\documentclass[UKenglish,cernpreprint,texlive=2011,txfonts]{atlasdoc} 
\pdfoutput=1

\usepackage{atlaspackage}
\usepackage{atlasphysics}
\usepackage{hyperref}
\usepackage{layouts}

\AtlasTitle{A measurement of the calorimeter response to single hadrons and determination of the jet energy scale uncertainty using LHC Run-1 $pp$-collision data with the ATLAS detector}

\AtlasJournalRef{Eur. Phys. J. C (2017) 77:26}
\AtlasDOI{10.1140/epjc/s10052-016-4580-0}

\PreprintIdNumber{CERN-EP-2016-149}

\AtlasAbstract{
A measurement of the calorimeter response to isolated charged hadrons in the ATLAS detector at the LHC is presented.
This measurement is performed with 3.2~nb$^{-1}$ of proton--proton collision data at $\sqrt{s}=7$~\TeV\ from 2010 and 0.1~nb$^{-1}$ of data at $\sqrt{s}=8$~\TeV\ from 2012.
A number of aspects of the calorimeter response to isolated hadrons are explored.
After accounting for energy deposited by neutral particles, there is a 5\% discrepancy in the modelling, using various sets of {\sc Geant4} hadronic physics models, of the calorimeter response to isolated charged hadrons in the central calorimeter region.
The description of the response to anti-protons at low momenta is found to be improved with respect to previous analyses.
The electromagnetic and hadronic calorimeters are also examined separately, and the detector simulation is found to describe the response in the hadronic calorimeter well.
The jet energy scale uncertainty and correlations in scale between jets of different momenta and pseudorapidity are derived based on these studies.
The uncertainty is 2--5\% for jets with transverse momenta above 2~\TeV, where this method provides the jet energy scale uncertainty for ATLAS.
}

\usepackage{graphicx}
\usepackage{caption}
\usepackage{xspace}

\newcommand{\pL}{\ensuremath{\Lambda}\xspace}
\newcommand{\pLB}{\ensuremath{\bar{\Lambda}}\xspace}
\newcommand{\pKS}{\ensuremath{K_\text{S}^{0}}\xspace}
\newcommand{\pKL}{\ensuremath{K_\text{L}^{0}}\xspace}
\newcommand{\pip}{\ensuremath{\pi^+}\xspace}
\newcommand{\pim}{\ensuremath{\pi^-}\xspace}
\newcommand{\ep}{\ensuremath{E/p}\xspace}
\newcommand{\epav}{\ensuremath{\langle E/p \rangle}\xspace}
\newcommand{\epcor}{\ensuremath{\langle E/p \rangle_{\mathrm{COR}}}\xspace}
\newcommand{\epbg}{\ensuremath{\langle E/p \rangle_{\mathrm{BG}}}\xspace}
\newcommand{\Ea}{\ensuremath{E_a}\xspace}
\usepackage{atlasbiblatex}
\begin{document}

\maketitle

\tableofcontents

\section{Introduction}
\label{sec-intro}
The proton--proton collisions measured by the ATLAS detector at the Large Hadron Collider (LHC) produce quarks and gluons that are observed as collimated sprays of hadrons, called jets.
The hadrons in jets are measured as charged-particle tracks and showers of particles in the calorimeters.
Uncertainties in the measurement of jet energies and the modelling of the calorimeter response to hadrons often dominate systematic uncertainties in measurements at the LHC.

The measurement of the calorimeter response to single charged hadrons provides an important validation of the modelling of hadronic showers in the calorimeters and of the detector geometry implemented in the ATLAS simulation~\cite{SOFT-2010-01}.
It is one of the few low-level measurements that can verify specific aspects of the modelling of the jet response.
It also allows a component-wise derivation of the jet energy scale uncertainty and the extension of the uncertainty to high jet transverse momentum ($\pt>1.8$~\TeV\ in 2012) where there are too few jets in the data for standard in situ calibration techniques (e.g. dijet or multi-jet balance techniques) to be applied~\cite{PERF-2012-01}.
The response to isolated charged hadrons was measured in ATLAS using data collected in 2009 and 2010~\cite{PERF-2011-05} and has been used to evaluate part of the standard ATLAS jet energy scale uncertainty since 2010~\cite{PERF-2012-01,PERF-2011-03}.
The response to charged hadrons also has been used for the calibration of the detector response to hadronically decaying $\tau$-leptons~\cite{PERF-2013-06}.

This paper describes the updated measurement of the response to isolated charged hadrons using data from both 2010 and 2012 with the most recent detector simulation.
Between 2010 and 2012, the centre-of-mass energy was increased from 7~\TeV\ to 8~\TeV, and the calorimeter conditions changed as the calorimeters were repaired and recalibrated.
In particular for comparisons sensitive to these changes, both 7 and 8~\TeV\ data are presented.  Generally, the conclusions are consistent between the two years.
The detector simulation includes significant improvements in the detector description~\cite{PERF-2011-08,PERF-2013-05} and makes use of new models of hadronic physics in {\sc Geant4}~\cite{GEANT4}.
Several variations of the inclusive response measurement are used to validate key aspects of the modelling of energy reconstruction in the calorimeter.
As in Ref.~\cite{PERF-2011-05}, the decays of identified particles are used to identify the type of particle entering the calorimeter (e.g. $\pi^\pm$, proton, or anti-proton), in order to further validate the details of the hadronic physics models.

The calibration of jets based on the energy deposited by individual particles involves a number of steps that can be separately tested.
Many particles are not sufficiently energetic to reach the calorimeter, and some particles interact before reaching the calorimeter and do not deposit a significant amount of energy.
The fraction of particles not depositing energy in the calorimeter is the first important test of the geometry (i.e. description of the detector material distribution) and simulation of hadrons, discussed in Section~\ref{sec:raw_ep}.
The energy deposited in the calorimeter is then grouped into topological clusters.
The procedure by which the clusters are constructed should not bias the energy measured in the calorimeter; this is explored in Section~\ref{sec-clusters_vs_cells}.
These topological clusters can be calibrated to the hadronic scale, and the way in which the calibration affects the calorimeter energy measurement is discussed in Section~\ref{sec:cluster_calibration}.
The construction of topological clusters involves an energy threshold, which differs between different data-taking periods at the LHC.
The effect of changing these thresholds on the measured response is explained in Section~\ref{sec-thresholds}.

A jet is a complex object, however, and good modelling of the average properties of jets does not always indicate that jets would be well described in more extreme configurations.
Some jets may be composed of more positively or more negatively charged particles, resulting in differences in response.
The separate modelling of positively and negatively charged particles is discussed in Section~\ref{sec:charge_dependence}.
Because hadrons may interact early or late in the calorimeter, jets may not be regularly distributed longitudinally.
The description of the separate calorimeter layers is discussed further in Sections~\ref{sec-tilecal} and~\ref{sec-emcal}.
A number of different hadron species can contribute to jets.  Their charged components are primarily charged pions, charged kaons, and (anti-)protons.
The responses of these individual species of hadrons are discussed further in Section~\ref{sec-resonances}.
These results primarily build confidence in the extrapolation from simple isolated-hadron configurations to complex jet configurations.

The studies of the calorimeter response to isolated charged hadrons are then used to construct a jet energy scale uncertainty in Section~\ref{sec-jets}.
The jet energy scale uncertainty, derived in this manner, is applicable only to the particular set of jets used in the derivation.
Just as in the case of the other in situ uncertainty estimations, additional uncertainties that depend on the details of the jet selection must be considered.
One of these is an uncertainty due to the modelling of additional proton--proton collisions (pile-up) simultaneous with the collision of interest.
Historically, although searches for new physics and measurements of the Standard Model are almost always performed in events with pile-up, isolated hadron response studies have always been performed in events with low or no pile-up.
In Section~\ref{sec:pileup}, the calorimeter response to isolated charged hadrons in events with pile-up is discussed.
These studies are sufficiently promising that future studies of the calorimeter response to isolated charged hadrons might be performed in larger data sets, including events with pile-up.

The paper is organised as follows.
The ATLAS detector is briefly introduced in Section~\ref{sec-detector}.
Section~\ref{sec-selection} describes the data and simulated event samples and event selection, as well as the reconstruction methodology.
Section~\ref{sec-inclusive} then details several features of the response to isolated charged hadrons, including the subtraction of neutral background particles.
The calorimeter response to specific species of charged hadrons identified using displaced decays is described in Section~\ref{sec-resonances}.
The calorimeter response to charged hadrons is used to understand the jet response and uncertainties in Section~\ref{sec-jets}.
Finally, Section~\ref{sec-conclusion} provides the conclusions of these studies.

\section{ATLAS detector}
\label{sec-detector}
The ATLAS detector~\cite{PERF-2007-01} is a general purpose particle detector covering almost $4\pi$ in solid angle\footnote{ATLAS uses a right-handed coordinate system with its origin at the nominal interaction point (IP) in the centre of the detector and the $z$-axis along the beam pipe. The $x$-axis points from the IP to the centre of the LHC ring, and the $y$-axis points upward. Cylindrical coordinates $(r,\phi)$ are used in the transverse plane, $\phi$ being the azimuthal angle around the $z$-axis. The pseudorapidity is defined in terms of the polar angle $\theta$ as $\eta=-\ln\tan(\theta/2)$.  The opening angle in $\phi$ ($\eta$) between two objects is denoted $\Delta\phi$ ($\Delta\eta$).} and consisting of an inner tracking detector (ID), a calorimeter, and a muon spectrometer.
The ID consists of silicon pixel and strip (SCT) tracking detectors covering $|\eta|<2.5$ and a straw-tube tracker (TRT) covering $|\eta|<2.0$, all immersed in an axial 2~T magnetic field provided by a superconducting solenoid.
A typical central track includes three measurements (hits) in the pixel detector, eight hits in the SCT, and $35$ hits in the TRT.
Below $|\eta|=0.6$, a particle passes through approximately 0.5 radiation lengths (0.2 interaction lengths) of material before reaching the calorimeter.
Between $|\eta|=0.6$ and $|\eta|=1.8$, the amount of material in the ID rises from 1.5 radiation lengths (0.4 interaction lengths) to a maximum of almost 2.5 radiation lengths (0.7 interaction lengths).
The sampling calorimeter is hermetic out to $|\eta|=4.9$ and is generally divided into barrel ($|\eta|<1.4$), endcap ($1.4\leq|\eta|<3.2$) and forward ($3.2\leq|\eta|<4.9$) regions.
The highly-segmented electromagnetic (EM) calorimeter uses liquid argon (LAr) with lead or copper absorber material and includes three longitudinal sampling layers in addition to a presampler for $|\eta|<1.8$.
The hadronic calorimeter uses scintillator tiles with steel absorber in the barrel ($|\eta|<1.7$) and LAr with copper (tungsten) absorber in the endcap (forward) region.

A three-level trigger system is used to select events for offline analysis.  
The first level is hardware-based, while the second two levels are implemented in software.
The minimum-bias trigger scintillators, used for selecting events in this analysis, are two sets of 16 thin scintillators covering $2.08<|\eta|<3.83$.
These scintillators are highly efficient for selecting events with charged particles in this $|\eta|$ range and are integrated into the first level of the trigger.

\section{Data sets and selection}
\label{sec-selection}
\subsection{Data samples}

The primary data sample consists of eight million proton--proton collision events corresponding to an integrated luminosity of 0.1~nb$^{-1}$ from a data taking period at the beginning of 2012 at $\sqrt{s} = 8$~\TeV.
Additionally, a sample of three million data events corresponding to an integrated luminosity of 3.2~nb$^{-1}$ recorded during 2010 at $\sqrt{s}=7$~\TeV\ are examined.
These data from 2010 were studied previously in Ref.~\cite{PERF-2011-05}, but they were subsequently reanalysed with improved understanding of the detector (e.g. improved knowledge of the detector material and alignment).
Both of these samples were collected during periods in which the fraction of events with pile-up was negligible.  In events with pile-up, the average number of simultaneous collisions is denoted $\langle\mu\rangle$.
To study issues related to pile-up, an additional data sample from 2012 is used, corresponding to an integrated luminosity of 551~nb$^{-1}$, which has approximately $15$ proton--proton collisions per event on average and collisions every 50~ns.  This data sample includes some effects from both in-time pile-up, collisions simultaneous to the collision of interest, and out-of-time pile-up, collisions in bunch crossings before or after the collision of interest.  Out-of-time pile-up primarily affects the calorimeter response due to the response time of the calorimeter and the bipolar signal pulse shaping in the LAr calorimeter~\cite{PERF-2007-01}.
All data samples are required to pass basic data-quality requirements.

\subsection{Monte Carlo simulation}

The primary 2012 data sample is compared to 20 million simulated single-, double-, and non-diffractive proton--proton collision events, generated using {\sc Pythia8.160}~\cite{PYTHIA8} using the A2 configuration of underlying event and hadronization parameters (tune)~\cite{AU2} and the MSTW 2008 leading-order parton distribution function set~\cite{MSTW,MSTW2}.  Throughout the paper, the $\pt$ spectrum of tracks in Monte Carlo (MC) simulation is weighted to match that of the data. A separate MC simulation sample is produced with conditions consistent with that of the 2010 data taking period for comparison to the 2010 data sample.

The simulated events are passed through the ATLAS detector simulation~\cite{SOFT-2010-01} based on {\sc Geant4} 9.4~\cite{GEANT4}.\footnote{Simulation with {\sc Geant4} version 9.6, which includes slightly different tunes of hadronic and electromagnetic physics models, was also tested, and the results were found to be compatible between the two versions.}  Two samples with different collections of hadronic physics models~\cite{G4hadronics} are used: one, called \texttt{QGSP\_BERT}, includes a quark--gluon string model~\cite{QGS15,QGS16,QGS17,QGS18,QGS19} with a pre-compound and evaporation model for hadron momenta above 12~\GeV, the parameterised low-energy proton inelastic model based on GHEISHA~\cite{GHEISHA20} from 9.5 to 25~\GeV, and the Bertini intra-nuclear cascade~\cite{BERT21,BERT22,BERT23} and nuclear de-excitation model below 9.9~\GeV.  In the regions where the models overlap, a smooth transition from one to the other is enforced.  For protons and neutrons, an additional quasi-elastic scattering model is applied.  The other set of hadronic physics models, called \texttt{FTFP\_BERT}, includes the Fritiof model~\cite{FTF24,FTF25,FTF26,FTF27} with a pre-compound model above 4~\GeV\ and the Bertini intra-nuclear cascade model below 5~\GeV.  These two sets of hadronic physics models also differ in the models applied to anti-hyperons and anti-baryons, which in particular leads to an expected difference in the modelling of the calorimeter response to anti-protons.

In all cases, the simulated detector conditions match those of the data-taking period, and the simulated events and data are passed through the same trigger and reconstruction software.  Where the data include pile-up, minimum-bias events generated with {\sc Pythia8} are overlaid on top of one another to mimic the simultaneous collisions in the detector, including the bipolar pulse shape of the calorimeter electronics.  The MC simulation samples with pile-up are only included using the \texttt{QGSP\_BERT} set of hadronic physics models.

\subsection{Event selection and reconstruction}
\label{sec-triggersel}

In order to be selected by the trigger system, events in the low-$\langle\mu\rangle$ data are required to have at least two hits in the minimum-bias trigger scintillators.  In the MC simulation, this trigger is more than $95$\% efficient for events passing the following offline selection. 

To collect data during data-taking periods with pile-up, three triggers are used.  Only a fraction of events passing the selection criteria of any of the triggers are written out, as the rates are above the maximum bandwidth for the trigger.  The first trigger is random and requires only crossing proton bunches in the detector. The events accepted by this trigger correspond to an integrated luminosity of 24 nb$^{-1}$.  Two triggers that require tracks that are isolated from other charged tracks at the front-face of the calorimeter and have at least 9~\GeV\ or 18~\GeV\ of $\pt$ are used to provide additional events.  The events accepted by these two triggers correspond to 499 nb$^{-1}$ and 551~nb$^{-1}$, respectively. In the MC simulation, because it is highly efficient, the trigger requirement has no significant impact on the results.

Each event is required to have a well-reconstructed vertex with at least four associated tracks with $\pT>400$~\MeV. In the low-$\langle\mu\rangle$ data set, the events are required to have exactly one vertex, to further suppress any residual contribution from pile-up. The tracks selected for the measurement are required to have $\pT>500$~\MeV\ and to have at least one hit in the pixel detector and six hits in the SCT, as well as small longitudinal and transverse impact parameters $|z_0\times\sin{\theta}| < 1.5$~mm and $|d_0| < 1.5$~mm with respect to the primary vertex~\cite{PERF-2011-05}. This reduces the contribution from spurious and poorly measured tracks to a negligible level. In order to ensure that the tracks are isolated, no other track extrapolated to the second longitudinal layer of the EM calorimeter is allowed within a cone of size $\Delta R = \sqrt{(\Delta\phi)^2 + (\Delta\eta)^2} < 0.4$ around the track of interest.  This criterion was shown previously to reduce the effect from other nearby charged particles on the measurement to a negligible level~\cite{PERF-2011-05}.

Although it does not provide the same level of precision tracking information as the pixel and SCT layers, the TRT provides additional information to efficiently reject the tracks originating from hadronic interactions in the ID material.
Tracks interacting in the ID produce a range of secondary particles, often including ions and neutrons, which can be difficult to model correctly.
Moreover, the kinematics and species of of hadrons resulting from these interactions is generally poorly known and may not be well modeled.
For most of the results in this paper, in the region $|\eta|<2.0$, more than 20 hits in the TRT are required to ensure that the particle producing the track reaches the calorimeter.
The impact of this selection criterion on the measured calorimeter response is carefully examined in Section~\ref{sec:TRTcutMotivation}.

In each event, the calorimeter cells' energies are topologically clustered using a 4--2--0 algorithm~\cite{TopoClusters}. This algorithm suppresses noise by forming energy clusters around seeds with at least four times larger (in absolute value) signal than the average noise, which includes both the electronic noise and the pile-up contributions.  The threshold is defined by the width of the energy distribution in a cell in an MC simulation sample containing a fixed amount of pile-up (e.g. $\langle\mu\rangle=30$).  All neighbouring cells with at least twice larger signal than the average noise are added to the clusters, and a final layer of cells at the boundary of the cluster is added without any noise threshold requirement. This final layer improved the energy resolution in single-particle studies with the ATLAS calorimeter test beam~\cite{Speckmayer}. After this procedure, clusters may be split if there are several local maxima of energy found within them.  Because the cell energy requirements are on the magnitude of the signal, negative energy clusters are possible.  The topological clusters are not calibrated beyond a correction for the sampling fraction of an electron shower in the calorimeter; no correction is applied for non-compensation or energy loss outside of the sampling portion of the calorimeter.  Thus, the topological clusters are calibrated only to the electromagnetic scale (EM scale).

The noise thresholds used in the clustering procedure for the low-$\langle\mu\rangle$ data and MC simulation are extracted from simulated events without pile-up.
In the data and MC simulation that include pile-up, the thresholds are re-calculated in simulated events with $\langle\mu\rangle=30$.
The difference between the two noise calculations is a factor of two in $|\eta|<2.0$, rising to more than a factor of 20 for $|\eta|>4.0$~\cite{PERF-2014-03}.
The impact of the difference between these thresholds are described further in Section~\ref{sec-thresholds}.
In the remainder of the paper, unless explicitly stated, the nominal data set includes the low-$\langle\mu\rangle$ data with calorimeter thresholds calculated in events with $\langle\mu\rangle=0$.

\section{Charged hadron response}
\label{sec-inclusive}
The calorimeter response to charged hadrons is studied using the ratio of the energy deposited by the isolated charged particle in the calorimeter, $E$, to the momentum of its track, $p$, as measured by the ID. The ratio is denoted \ep, and the average ratio is denoted \epav. As the track momentum measurement has a small uncertainty in for the range $0.5<p/\GeV<30$ considered in this paper, which is negligible below 5~\GeV\ and is taken as a conservative 1\% above this value~\cite{PERF-2011-05}, it is an excellent proxy for the energy measurement of the isolated charged hadron.

The energy of a topological cluster in a certain layer of the calorimeter is matched to the track if the energy-weighted position of the cells associated with the topological cluster in that layer is situated within $\Delta R = 0.2$ of the extrapolated position of the track in that particular layer.  The cone size of $\Delta R = 0.2$ around the track was optimised such that, on average, about 90\% of the energy of the charged hadron is included, while the contribution from the neutral-particle background is kept to a low level~\cite{PERF-2011-05}.

\begin{figure}[htbp]
\centering
\subfloat[]{
  \includegraphics[width=.4\textwidth]{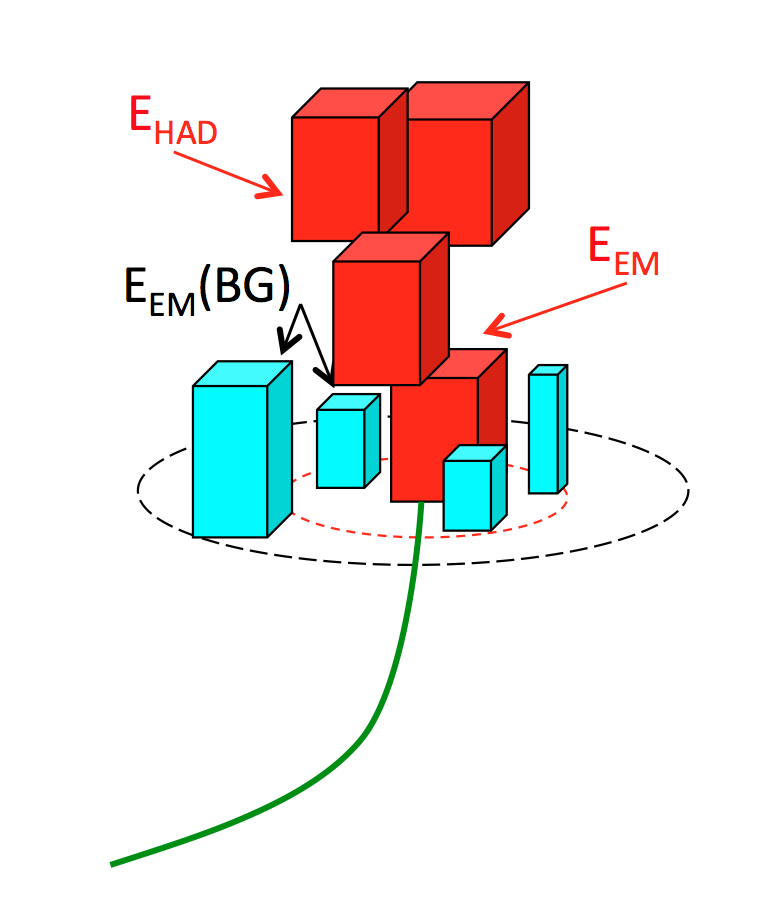}
}
\subfloat[]{
  \includegraphics[width=.4\textwidth]{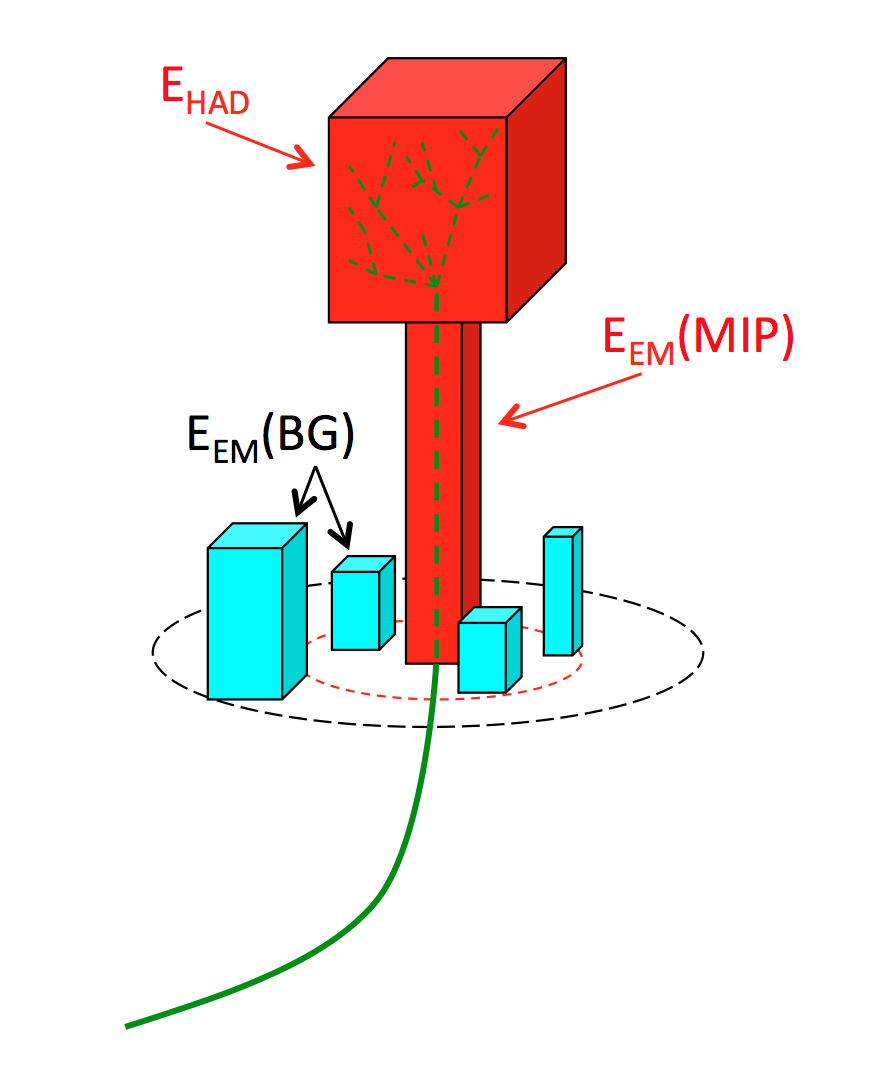}
}
\caption{An illustration (a) of the construction of the \ep variable used throughout this paper.  The particle is identified by matching a track (green) with momentum $p$ to topological energy clusters in the EM and hadronic calorimeters (red), while nearby topological energy clusters from neutral-particle background (light blue) must be removed. The red (black) dashed circle on the horizontal plane has a radius $\Delta R = 0.1$ ($0.2$) around the track.  The same diagram is shown for the neutral-particle background selection (b) using late-showering hadrons, described in Section~\ref{sec:neutral_bg}.  The construction is similar, but the energy deposited in the EM calorimeter by the particle is required to be consistent with a minimally-ionising particle (MIP).  No attempt is made to identify individual clusters as originating from background particles. The subtraction is done on average}
\label{fig:Eoverp_cartoon}
\end{figure}

The construction of the \ep variable is illustrated in Figure~\ref{fig:Eoverp_cartoon}.

\subsection{\ep distributions}
\label{sec:raw_ep}

\begin{figure}[htbp]
\centering
\subfloat[]{
  \includegraphics[width=0.48\textwidth]{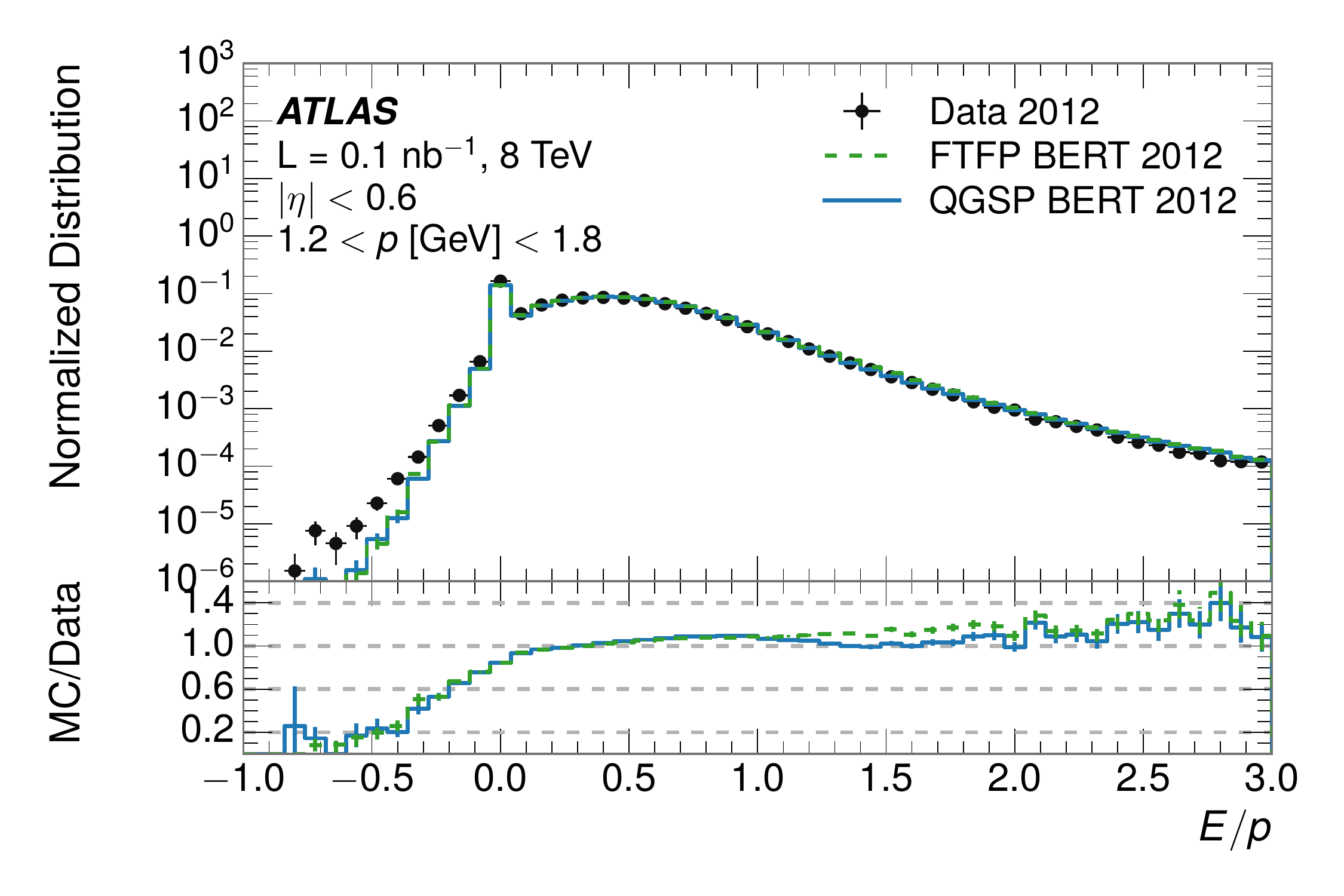}
}
\subfloat[]{
  \includegraphics[width=0.48\textwidth]{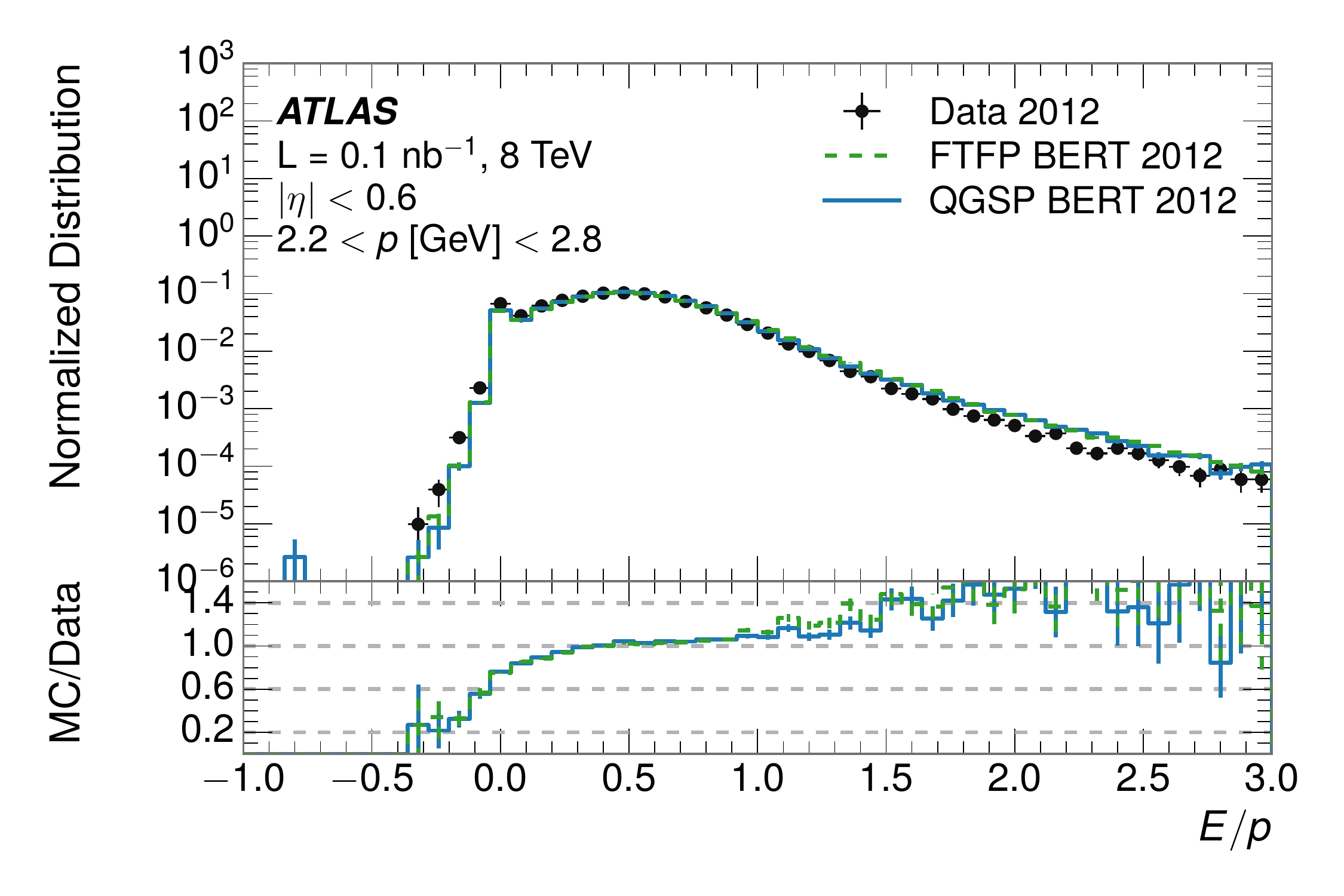}
  \label{fig-eoverp-highp}
}
\\
\subfloat[]{
  \includegraphics[width=0.48\textwidth]{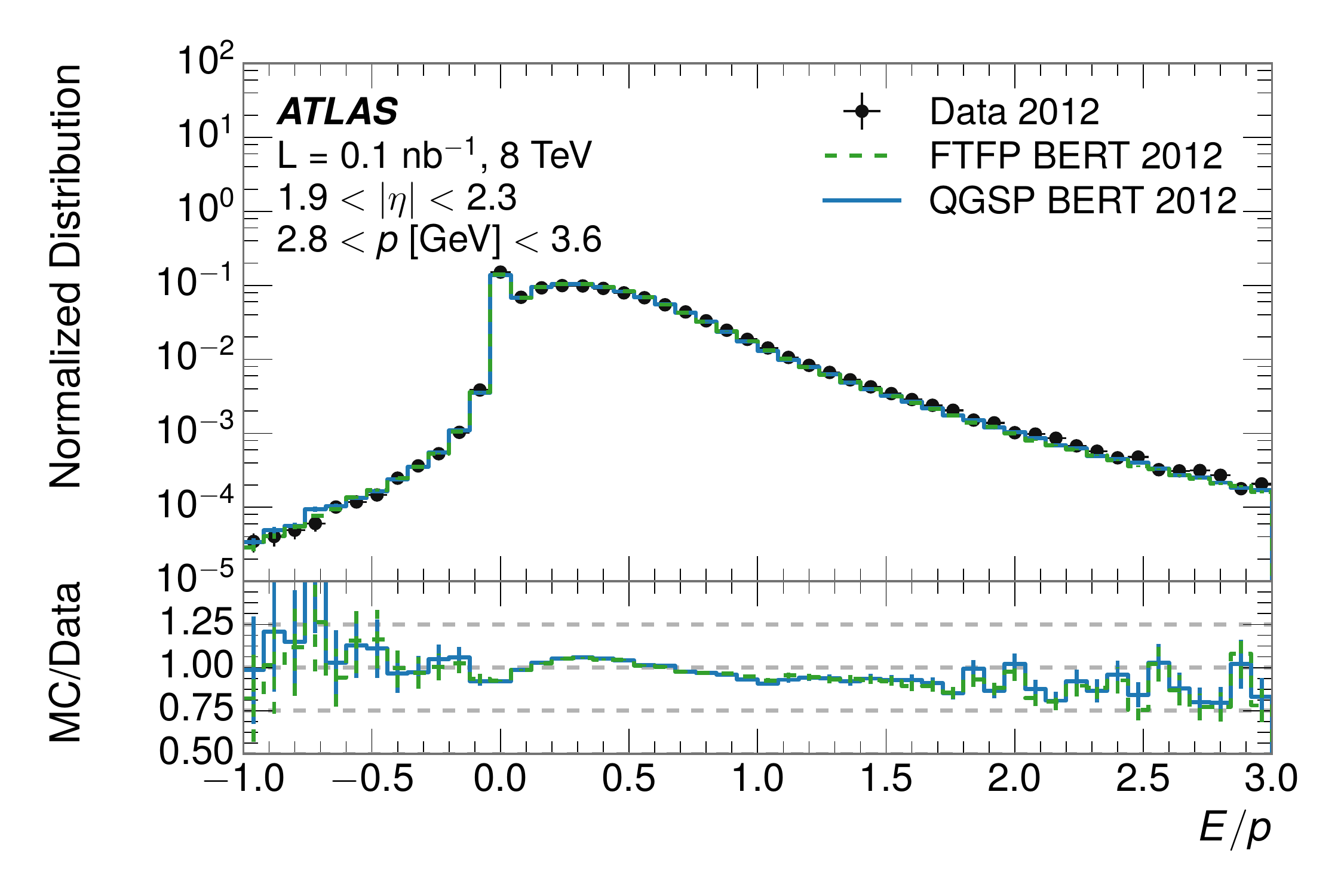}
}
\subfloat[]{
  \includegraphics[width=0.48\textwidth]{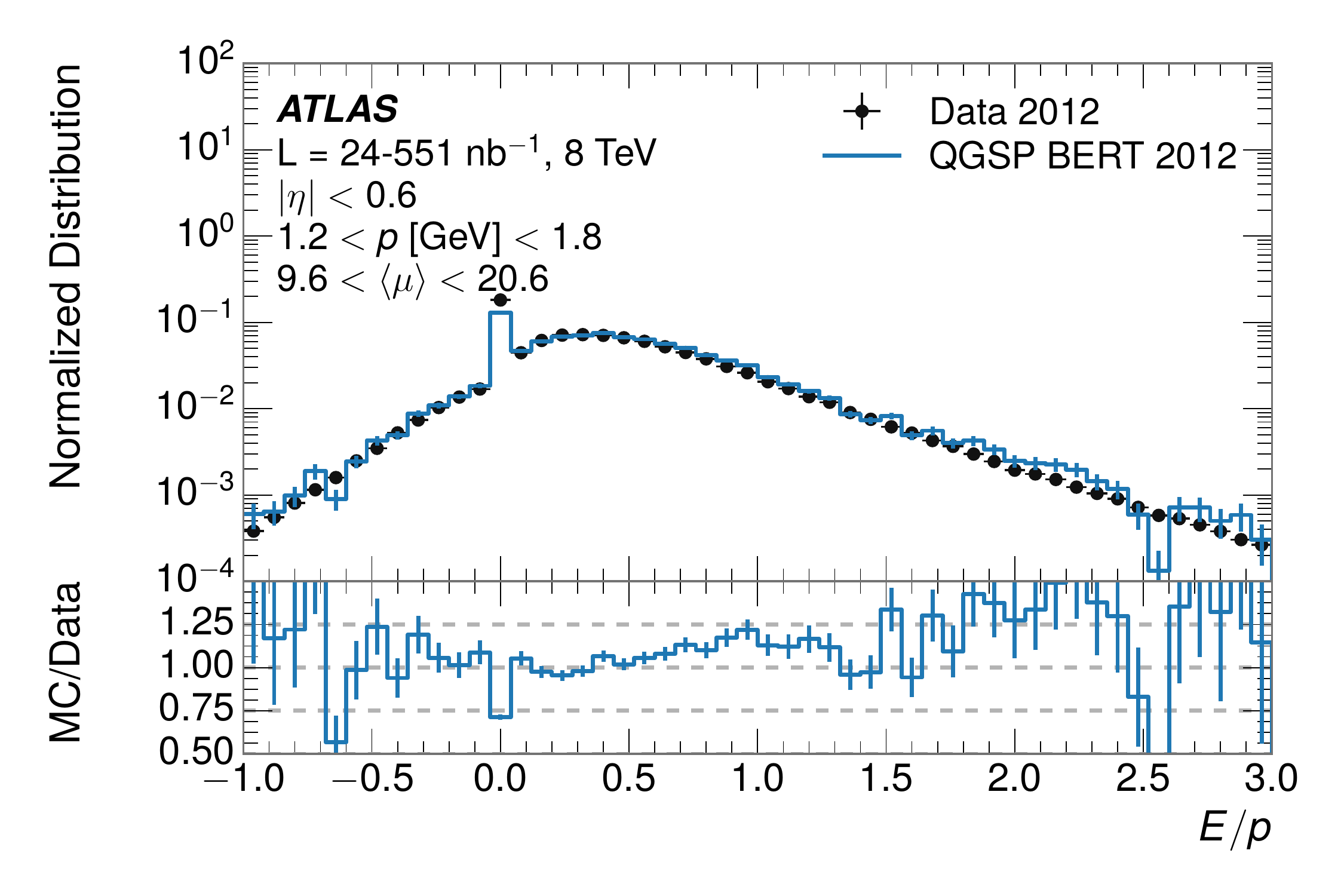}
  \label{fig-eoverp-highmu}
}
\caption{The \ep distribution for isolated tracks with (a) $|\eta| < 0.6$ and $1.2 < p /\GeV < 1.8$; (b) $|\eta| < 0.6$ and $2.2 < p /\GeV < 2.8$; (c) $1.9 < |\eta| < 2.3$ and $2.8 < p / \GeV < 3.6$; (d) $|\eta| < 0.6$ and $1.2 < p /\GeV < 1.8$ and high $\langle\mu\rangle$ ($9.6 < \langle\mu\rangle < 20.6$). The bottom portion of each panel shows the ratio of MC simulation to data, separately for the two sets of hadronic physics models. The error bars represent statistical uncertainties.}
\label{fig-eoverp}
\end{figure}

Figure~\ref{fig-eoverp} shows several examples of the \ep distributions for data and MC simulations with both sets of hadronic physics models. The distributions are normalised to have unit area. Examples are given for two track momentum bins in the central region of the calorimeter, for data with low $\langle\mu\rangle$ and a higher $|\eta|$ region, and for data in the central region of the calorimeter with higher $\langle\mu\rangle$.  In these distributions, no requirement is made on the number of TRT hits associated to the track. The mean of the distribution is significantly less than one because of the loss of some energy outside of the clusters included in the definition of $E$ and the non-compensating response of the calorimeter.  The large fraction of tracks with $\ep = 0$ corresponds to tracks without an associated topological cluster in the calorimeter. This can happen if either a particle interacts hadronically before reaching the calorimeter or no single energy deposit is large enough to seed a topological cluster~\cite{PERF-2011-05}. The negative tail of these distributions is caused by noise in the calorimeter, which for data taking conditions with low $\langle\mu\rangle$ consists mostly of electronics noise, while the long positive tail is caused by the background of neutral particles, since these particles add to the measured $E$ but not to $p$. In events with significant amounts of pile-up, such as those shown in Figure~\ref{fig-eoverp-highmu}, the positive tail from in-time pile-up can be quite large, and the negative tail is enhanced by several orders of magnitude due to the impact of out-of-time pile-up.  At low $|\eta|$, the MC simulation underestimates the negative $E$ tail from noise; however, this tail is only a very small fraction of the distribution.  In the same $|\eta|$ region but at higher momenta, the amount of energy from neutral-particle backgrounds is overestimated by the MC simulation, as can be seen from the high \ep region in Figure~\ref{fig-eoverp-highp}.

\begin{figure}[htbp]
\centering
\subfloat[]{
  \includegraphics[width=0.48\textwidth]{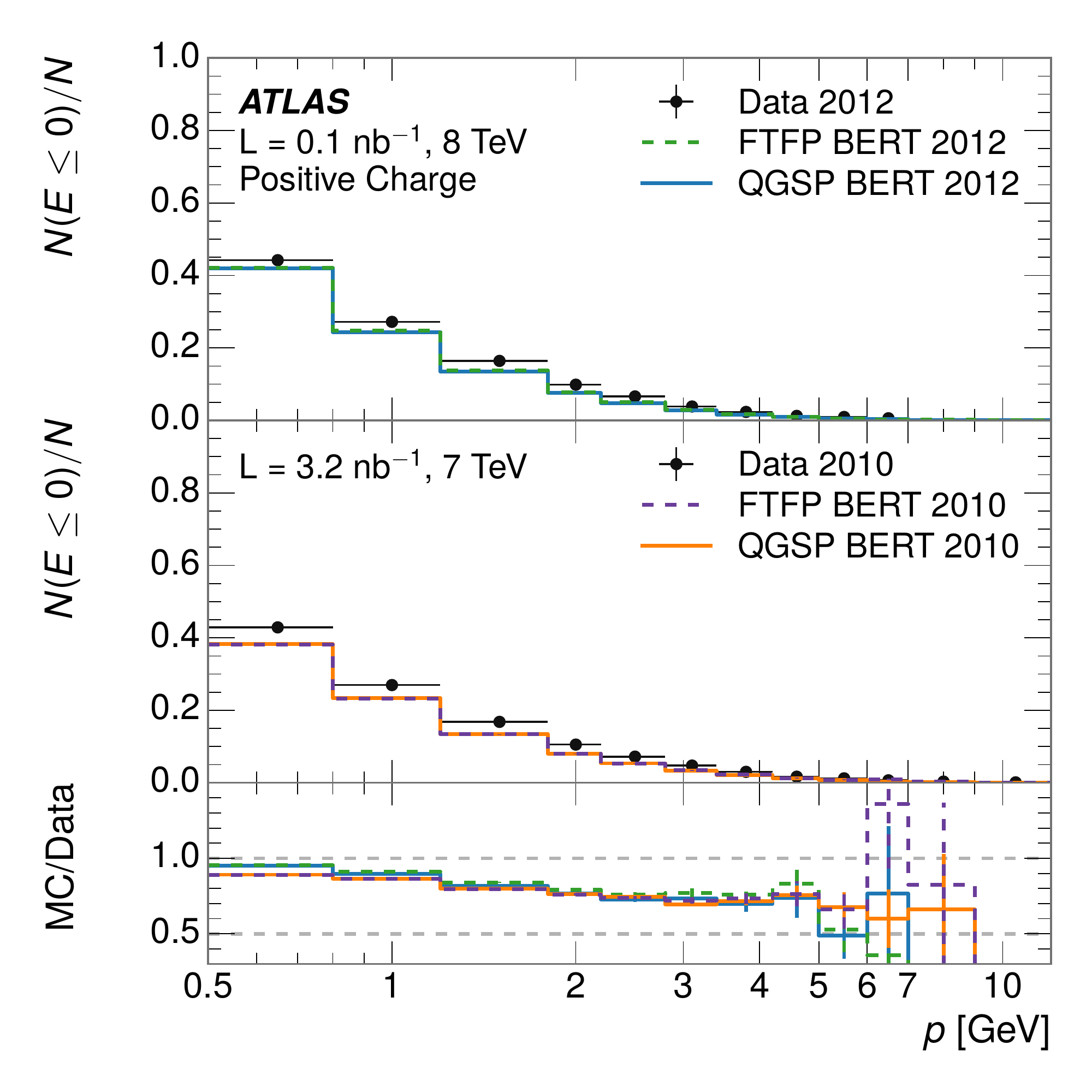}
}
~
\subfloat[]{
  \includegraphics[width=0.48\textwidth]{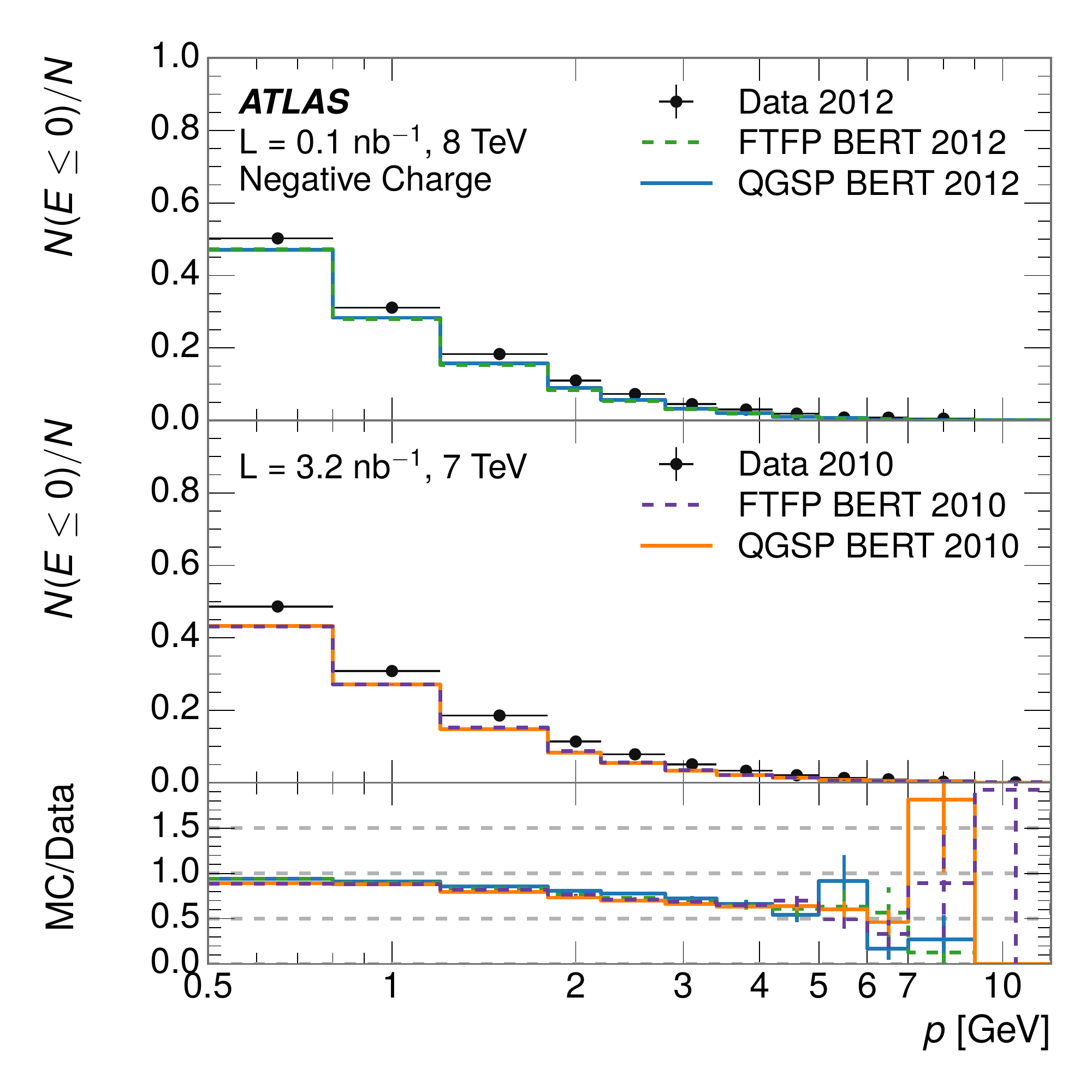}
}
\\
\subfloat[]{
  \includegraphics[width=0.48\textwidth]{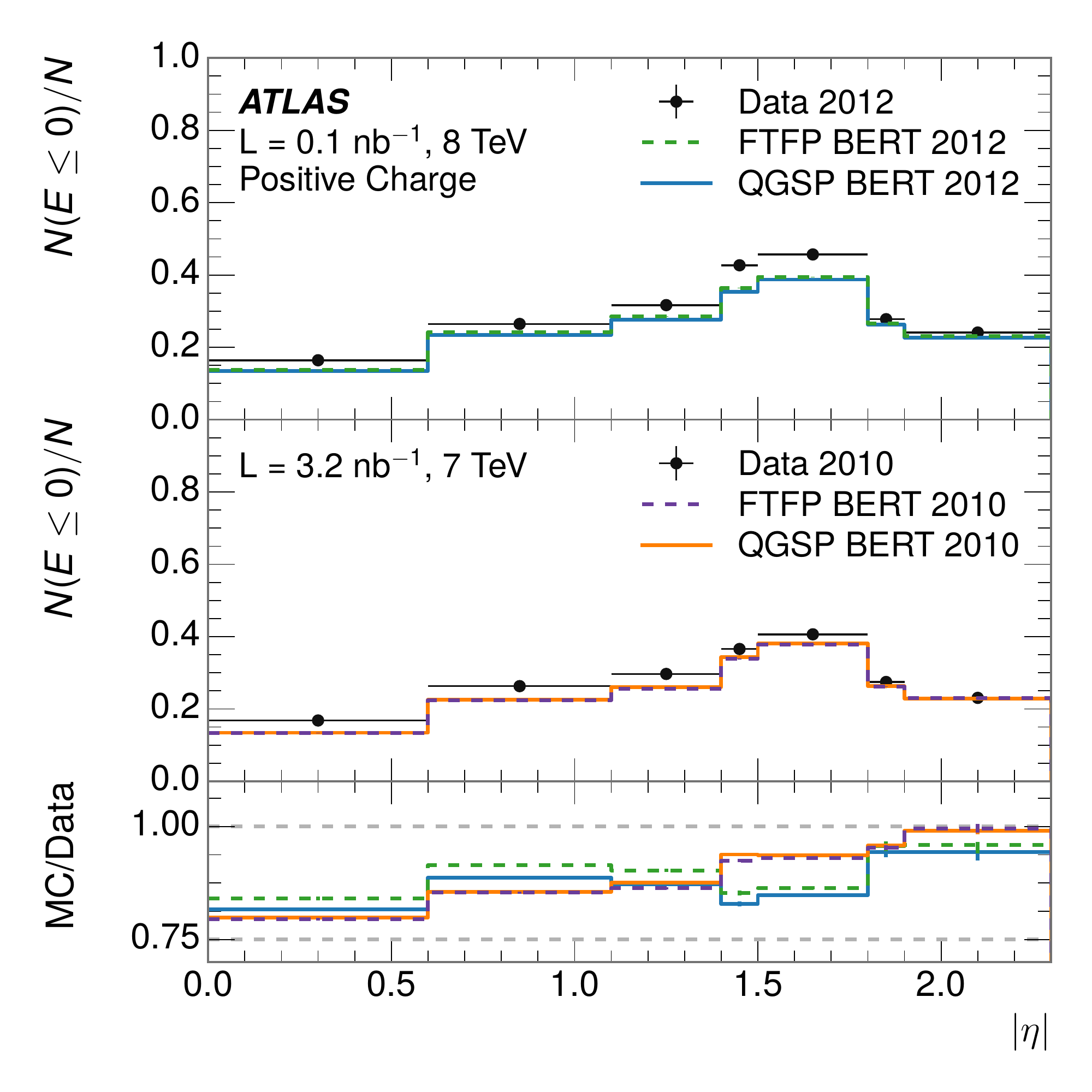}
}
~
\subfloat[]{
  \includegraphics[width=0.48\textwidth]{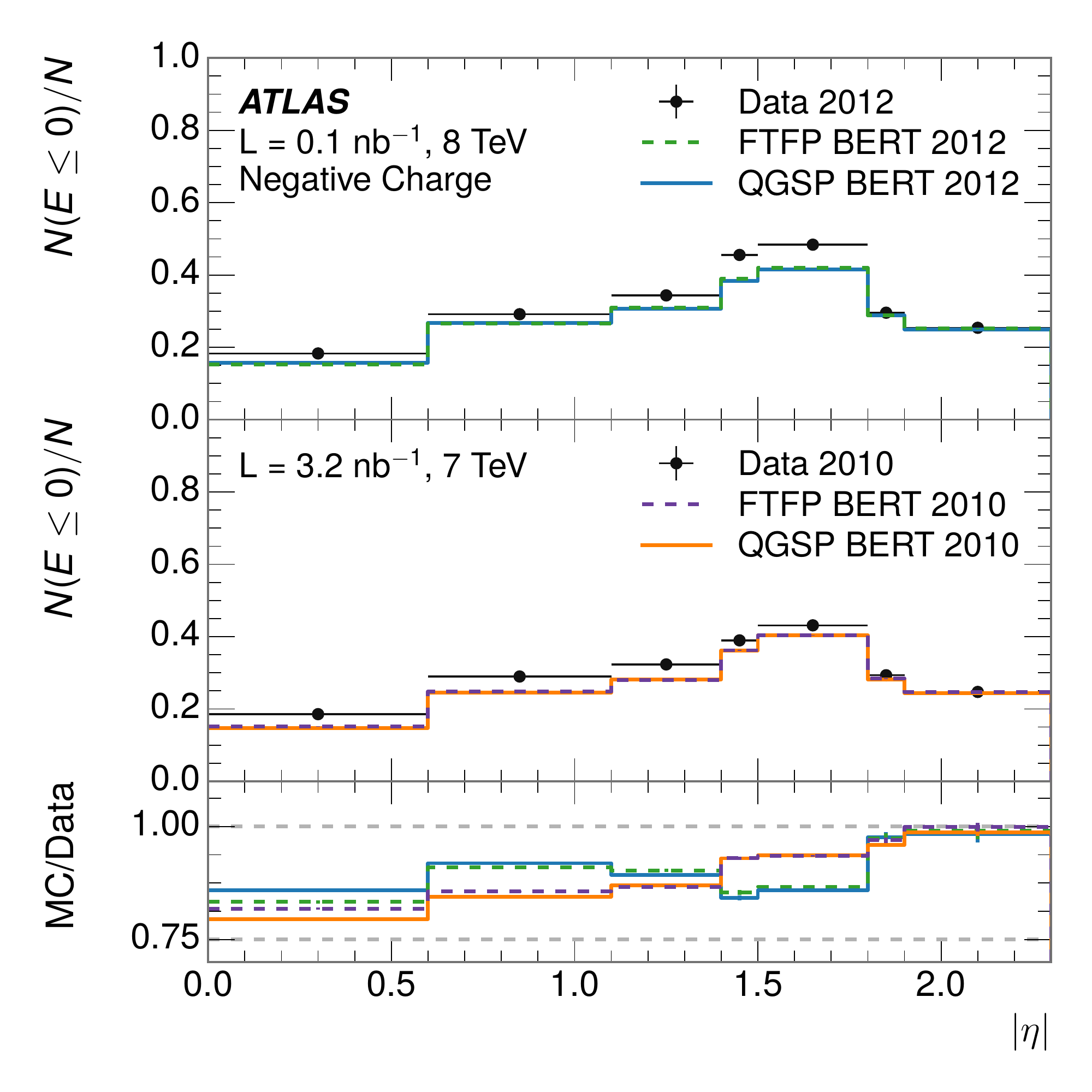}
}
\caption{The fraction of tracks as a function (a, b) of momentum and (c, d) of $|\eta|$ with $E \leq 0$ for tracks with positive (a, c) and negative (b, d) charge.  The bottom portion of each panel shows the ratio of MC simulation to data, separately for 2010 and 2012, and separately for the two sets of hadronic physics models.  The error bars represent statistical uncertainties.}
\label{fig-zerofracincl}
\end{figure}

\begin{figure}[htbp]
\centering
\subfloat[]{
  \includegraphics[width=0.48\textwidth]{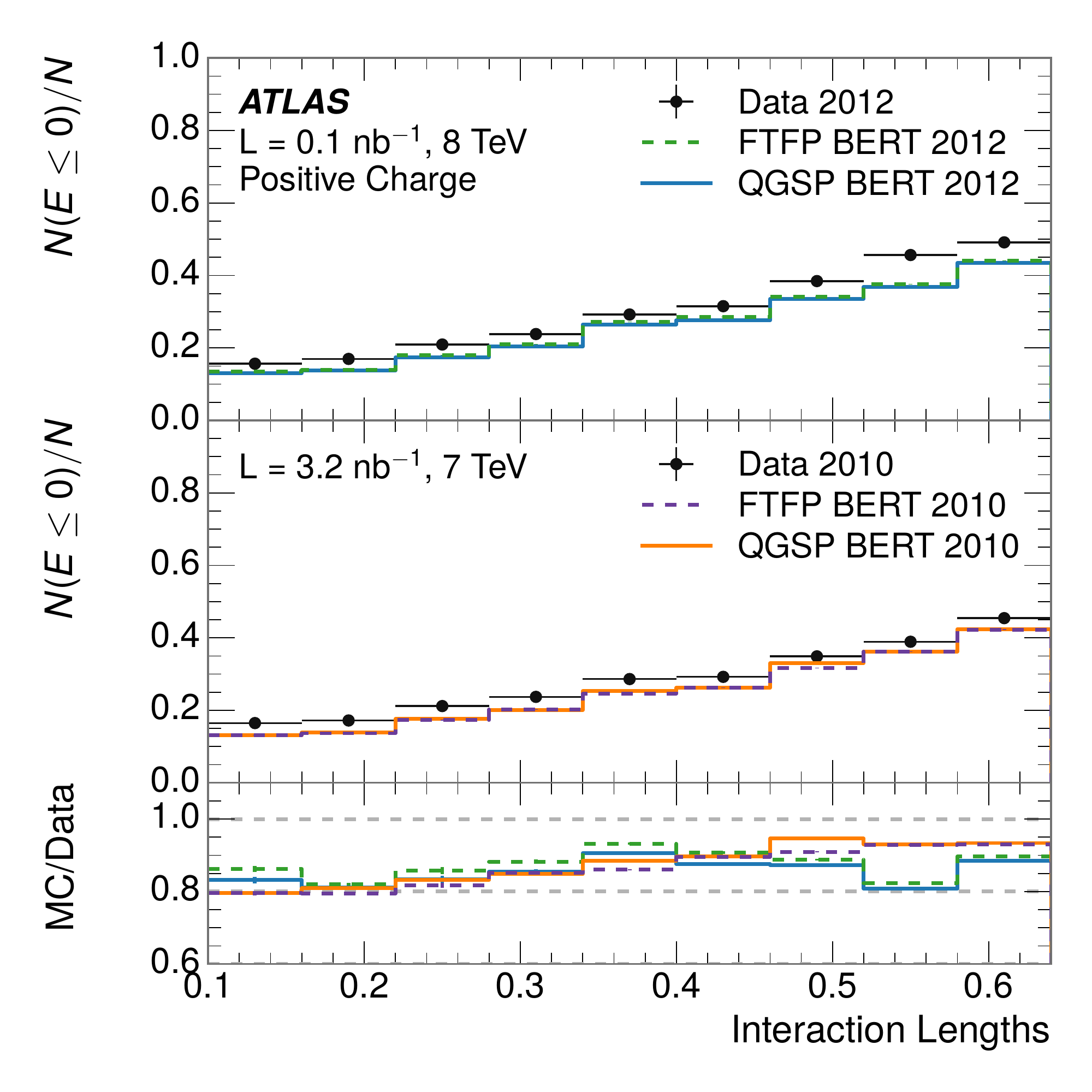}
}
~
\subfloat[]{
  \includegraphics[width=0.48\textwidth]{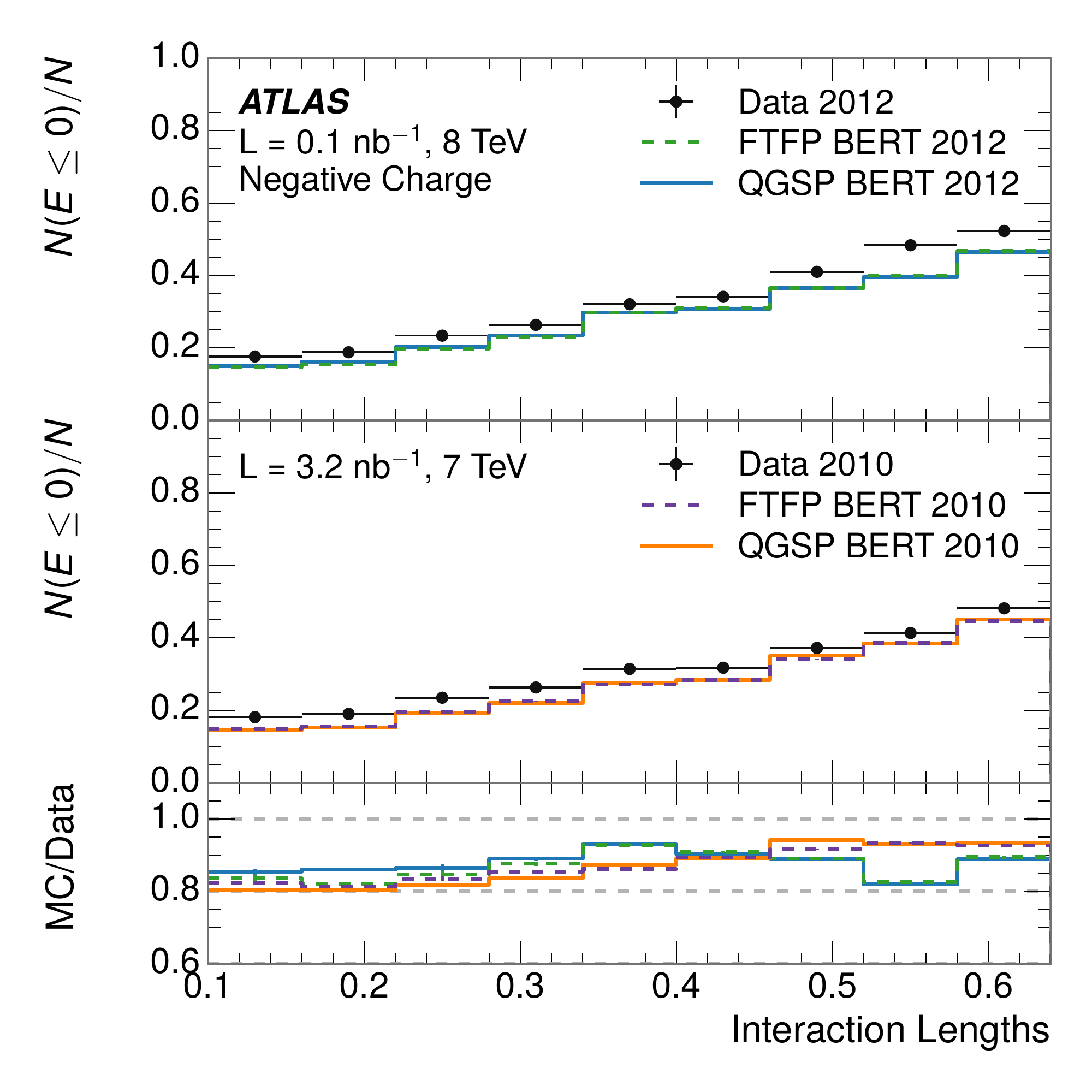}
}
\caption{ The fraction of tracks as a function of interaction lengths of material in front of the detector with $E \leq 0$ for tracks with $1.2 < p /\GeV < 1.8$ and (a) positive or (b) negative charge.  The bottom portion of each panel shows the ratio of MC simulation to data, separately for 2010 and 2012, and separately for the two sets of hadronic physics models.  The error bars represent statistical uncertainties.}
\label{fig-zerofracincl2}
\end{figure}

The fraction of the distribution with $E\leq 0$ can be further examined to understand features of the geometry, hadronic interaction models, and noise modelling.
No requirement is placed on the number of TRT hits associated to the track for these distributions in order to include particles that may have undergone a hadronic interaction earlier in the ID.
This fraction for inclusive charged particles is shown in Figure~\ref{fig-zerofracincl} as a function of track momentum and $|\eta|$ separately for tracks of positive and negative charges. 
The bin edges in $|\eta|$ in these distributions follow geometric features of the calorimeters.
The 2010 and 2012 data are shown separately and display similar features.
This fraction is directly displayed as a function of the number of traversed interaction lengths of material as described by the geometry of the simulation in Figure~\ref{fig-zerofracincl2} for tracks with $1.2 < p /\GeV < 1.8$.
The fraction of the distribution with $E\leq 0$ increases with $|\eta|$ and interaction lengths, as the detector material increases, and decreases with momentum.
Differences between the two charge distributions are clearly visible, particularly at low momenta.
These differences are closely related to the population of particle species present in the two samples and is discussed further in Section~\ref{sec:pid_response}.
The data and MC simulation are discrepant across a large range of interaction lengths and $|\eta|$ regions, indicating that the modelling of hadronic interactions, rather than of geometry, is primarily responsible for this discrepancy.

\begin{figure}[ht]
\centering
\subfloat[]{
  \includegraphics[width=0.48\textwidth]{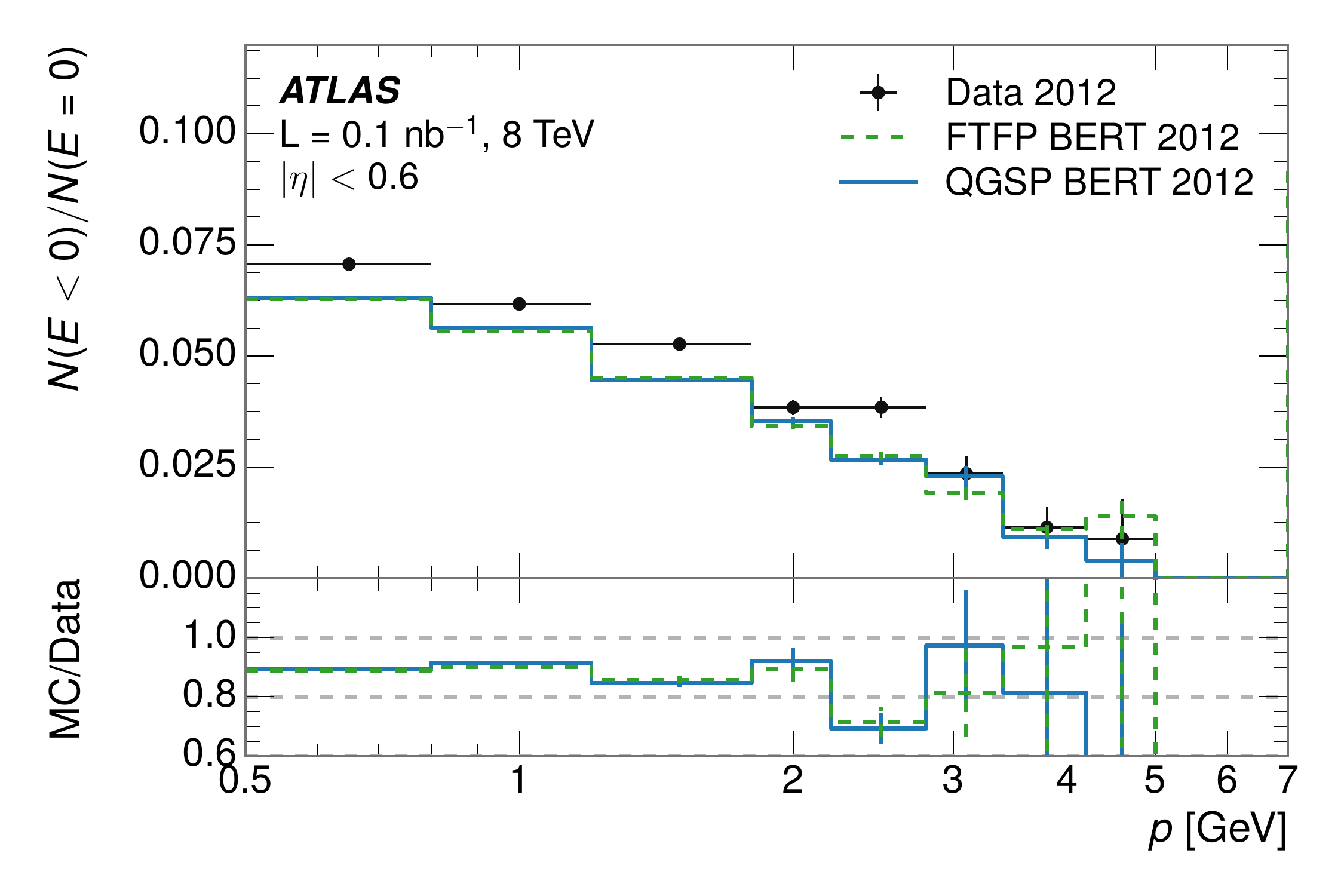}
}
~
\subfloat[]{
  \includegraphics[width=0.48\textwidth]{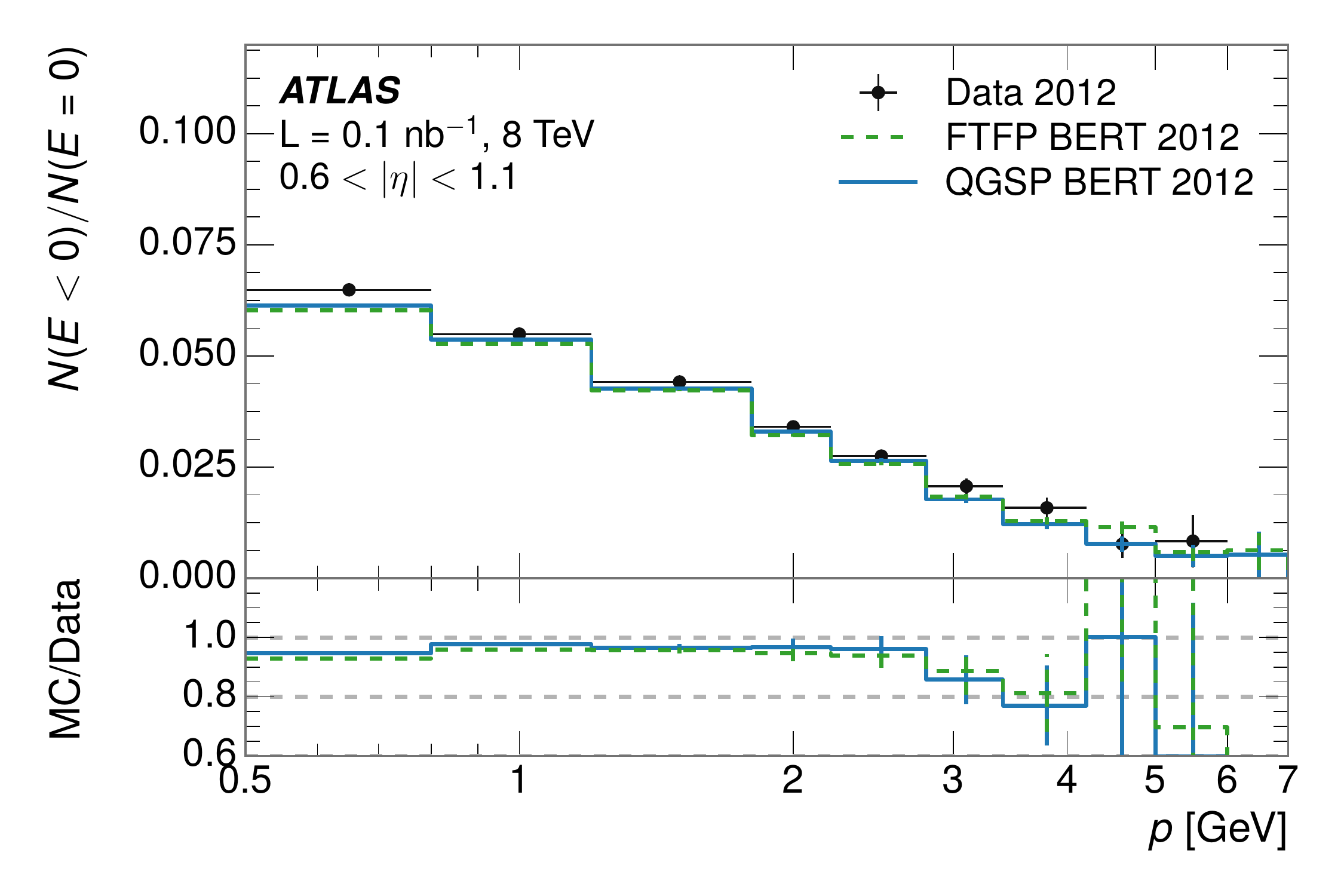}
}
\caption{ Ratio of the number of tracks with $E<0$ to the number with $E=0$ as a function of track momentum, for tracks with (a) $|\eta|<0.6$ and (b) $0.6 < |\eta| < 1.1$.  The bottom portion of each panel shows the ratio of MC simulation to data, separately for the two sets of hadronic physics models.  The error bars represent statistical uncertainties.}
\label{fig-zerofracLTZ}
\end{figure}

Detector noise, which is largely symmetric, drives the portion of the response distribution with $E<0$.
This region is dominated by particles that did not have any significant energy deposited in the calorimeter.
Thus, the tail can be used to further examine the modelling of calorimeter noise by the simulation.
Figure~\ref{fig-zerofracLTZ} shows the ratio of the number of tracks with associated $E<0$ to those with no associated clusters of energy as a function of track momentum -- this is an approximation of the relative rate at which particles with low momenta, or which have scattered before reaching the calorimeter, coincide with a sufficiently large amount of noise that a negative-energy topological cluster is formed.
In general, additional detector material in the simulation should manifest itself as a higher $E=0$ rate in the simulation than in the data, but this effect is cancelled in the ratio.
The ratio shows a disagreement at the 10\% level in the central region of the calorimeter, but the data and MC simulation are consistent in a more forward region $0.6< |\eta|<1.1$.  Further forward $|\eta|$ bins indicate 10\%-level disagreements.

\subsection{Neutral background subtraction}
\label{sec:neutral_bg}

Energy deposits from close-by particles bias the calorimeter measurement of $E$. While the isolation requirement on the charged-hadron track is efficient at reducing the effect from other charged particles to negligible levels, there is no equivalent method for eliminating the contribution from neutral particles, such as neutral hadrons or photons from $\pi^0 \to \gamma \gamma$ decays.

Since neutral particles, which are mostly photons with some low-energy hadrons, deposit their energy mostly in the EM calorimeter, it is possible to measure the background in situ by considering late showering charged hadrons, which behave like minimally-ionising particles (MIP) in the EM layers of the calorimeter. Such late showering hadrons are selected by requiring that they leave less than 1.1~\GeV\ in the EM calorimeter inside a cone of size $\Delta R = 0.1$ around their track.  They are further required to have energy deposited in the same cone in the hadronic calorimeter that is at least 40\% and less than 90\% of the track momentum. The energy deposited by close-by neutral particles is then measured in the EM calorimeter in the region $0.1 < \Delta R < 0.2$ around the MIP particle's track. A geometric factor of $4/3$ is applied to estimate the energy deposits from the neutral-particle background in the whole $\Delta R = 0.2$ cone. The mean of this background distribution over many events in a given track momentum and pseudorapidity bin, \epbg, estimates the energy deposited by photons and neutral hadrons in the EM calorimeter. This selection is illustrated in Figure~\ref{fig:Eoverp_cartoon}.  Using a similar method with the individual layers of the hadronic calorimeter, the energy deposited by the background particles in the hadronic calorimeter was found to be negligible. As described in Ref.~\cite{PERF-2011-05}, an alternative method that used information about the shape of the hadronic shower was used to estimate the neutral-particle background.  The difference between this method and the nominal method of about 10\% of the background, which itself is generally less than 25\% of the measured response, is taken as a systematic uncertainty.  

The \epav is corrected by the average background to give the corrected average response: $\epcor = \epav - \epbg$.
This corrected response is the primary observable used in the studies of calorimeter response to isolated charged hadrons in the remainder of this paper.

While this method accounts for the average energy deposited by the neutral-particle background, it cannot account for per-event background fluctuations.
This is particularly important when considering threshold effects, since a small background energy deposit might be sufficient to raise the signal in a cell above the threshold required to seed a topological cluster.
In events with large background, this can lead to a positive bias in \epcor.
Even if the hadronic shower and calorimeter response to the charged hadron were perfectly modeled, significant mis-modelling of the neutral-particle background can lead to different rates of this signal promotion.
Above $p\approx 4$~\GeV, when the fraction of tracks with $E=0$ is small, this effect is negligible.

\begin{figure}[htbp]
\centering
\subfloat[]{
  \includegraphics[width=0.48\textwidth]{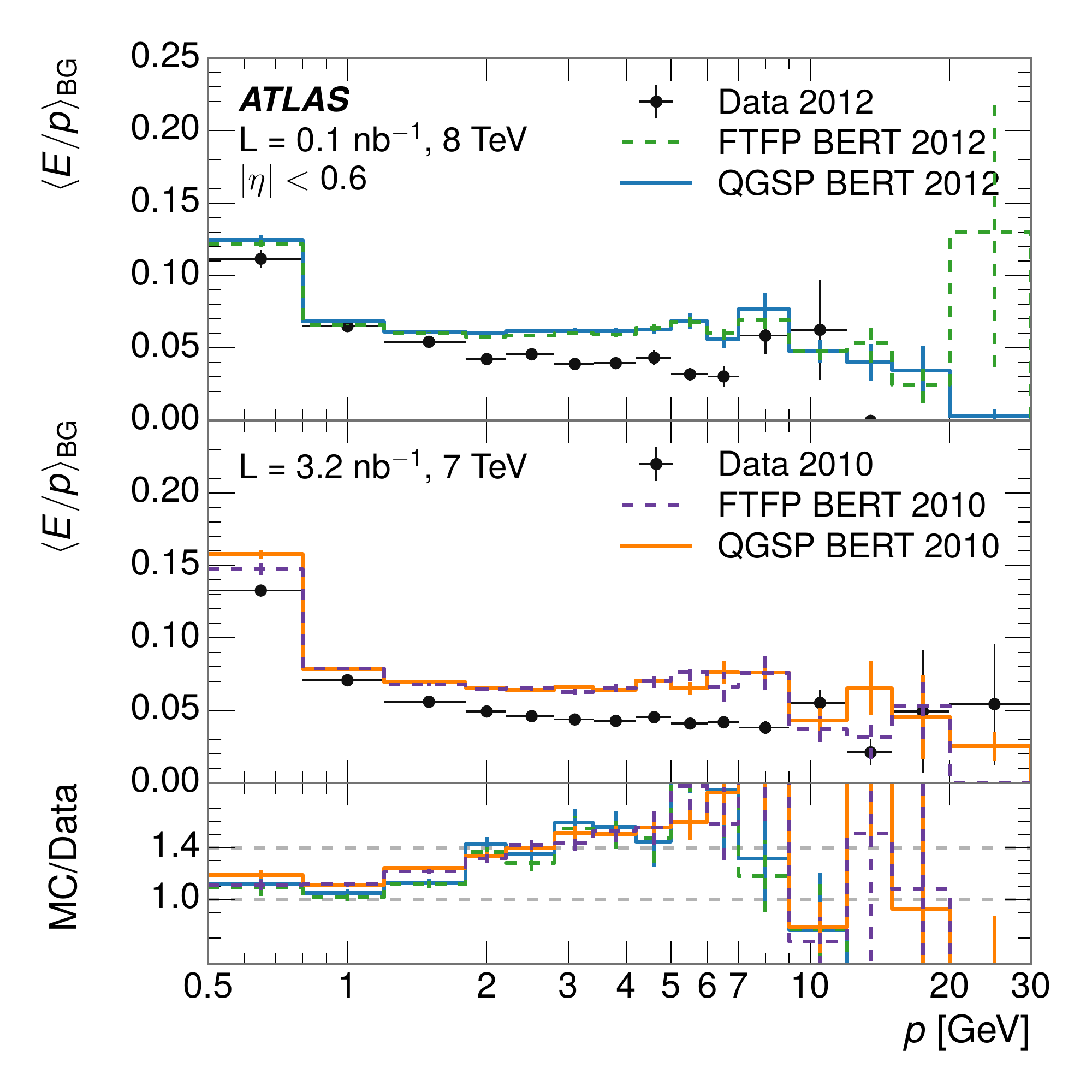}
}
~
\subfloat[]{
  \includegraphics[width=0.48\textwidth]{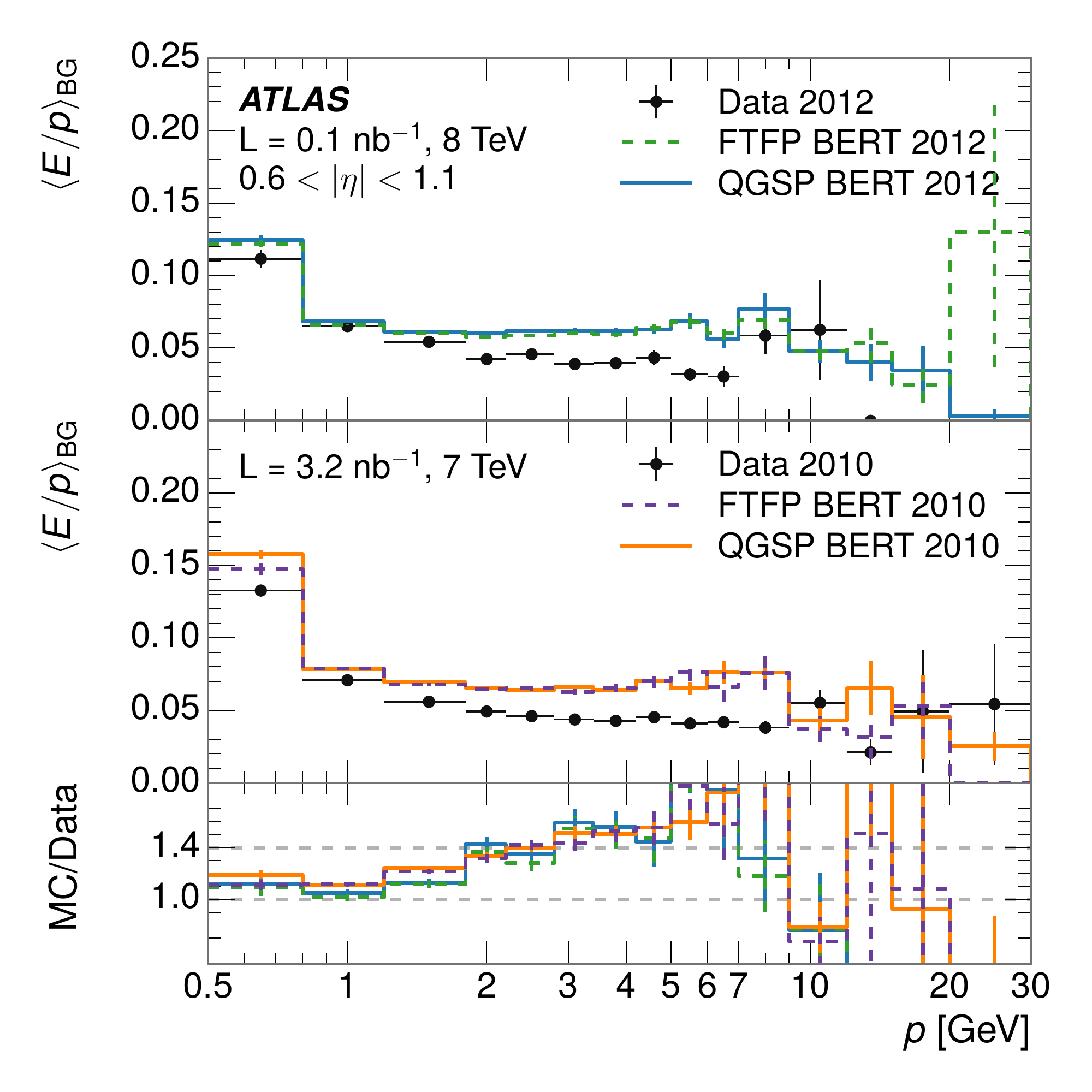}
}
\\
\subfloat[]{
  \includegraphics[width=0.48\textwidth]{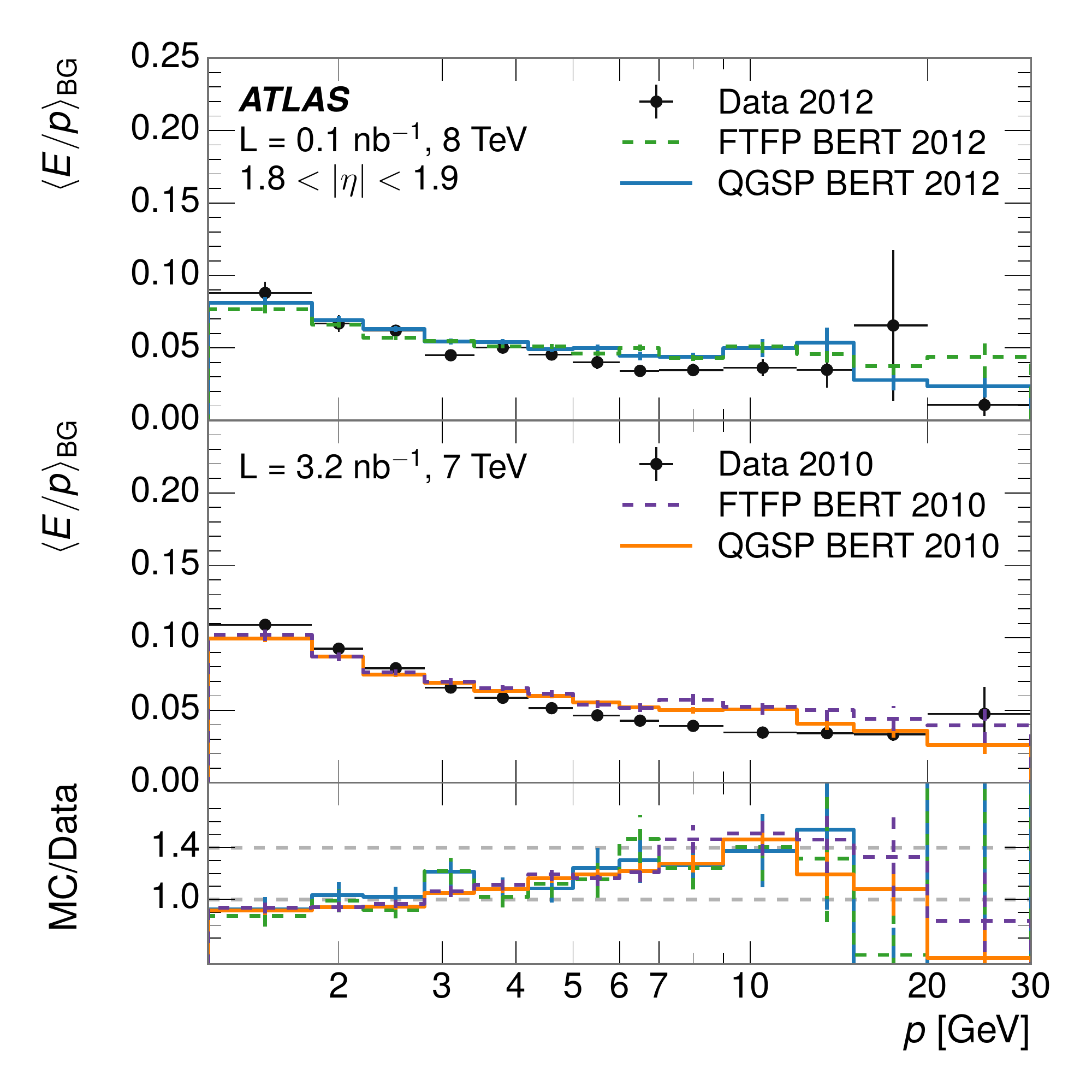}
}
~
\subfloat[]{
  \includegraphics[width=0.48\textwidth]{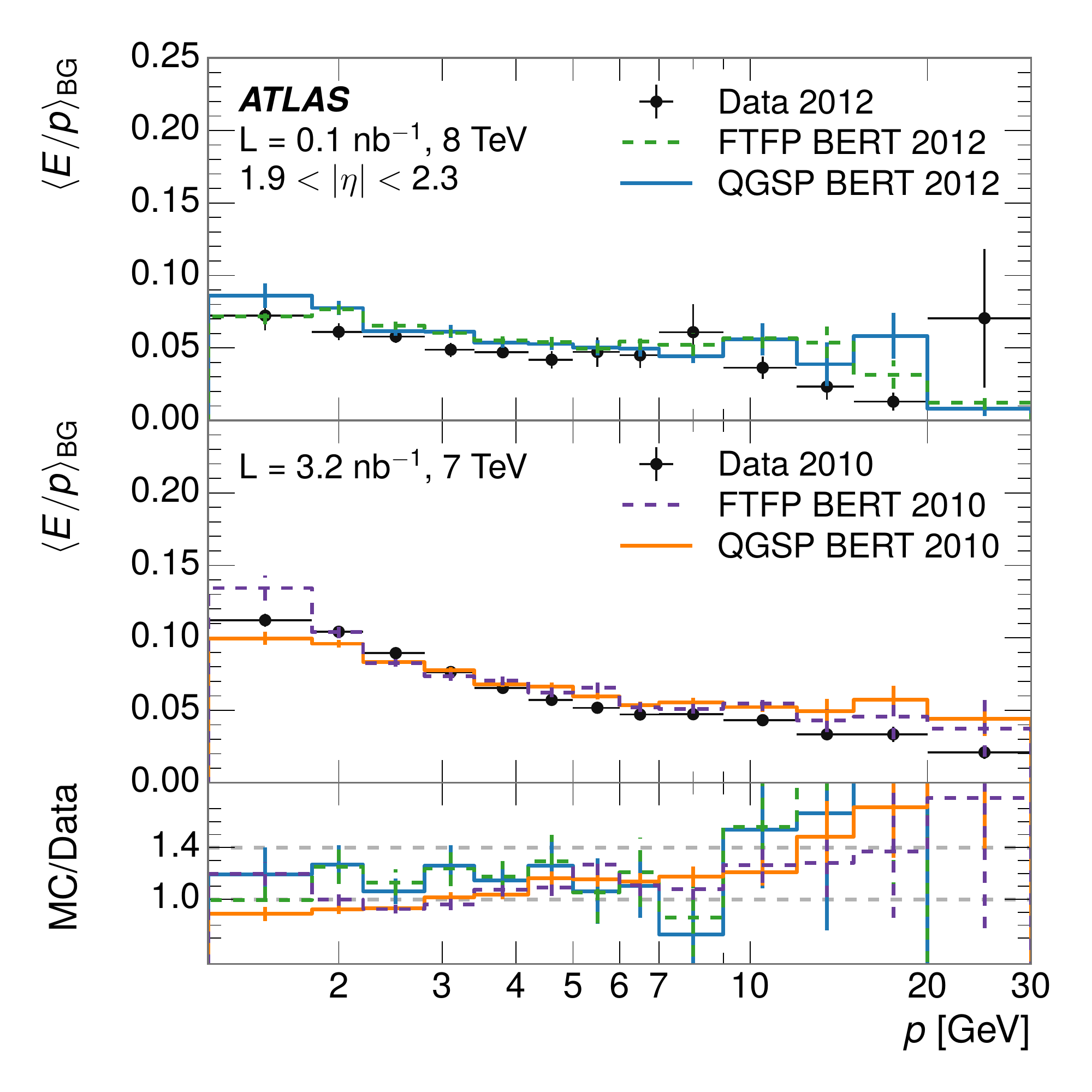}
}
\caption{\epbg as a function of the track momentum, for tracks with at least 20 TRT hits and (a) $|\eta| < 0.6$, (b) $0.6 < |\eta| < 1.1$,
(c) $1.8 < |\eta| < 1.9$, and (d) $1.9 < |\eta| < 2.3$.  The bottom portion of each panel shows the ratio of MC simulation to data.  The error bars represent statistical uncertainties.}
\label{fig-eoverp_background}
\end{figure}

\begin{figure}[htbp]
\centering
\subfloat[]{
  \includegraphics[width=0.48\textwidth]{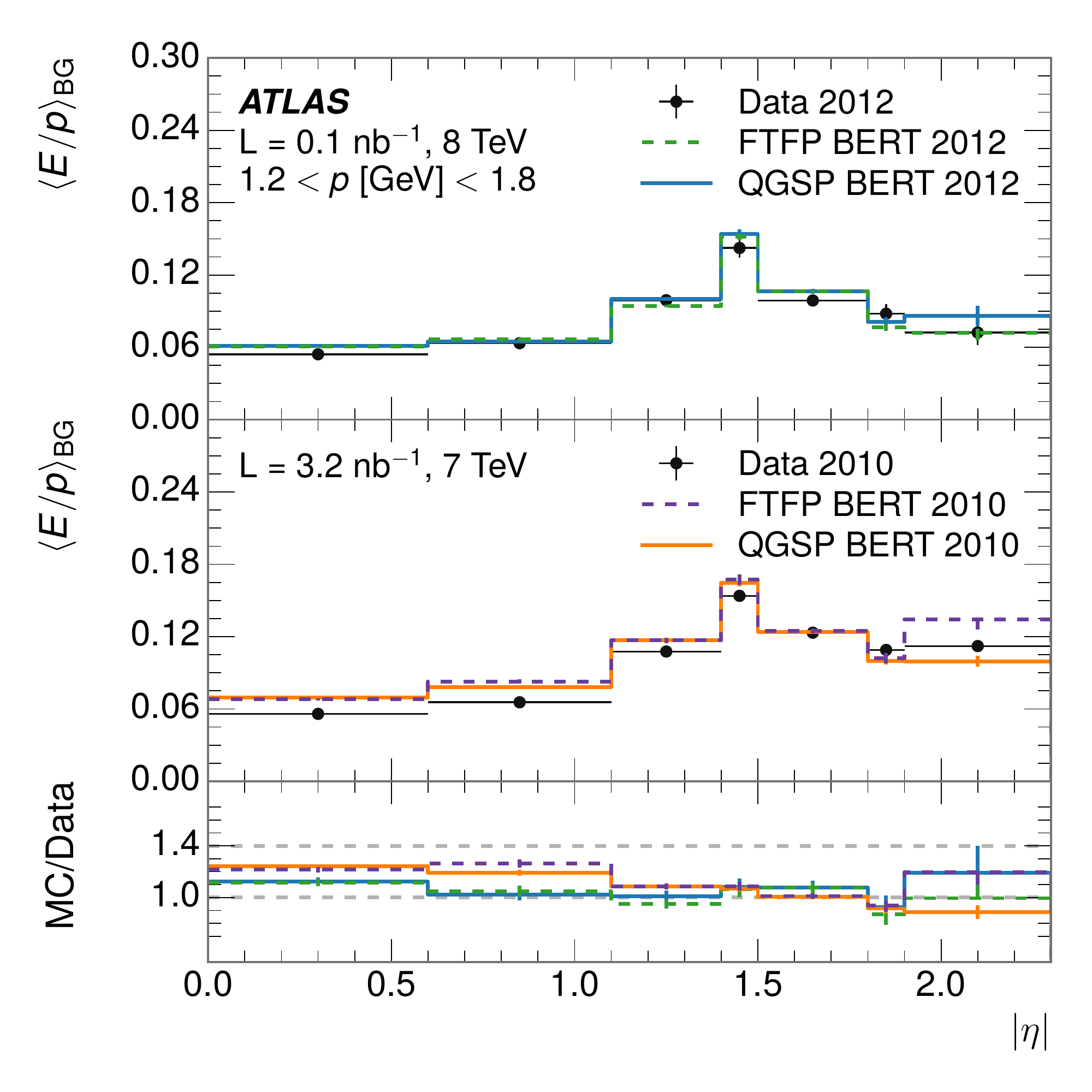}
}
~
\subfloat[]{
  \includegraphics[width=0.48\textwidth]{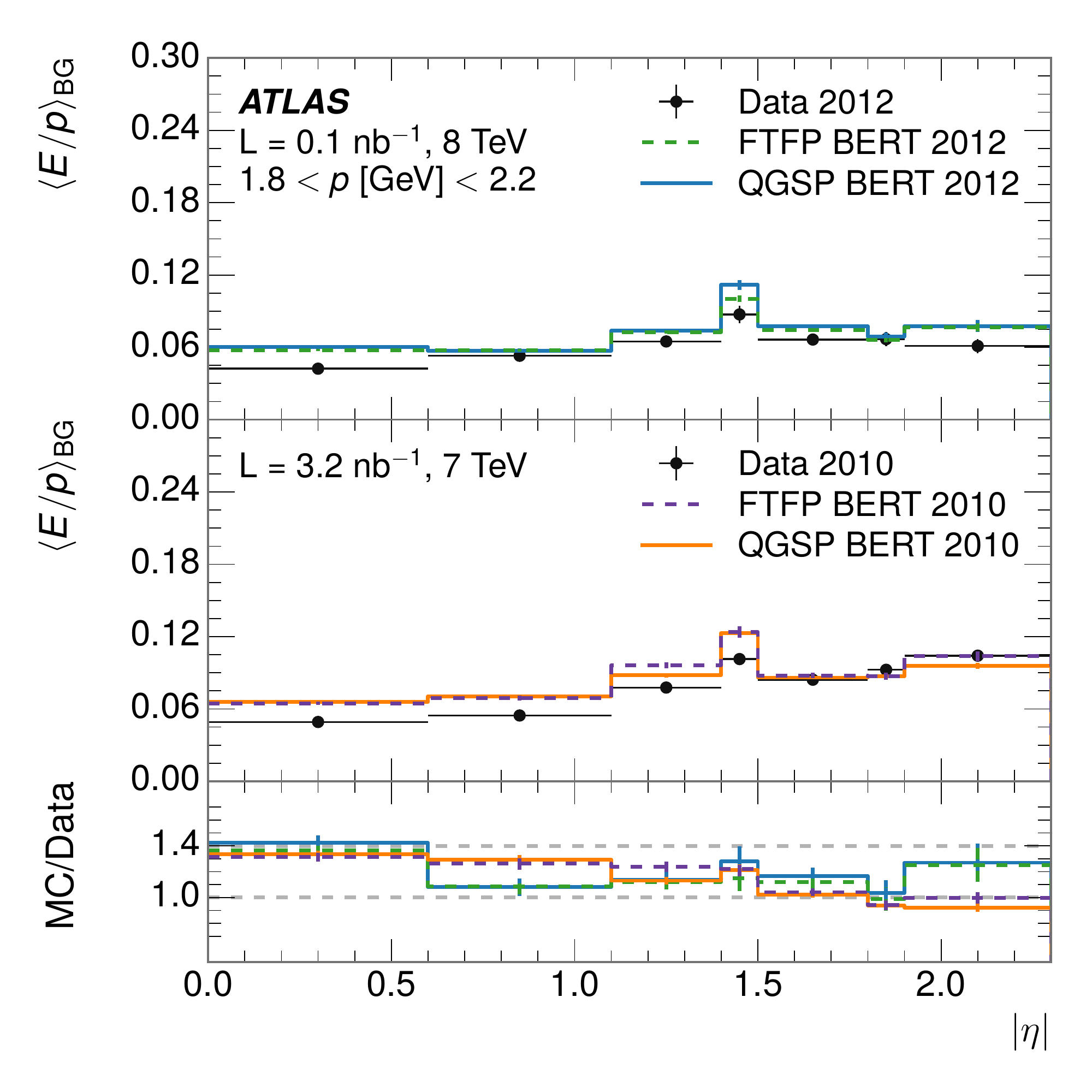}
}
\\
\subfloat[]{
  \includegraphics[width=0.48\textwidth]{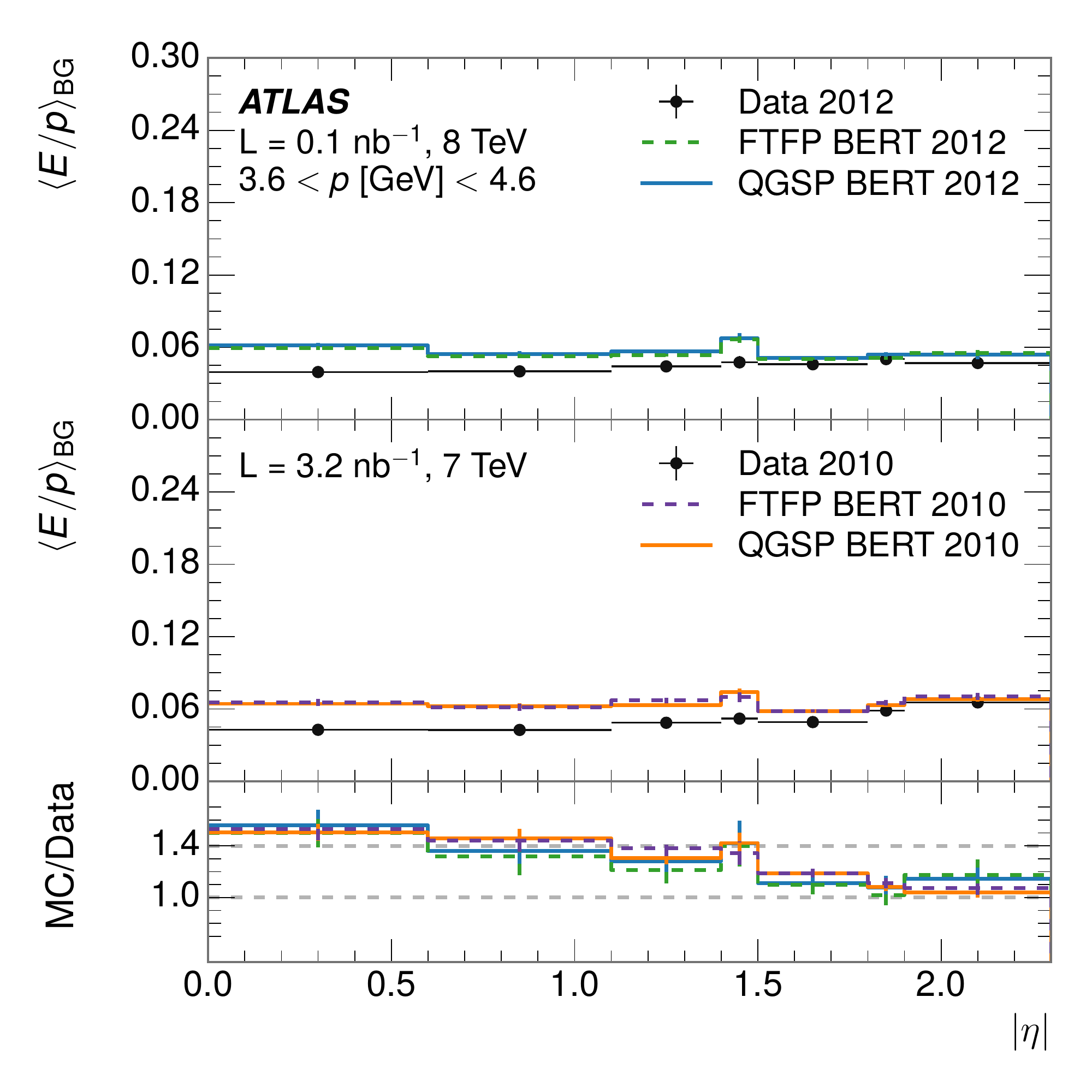}
}
~
\subfloat[]{
  \includegraphics[width=0.48\textwidth]{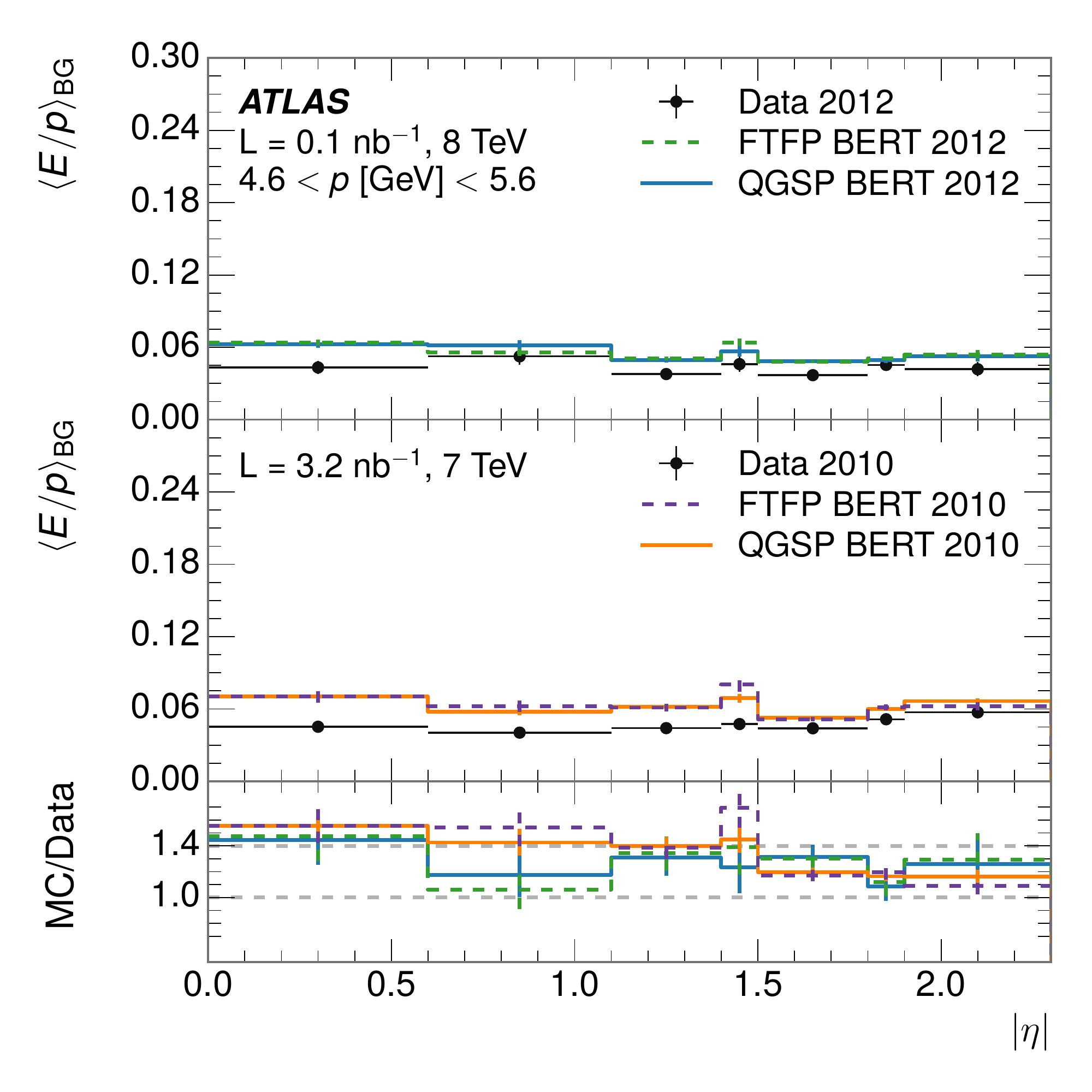}
  \label{fig-eoverp_background_eta_highp}
}
\caption{\epbg as a function of the track pseudorapidity, for tracks with at least 20 TRT hits and (a) $1.2 < p / \GeV < 1.8$, (b) $1.8 < p / \GeV < 2.2$,
(c) $3.6 < p / \GeV < 4.6$, and (d) $4.6 < p / \GeV < 5.6$.  The bottom portion of each panel shows the ratio of MC simulation to data.  The error bars represent statistical uncertainties.}
\label{fig-eoverp_background_eta}
\end{figure}

Figure~\ref{fig-eoverp_background} and Figure~\ref{fig-eoverp_background_eta} show the measured \epbg in data and MC simulation as a function of track momentum and pseudorapidity, respectively.  The general shape of the background is reasonably well modeled by the simulations, but important discrepancies exist between the two, such as the simulation's overestimation of the background at intermediate ($2 < p / \GeV < 8$) track momentum in the central ($|\eta|<1.1$) region of the detector. These differences are attributed to the phenomenological models used to describe non-perturbative QCD processes in {\sc Pythia8}, as well as the difficulty of correctly modelling the calorimeter response to low-momentum neutral particles. They do not strongly indicate a deficiency in the detector description, which would typically be isolated in a narrow region of $|\eta|$ away from the well-understood central ID region. The broad discrepancy as a function of $|\eta|$ shown in Figure~\ref{fig-eoverp_background_eta_highp} is consistent with a deficiency in the modelling of coherent neutral particle radiation in {\sc Pythia8}, as the background is consistently and significantly higher in the MC simulation than in the data. 
Provided the neutral-particle background is correctly accounted for separately in data and MC simulation, however, any imperfection in the modelling of neutral particles can be removed from the comparison of calorimeter response.
The neutral-particle background is calculated separately for each dataset and calorimeter configuration considered in this paper.

\subsection{Reduction of hadronic interactions in the ID}
\label{sec:TRTcutMotivation}

\begin{figure}[tb]
\centering
\subfloat[]{
  \includegraphics[width=0.48\textwidth]{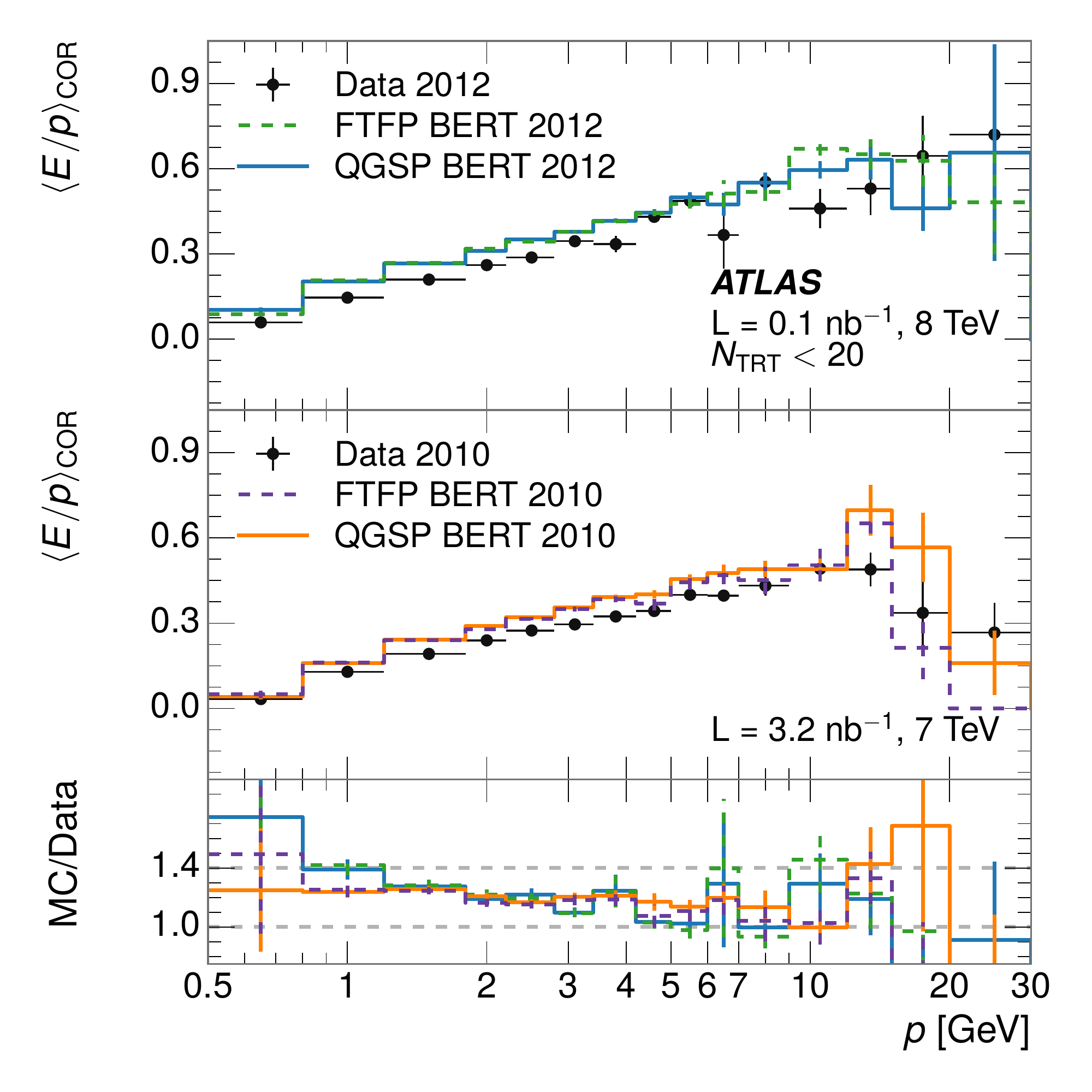}
}
~
\subfloat[]{
  \includegraphics[width=0.48\textwidth]{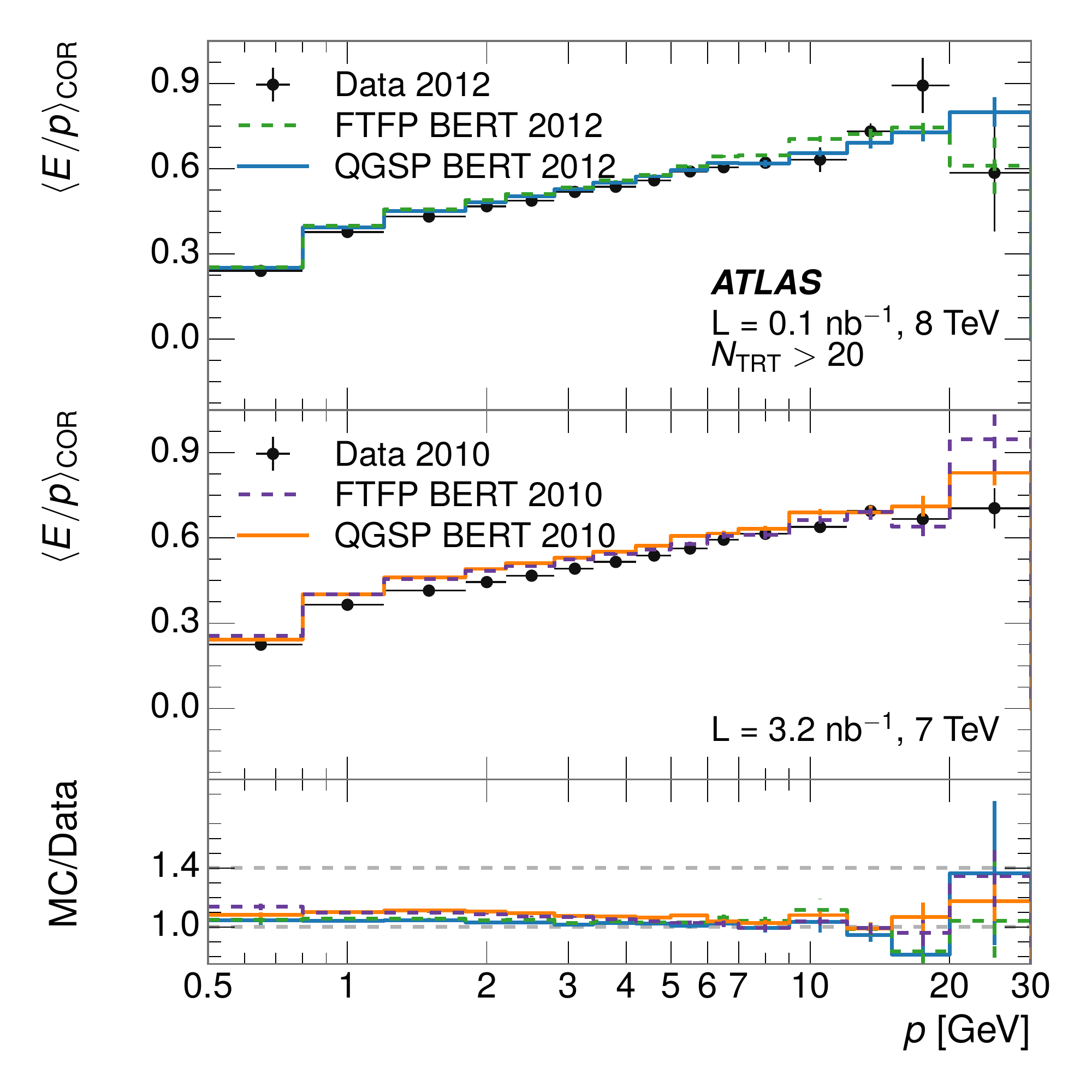}
}
\caption{Comparison of the \epcor for tracks with (a) less than and (b) greater than 20 hits in the TRT.  The bottom portion of each panel shows the ratio of MC simulation to data, separately for the two sets of hadronic physics models, and separately for 2010 and 2012 data. The error bars represent statistical uncertainties.}
\label{fig-zerofracTRT}
\end{figure}

Part of the difference between the rate of tracks with no associated energy in the simulation and the data (e.g. in Figure~\ref{fig-zerofracincl2}) may be due to geometry differences, since additional material tends to increase the rate of particles that do not reach the calorimeter and deposit significant energy.
Another part may be due to poor modelling of secondary particles from hadronic interactions that occur before the calorimeter, as suggested in Section~\ref{sec-selection}.
Tracks that do not have a large number of hits in the TRT are likely to result from particles that have undergone hadronic interactions while propagating through the ID.
The large scattering angles of secondary charged particles, as well as the rate of secondary neutral particles, both contribute to the original track not being extended to the face of the calorimeter.
Thus, examining tracks with a small number of hits in the TRT provides information about particles that likely underwent hadronic interactions.
A comparison of the \epcor for tracks that do not have a large number of hits in the TRT with those that do is shown in Figure~\ref{fig-zerofracTRT}.
The \epcor for tracks without a large number of TRT hits is very poorly modeled by the simulation, showing 25\%-level discrepancies at low momenta, suggesting problems with the description of secondary particles from these relatively low-energy nuclear interactions.  For tracks with a large number of TRT hits, there remains a 5--10\% discrepancy.  This discrepancy, which is smaller for 2012 data than in the case of 2010 data, is explored in subsequent sections of the paper.

For the remainder of the paper, more than 20 hits are required in the TRT, in order to suppress tracks from particles that undergo nuclear interactions before the calorimeter.

\subsection{Background-corrected single-hadron response}
\label{sec-final_eoverp}

\begin{figure}[p!]
\centering
\subfloat[]{
  \includegraphics[width=0.48\textwidth]{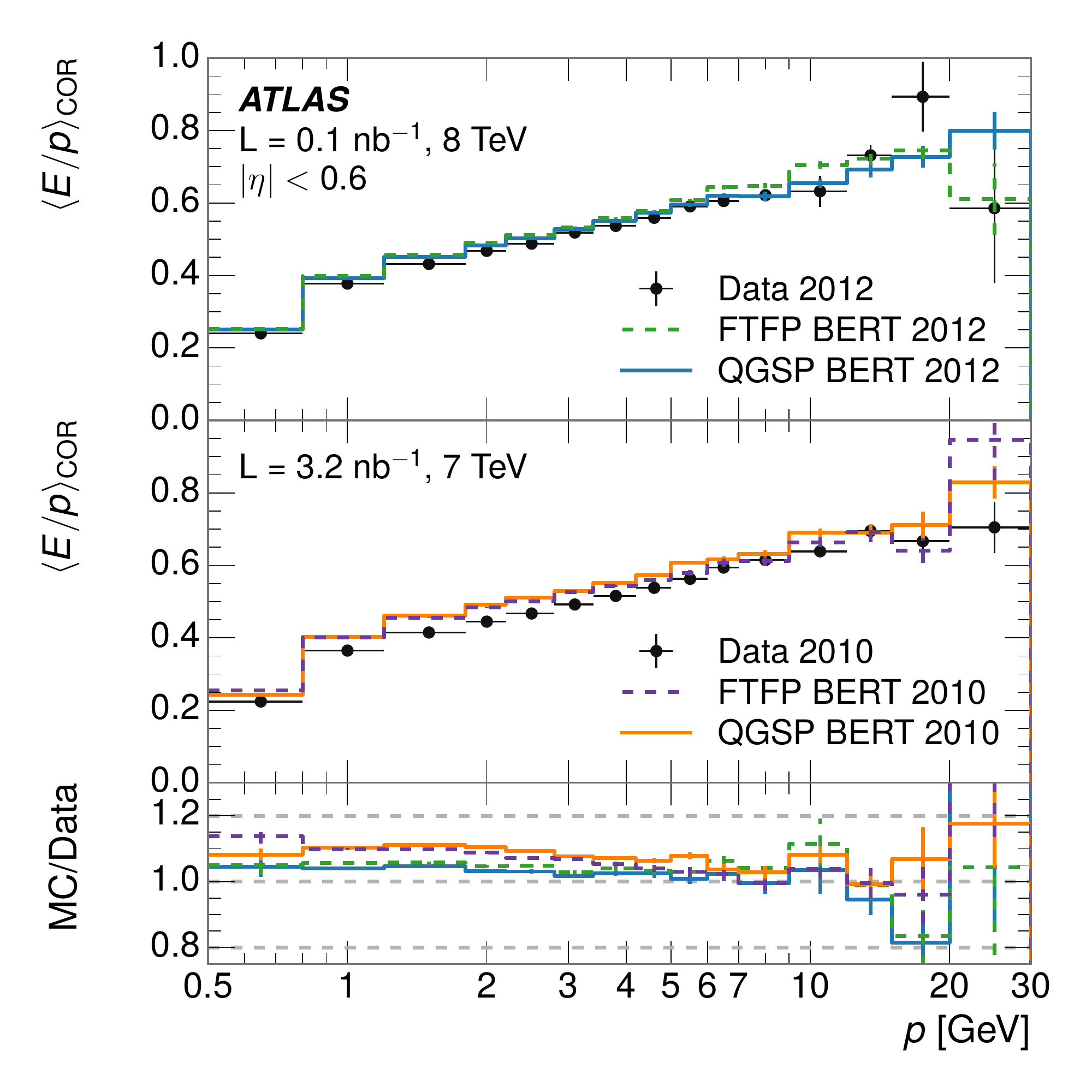}
  \label{fig-eoverp_corrected_smallEta}
}
~
\subfloat[]{
  \includegraphics[width=0.48\textwidth]{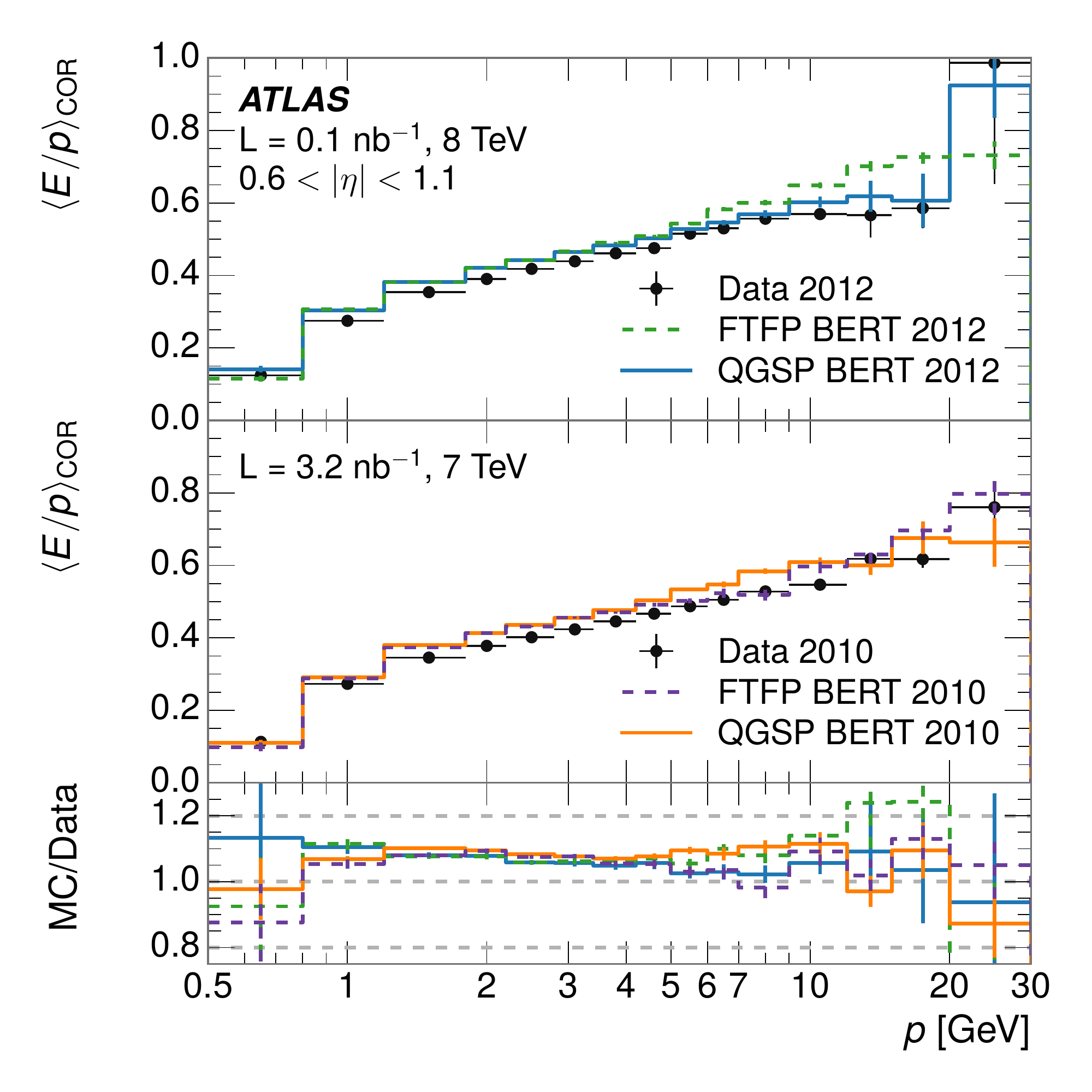}
}
\\
\subfloat[]{
  \includegraphics[width=0.48\textwidth]{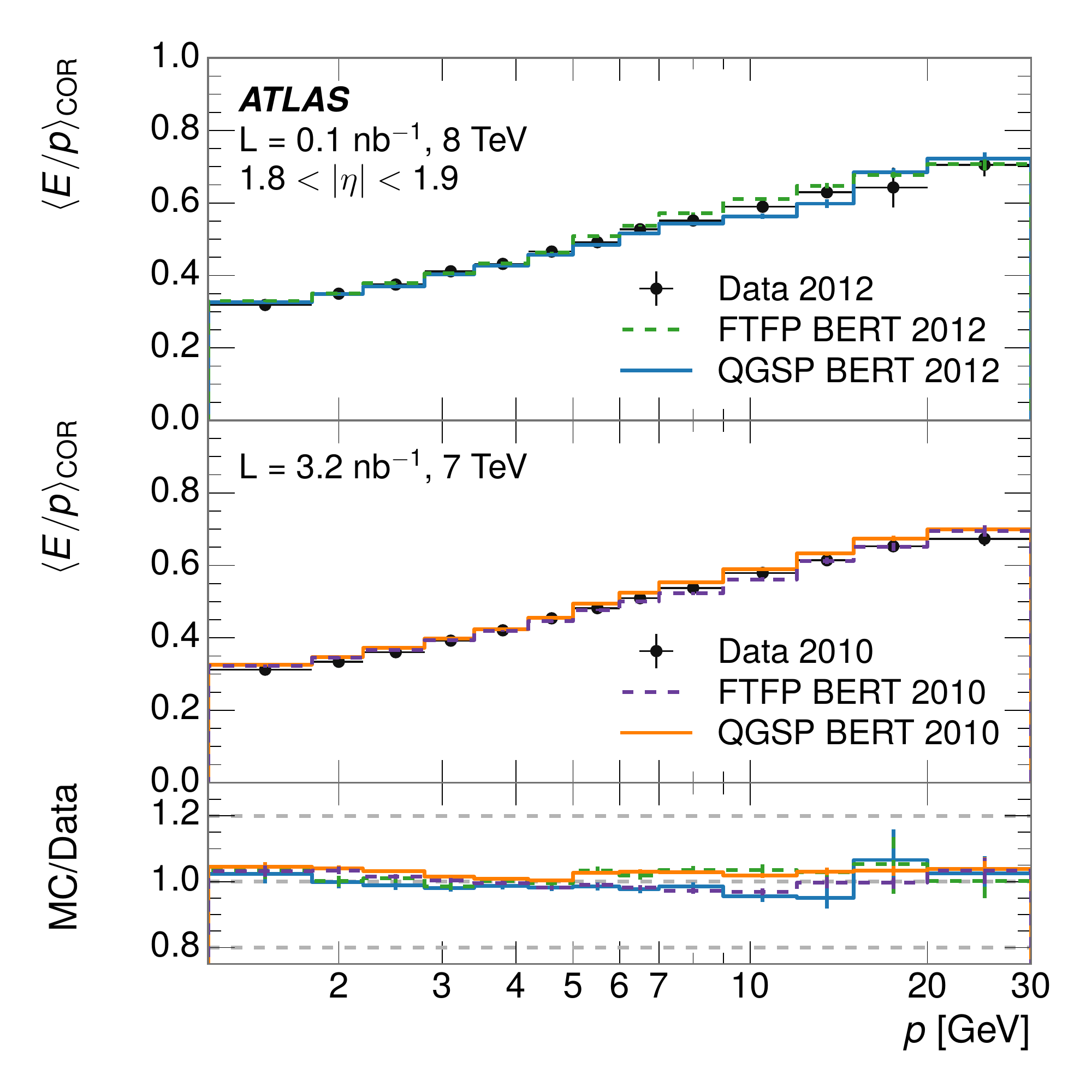}
}
~
\subfloat[]{
  \includegraphics[width=0.48\textwidth]{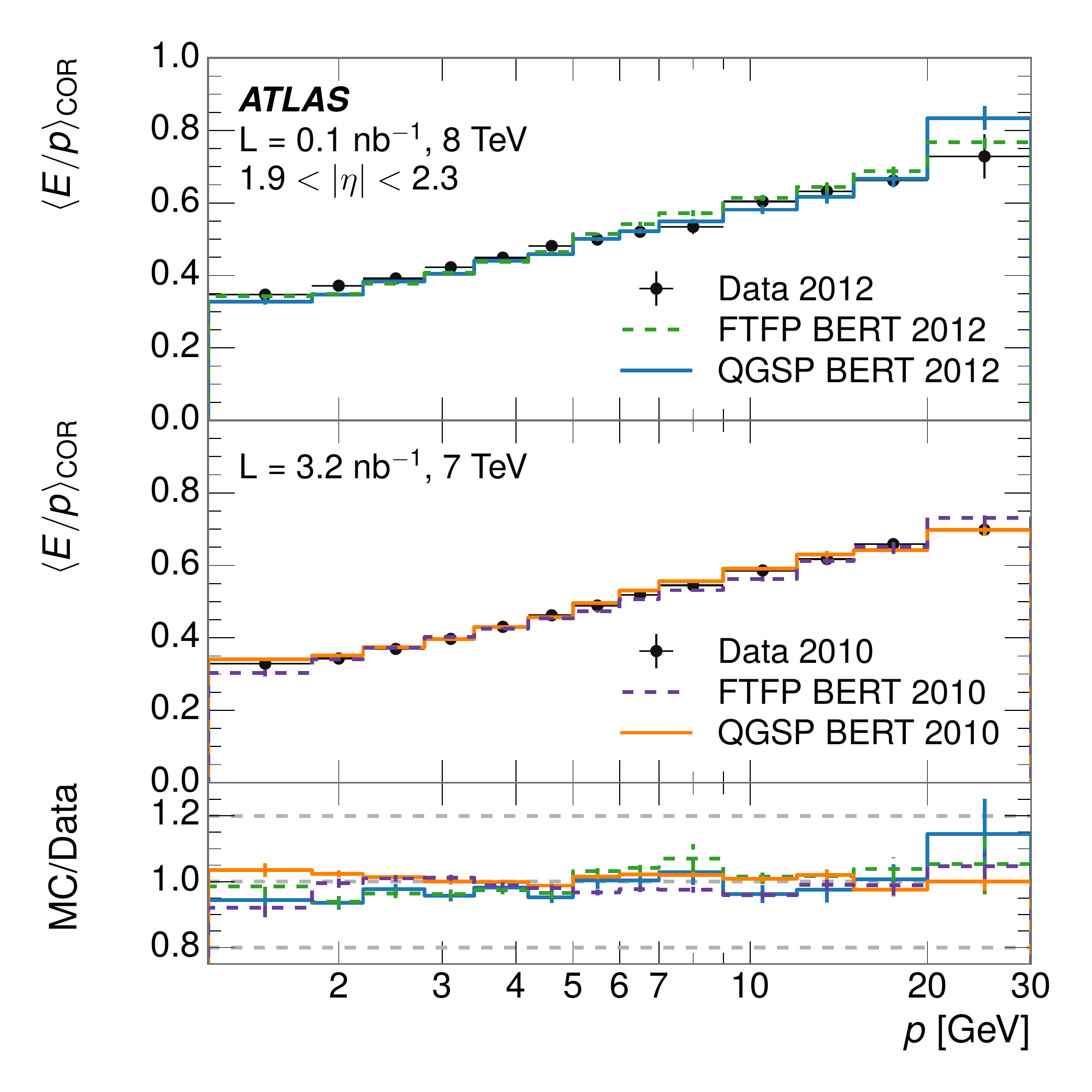}
}
\caption{\epcor as a function of track momentum, for tracks with (a) $|\eta| < 0.6$, (b) $0.6 <  |\eta| < 1.1$,
(c) $1.8 <  |\eta| < 1.9$, and (d) $1.9 <  |\eta| < 2.3$.  Tracks not matching any topological energy clusters in the calorimeter are included.  The bottom portion of each panel shows the ratio of MC simulation to data.  The error bars represent statistical uncertainties.}
\label{fig-eoverp_corrected}
\end{figure}

The corrected \epav variable (\epcor), in which the average neutral-particle background is subtracted, is shown in Figure~\ref{fig-eoverp_corrected}, with statistical uncertainties, for several bins of pseudorapidity. Here, the background estimated in data is subtracted from the raw data \epav, and the background estimated in MC simulation is subtracted from the raw MC simulation \epav. The maximum momentum that can be effectively probed with the available data is about 30~\GeV, and the measurement has large statistical uncertainties above 20~\GeV, limiting the comparison. Both the \texttt{QGSP\_BERT} and \texttt{FTFP\_BERT} MC simulation event samples overestimate \epcor at low momentum, by approximately 5\% in the most central $|\eta|$ region. In more forward regions (e.g. beyond $|\eta|=1.8$), where the background is well modeled for the same momenta but the material in front of the calorimeter is substantially larger, the MC simulation describes the data to within the uncertainties.  Tracks that are assigned calorimeter energy $E=0$ are included in these distributions.

\subsubsection{Use of clusters or cells in response measurement}
\label{sec-clusters_vs_cells}

\begin{figure}[ht]
\centering
\subfloat[]{
  \includegraphics[width=0.48\textwidth]{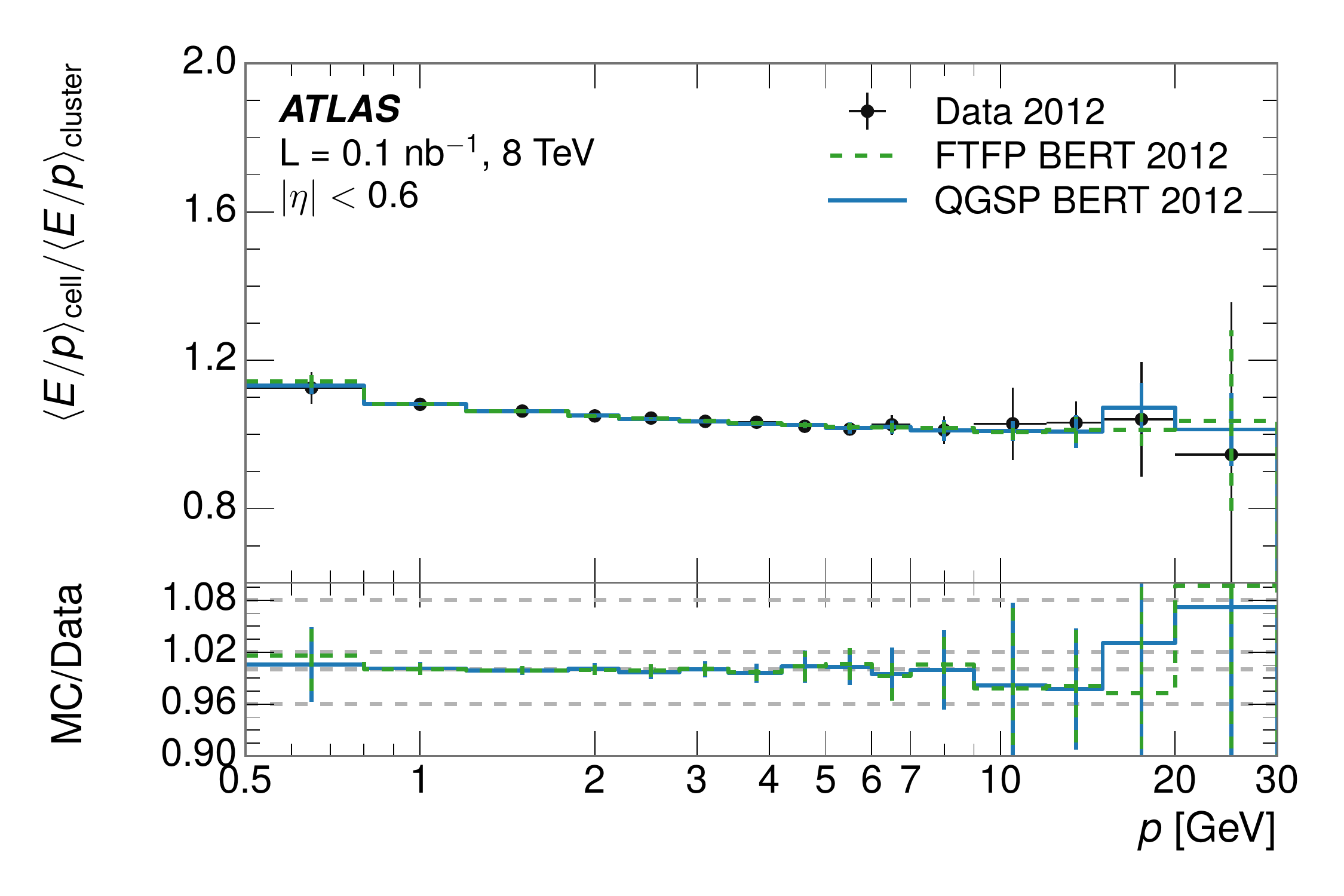}
}
~
\subfloat[]{
  \includegraphics[width=0.48\textwidth]{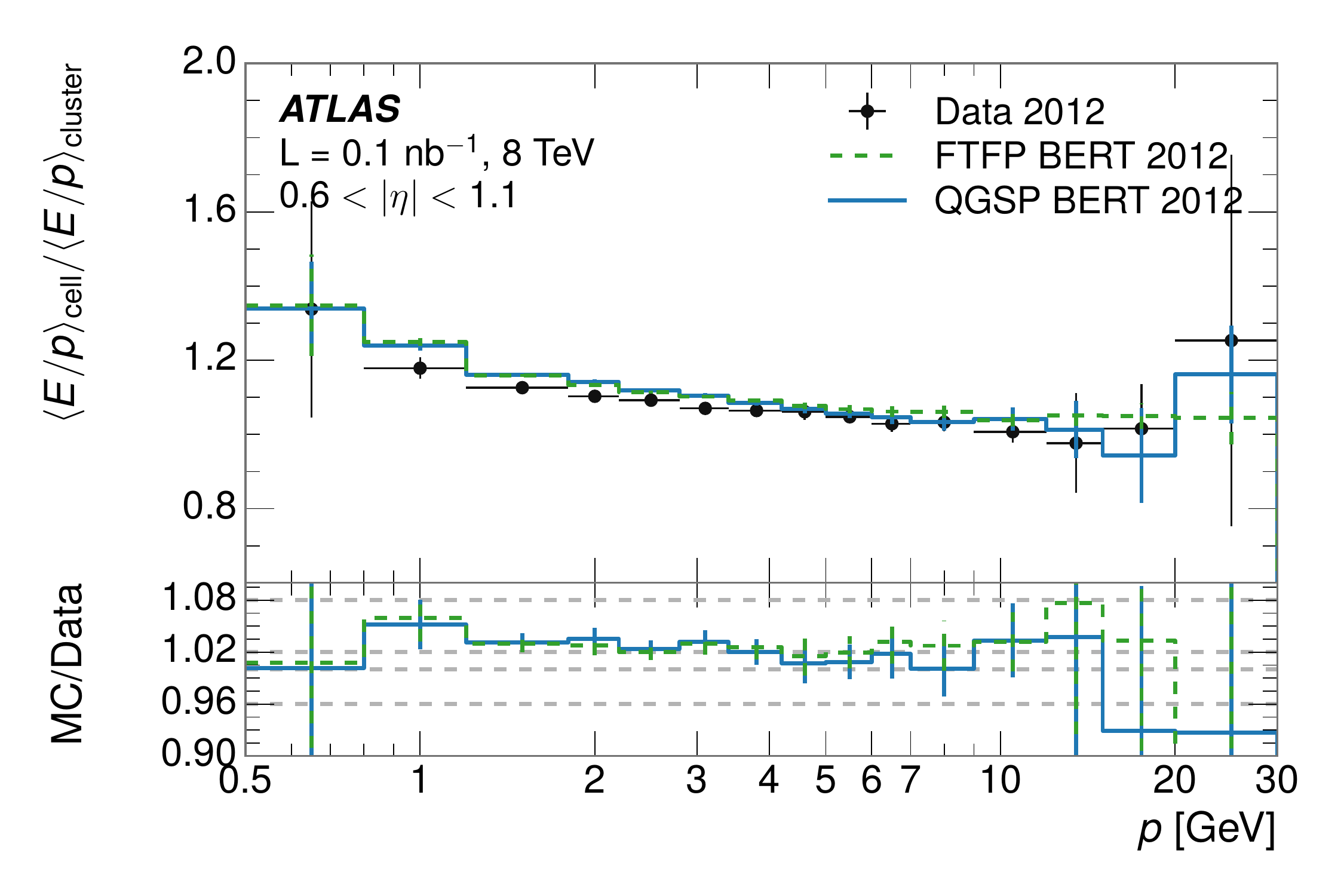}
}
\caption{ Ratio of \epcor calculated with cells to \epcor calculated with topological clusters as a function of track momentum, for tracks with (a) $|\eta|<0.6$ and (b) $0.6 < |\eta| < 1.1$. The bottom portion of each panel shows the ratio of MC simulation to data.  The error bars represent statistical uncertainties.}
\label{fig-eoverp_corrected_cells}
\end{figure}

The calorimeter response is normally calculated using topological clusters of energy in the calorimeter.
Figure~\ref{fig-eoverp_corrected_cells} shows a ratio of the \epcor derived directly from the energy deposition in the calorimeter cells near the extrapolated track position, $\epav_\text{cell}$, to the \epcor calculated using topological energy clusters, here labelled $\epav_\text{cluster}$.
For this comparison, all cells within $\Delta R=0.2$ of the extrapolated track position are included in calculating the cell-level energy.
In order to provide a subtraction of the appropriate background contribution, the background is also calculated using cells instead of clusters.
This comparison provides a useful test of the modelling of topological clustering effects and the hadronic shower width by the simulation.
These distributions agree within the statistical uncertainties for central $|\eta|$, demonstrating an excellent modelling of the effect of topological clustering on the calorimeter response distribution.
In the more forward region, there are percent-level disagreements.
It is also clear that the cell calculation provides a response up to 15\% higher in the central region at low momentum, which is expected because of the effect of the threshold on the calorimeter cells applied during the clustering.

\subsubsection{Effect of cluster calibration on response measurement}
\label{sec:cluster_calibration}

\begin{figure}[ht]
\centering
\subfloat[]{
  \includegraphics[width=0.48\textwidth]{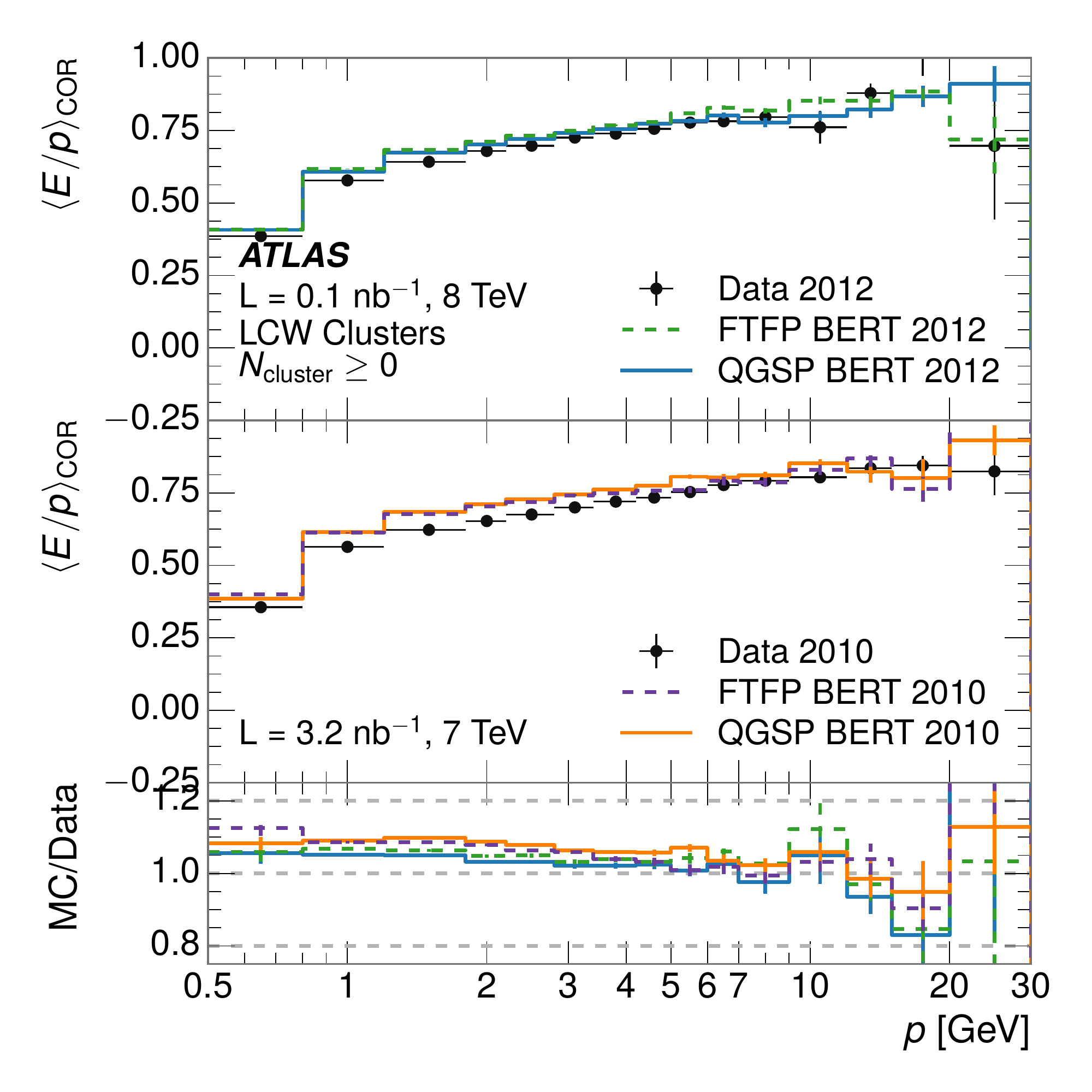}
  \label{fig-eoverp_corrected_lcw_smallEta}
}
~
\subfloat[]{
  \includegraphics[width=0.48\textwidth]{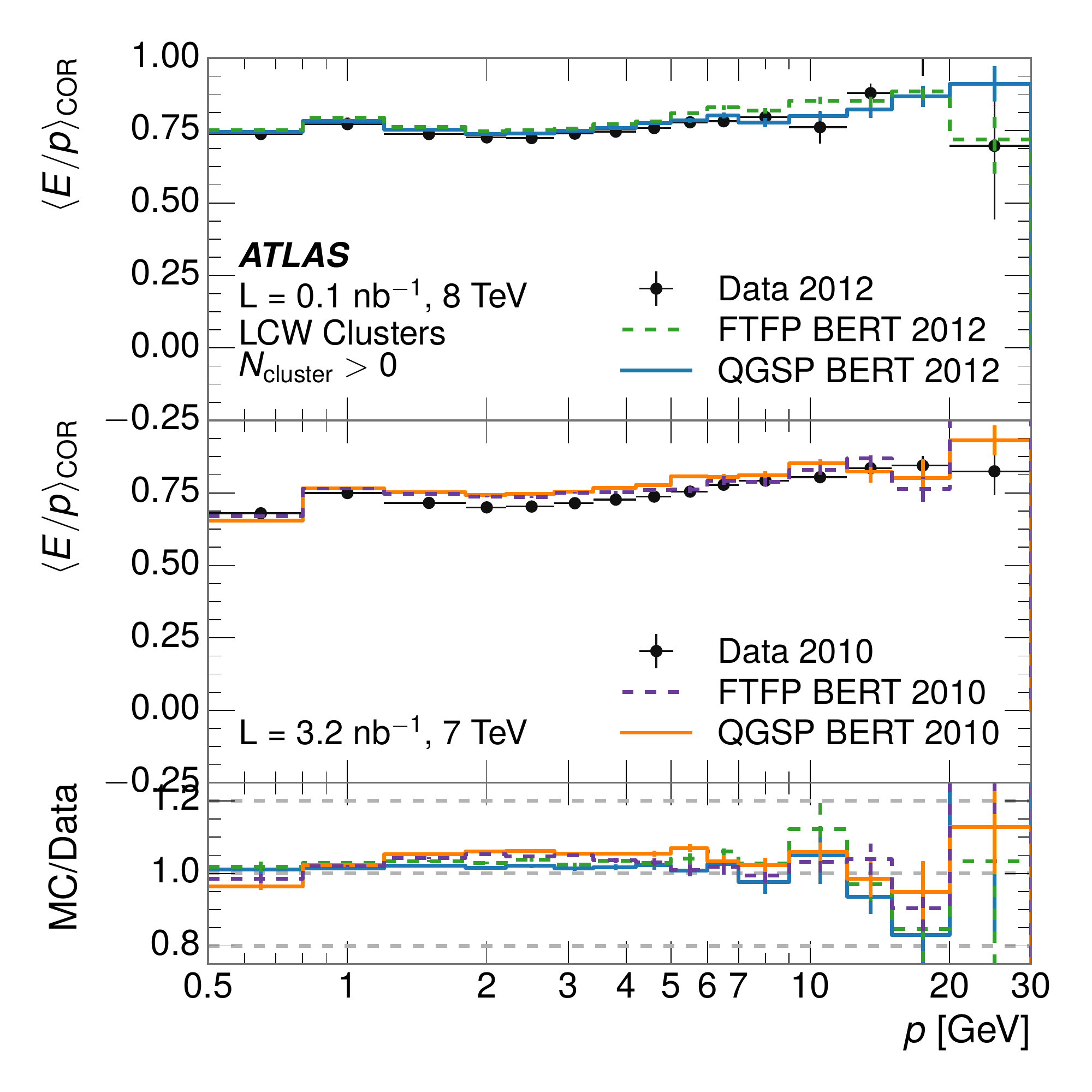}
}
\caption{Comparison of the \epcor calculated using LCW-calibrated topological clusters as a function of track momentum, corrected for the neutral-particle background, for tracks with $|\eta|<0.6$, and (a) zero or more associated topological clusters or (b) one or more associated topological clusters. Figure~\ref{fig-eoverp_corrected_smallEta} shows the same quantity as Figure~\ref{fig-eoverp_corrected_lcw_smallEta}, calculated using EM-scale topological clusters. The bottom portion of each panel shows the ratio of MC simulation to data, separately for the two sets of hadronic physics models, and separately for 2010 and 2012 data. The error bars represent statistical uncertainties.}
\label{fig-eoverp_corrected_lcw}
\end{figure}

The topological clusters used for the calorimeter response comparison are measured at the EM scale, meaning that the calibration does not attempt to compensate for energy depositions measured by the calorimeter outside of the topological cluster, energy losses in uninstrumented material inside and outside of the topological cluster, or the non-compensating response of the calorimeter.
The local cluster weighting (LCW) technique~\cite{PERF-2012-01} applies a calibration to the topological cluster energy according to the position and properties of the energy depositions in the topological cluster (e.g. depth in the calorimeter and energy density) in order to account for these effects.
Figure~\ref{fig-eoverp_corrected_lcw} shows a comparison of the LCW-calibrated \epcor in data and simulation, both including and omitting tracks with $E=0$.
When calculating \epcor with the LCW calibration, the same calibration method is applied to the clusters used to determine the background.
The response is significantly higher than that of Figure~\ref{fig-eoverp_corrected_smallEta} due to the calibration, since the calibration raises the average response for the clusters.
Agreement between data and MC simulation is almost identical with both calibrations, implying no gross mis-modelling of the hadronic shower properties on which the LCW calibration depends.
Agreement is marginally better when considering only tracks with at least one associated topological cluster in the calorimeter, again suggesting a discrepancy in the description of hadronic interactions before the sampling portion of the calorimeter.
The calibrated response to single particles, which is not unity with either of these selections, is a result of imperfections in the calibration procedure.
Nonetheless, the momentum dependence of the response is almost completely removed by the LCW calibration when considering tracks with at least one associated topological cluster.
As the discrepancies between MC simulation and data are most critical for the studies presented here and the LCW calibration does not affect these discrepancies significantly, in the remainder of this paper the EM-scale response is used for almost all comparisons.

\subsubsection{Charge dependence of response}
\label{sec:charge_dependence}
\begin{figure}[ht]
\centering
\subfloat[]{
  \includegraphics[width=0.48\textwidth]{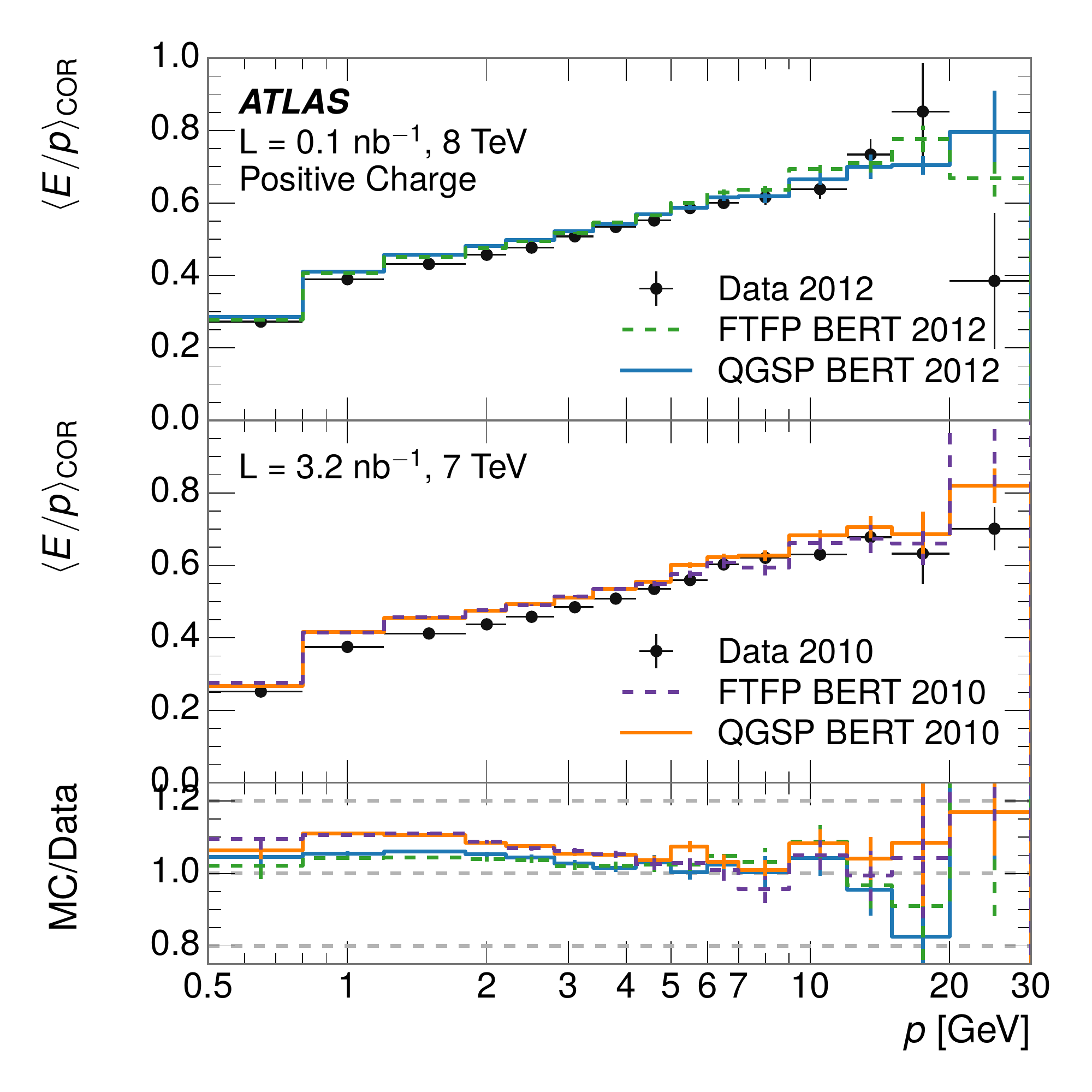}
}
~
\subfloat[]{
  \includegraphics[width=0.48\textwidth]{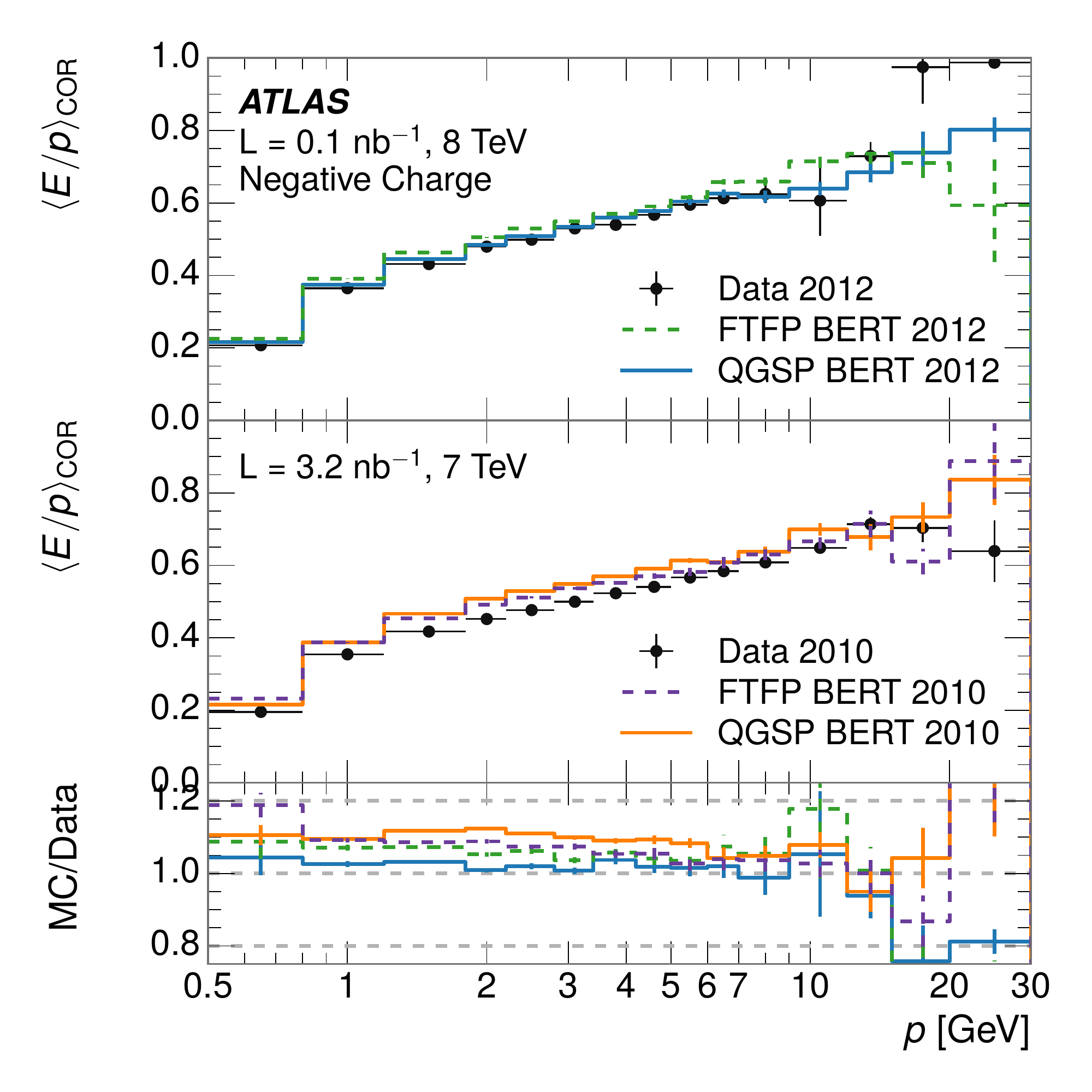}
}
\caption{Comparison of the \epcor for (a) positive and (b) negative tracks as a function of track momentum, corrected for the neutral-particle background, for tracks with $|\eta|<0.6$, in simulation with the \texttt{FTFP\_BERT} and \texttt{QGSP\_BERT} sets of hadronic physics models. The bottom portion of each panel shows the ratio of MC simulation to data, separately for the two sets of hadronic physics models, and separately for 2010 and 2012 data.  The error bars represent statistical uncertainties.}
\label{fig-eoverp_corrected_ftfp}
\end{figure}

The \epcor for positive and negative tracks, including tracks that do not match a topological cluster, is shown in Figure~\ref{fig-eoverp_corrected_ftfp}.
The two sets of hadronic physics models provide an almost identical result for positively charged tracks, which are dominated by \pip, $K^+$, and protons.
At low momenta the models are identical, and at high momenta they are tuned to the same thin-target data.
For negatively charged tracks, some difference between \texttt{QGSP\_BERT} and \texttt{FTFP\_BERT} is observed.

\begin{figure}[ht]
\centering
\subfloat[]{
  \includegraphics[width=0.48\textwidth]{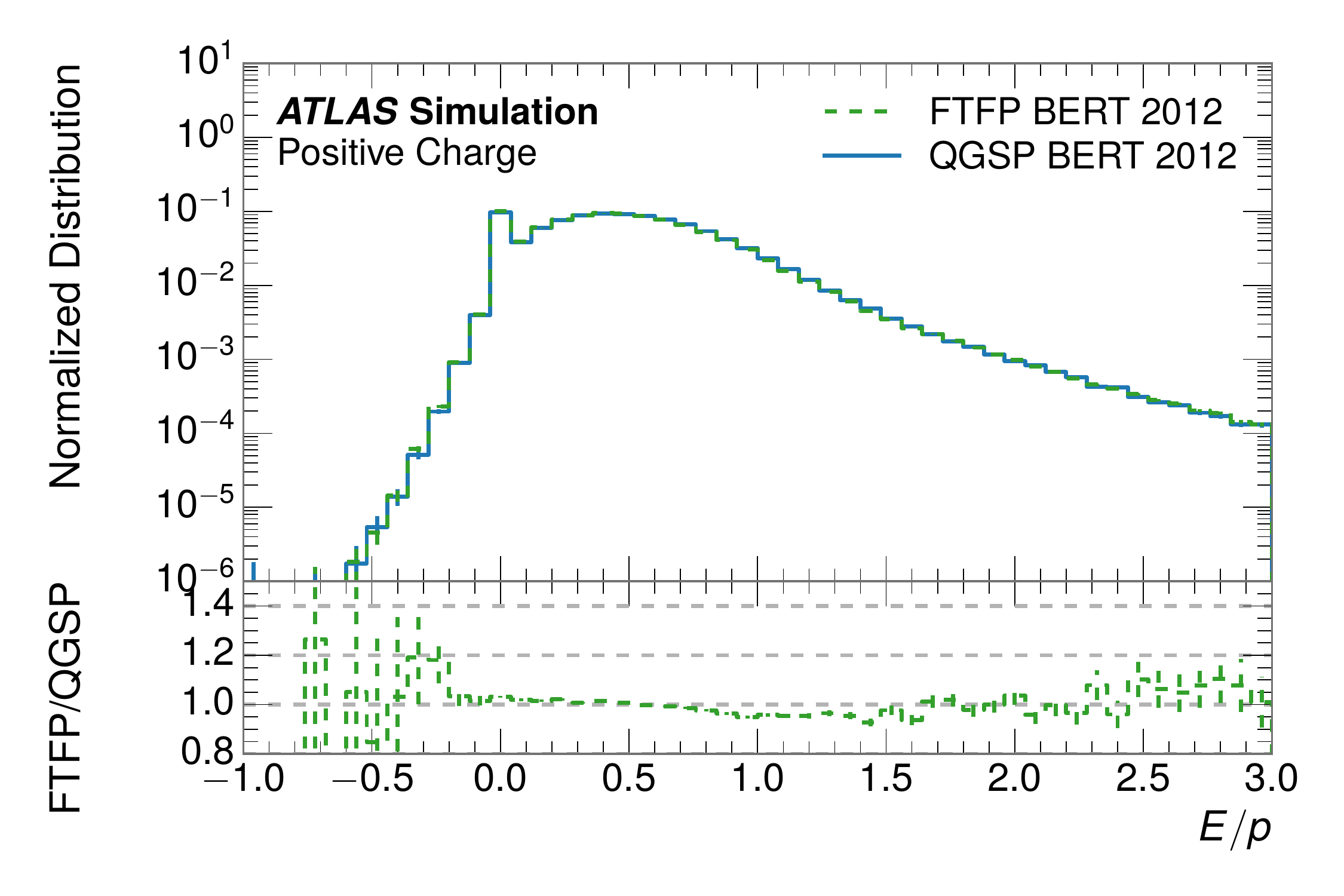}
}
~
\subfloat[]{
  \includegraphics[width=0.48\textwidth]{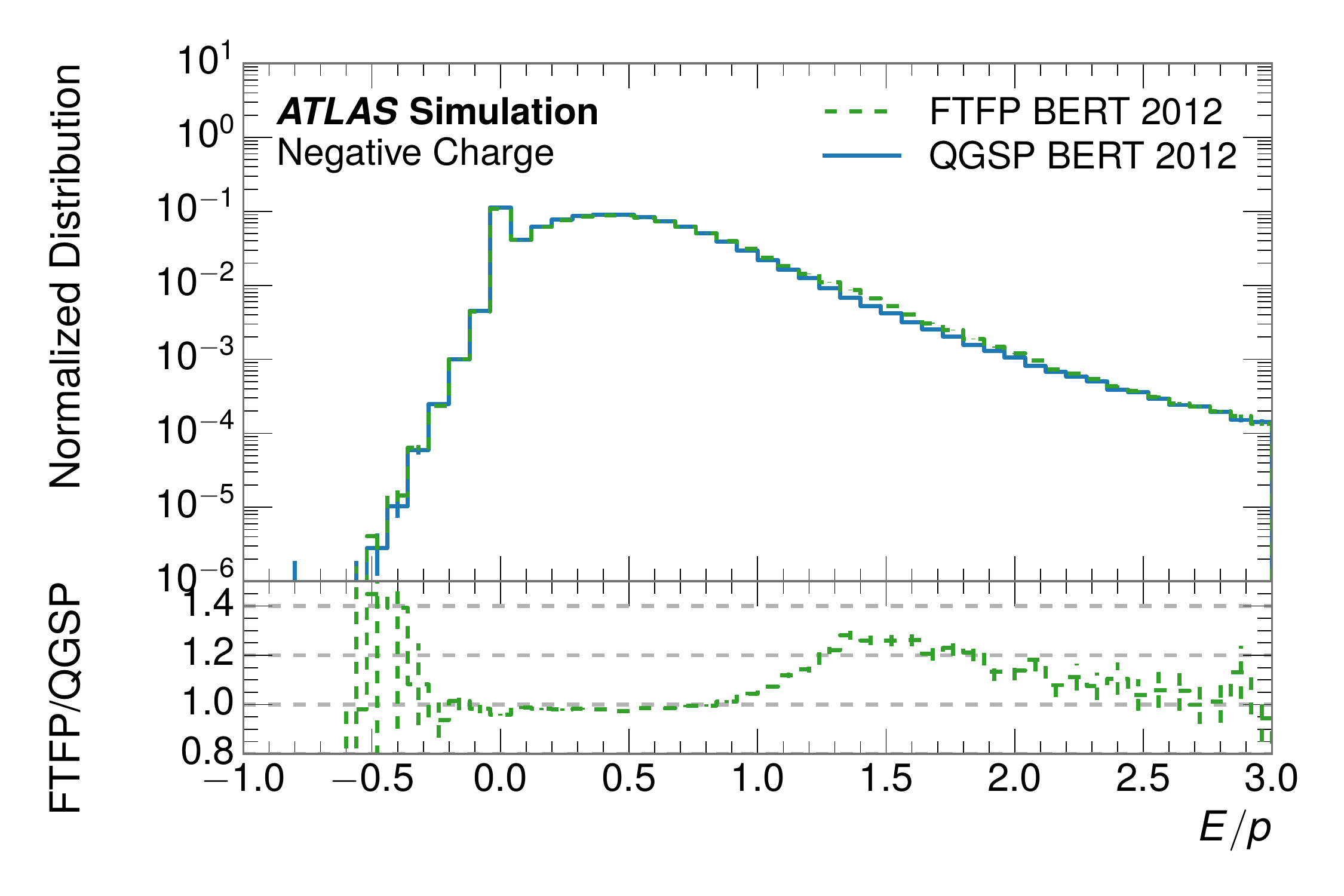}
}
\caption{ Comparison of the \ep distributions for (a) positive and (b) negative tracks with $0.8< p/\GeV<1.2$ and $|\eta|<0.6$, in simulation with the \texttt{FTFP\_BERT} and \texttt{QGSP\_BERT} sets of hadronic physics models. The bottom portion of each panel shows the ratio of the two sets of hadronic physics models.  The error bars represent statistical uncertainties.}
\label{fig-MCcomparison_plus_minus}
\end{figure}

This difference is primarily due to the difference in the modelling of the response to anti-protons, as is suggested by Figure~\ref{fig-MCcomparison_plus_minus}, which compares the \ep distributions for negatively charged and positively charged tracks in a low-momentum bin.  The two sets of hadronic physics models show identical distributions for positively charged tracks and show a clear discrepancy for negatively charged tracks with $E\approx 1.5\times p$.  In this momentum bin, the average calorimeter response is around 0.4, as seen in Figure~\ref{fig-eoverp_corrected_smallEta}.  Anti-protons, however, also contribute their annihilation energy in the calorimeter.  This additional $2$~\GeV~(938~\MeV\ for each of the anti-proton and the proton with which it annihilates), after including the effect of the non-compensating response of the calorimeter (roughly 50\%), gives an extra $1~\GeV$ to the energy measured in the calorimeter.  This difference is explored further in Section~\ref{sec:pid_response}.

\subsubsection{Single-hadron response in events with pile-up}
\label{sec:pileup}

Historically, the calorimeter response to isolated single particles has been measured using events with only a single proton--proton collision in the event.
Pile-up contributes additional neutral-particle background to the event that is normally only removed on average from the topological clusters.
The charged-particle background from pile-up can still be removed using the track isolation requirement.
Moreover, fluctuations of the neutral-particle background significantly worsen the energy resolution for low-momentum particles.
Nonetheless, as the background subtraction technique in this paper depends only on the average background distributions, the (pile-up and background-corrected) \epcor can still be measured in events with pile-up, in this case also binned in $\langle\mu\rangle$ and the number of reconstructed vertices.
To ensure a fair comparison, all data and MC simulation samples used in these comparisons are reconstructed with consistent calorimeter thresholds corresponding to $\langle\mu\rangle=30$.

\begin{figure}[ht]
\centering
\subfloat[]{
  \includegraphics[width=0.48\textwidth]{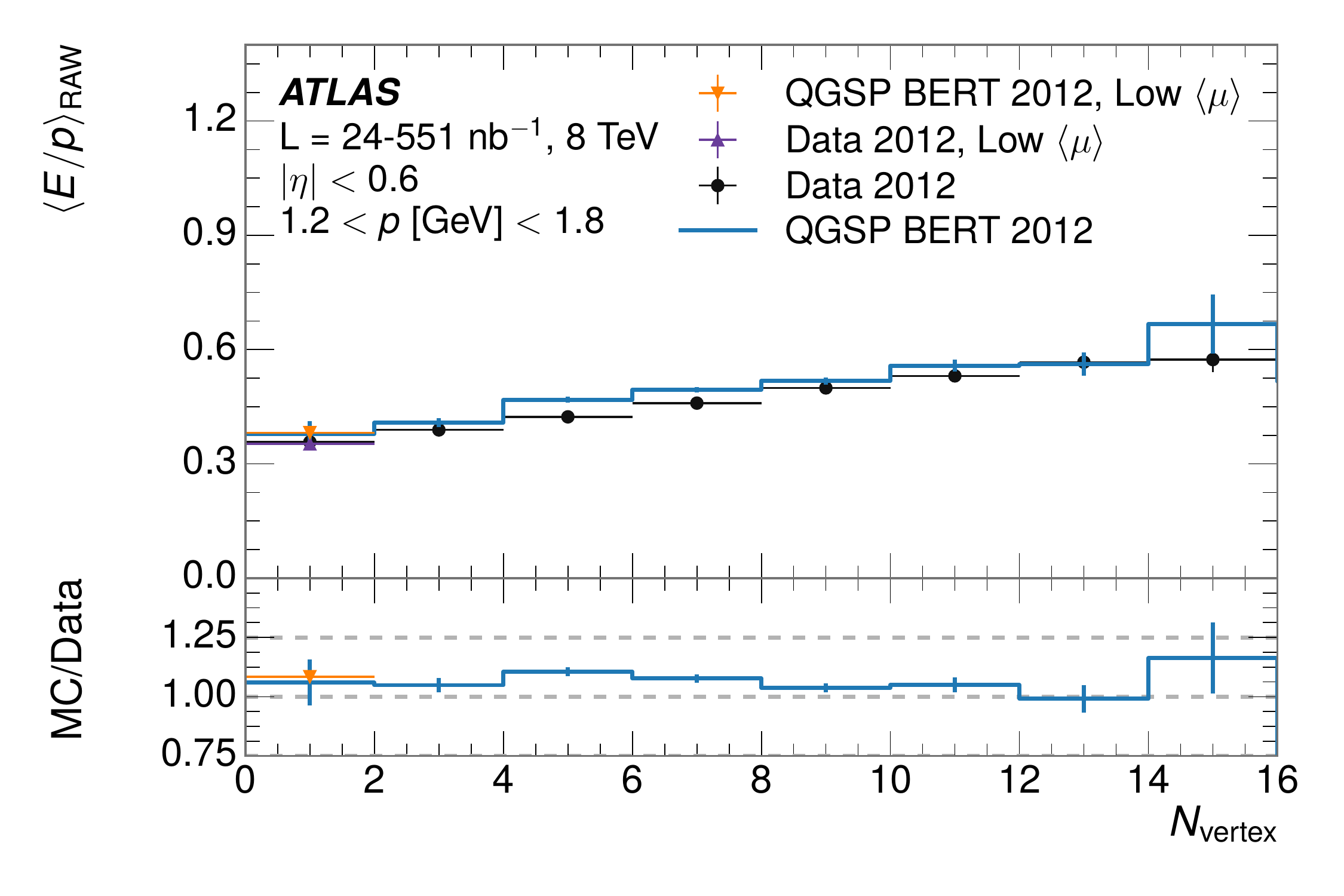}
}
~
\subfloat[]{
  \includegraphics[width=0.48\textwidth]{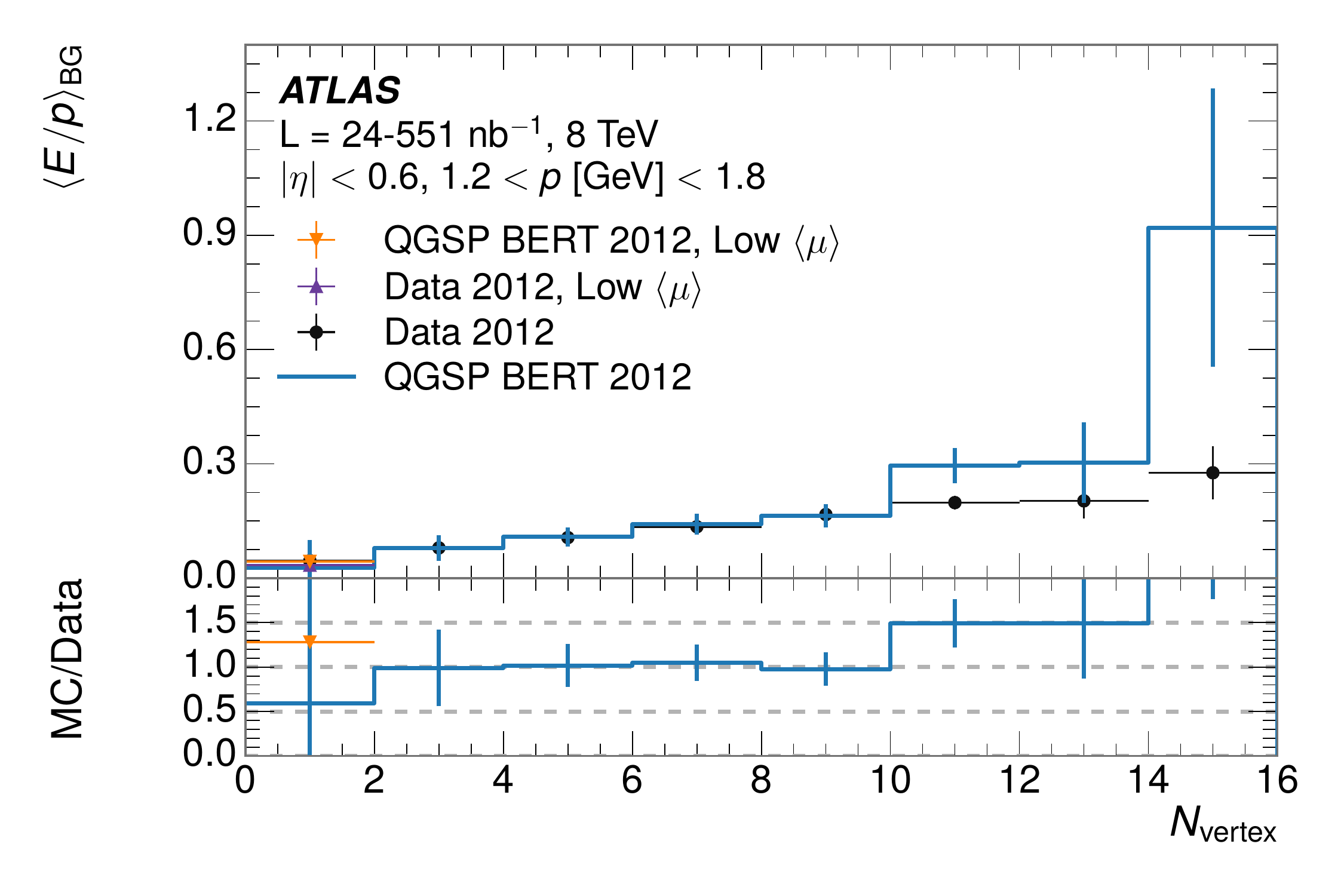}
}
\\
\subfloat[]{
  \includegraphics[width=0.48\textwidth]{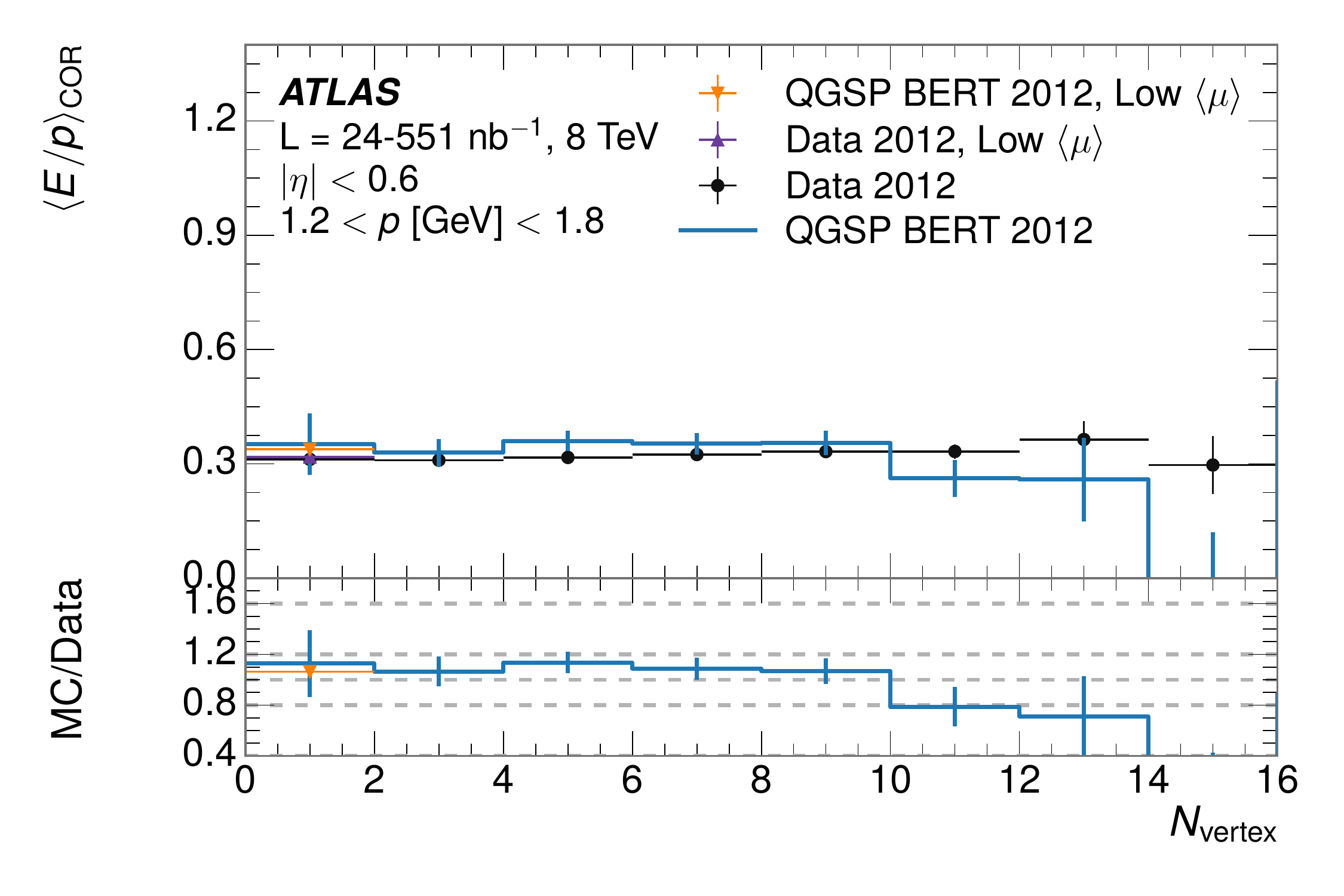}
}
~
\subfloat[]{
  \includegraphics[width=0.48\textwidth]{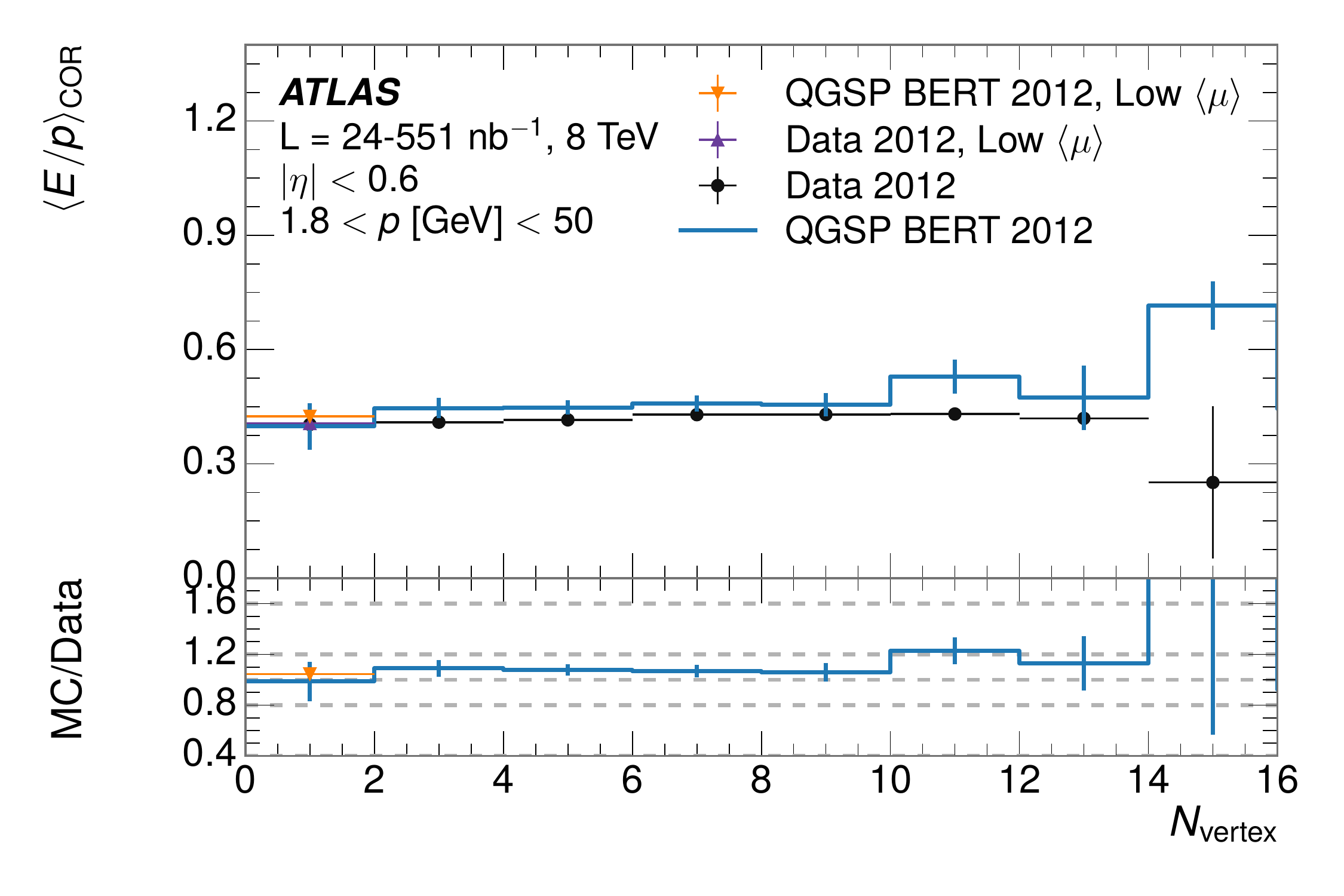}
}
\caption{ The (a) $\epav_\text{RAW}$, (b) $\epbg$, and (c) $\epcor$ with $1.2 < p / \GeV < 1.8$ and (d) $\epcor$ with $1.8 < p / \GeV < 50$ as a function of the number of reconstructed primary vertices, for tracks with $|\eta|<0.6$ and for $\langle\mu\rangle < 20.6$.  Here, low-$\langle\mu\rangle$ refers to data taken with average pile-up $\langle\mu\rangle\ll 1$. The bottom portion of each panel shows the ratio of MC simulation to data.  The error bars represent statistical uncertainties.}
\label{fig-eoverp_corrected_vtx}
\end{figure}

There are two response issues to be addressed in these events.
The first is the dependence of the response on the number of reconstructed vertices, which is an excellent proxy for the in-time pile-up.
The raw response, background, and background-corrected response to isolated charged hadrons as a function of the number of reconstructed vertices is shown in Figure~\ref{fig-eoverp_corrected_vtx}.
There is a clear dependence in both the raw and background distributions.
The difference is almost completely removed, however, in the \epcor distribution.
After the background subtraction, the values are also in good agreement with those calculated in the low-$\langle\mu\rangle$ dataset.
In each case, similar trends are present in both data and MC simulation.
In the MC simulation with pile-up, the events are weighted such that the $\langle\mu\rangle$ distribution matches that of the data, in order to ensure that out-of-time pile-up contributions are well modeled.  Both samples are required to have $\langle\mu\rangle<20.6$.

\begin{figure}[ht]
\centering
\subfloat[]{
  \includegraphics[width=0.48\textwidth]{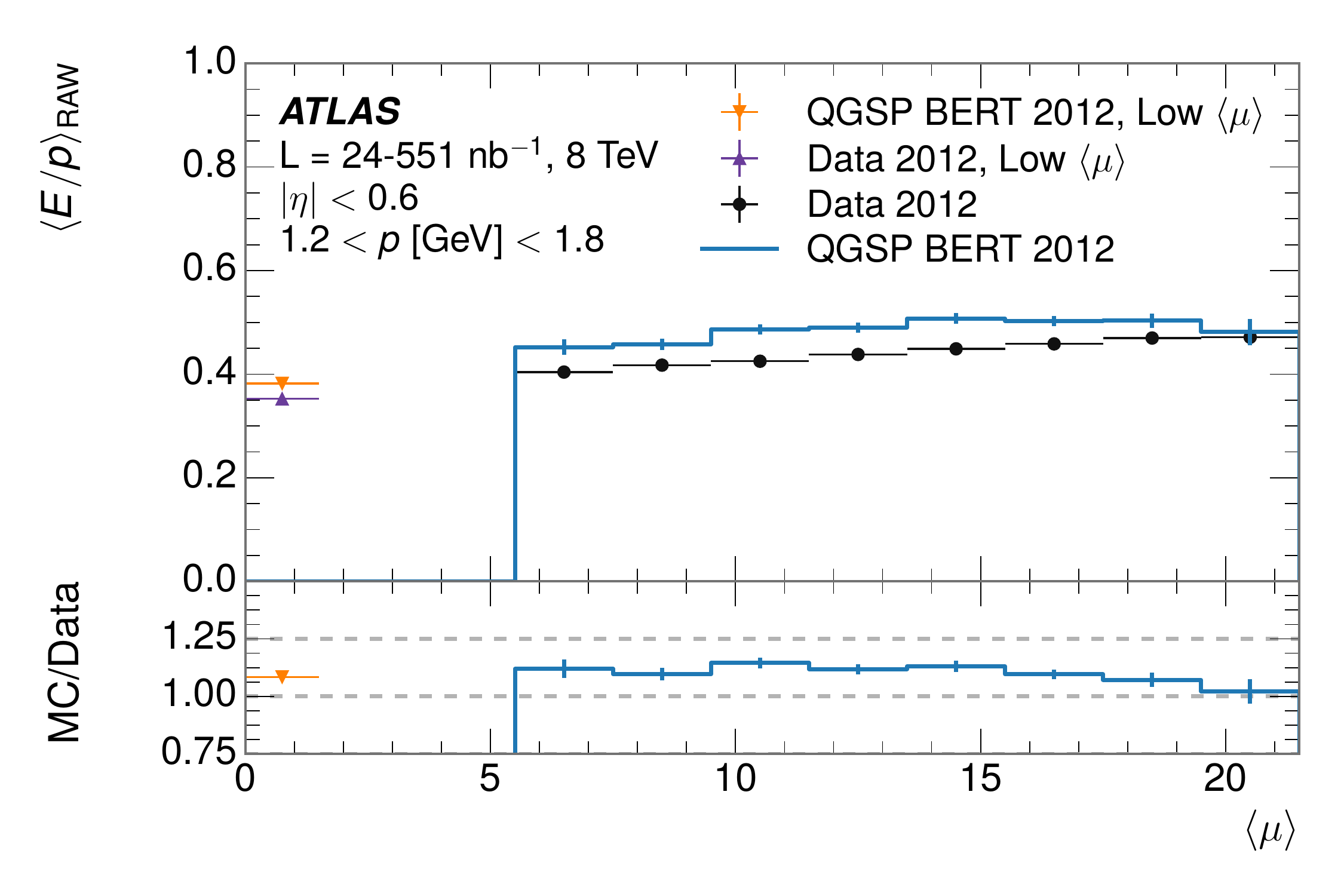}
}
~
\subfloat[]{
  \includegraphics[width=0.48\textwidth]{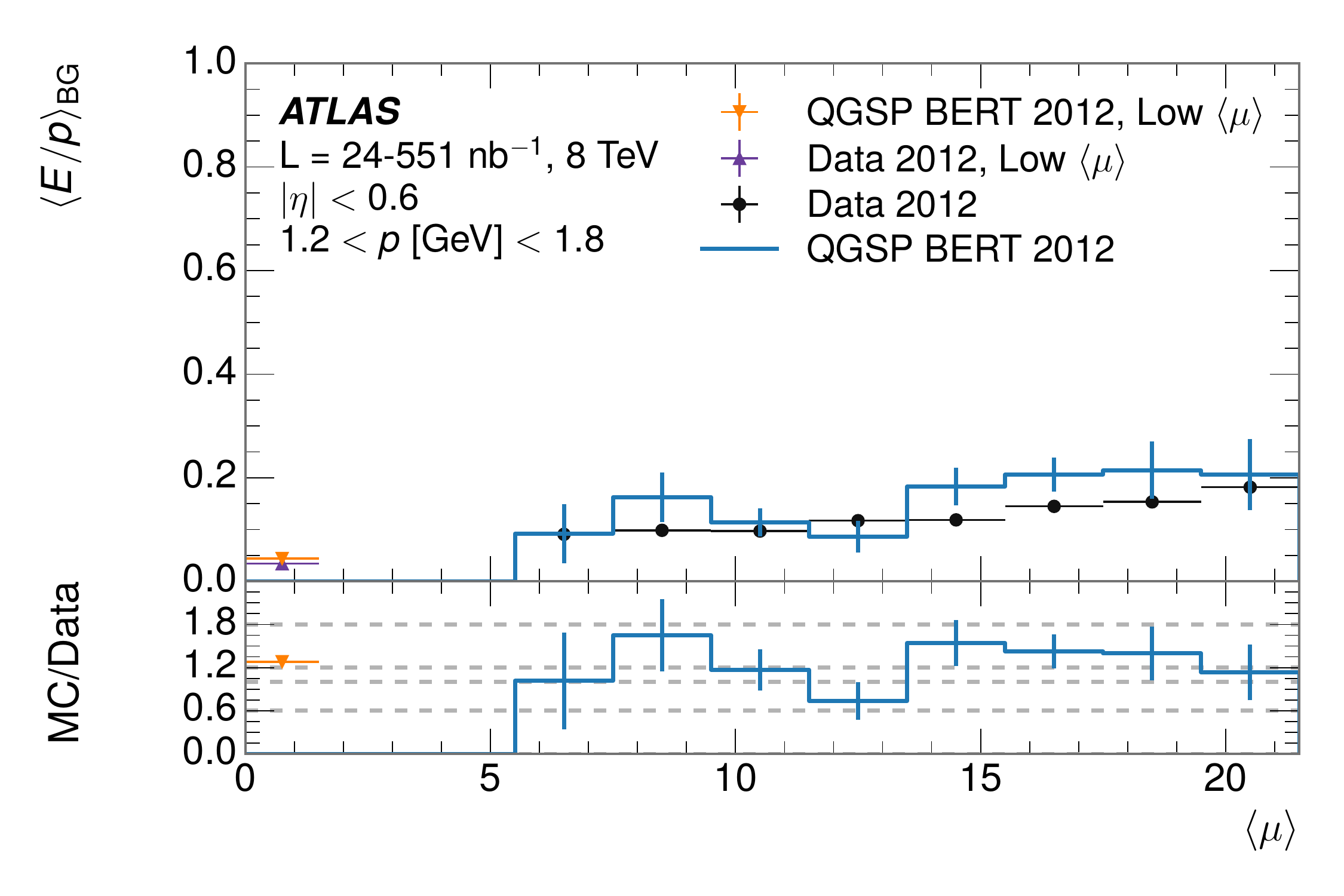}
}
\\
\subfloat[]{
  \includegraphics[width=0.48\textwidth]{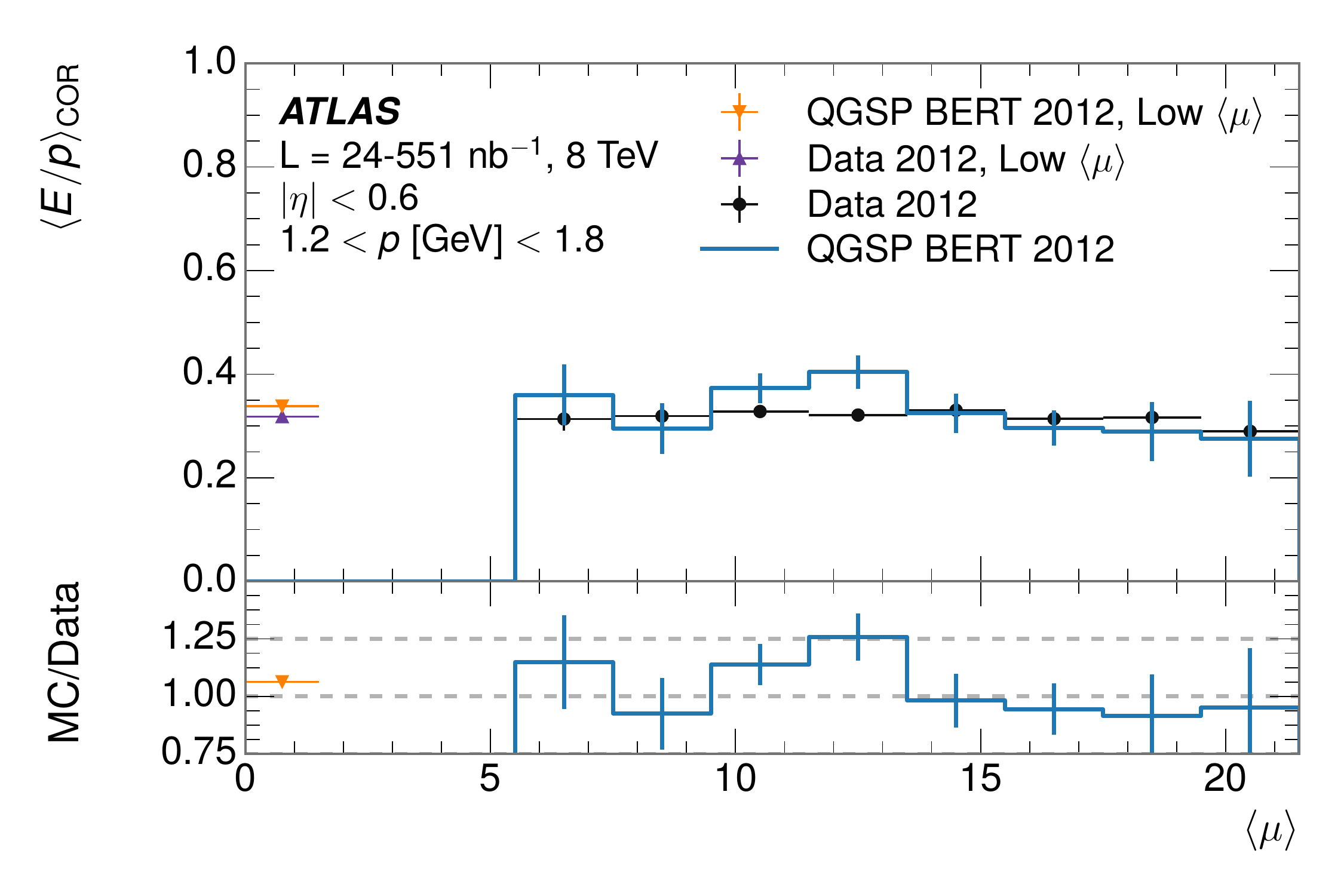}
}
~
\subfloat[]{
  \includegraphics[width=0.48\textwidth]{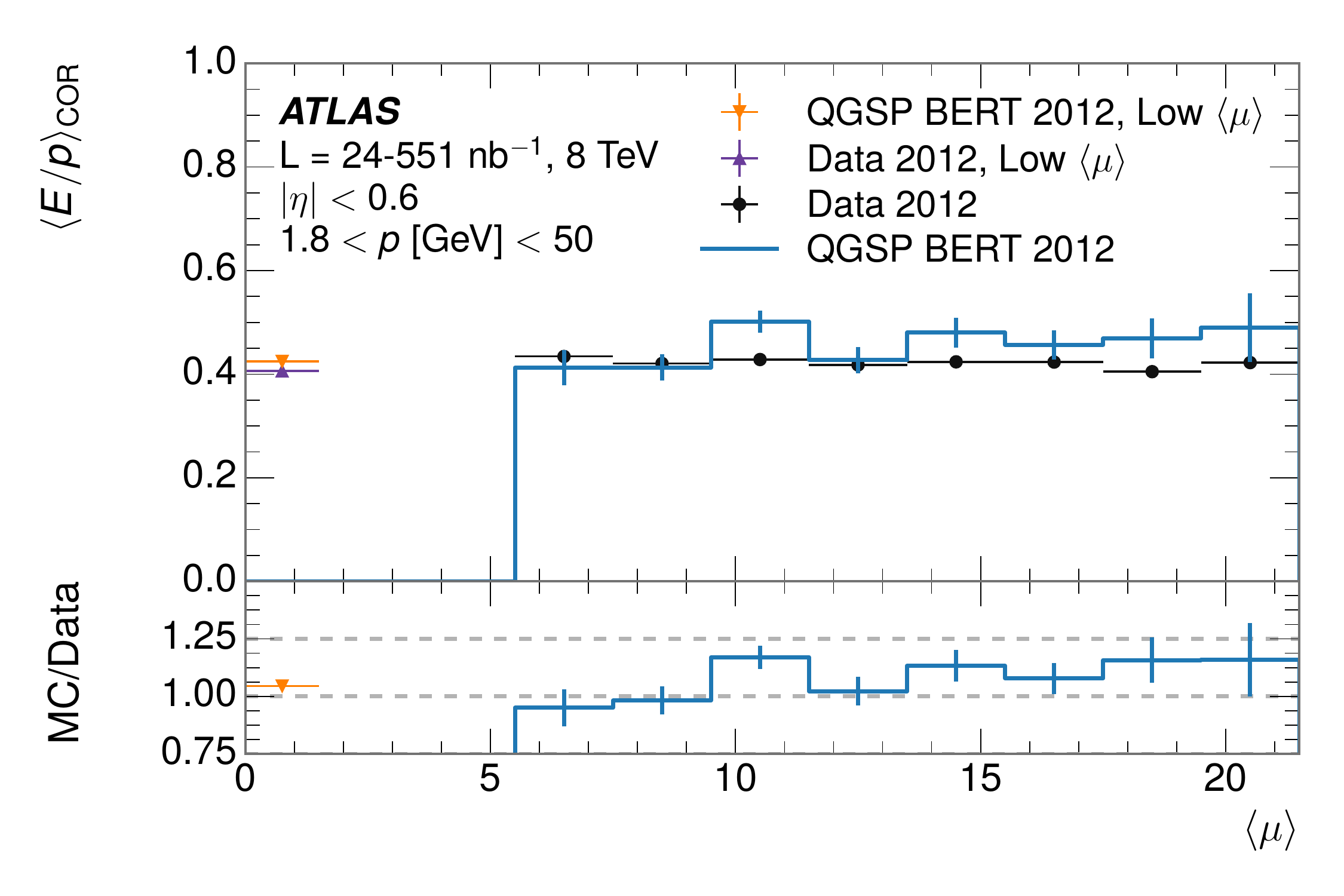}
}
\caption{ The (a) $\epav_\text{RAW}$, (b) $\epbg$, and (c) $\epcor$ with $1.2 < p / \GeV < 1.8$ and (d) $\epcor$ with $1.8 < p / \GeV < 50$ as a function of $\langle\mu\rangle$ for tracks with $|\eta|<0.6$.  Here, low-$\langle\mu\rangle$ refers to data taken with average pile-up $\langle\mu\rangle\ll 1$. The bottom portion of each panel shows the ratio of MC simulation to data.  The error bars represent statistical uncertainties.}
\label{fig-eoverp_corrected_pu}
\end{figure}

The ATLAS calorimeter is additionally sensitive to out-of-time pile-up, collisions in bunch crossings close in time to the one that was selected by the trigger, although this dependence is mitigated somewhat by the bipolar pulse shape of the calorimeter electronics.
A bunch-dependent correction is applied to the calorimeter energy measured in each calorimeter cell to correct for the residual average energy shift per bunch due to the bunch train structure and the fluctuations in the luminosity per bunch crossing.
Energy deposits up to 100~ns after, and up to 800~ns before the collision of interest may affect the energy measured in a calorimeter cell.
Thus, an equally important test is the stability of the response to isolated charged hadrons against the average number of proton--proton collisions per bunch crossing, $\langle\mu\rangle$.
This dependence is shown in Figure~\ref{fig-eoverp_corrected_pu}.
Again, there is a dependence in both the raw and background distributions, while the \epcor distribution shows that the pile-up is well compensated for by the background subtraction scheme.
As shown in the figure, the low-$\langle\mu\rangle$ values of \epcor are consistent with those at higher values of $\langle\mu\rangle$.

\subsubsection{Single-hadron response in the hadronic calorimeter}
\label{sec-tilecal}

To measure the response of the hadronic calorimeter, only tracks that deposit an amount of energy in the EM calorimeter consistent with a MIP are selected. The criteria for selecting a MIP are identical to those described in Section~\ref{sec:neutral_bg}. The measured energy corresponds to the energy of the topological clusters in the hadronic calorimeter within $\Delta R = 0.2$ of the extrapolated track.

\begin{figure}[htbp]
\centering
\subfloat[]{
  \includegraphics[width=0.48\textwidth]{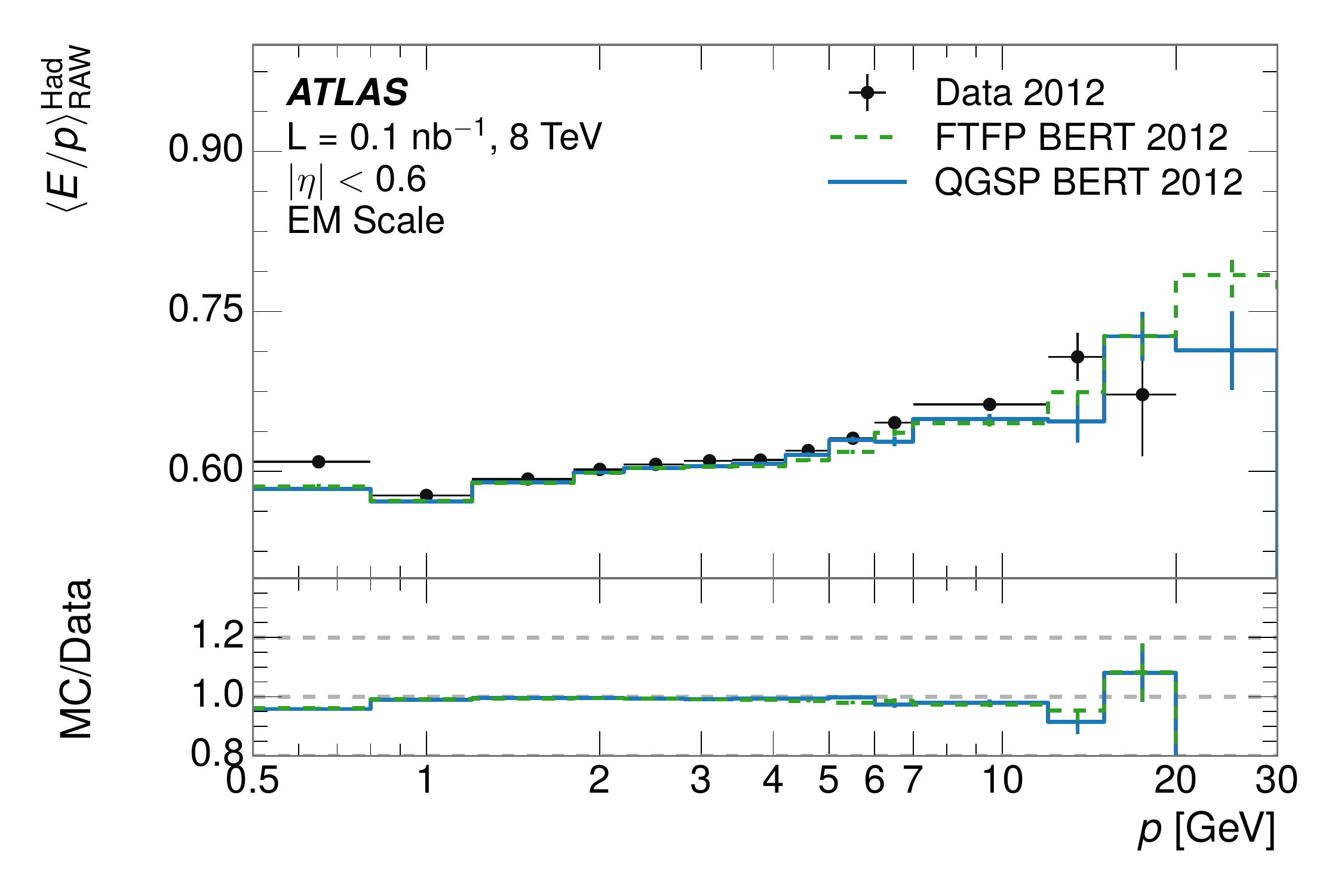}
}
~
\subfloat[]{
  \includegraphics[width=0.48\textwidth]{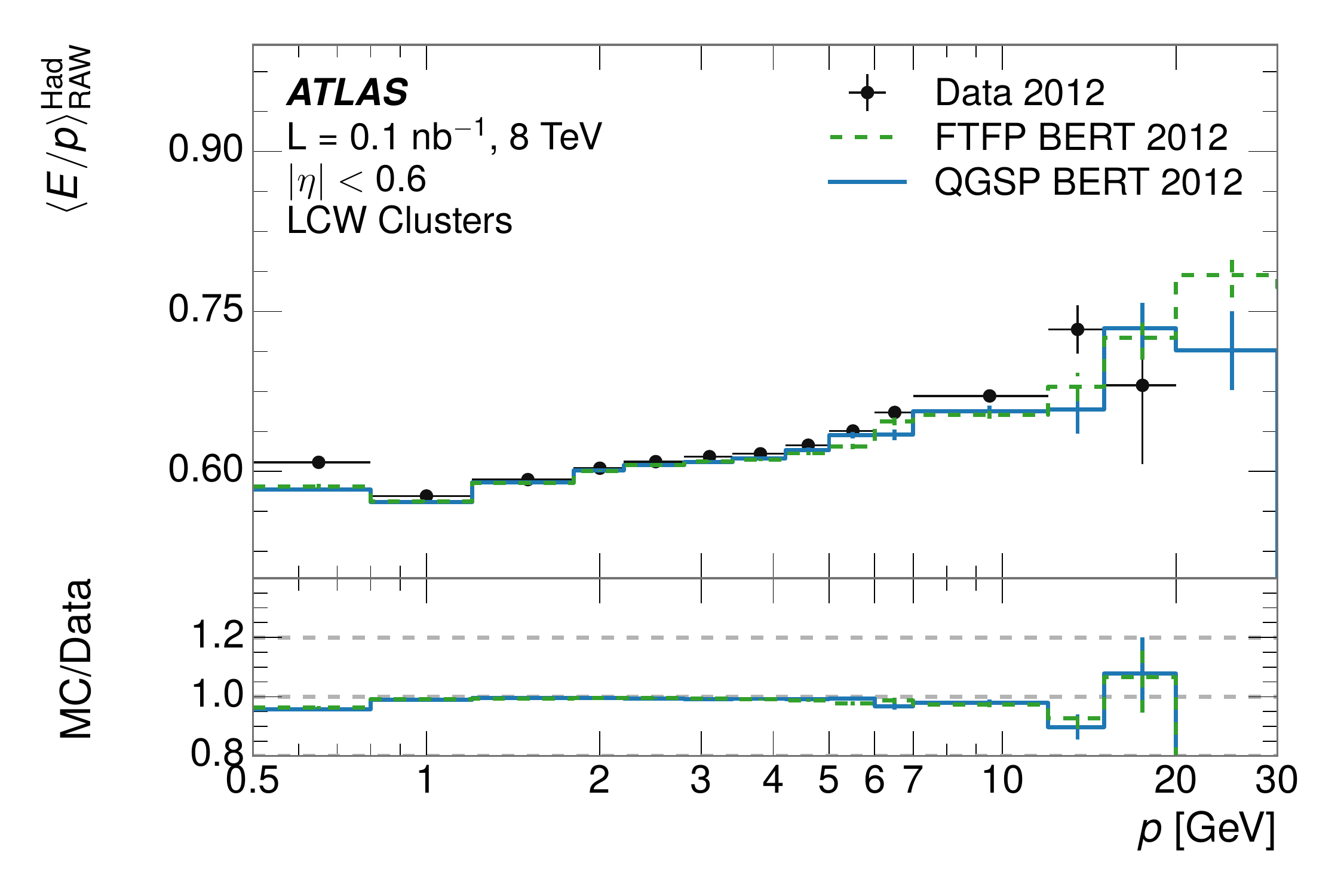}
}
\caption{Comparison of the response of the hadronic calorimeter as a function of track momentum between the data and MC simulation in $|\eta| < 0.6$ (a) at the EM-scale and (b) after the LCW calibration. The bottom portion of each panel shows the ratio of MC simulation to data, separately for the two sets of hadronic physics models.  The error bars represent statistical uncertainties.}
\label{fig-HCAL}
\end{figure}

\begin{figure}[htbp]
\centering
\subfloat[]{
  \includegraphics[width=0.48\textwidth]{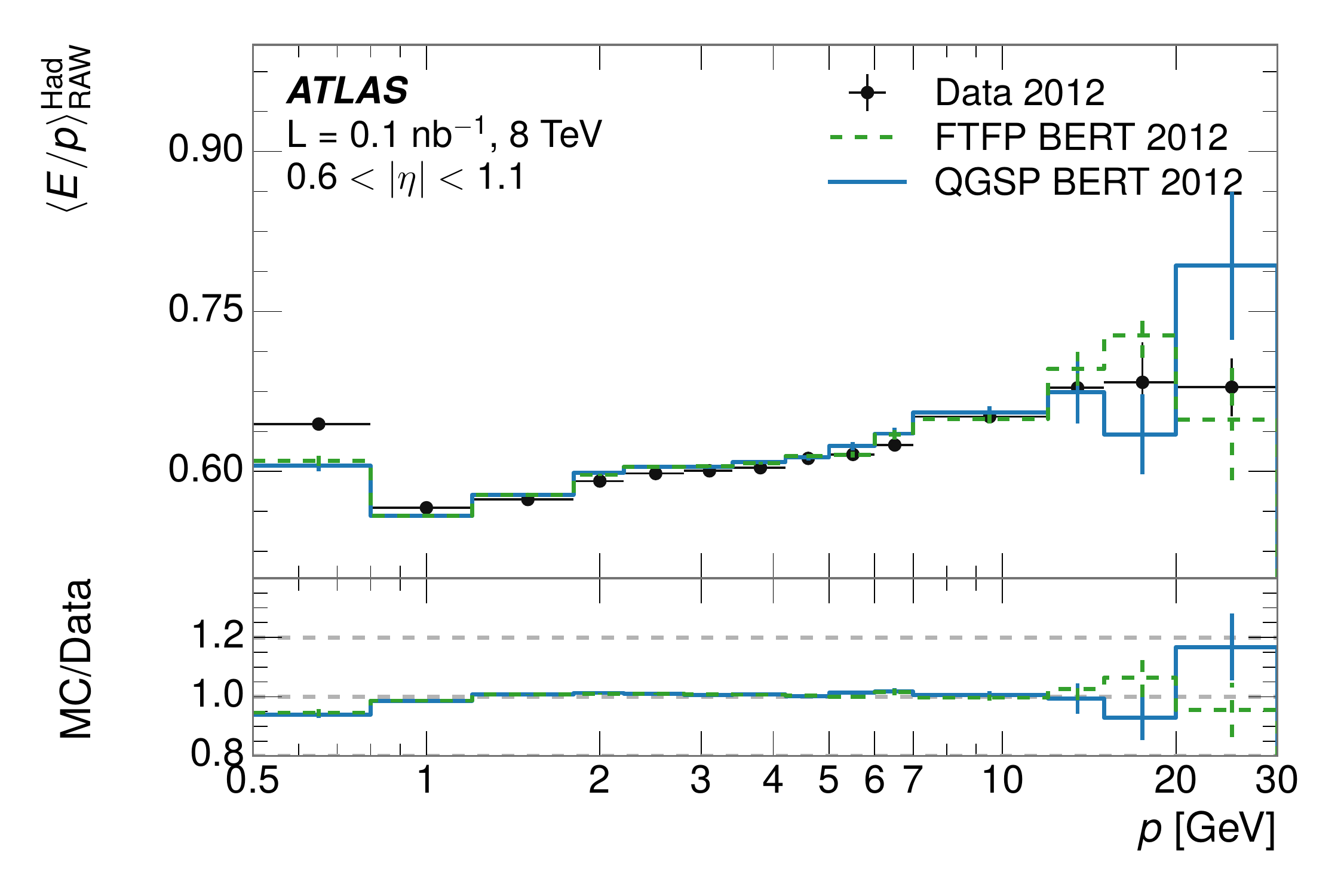}
}
~
\subfloat[]{
  \includegraphics[width=0.48\textwidth]{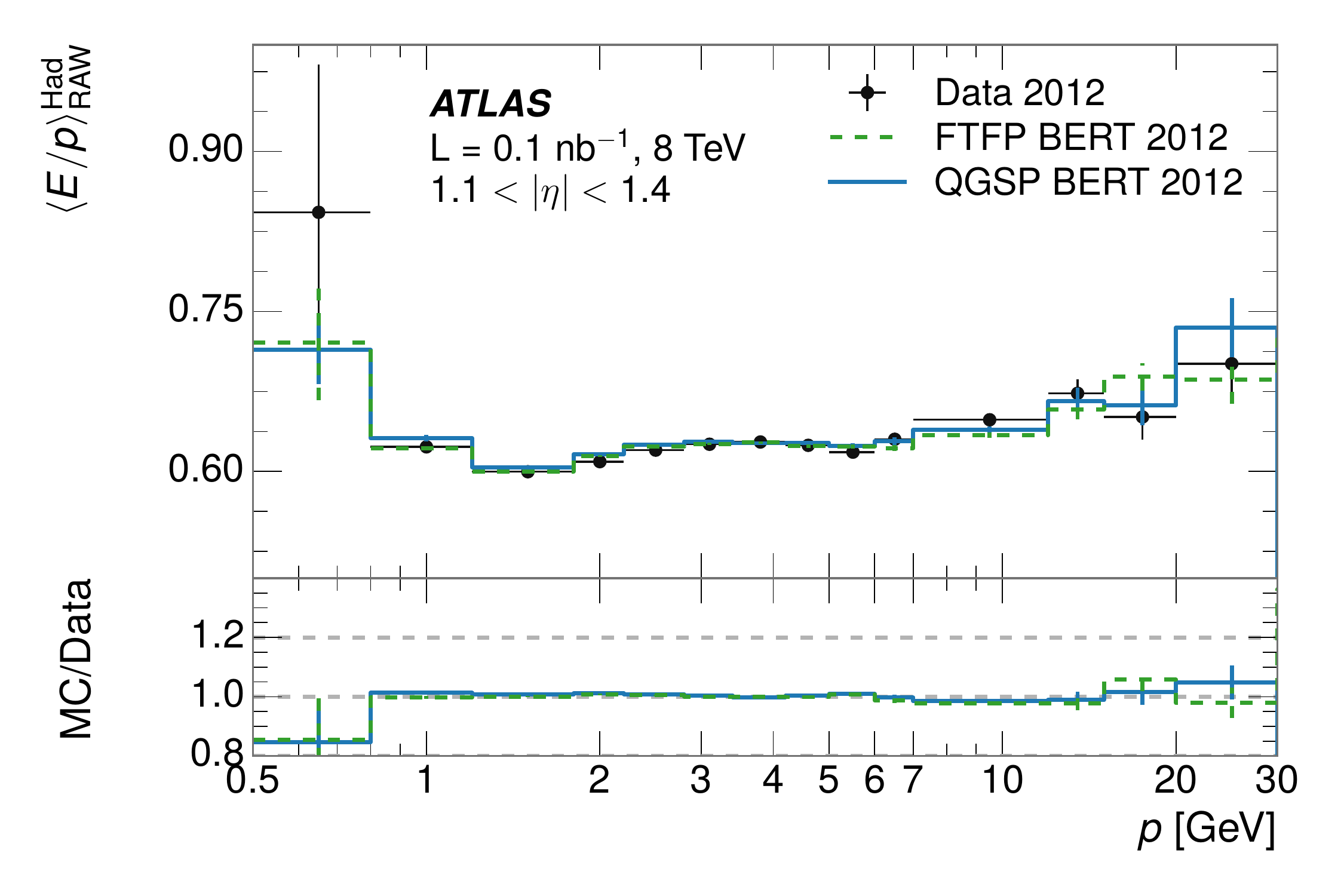}
}
\caption{Response of the hadronic calorimeter as a function of track momentum in (a) $0.6 < |\eta| < 1.1$ and (b) $1.1 < |\eta| < 1.4$ at the EM-scale. The bottom portion of each panel shows the ratio of MC simulation to data, separately for the two physics sets of hadronic physics models.  The error bars represent statistical uncertainties.}
\label{fig-HCAL-eta}
\end{figure}

Figure~\ref{fig-HCAL} shows a comparison of the data to the MC simulation for tracks passing this MIP selection of \epav$_\text{RAW}^\text{Had}$, built using topological clusters in the hadronic calorimeter, calibrated at the EM scale and after the LCW calibration, in the central region, $|\eta| < 0.6$. The raw and corrected values are identical because the background in the tile calorimeter is negligible. Agreement of the data and the simulation is better than in the inclusive \epcor shown in the previous section.  Any residual neutral-particle background effects that might be present in the response to inclusive single particles are negligible in this comparison, but particles are selected that had a particularly late shower, weakening the dependence on the distribution of secondary particles from hadronic interactions.  This measure of the response is repeated for different detector regions. Figure \ref{fig-HCAL-eta} shows the response of the hadronic calorimeter for $0.6 < |\eta| < 1.1$ and $1.1 < |\eta| < 1.4$.

\subsubsection{Single-hadron response in the EM calorimeter}
\label{sec-emcal}

\begin{figure}[htbp]
\centering
\subfloat[]{
  \includegraphics[width=0.48\textwidth]{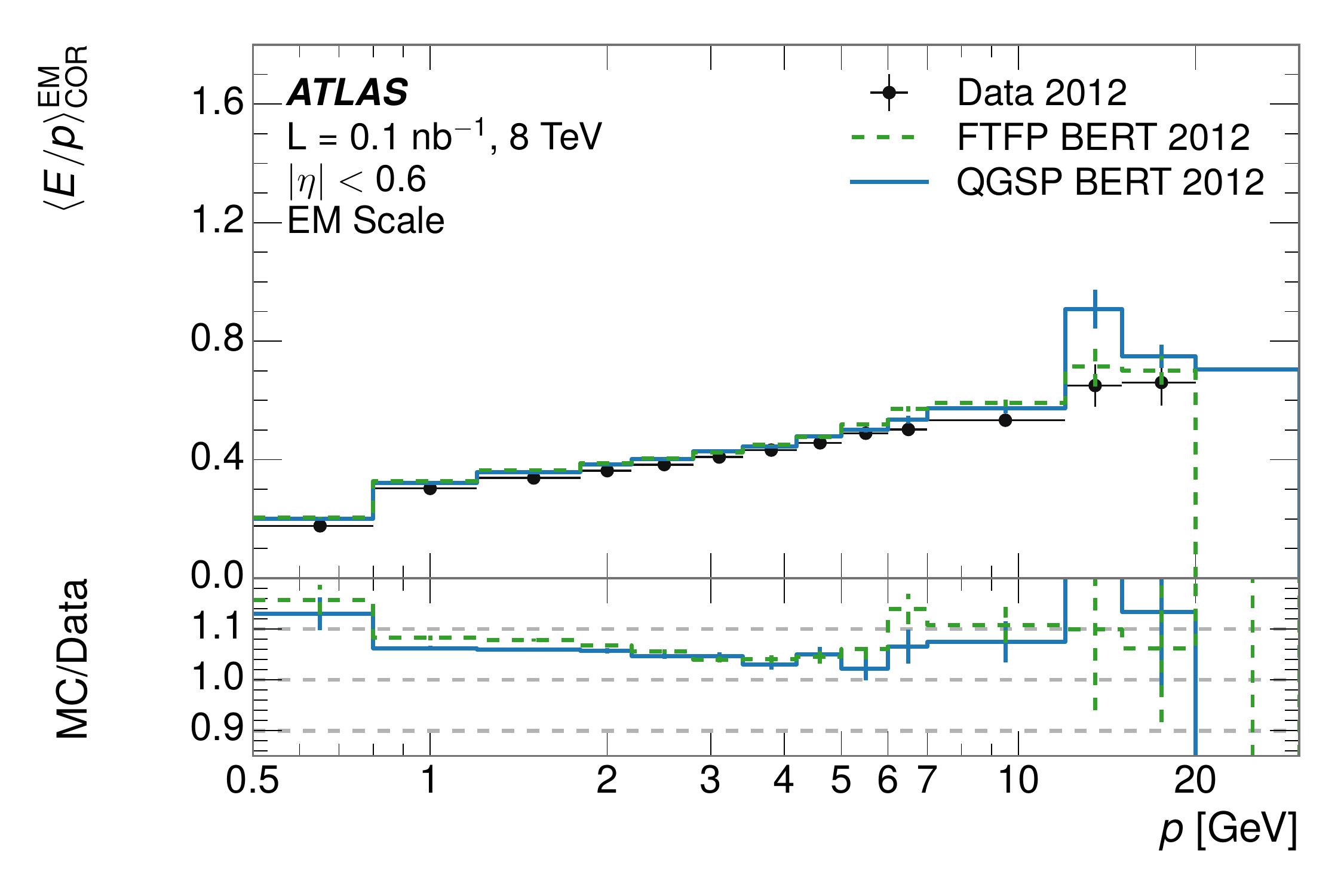}
}
~
\subfloat[]{
  \includegraphics[width=0.48\textwidth]{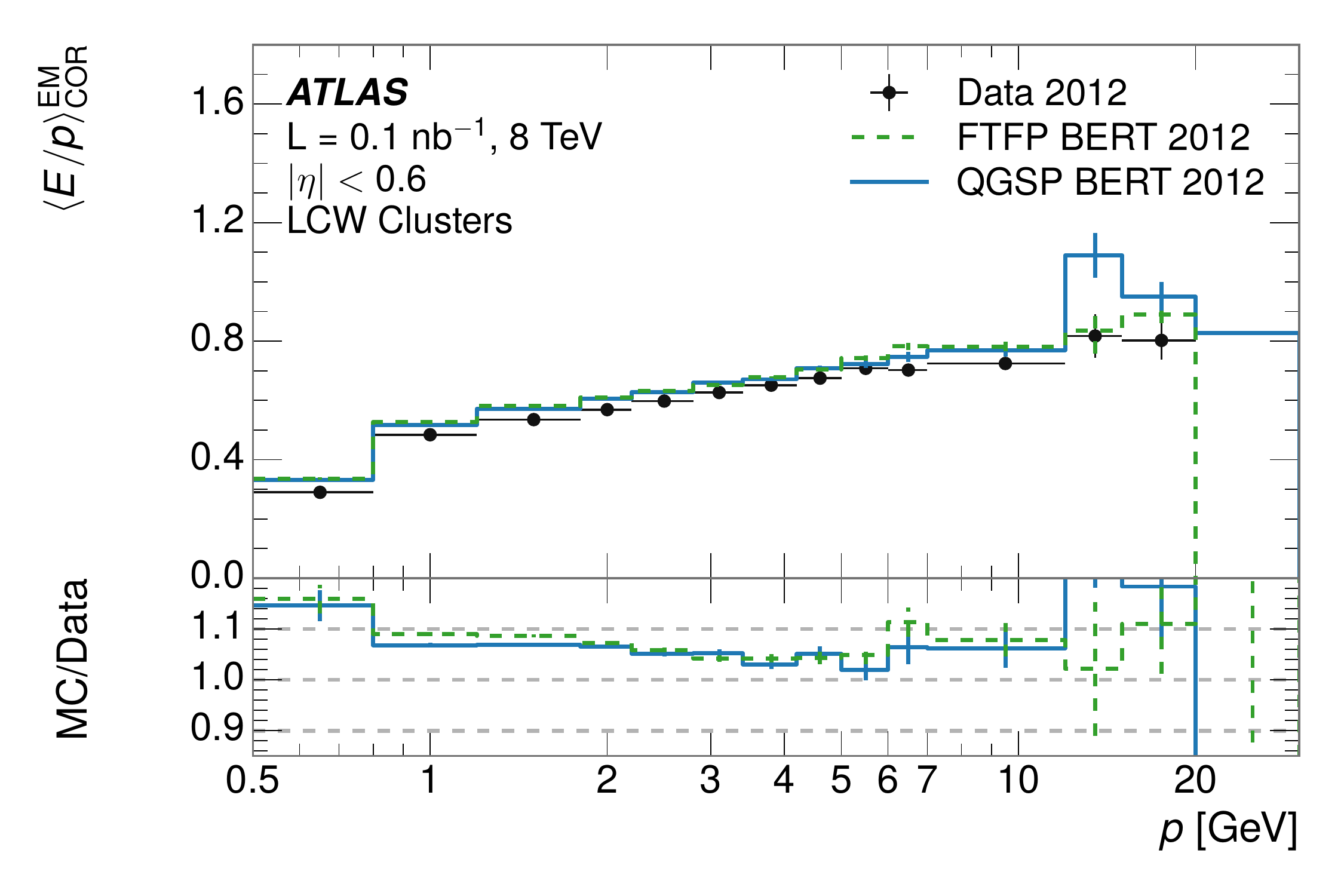}
}
\caption{Comparison of the response of the EM calorimeter as a function of track momentum between the data and MC simulation in $|\eta| < 0.6$, (a) at the EM-scale and (b) with the LCW calibration. The bottom portion of each panel shows the ratio of MC simulation to data, separately for the two physics sets of hadronic physics models. The error bars represent statistical uncertainties.}
\label{fig-EMCal}
\end{figure}

In order to examine the response of the EM calorimeter alone, particles are selected that deposit most of their energy in the EM calorimeter.  In this case, tracks are required to have no associated energy in the hadronic calorimeter.  Such a selection is inherently more sensitive to neutral-particle backgrounds, which deposit most of their energy in the EM calorimeter.  A comparison of \epav$^\text{EM}_\text{COR}$, the \epcor built only from energy deposits in the EM calorimeter, between data and MC simulation is shown in Figure~\ref{fig-EMCal} for EM scale response and after the LCW calibration is applied.  These distributions show disagreement at the 5\% level over much of the momentum range, for both topological cluster calibrations.  This is consistent with the description of the response of this calorimeter component being the main cause of the discrepancy observed in the inclusive distributions.

\subsubsection{Modelling of response with modified calorimeter noise thresholds}
\label{sec-thresholds}

\begin{figure}[htbp]
\centering
\subfloat[]{
  \includegraphics[width=0.48\textwidth]{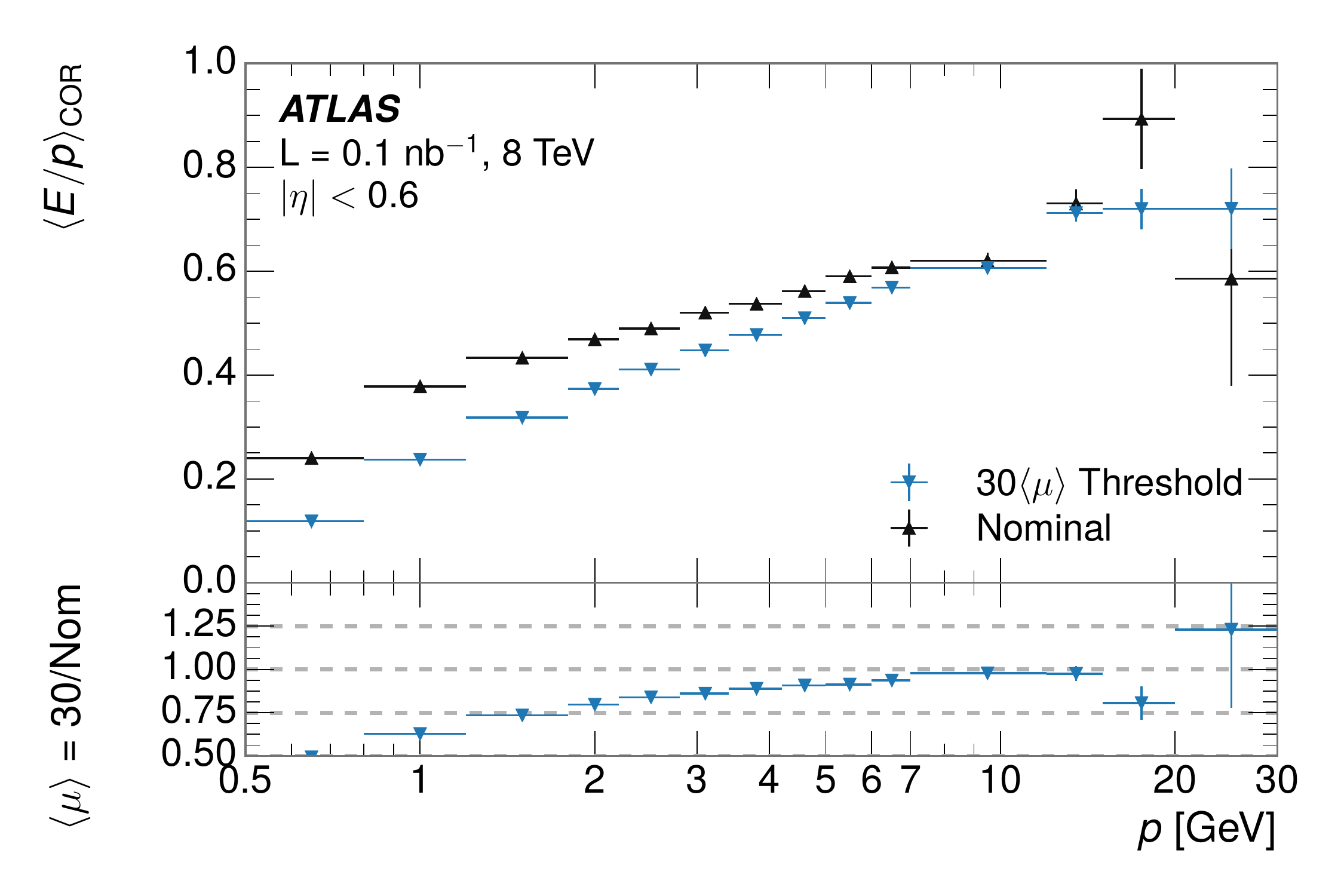}
}
~
\subfloat[]{
  \includegraphics[width=0.48\textwidth]{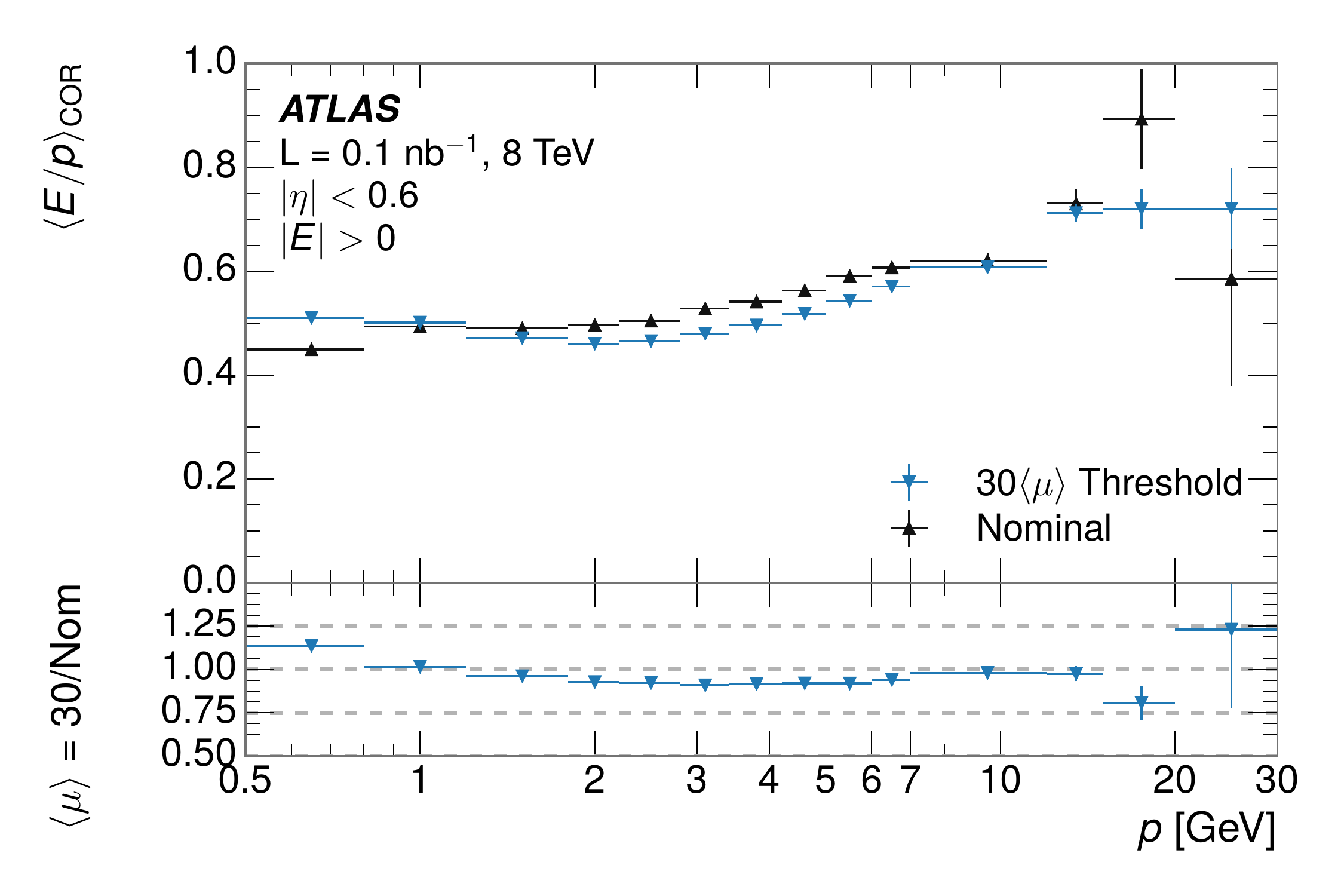}
}
\caption{Comparison of the response of the calorimeter between the nominal topological cluster threshold and the threshold corresponding to $\langle\mu\rangle=30$ with $|\eta|<0.6$. (a), with no requirement on $E/p$, and (b), with $E/p > 0$. The bottom portion of each panel shows the ratio of the response with the different thresholds. The error bars represent statistical uncertainties.}
\label{fig-threshold}
\end{figure}

\begin{figure}[htbp]
\centering
\subfloat[]{
  \includegraphics[width=0.48\textwidth]{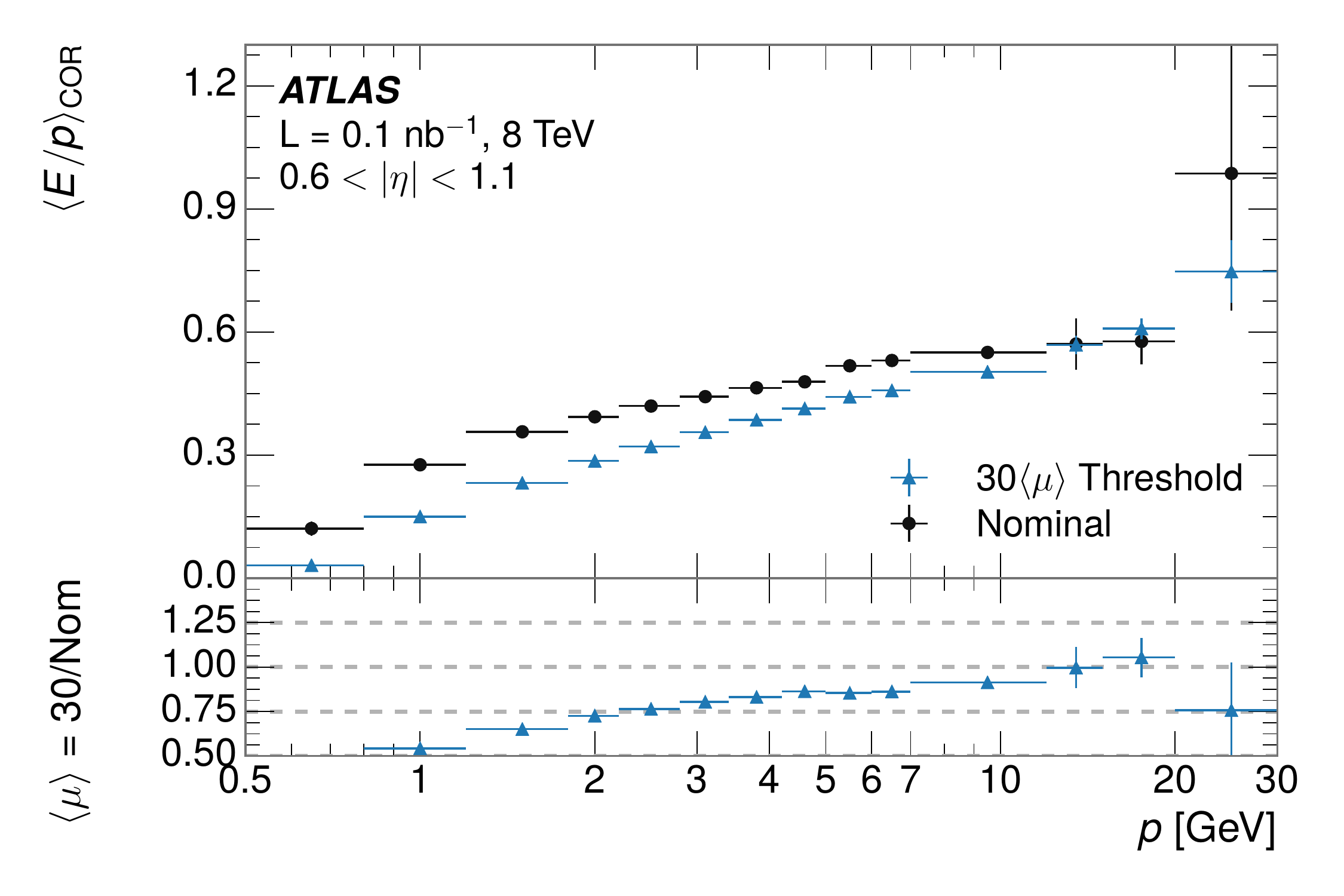}
}
~
\subfloat[]{
  \includegraphics[width=0.48\textwidth]{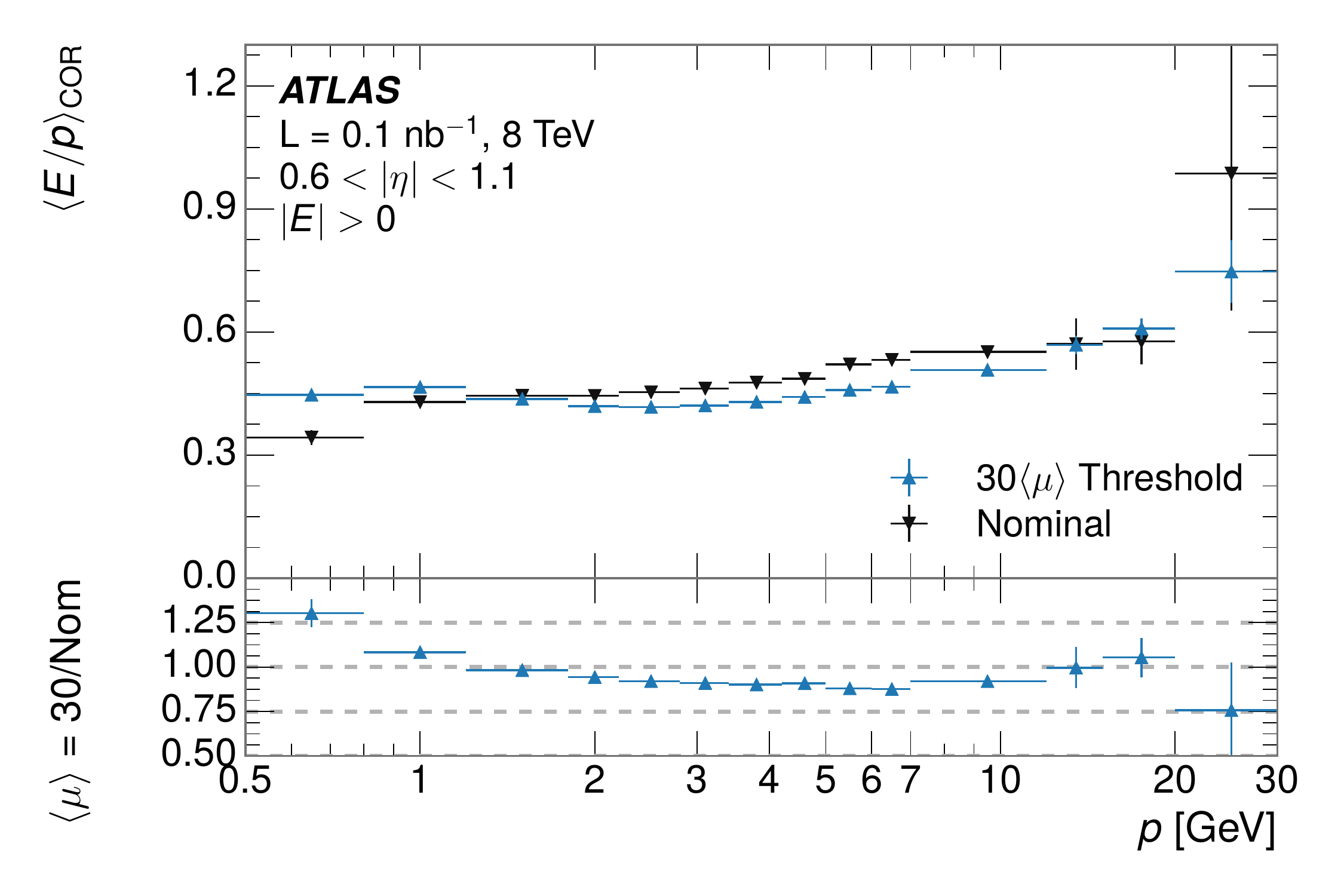}
}
\caption{Comparison of the response of the calorimeter between the nominal topological cluster threshold and the threshold corresponding to $\langle\mu\rangle=30$ with $0.6< |\eta|<1.1$. (a), with no requirement on $E/p$, and (b), with $E/p > 0$. The bottom portion of each panel shows the ratio of the response with the different thresholds. The error bars represent statistical uncertainties.}
\label{fig-threshold-eta}
\end{figure}

During the low-$\langle\mu\rangle$ data-taking period, the noise threshold used for clustering of energy included only electronics noise.  During most of the data-taking period in 2012, however, a different calorimeter noise threshold setting was applied when defining topological clusters.  This higher threshold serves to suppress pile-up, while lowering the clustering efficiency for true energy deposits.  A comparison of the raised threshold used for most of 2012 (corresponding to $\langle\mu\rangle=30$) to the threshold used during the low-$\langle\mu\rangle$ data-taking period (corresponding to $\langle\mu\rangle=0$) in the same dataset is shown in Figure~\ref{fig-threshold}.  When including a higher threshold, as expected, a higher fraction of tracks are not associated with a topological cluster because a more significant energy deposit is required to seed a cluster.  This manifests as a significant drop in the average response at low momentum.  At high momentum ($p>6$~\GeV), however, and when considering only tracks that match to at least one topological cluster, agreement between the two threshold settings is typically within 10\%.  As most pile-up consists of low-momentum particles, this is an indication that the higher threshold setting is successful at rejecting pile-up, while keeping and not altering the high-energy signals typically associated with energetic jets.  When excluding tracks that do not match any cluster, at low momentum the higher minimum cluster energy increases the average response, because the majority of tracks match only one cluster.  At moderate momenta ($2<p/\GeV<7$), most tracks match more than one cluster, and a low-energy cluster is cut away by the higher threshold, resulting in a reduction in \epcor. Figure~\ref{fig-threshold-eta} shows the same comparison for a higher $|\eta|$ range. 

\begin{figure}[htbp]
\centering
\subfloat[]{
  \includegraphics[width=0.48\textwidth]{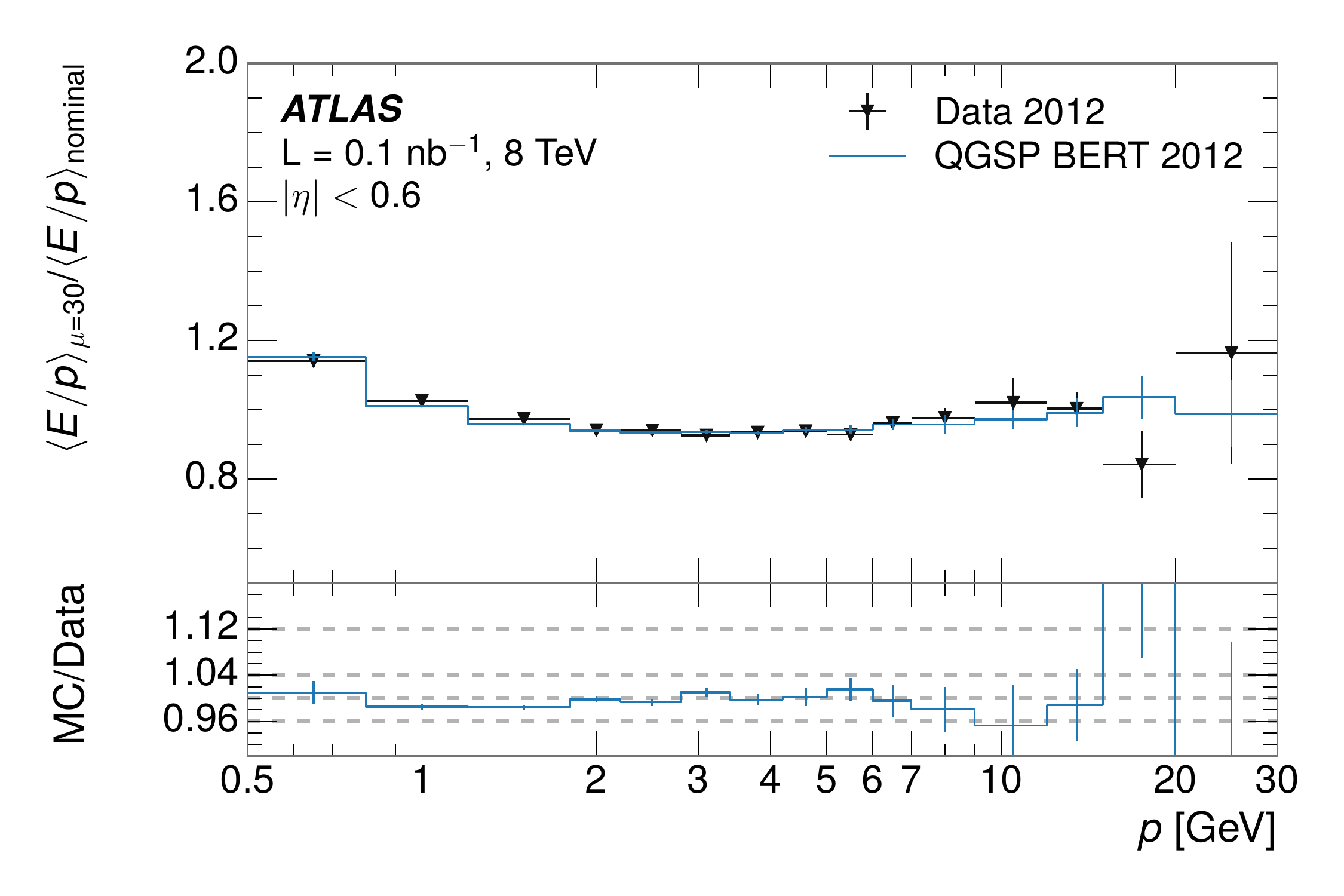}
}
~
\subfloat[]{
  \includegraphics[width=0.48\textwidth]{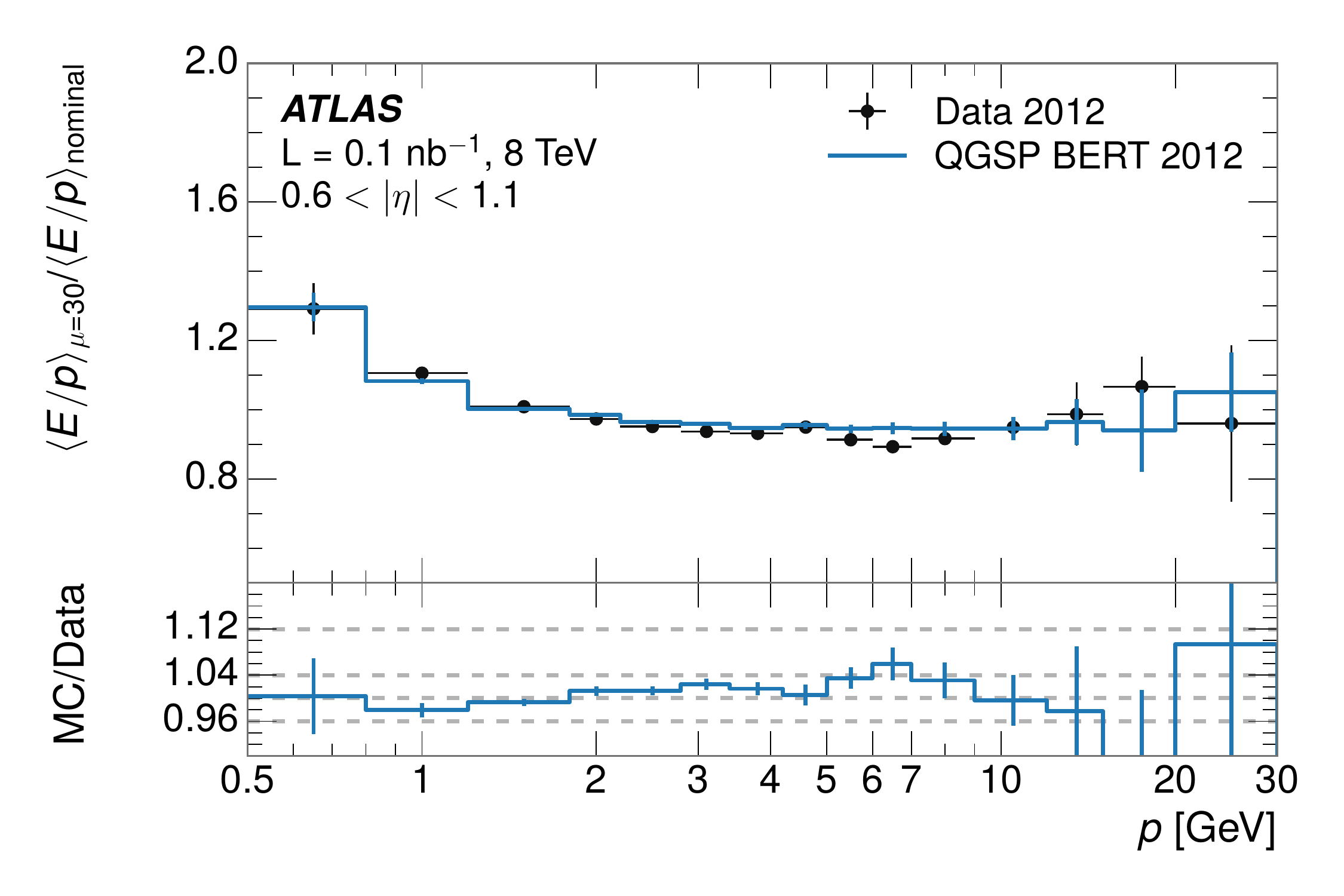}
}
\caption{Ratio of the response of the calorimeter between the threshold corresponding to $\langle\mu\rangle=30$ and the nominal topological cluster threshold  with (a) $|\eta|<0.6$ and (b) $0.6< |\eta|<1.1$, excluding tracks with $E\leq 0$. The bottom portion of each panel shows the ratio of MC simulation to data.  The error bars represent statistical uncertainties.}
\label{fig-threshold-dvmc}
\end{figure}

Figure~\ref{fig-threshold-dvmc} shows the ratio of \epcor with higher threshold to \epcor with the nominal threshold for data and MC simulation, where tracks with $E\leq 0$ have been excluded. The data and MC simulation agree over the entire range of momentum.

\begin{figure}[htbp]
\centering
\subfloat[]{
  \includegraphics[width=0.48\textwidth]{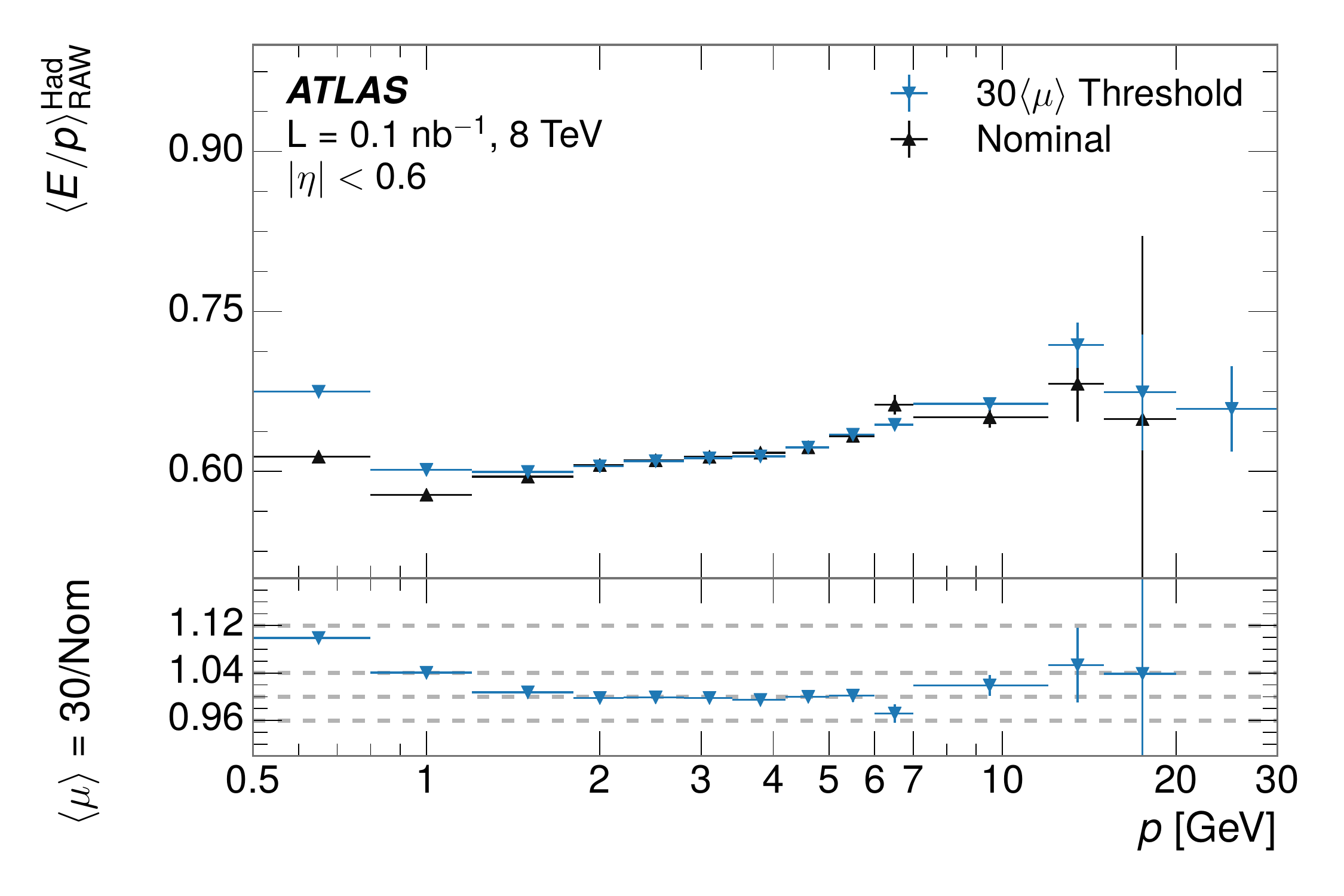}
}
~
\subfloat[]{
  \includegraphics[width=0.48\textwidth]{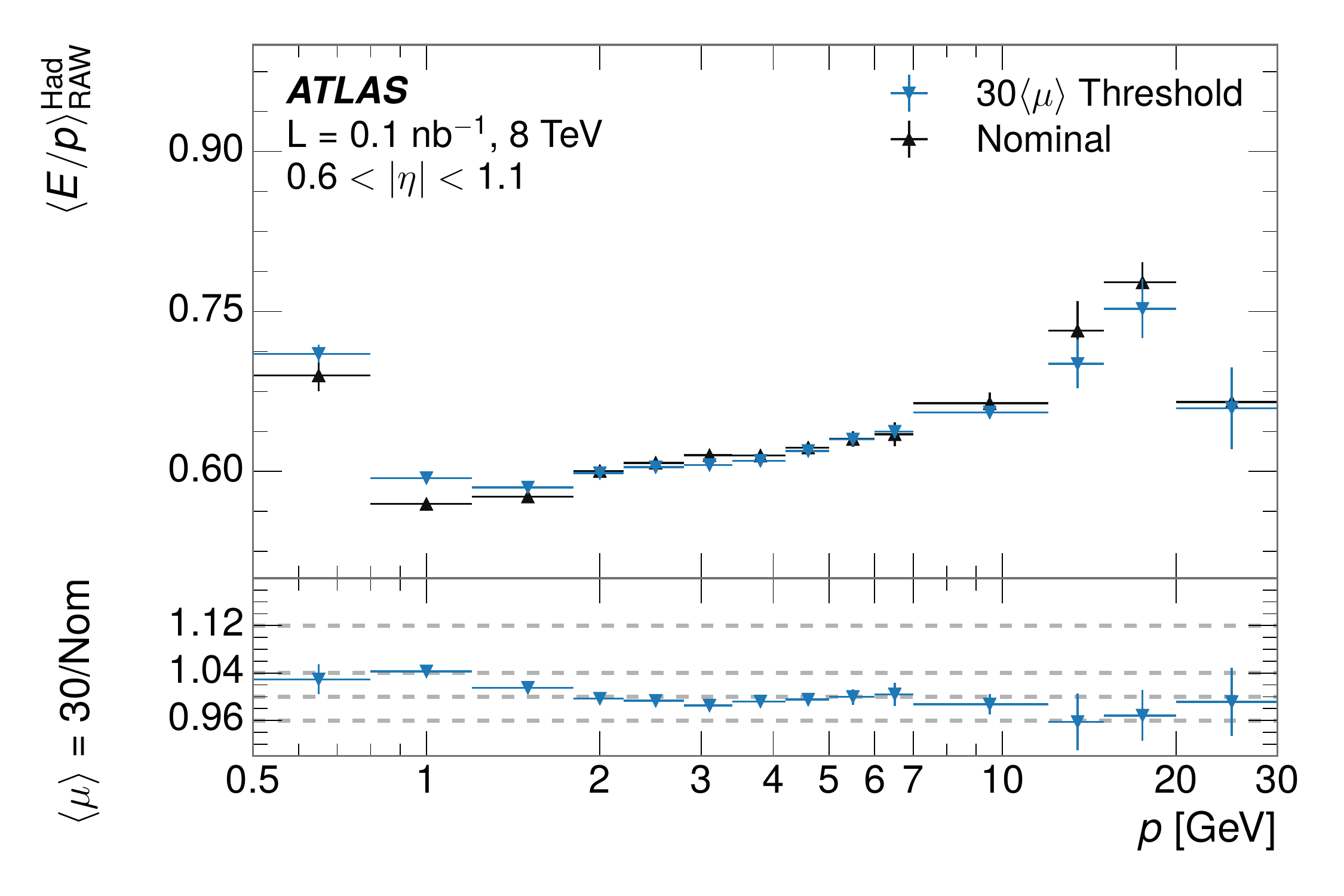}
}
\caption{Comparison of the response of the hadronic calorimeter with the nominal topological cluster threshold to that with the threshold corresponding to $\langle\mu\rangle=30$ for (a) $|\eta| < 0.6$ and (b) $0.6 < |\eta| < 1.1$. The bottom portion of each panel shows the ratio of the response with the different thresholds.  The error bars represent statistical uncertainties.}
\label{fig-HCAL-thresholds}
\end{figure}

The change in threshold settings affects the EM calorimeter in particular, because particles from pile-up tend to be low-energy and deposit most of their energy in the EM calorimeter, leading to more similar threshold settings in the hadronic calorimeter when calculated with and without pile-up.  For tracks leaving significant energy in the tile calorimeter and only minimal energy in the EM calorimeter, therefore, the two topological cluster settings are expected to provide comparable results.  This comparison is shown in Figure~\ref{fig-HCAL-thresholds} for two different ranges of $|\eta|$. As expected, agreement is better than 5\% above 800~\MeV, and the distributions are statistically consistent over most of the range. 

\clearpage
\section{Identified particle response}
\label{sec-resonances}
In addition to the calorimeter response to an inclusive set of isolated charged hadrons, uncertainties in jet energy scale calibration rely on an understanding of the modelling of the detector response to specific particle species. A jet includes a variety of hadrons that need to be modelled. A measurement of the response to individual hadron species can be used to validate that the MC simulation correctly models each component of the jet shower. This study uses decays of identified particles that have long enough lifetimes to be identified via a secondary vertex.  A sample of individual particle species is extracted to provide separate measurements of the calorimeter response to each. Only the 2012 data are used for these comparisons, as the 2010 and 2012 data show consistent features.

Decays of \pL, \pLB, and \pKS\ hadrons are used to identify individual protons, anti-protons, and pions respectively. These hadrons are required to be isolated from all other tracks in the event. The calorimeter response to these particles is expected to vary, particularly for the anti-proton because of its eventual annihilation. These differences can be measured at low energy, where they have a significant effect.

\subsection{Event selection}

In addition to the event selection required for the inclusive tracks listed in Section~\ref{sec-selection}, events are required to have at least one secondary vertex. 
The same selection used for inclusive tracks is applied to the identified hadrons, except for the impact parameter requirement.

To match the energy available to be deposited in the calorimeter, the ratio \ep is measured as a function of the available energy, \Ea, calculated using information in the tracker. For pions, the available energy is the total energy: $\Ea = \sqrt{p^2 + m^2}$. For protons, the available energy is the kinetic energy: $\Ea = \sqrt{p^2 + m^2} - m$. For anti-protons, the available energy is the sum of the total energy and the rest mass, to account for annihilation: $\Ea = \sqrt{p^2 + m^2} + m$.  For anti-protons, $\Ea\gtrsim 2$~\GeV.

\subsection{Reconstruction of particle candidates}
\label{sec:candidate_reco}

The selection of particle decays in the ID is based on previous ATLAS results~\cite{STDM-2010-09}. The decay $\pKS \rightarrow \pip\pim$, the dominant \pKS\ decay to charged particles, is used to select pions. Similarly, the decays $\pL \rightarrow \pim p$ and $\pLB \rightarrow \pip\bar{p}$, also the dominant \pL\ and \pLB\ decays to charged particles, are used to identify protons and anti-protons respectively, by selecting the higher-momentum track associated with the decay vertex. In \pL\ and \pLB\ decays, because of the boost of the \pL\ or \pLB, it is kinematically more likely for the proton or anti-proton to have greater momentum than the pion. Considering the two tracks associated with the decay vertex, a positively charged higher-momentum track indicates that the candidate is a \pL, while a negatively charged higher-momentum track indicates that the candidate is a \pLB. The decay-product tracks are both required to have $\pt > 500$~\MeV.  The tracks used to measure the \epav distributions are divided into two bins of pseudorapidity, $|\eta| < 0.6$ and $0.6 < |\eta| < 1.1$. Agreement between data and MC simulation for tracks with larger pseudorapidity is consistent with those at lower $|\eta|$, but has significantly larger statistical uncertainty. 

\begin{figure}[htbp]
\centering
\subfloat[]{
  \includegraphics[width=0.48\textwidth]{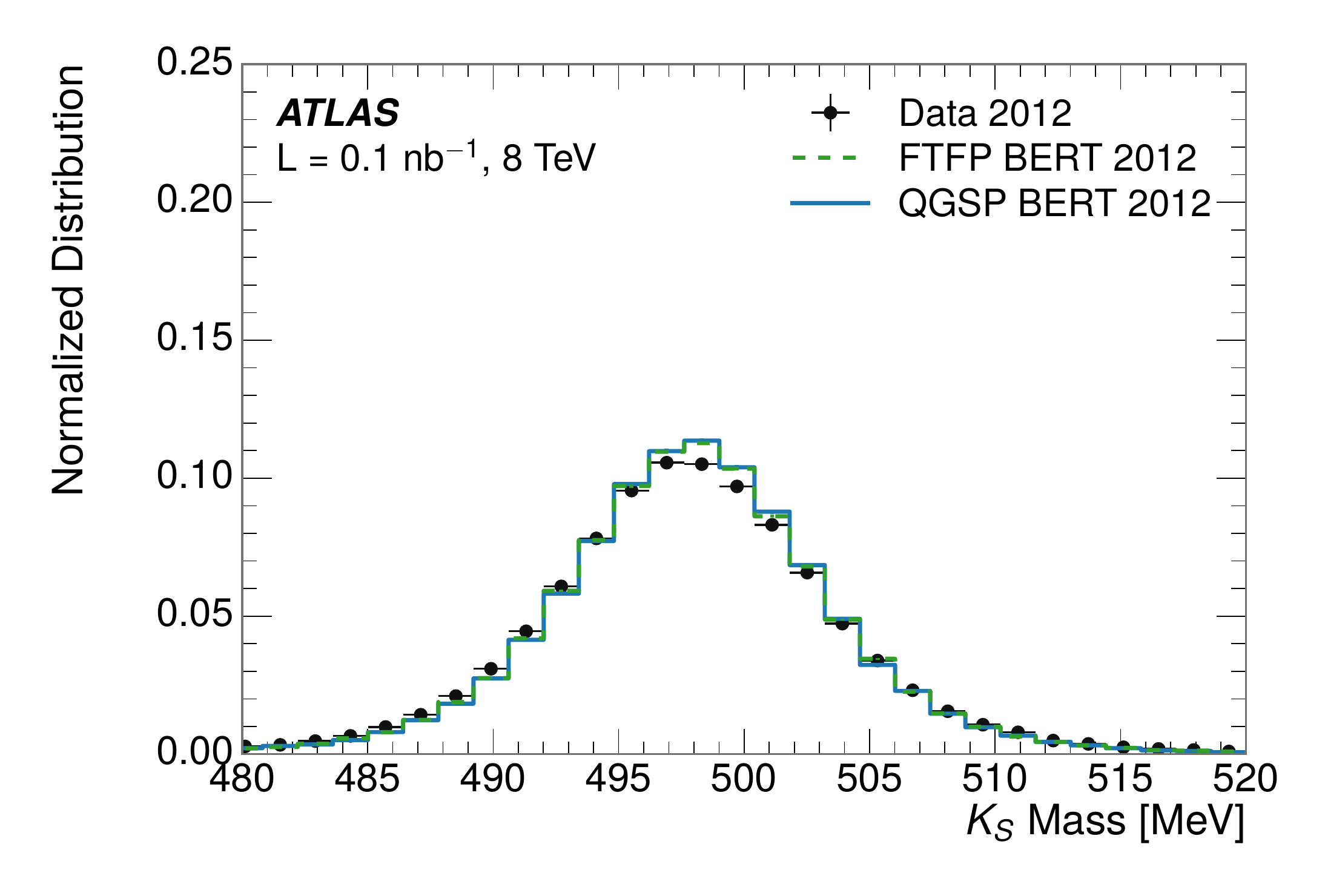}
}
\\
\subfloat[]{
  \includegraphics[width=0.48\textwidth]{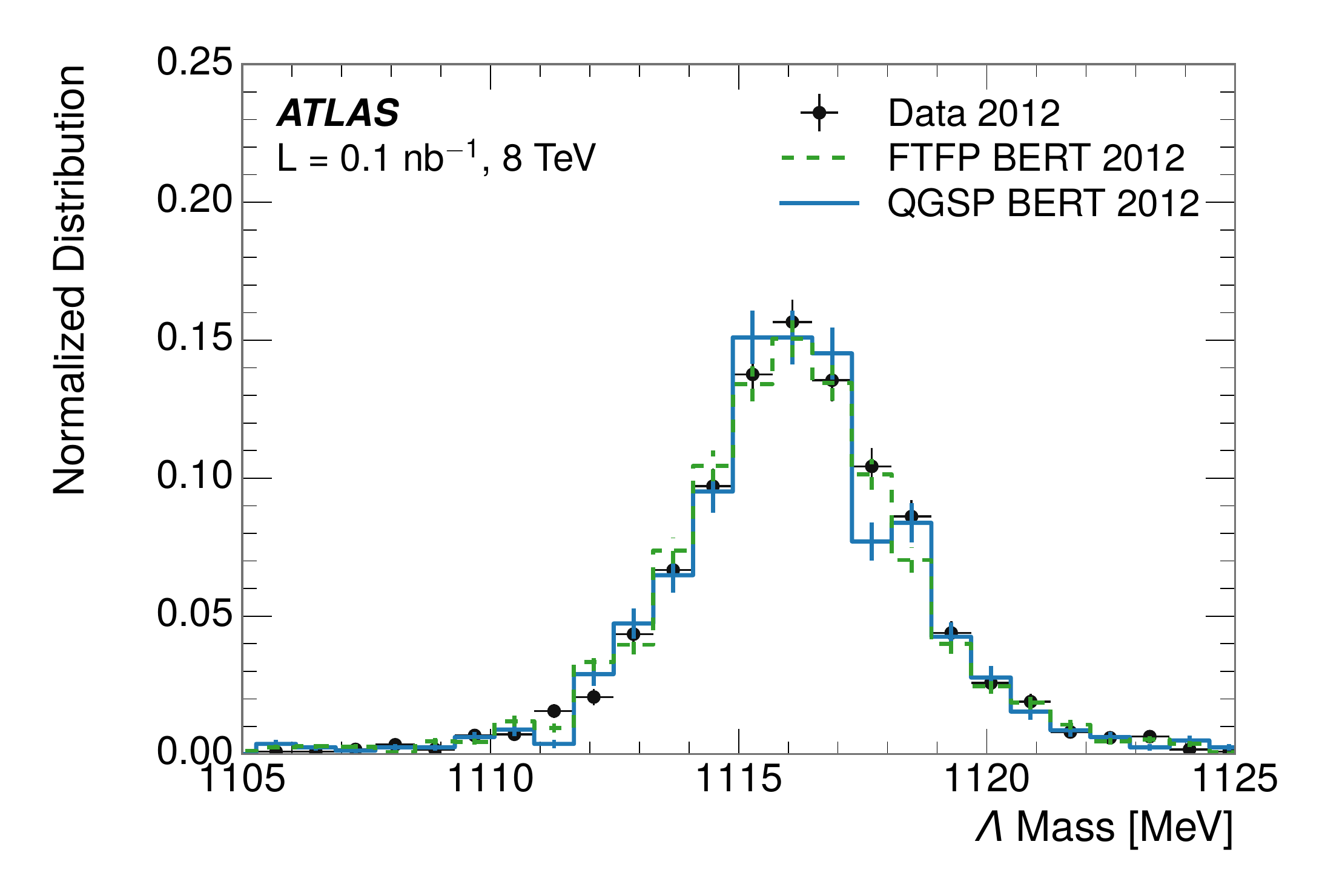}
}
~
\subfloat[]{
  \includegraphics[width=0.48\textwidth]{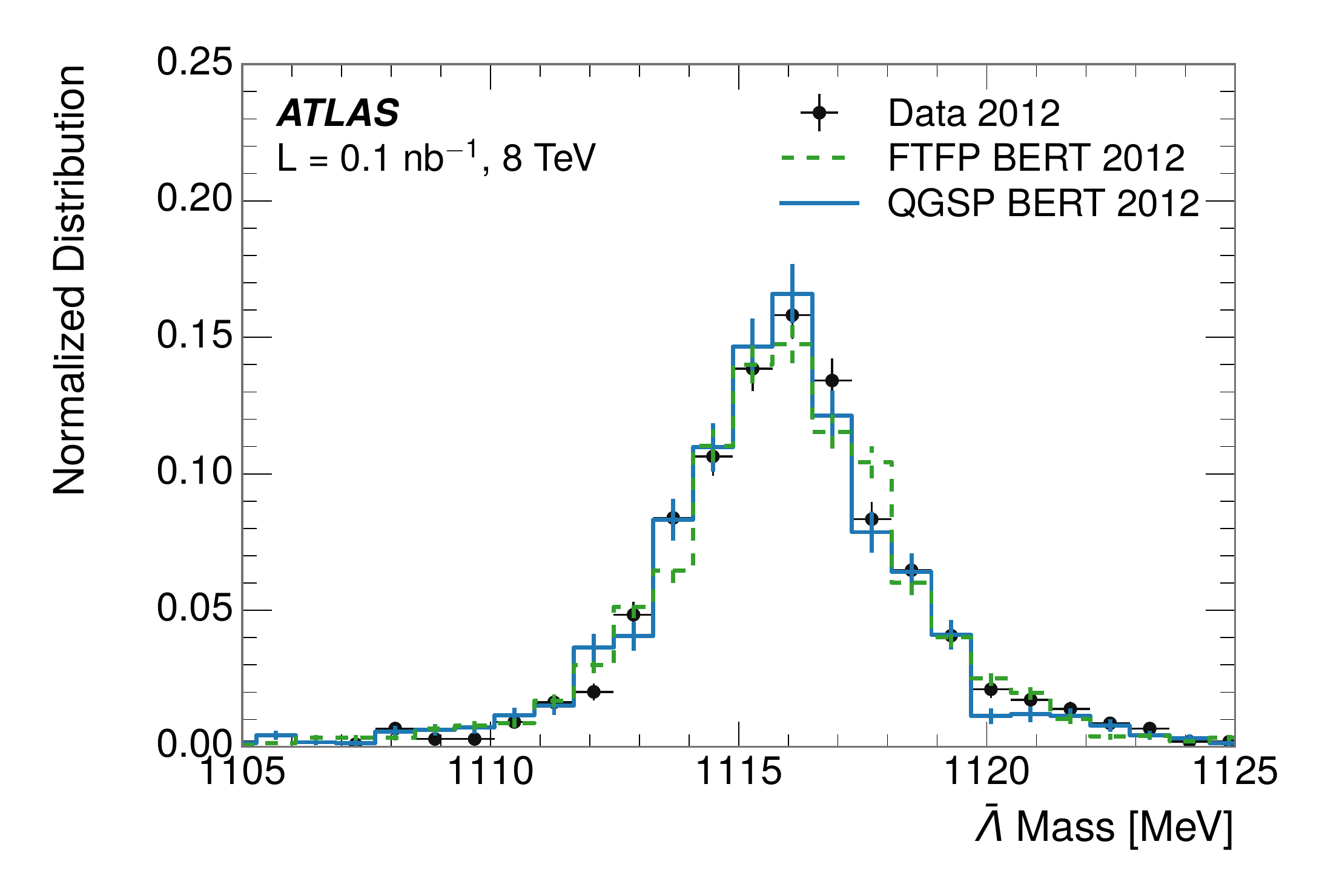}
}
\caption{The reconstructed mass peaks of (a) \pKS, (b) \pL, and (c) \pLB candidates in data and MC simulation with the \texttt{QGSP\_BERT} and \texttt{FTFP\_BERT} sets of hadronic physics models, for candidates with at least one track with $|\eta| < 0.6$. The distributions are normalised to unit area, and the error bars represent statistical uncertainties.}
\label{fig-mass}
\end{figure}

Example mass distributions for reconstructed \pKS, \pL, and \pLB\ candidates with at least one central track ($|\eta| < 0.6$) are shown in Figure~\ref{fig-mass}. These mass distributions are fitted to a modified Gaussian signal function and a polynomial background in bins of pseudorapidity as described in Ref.~\cite{PERF-2011-05}. The MC simulation and data distributions in these figures are normalised to unit area so that their shapes can be compared.  For each candidate type and each bin of pseudorapidity, the fits are used to construct an acceptance window to minimize background while retaining the majority of candidates. The windows are centred on the fitted mean and contain three standard deviations of the fitted signal function around the mean value. The number of candidates found in data and each MC simulation sample after passing the pseudorapidity dependent mass cuts and the remaining selection are shown in Table~\ref{tab-candidates}. These raw counts show clearly that roughly twice as many events were generated with \texttt{FTFP\_BERT} as with \texttt{QGSP\_BERT}. The ratio of \pL\ and \pLB\ candidates to \pKS\ candidates is 40\% higher in the data than in the MC simulation. A similar difference in the relative yields between data and MC simulation was observed in 2010~\cite{PERF-2011-05}.

\begin{table}[htbp]
\centering
\begin{tabular} {l c c c}
\hline
Candidate & Data & \texttt{QGSP\_BERT} & \texttt{FTFP\_BERT} \\
\hline
\pKS & $2.3 \times 10^{5}$ & $2.2 \times 10^{5}$ & $4.4 \times 10^{5}$\\
\pL  & $1.1 \times 10^{4}$ & $7.9 \times 10^{3}$ & $1.6 \times 10^{4}$\\
\pLB & $1.0 \times 10^{4}$ & $7.1 \times 10^{3}$ & $1.5 \times 10^{4}$\\
\hline
\end{tabular}
\caption{The number of signal candidates of each type found in data and each MC simulation sample.
}
\label{tab-candidates}
\end{table}

\subsection{Background estimation}

There are three primary sources of charged backgrounds in the identified particle \ep distributions. First, nuclear interactions in the ID can fake particle decays. The narrow mass window for decay candidates suppresses this background significantly. The particles can also undergo nuclear interactions before entering the calorimeter. These types of interactions are suppressed by requiring that the daughter tracks have many hits in the TRT. 

Another charged background for identified candidates comes from combinatoric sources. The purity of \pKS\ candidates is found to be high ($>99$\%), and the majority of tracks result from pions, so no correction is applied to the pion \ep distributions. The \pL\ and \pLB\ candidates can be faked by pions from \pKS\ decays by falsely treating them as protons. To remove this background, \pL\ and \pLB\ candidates which fall within the \pKS\ mass window when the invariant mass is calculated using the pion hypothesis are vetoed. After applying this veto as well as the remaining selection, the combinatoric background for \pL\ and \pLB\ is found to be small ($<1$\%).

The final charged background for the proton (anti-proton) \ep distributions occurs when a \pL\ (\pLB) decay fakes a \pLB\ (\pL) decay because the pion is actually the higher momentum track. This is most common for low energy \pL\ or \pLB, and is governed by well-understood two-body decay kinematics. These fakes are suppressed by the momentum requirements on the candidates, but are still present at the percent level. However, since two-body decay kinematics are straightforward to describe, it is accurately modeled by the MC simulation and is taken into account through MC simulation predictions.

There is also a contribution to the \ep distributions from the neutral background, as discussed in Section~\ref{sec-inclusive}. Where the distributions are presented as differences between particle species, the neutral background should cancel in the difference. The isolation from charged particles ensures that this background is small. This cancellation was tested using a simulation of single particles and was found to be valid to within statistical uncertainties~\cite{PERF-2011-05}. No additional correction or uncertainty is added for the neutral background. 

Thus, all of the systematic uncertainties arising from backgrounds are found to be negligible compared to the statistical errors of the available data sample.

\subsection{Identified particle response}
\label{sec:pid_response}

\begin{figure}[tbp]
\centering
\subfloat[]{
  \includegraphics[width=0.48\textwidth]{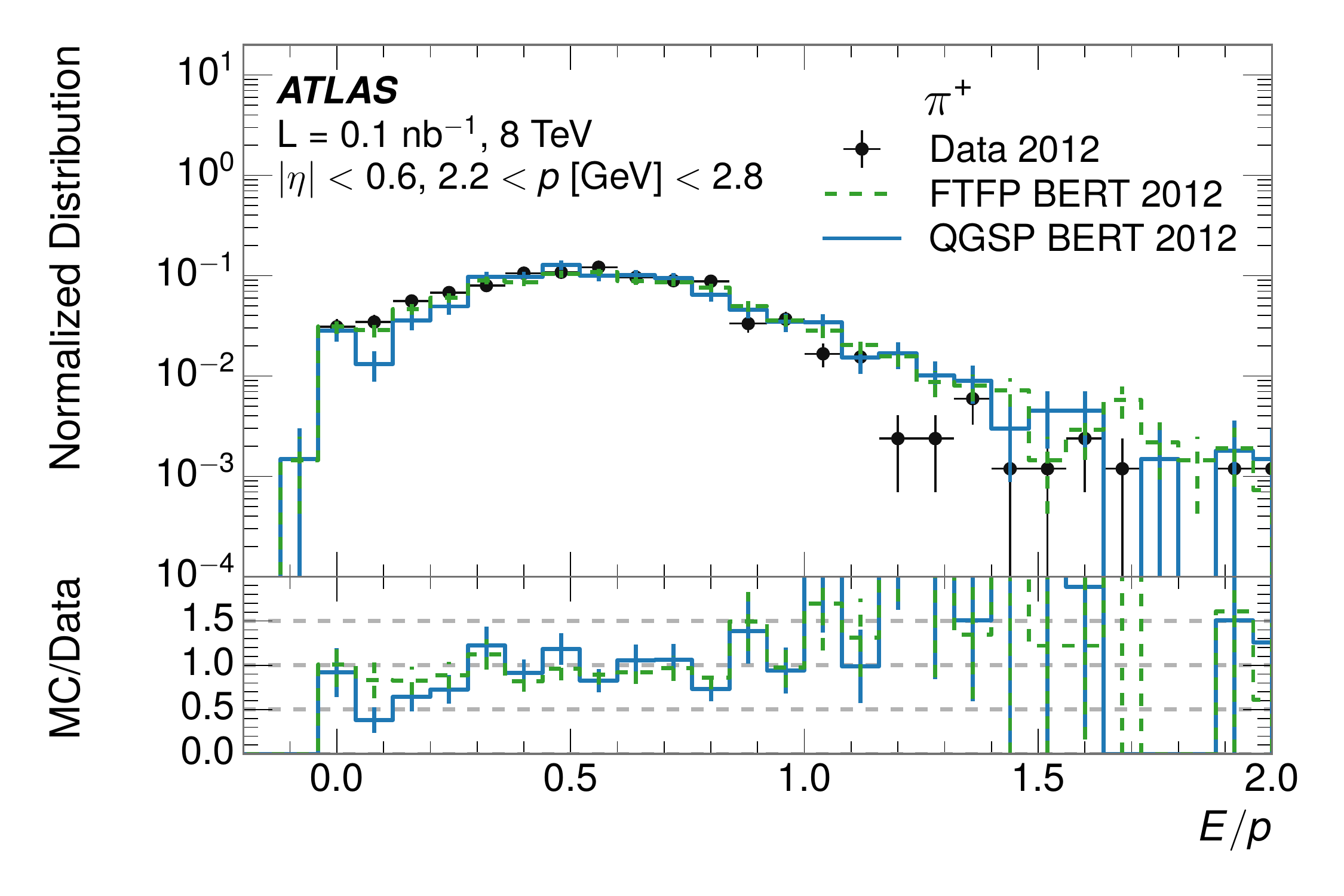}
}
~
\subfloat[]{
  \includegraphics[width=0.48\textwidth]{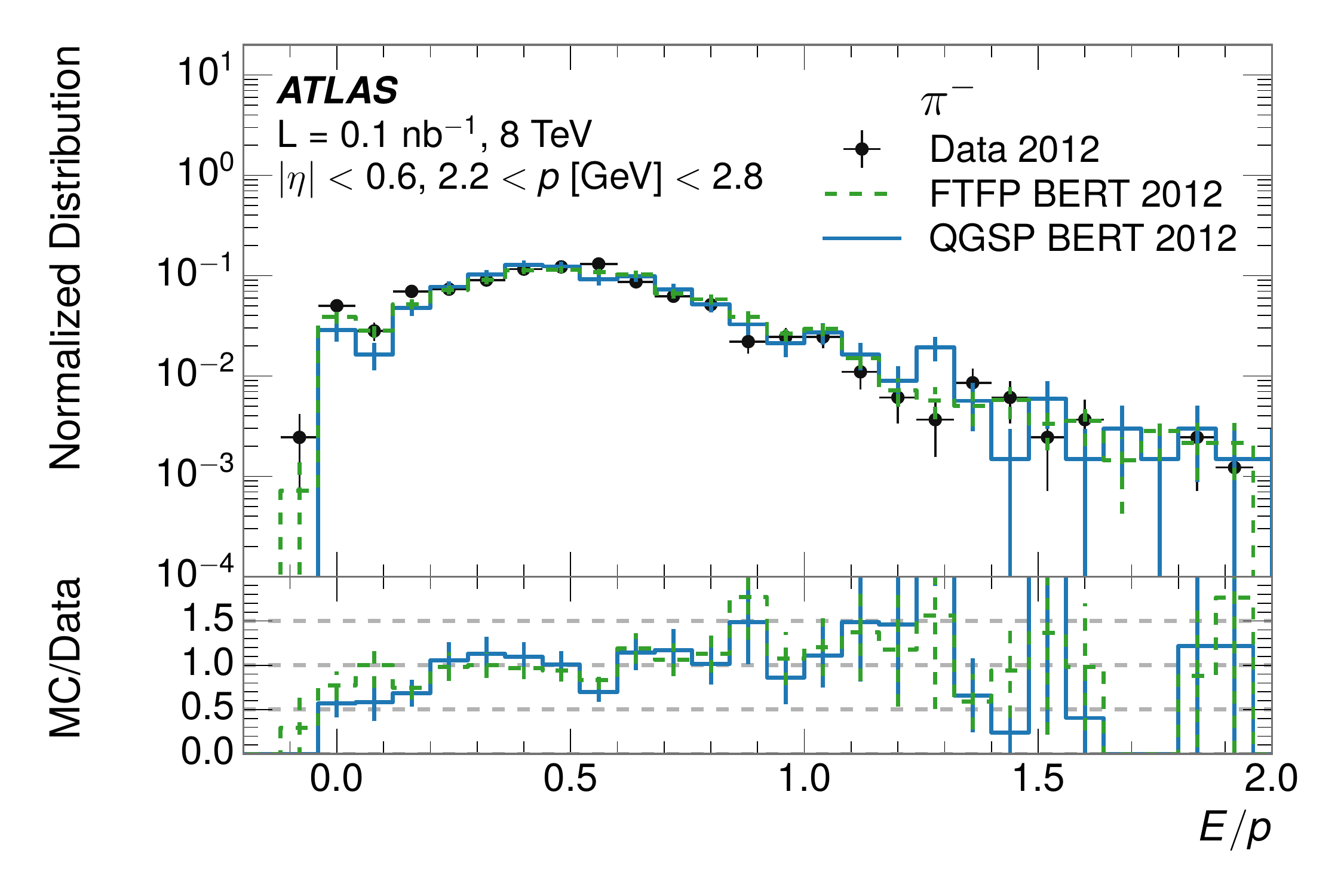}
}
\\
\subfloat[]{
  \includegraphics[width=0.48\textwidth]{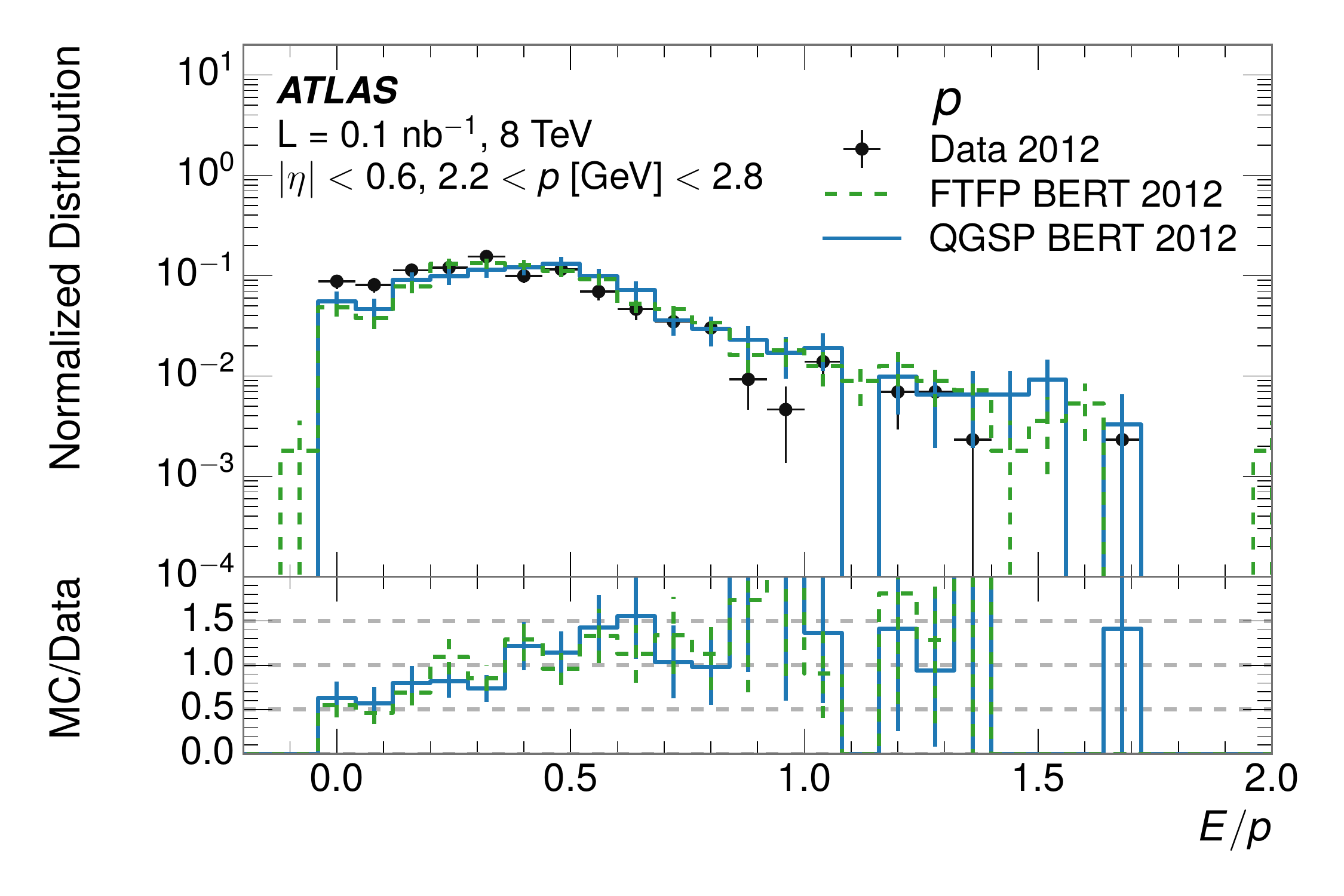}
}
~
\subfloat[]{
  \includegraphics[width=0.48\textwidth]{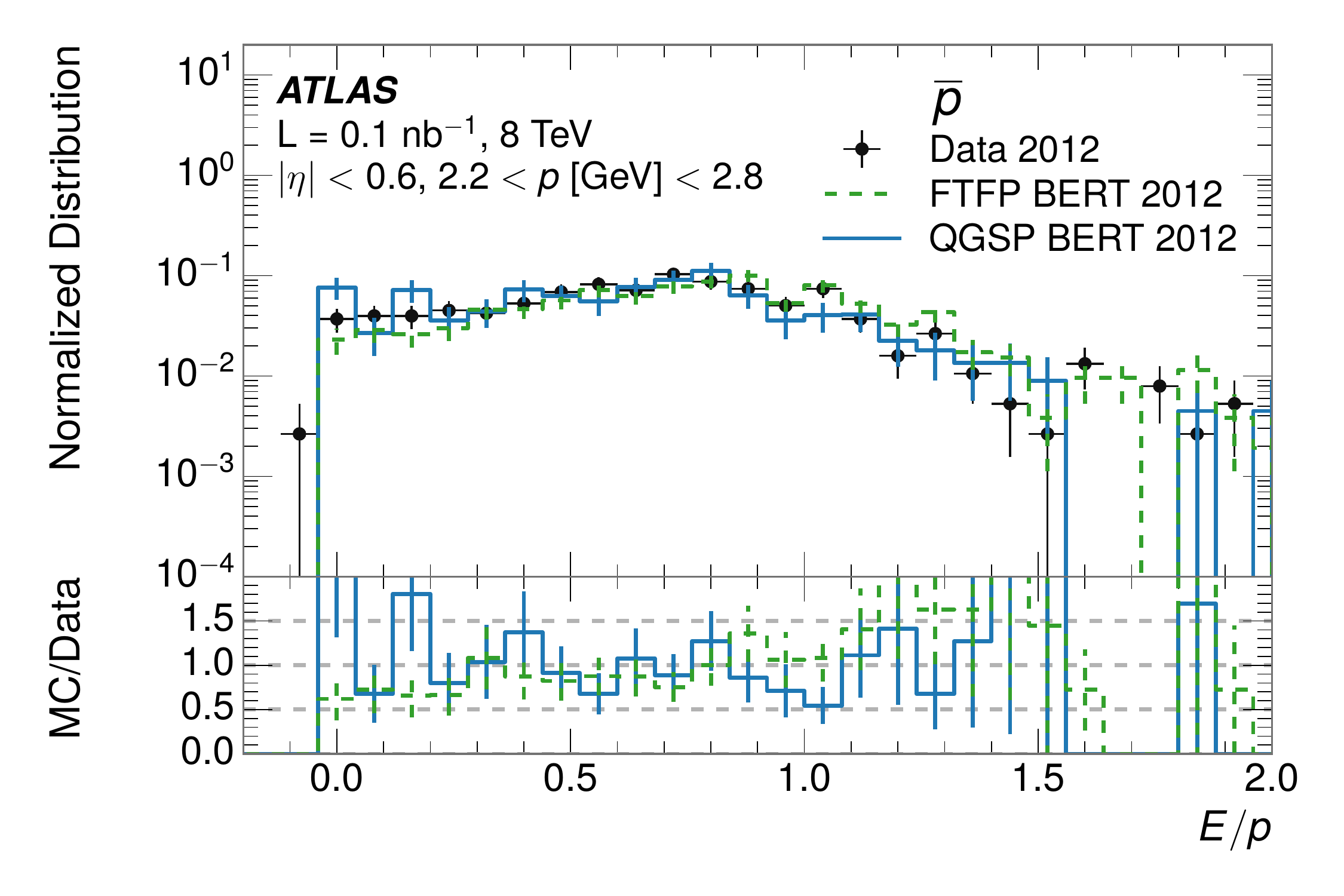}
}
\caption{The \ep distribution for isolated (a) \pip, (b) \pim, (c) proton, and (d) anti-proton tracks with $|\eta| < 0.6$ and $2.2 < p / \GeV < 2.8$. The bottom portion of each panel shows the ratio of MC simulation to data. The error bars represent statistical uncertainties.}
\label{fig-res_eoverp}
\end{figure}

Examples of the uncorrected distributions of \ep for \pip, \pim, protons, and anti-protons are shown in Figure~\ref{fig-res_eoverp}, for a single bin of available energy and pseudorapidity: $2.2 < E_a / \GeV < 2.8$ and $|\eta| < 0.6$. These distributions are normalised, so that their shapes can be compared without regard to the yield differences that are discussed in Section~\ref{sec:candidate_reco}.  As in the inclusive hadron response studies, a small fraction of the identified tracks have negative values of \ep. The energy in this distribution has a long positive tail due to the neutral background. The population of identified anti-protons with $\ep>1$ is much more prominent because of the annihilation of the anti-proton, which leads to a significantly greater calorimeter response to anti-protons than to pions or protons. The response distribution is well reproduced by the MC simulation to within the statistical precision. 

\begin{figure}[htbp]
\centering
\subfloat[]{
  \includegraphics[width = 0.48\textwidth]{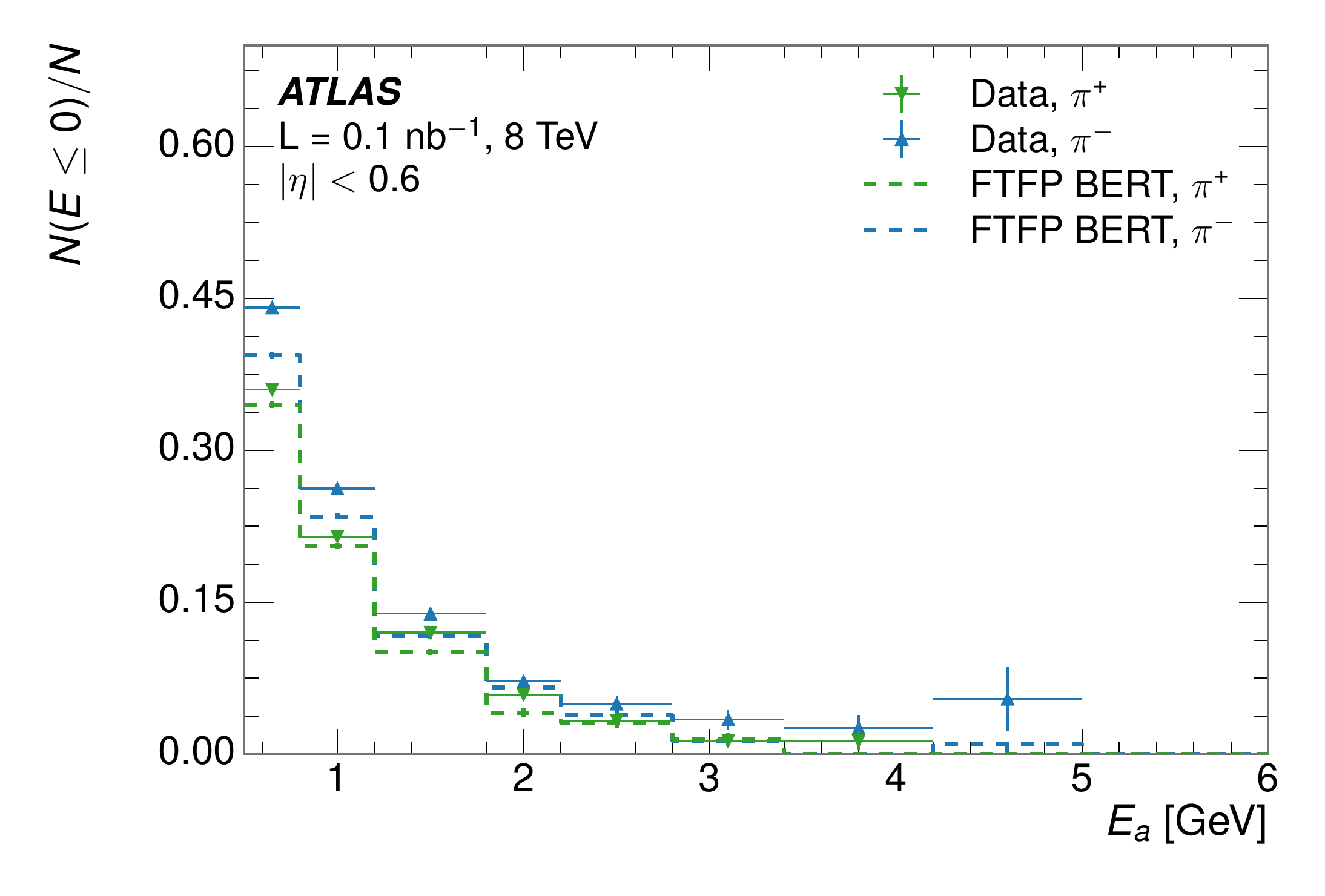}
}
~
\subfloat[]{
  \includegraphics[width = 0.48\textwidth]{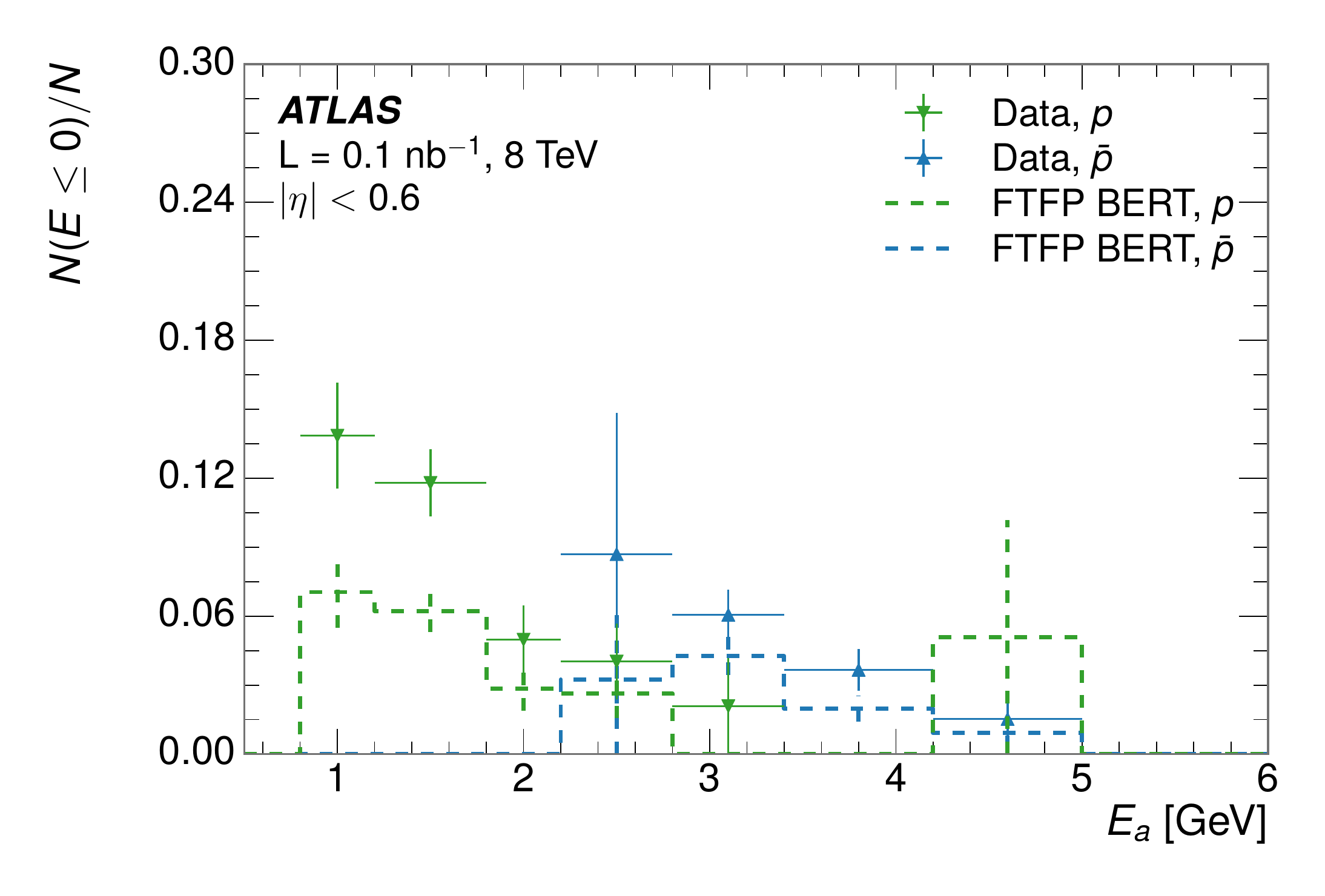}
}
\caption{ The fraction of tracks with $E \leq 0$ for identified (a) \pip and \pim, and (b) proton and anti-proton tracks with $|\eta| < 0.6$.
For anti-protons, $\Ea\gtrsim 2$~\GeV. The uncertainties shown are statistical only.}
\label{fig-res_zero_fraction}
\end{figure}

A significant feature of the \ep distributions is the fraction of tracks with $E \leq 0$, as discussed in Section~\ref{sec:raw_ep}. The fraction is large at low available energy, and the level of agreement between data and MC simulation reflects the modelling of the material in front of the calorimeter. This fraction compared between particle species is shown in Figure~\ref{fig-res_zero_fraction}, in data and simulation with the \texttt{FTFP\_BERT} set of hadronic physics models. The \texttt{QGSP\_BERT} set of hadronic physics models provides a similar description of the data. This fraction is underestimated by the MC simulation by approximately 10\% at low \Ea, with larger discrepencies for protons and anti-protons, although the available statistics in data and MC simulation are limited. This suggests that the source of the discrepancy in the fraction of tracks with $E \leq 0$ is \emph{not} from the hadronic-interaction model for one particle species, but is caused by an effect present for all particle species.

\subsubsection{Differences in calorimeter response between particles of different species}

\begin{figure}[htbp]
\centering
\subfloat[]{
  \includegraphics[width=0.48\textwidth]{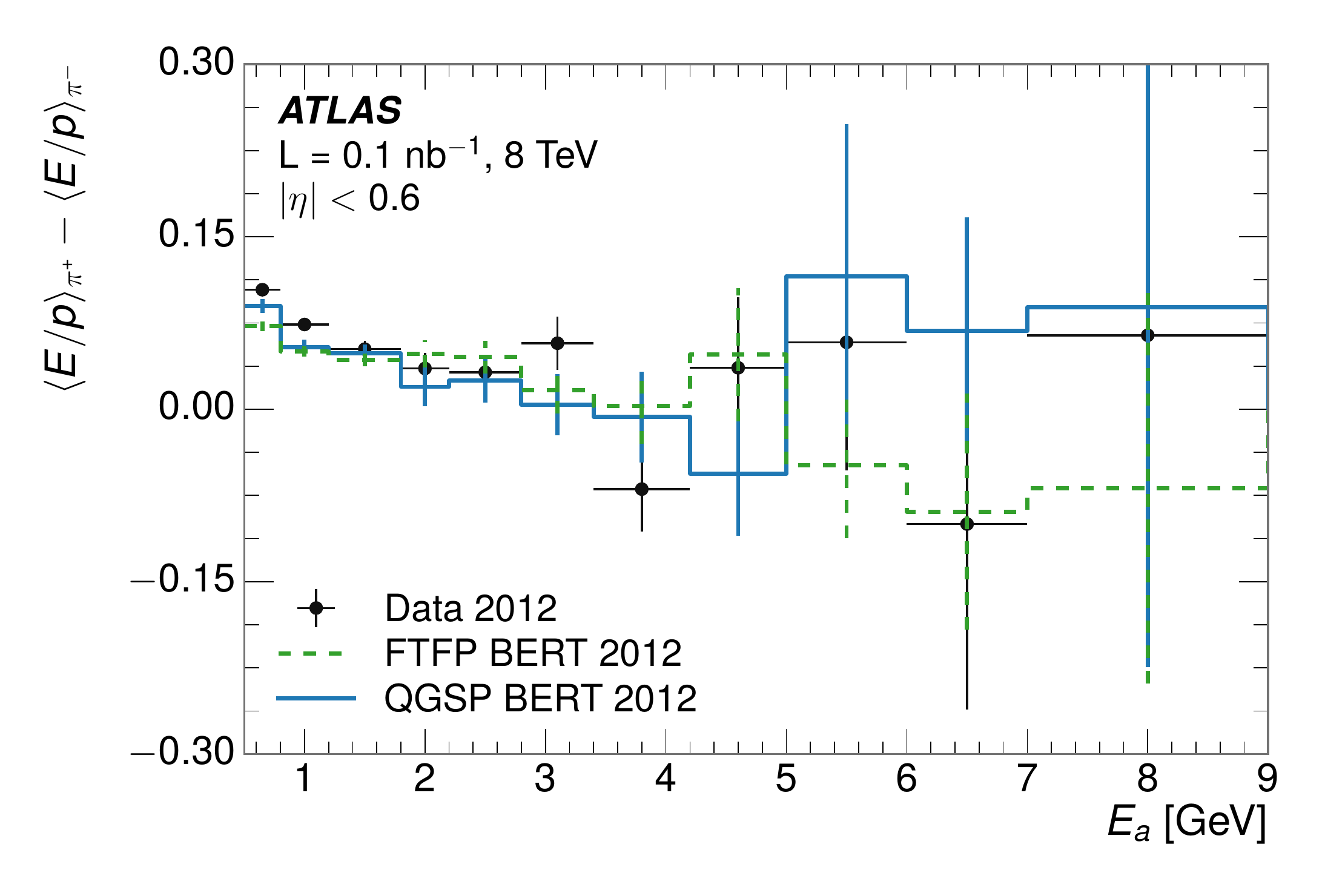}
}
~
\subfloat[]{
  \includegraphics[width=0.48\textwidth]{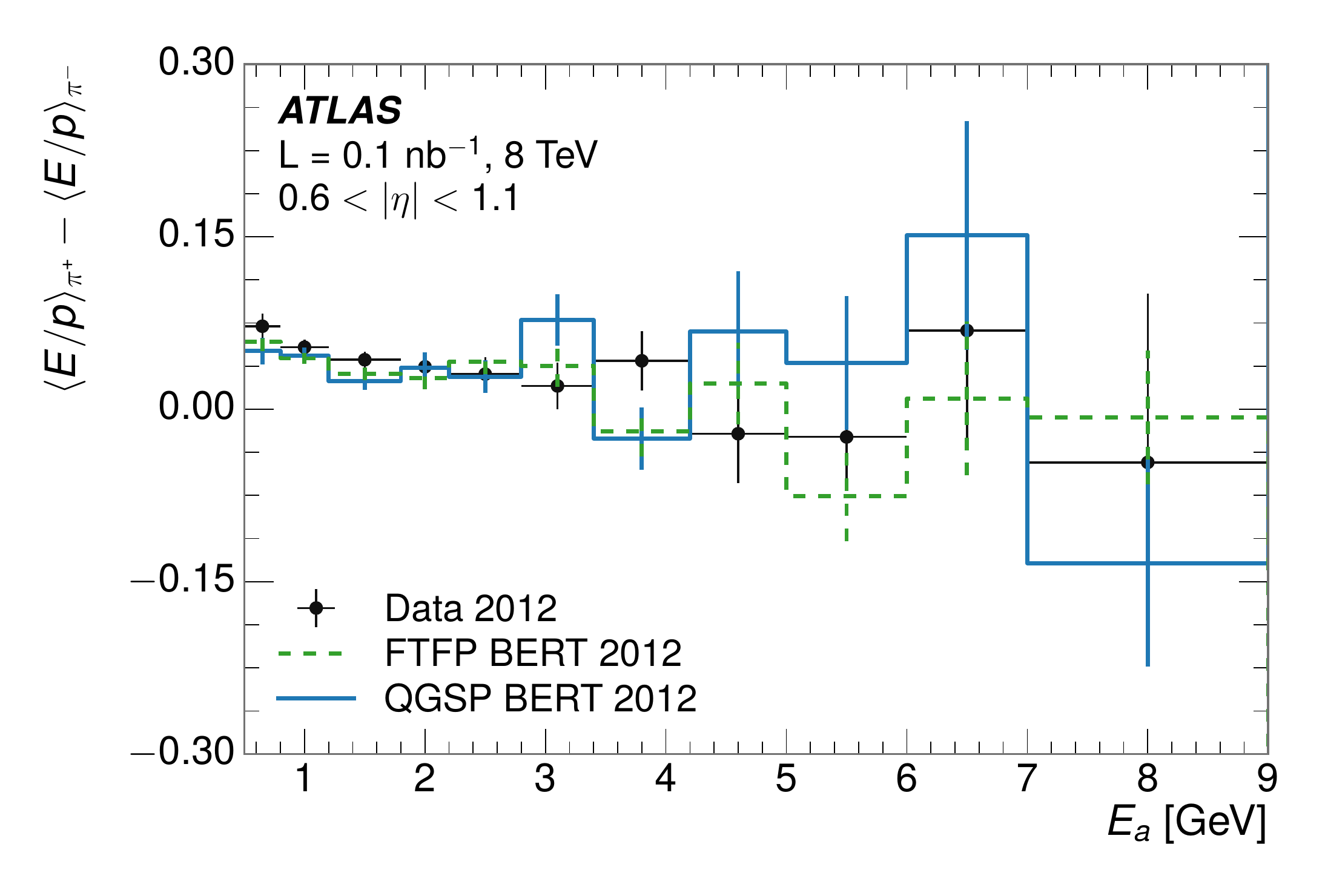}
}
\caption{ The difference in \epav between \pip\ and \pim\ with (a) $|\eta| < 0.6$ and (b) $0.6 < |\eta| < 1.1$. The error bars represent statistical uncertainties.}
\label{fig-epdiff_pions}
\end{figure}

In order to reduce the effect of the neutral background, the average responses, \epav, are measured as differences between pairs of particle species. The averages are just the arithmetic means of the \ep distributions. The difference between \pip\ and \pim\ is shown in Figure~\ref{fig-epdiff_pions}, for two bins of pseudorapidity ($|\eta| < 0.6$ and $0.6 < |\eta| < 1.1$). The response to \pip\ is greater on average than the response to \pim\ at low energy, which agrees with Ref.~\cite{particlebeam}, where the difference is attributed to a charge-exchange effect (i.e. the production of additional neutral pions in showers initiated by $\pi^+$). The simulation models the data well, with some trend to underestimate the difference, although there are large statistical uncertainties at high available energy. 

\begin{figure}[htbp]
\centering
\subfloat[]{
  \includegraphics[width=0.48\textwidth]{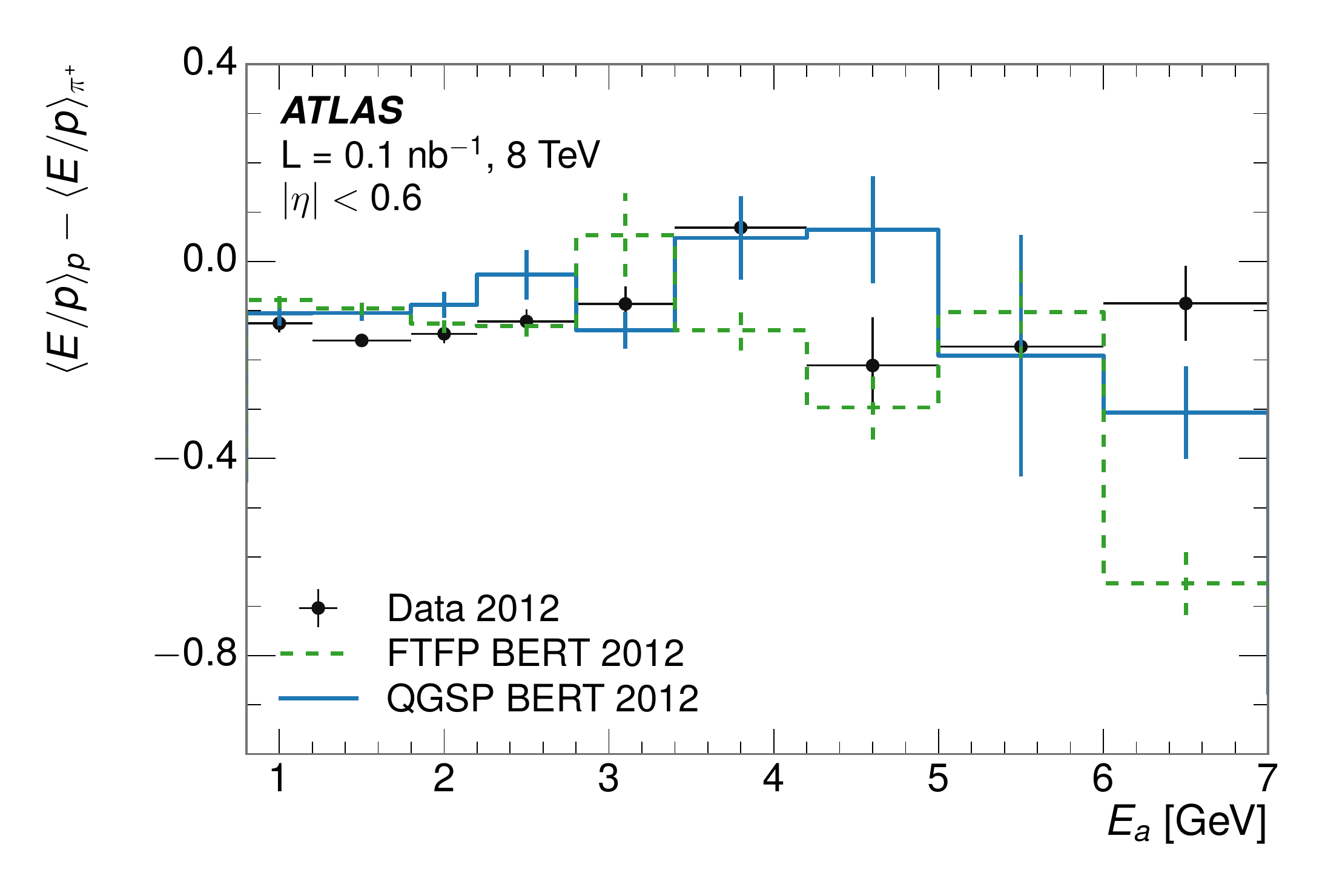}
}
~
\subfloat[]{
  \includegraphics[width=0.48\textwidth]{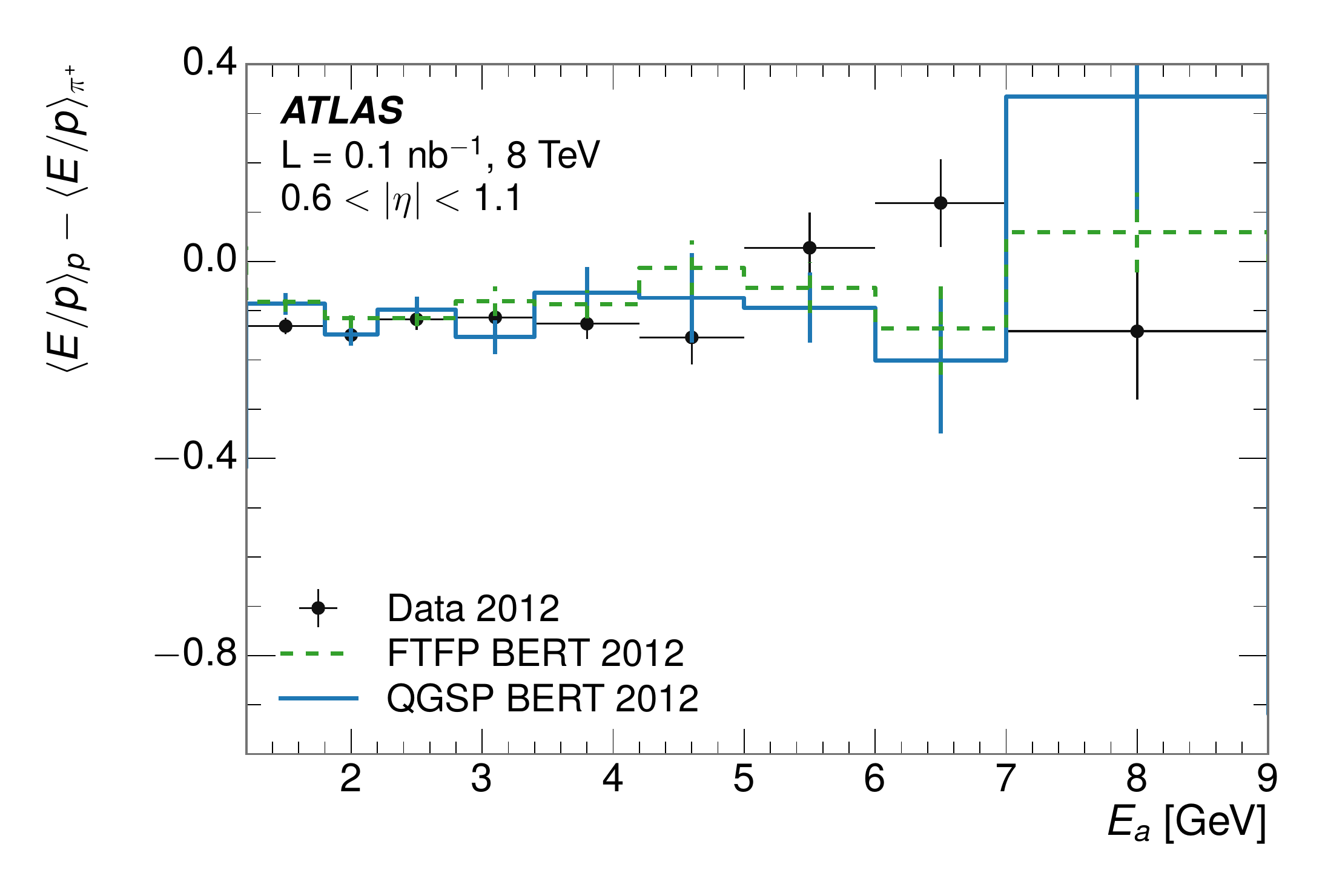}
}
\caption{ The difference in \epav between protons and \pip\ with (a) $|\eta| < 0.6$ and (b) $0.6 < |\eta| < 1.1$. The error bars represent statistical uncertainties.}
\label{fig-epdiff_pion_proton}
\end{figure}

Figure~\ref{fig-epdiff_pion_proton} shows the difference between the response to protons and \pip\ for two pseudorapidity bins. The response to protons is lower than the response to pions on average because a larger fraction of the initial energy is converted into an electromagnetic shower for pions~\cite{physicsg,TileTB}. This is evident in the \epav difference, where the response to pions is about 10\% greater on average. The data and MC simulation are fairly consistent, though there are large statistical uncertainties because of the low number of identified protons.

\begin{figure}[htbp]
\centering
\subfloat[]{
  \includegraphics[width=0.48\textwidth]{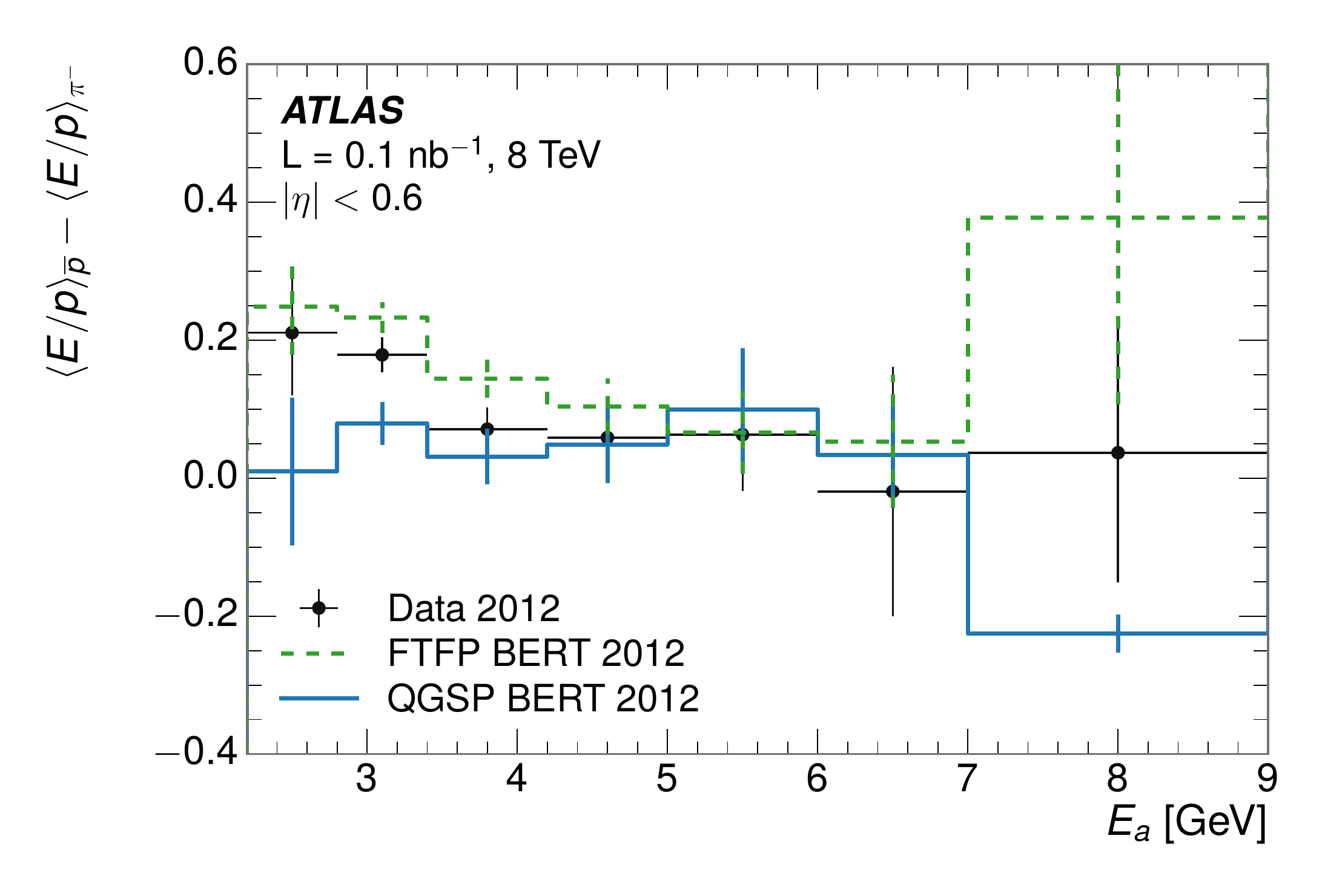}
}
~
\subfloat[]{
  \includegraphics[width=0.48\textwidth]{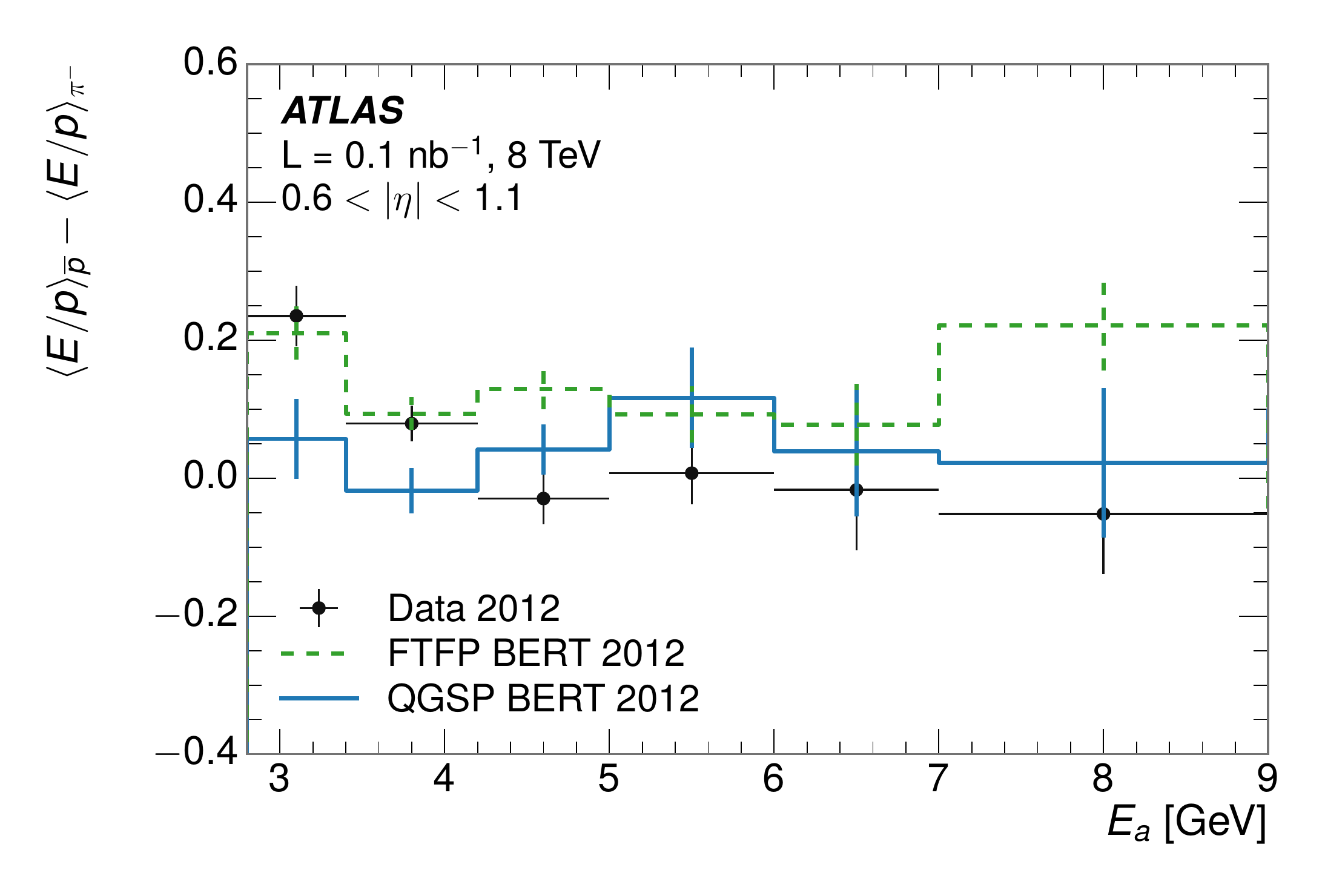}
}
\caption{ The difference in \epav between anti-protons and \pim\ with (a) $|\eta| < 0.6$ and (b) $0.6 < |\eta| < 1.1$. The error bars represent statistical uncertainties.}
\label{fig-epdiff_pion_antiproton}
\end{figure}

Figure~\ref{fig-epdiff_pion_antiproton} shows the difference between the response to anti-protons and \pim\ for two pseudorapidity bins. The response to anti-protons is expected to be significantly larger than the response to pions at low available energy, because of the annihilation of the anti-proton in the calorimeter. While the difference in response at low \Ea is small for the \texttt{QGSP\_BERT} set of hadronic physics models, because of a different model used to estimate anti-baryon nuclear interactions in the \texttt{FTFP\_BERT} set of hadronic physics models~\cite{G4hadronics}, the response to anti-protons is about 20\% greater. Here, the \texttt{FTFP\_BERT} set of hadronic physics models provides a slightly better description of the data. The imperfections in the \texttt{QGSP\_BERT} set of hadronic physics models were also reported in Ref.~\cite{PERF-2011-05}.

\subsubsection{Background-corrected isolated identified hadron response}

\begin{figure}[tbp]
\centering
\subfloat[]{
  \includegraphics[width=0.48\textwidth]{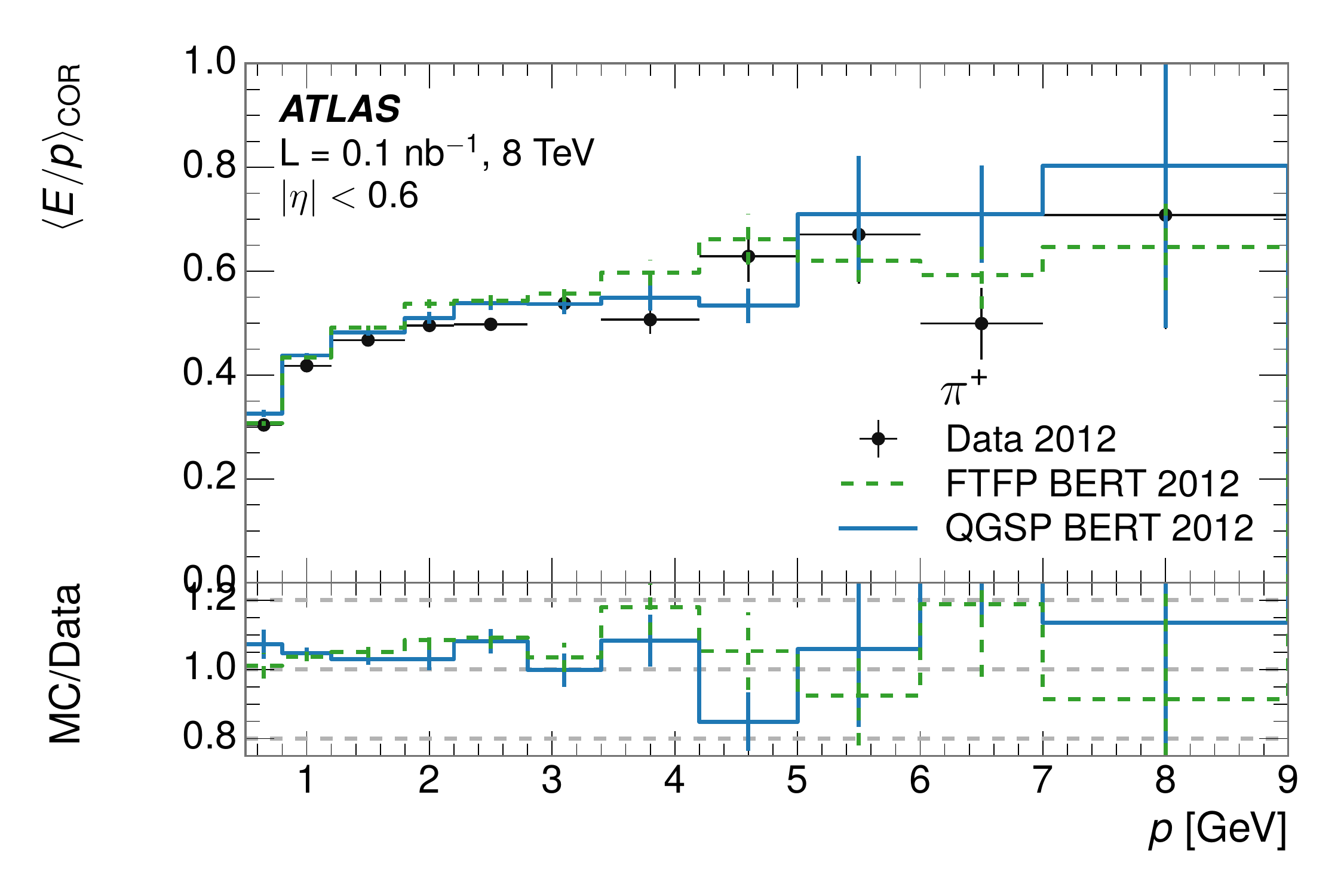}
}
\subfloat[]{
  \includegraphics[width=0.48\textwidth]{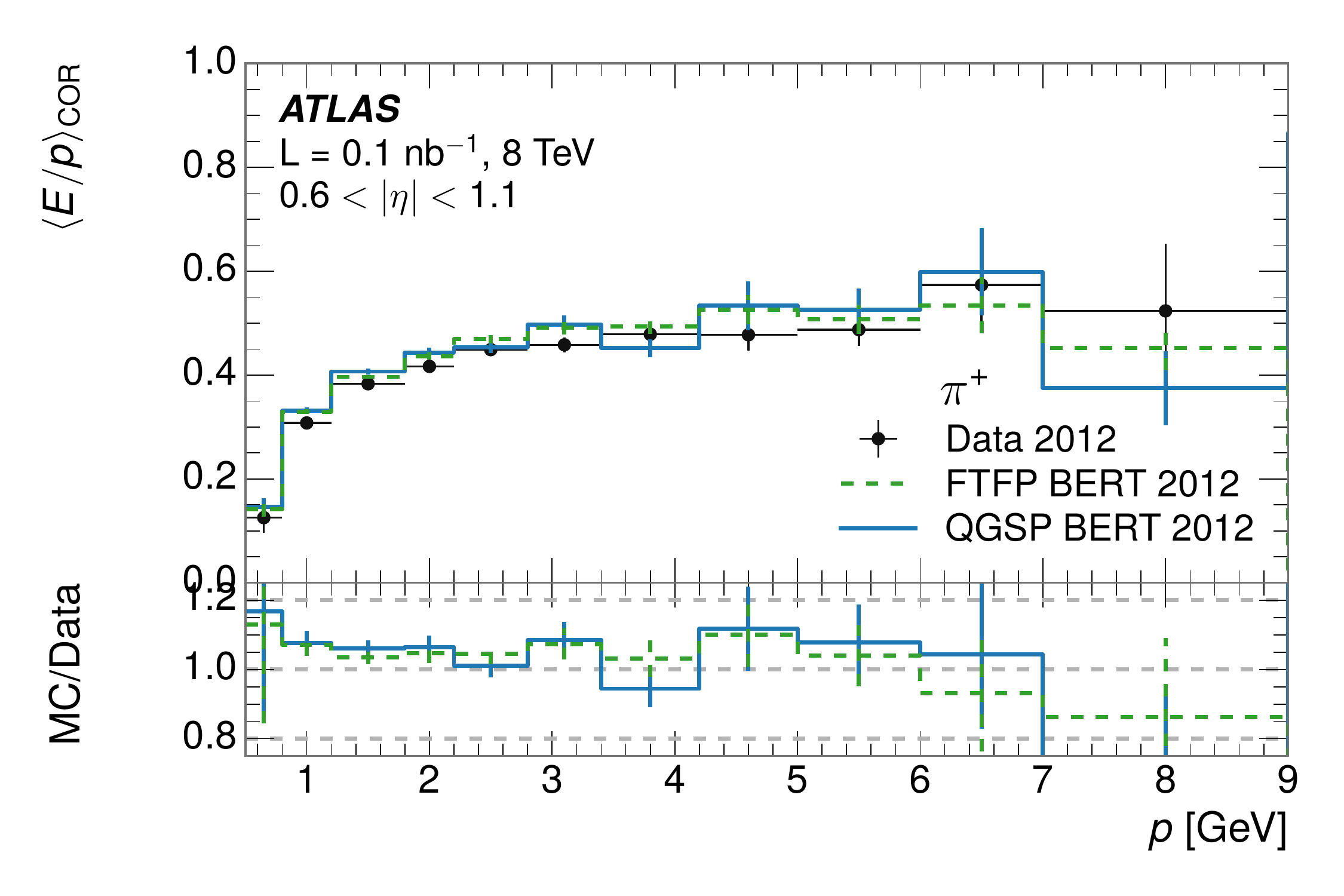}
}
\\
\subfloat[]{
  \includegraphics[width=0.48\textwidth]{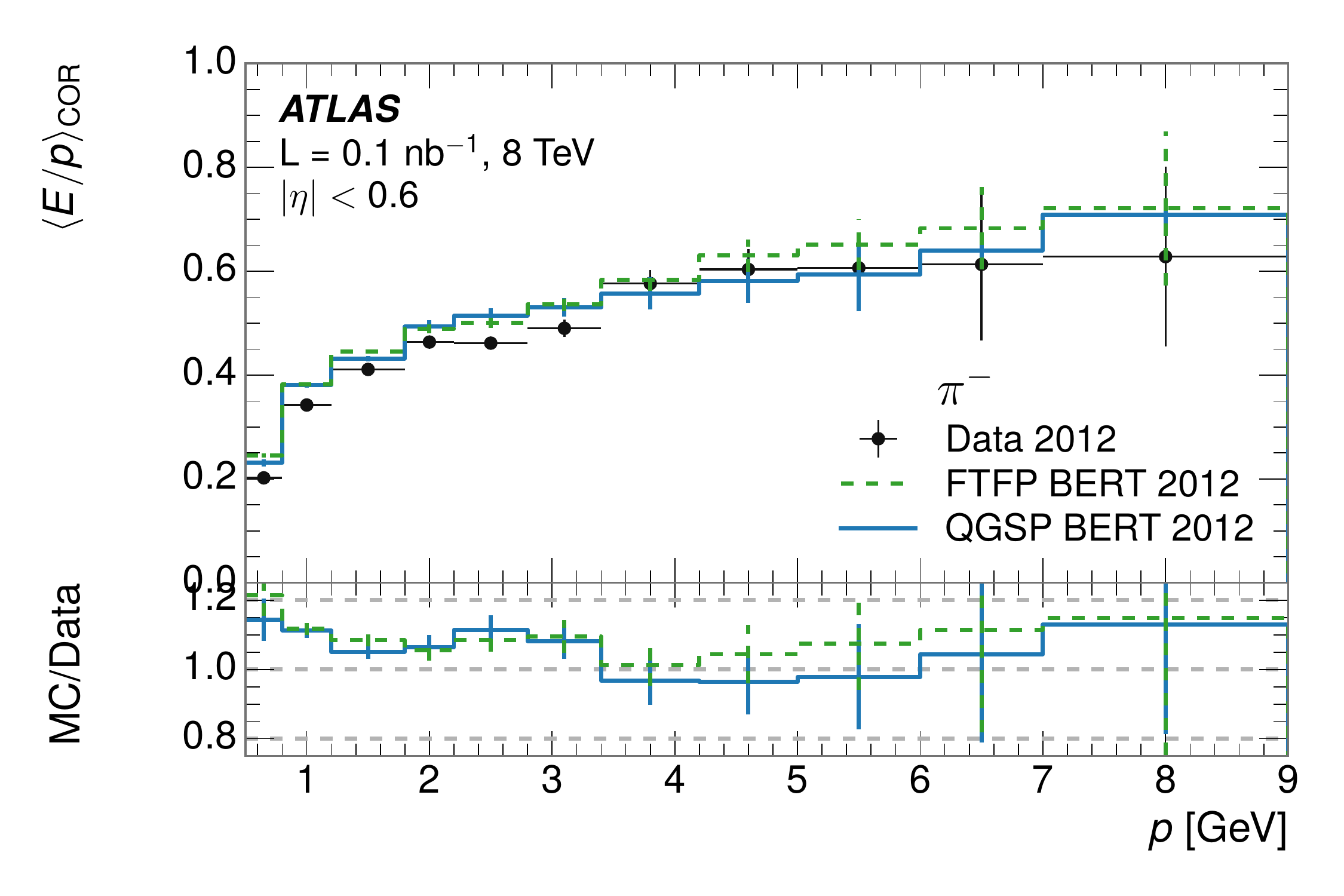}
}
\subfloat[]{
  \includegraphics[width=0.48\textwidth]{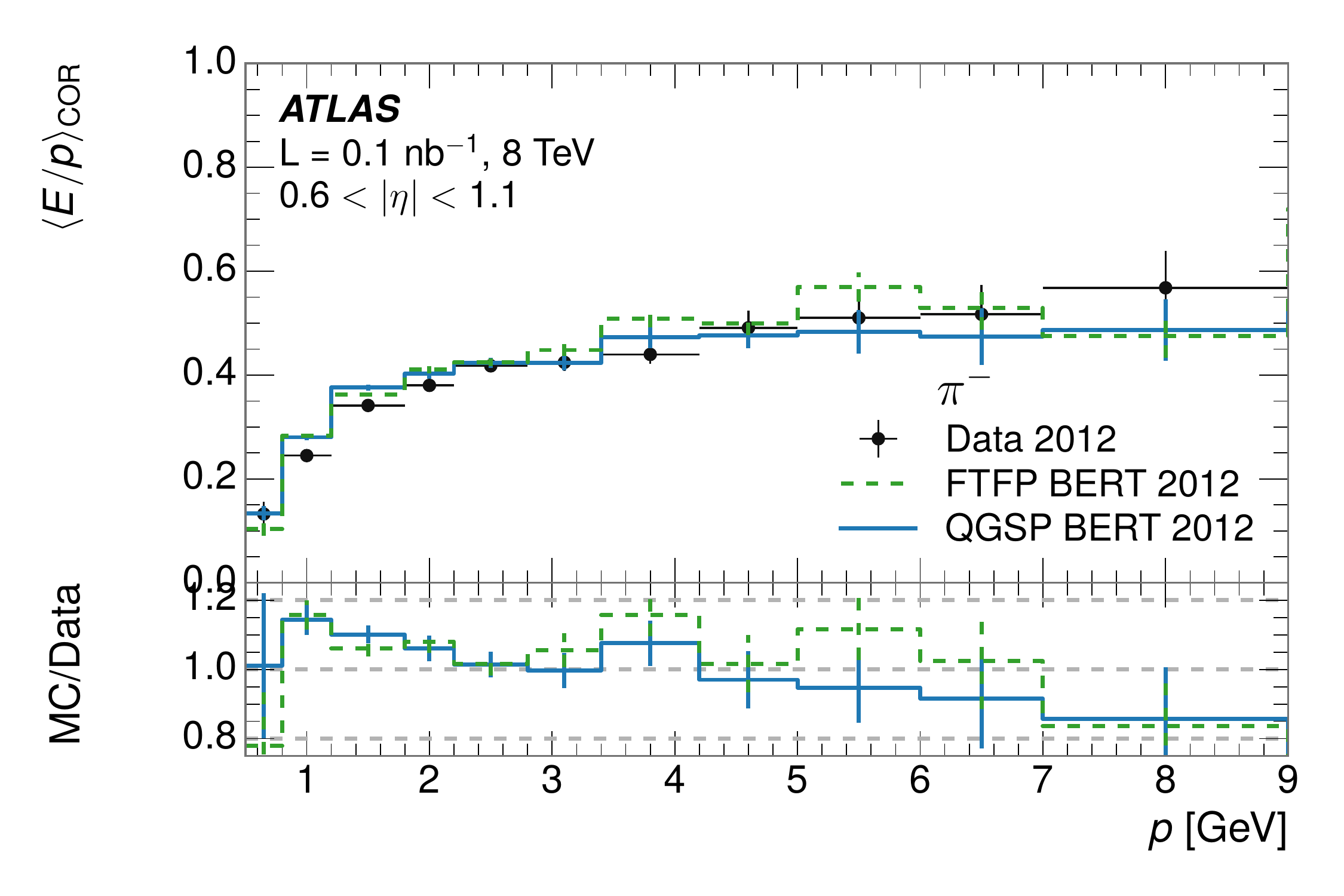}
}
\caption{\epcor as a function of track momentum, corrected for the neutral background, for \pip\ tracks with (a) $|\eta| < 0.6$ and (b) $0.6 < |\eta| < 1.1$ and for \pim\ tracks with (c) $|\eta|<0.6$ and (d) $0.6 < |\eta| < 1.1$. The bottom portion of each panel shows the ratio of MC simulation to data, separately for the two sets of hadronic physics models. The error bars represent statistical uncertainties.}
\label{fig-epcor_pim}
\end{figure}

Insofar as the neutral background is independent of the species of the particle of interest, the neutral background estimate described in Section~\ref{sec-inclusive} is equally applicable to identified particles from displaced decays. Figure~\ref{fig-epcor_pim} shows the \epcor distributions for \pip\ and \pim as a function of track momentum in two bins of pseudorapidity. The mean values of the \ep distributions for identified \pip\ and \pim\ tracks are similar to the distributions for inclusive tracks as expected. The difference between the response to \pip\ and \pim\ is also apparent here; the response to \pip\ is greater than the response to \pim\ for a given range of $|\eta|$ and $p$. Although the accessible range of momenta of identified tracks is limited, these distributions suggest that the data and MC simulation differ in the \pim\ distributions, where the difference is consistent with a difference of about 10\% for momenta below 2~\GeV.  The data and MC simulation are more consistent in the \pip\ distributions. This difference may explain the approximately 5\% discrepancy at low momentum in the inclusive \epcor distributions, shown in Figure~\ref{fig-eoverp_corrected}.

\subsubsection{Estimation of charged-kaon calorimeter response}

The charged-particle content of jets and of inclusive minimum-bias events generated using {\sc Pythia8} is dominated by charged pions (60--70\%), with lesser components from charged kaons (15--20\%) and protons and anti-protons (5--15\%). The largest component of this composition that has not been measured in ATLAS is the charged-kaon component.  Discriminants based on ionisation energy loss in thin detectors are only applicable at low momenta ($<2$~\GeV), and most relevant resonances ($\phi$, $K^{*}$) are too rare or suffer from too small a purity to construct a useful sample. It is possible to test the response to kaons using template subtraction, so that for each $p$ and $\eta$ bin the response to kaons is calculated from:

\begin{equation}
\epav^\text{inclusive} = f_{K^\pm} \times \epav^{K^\pm} + f_{\pip} \times \epav^{\pip} + f_{\pim} \times \epav^{\pim} + f_p \times \epav^{p} + f_{\bar{p}} \times \epav^{\bar{p}}
\end{equation}

\noindent where $f_X$ is the fraction of species $X$ as estimated in MC simulated event samples and $\epav^X$ is the average corrected response to particle species $X$. Uncertainties in the species rates are estimated using several different MC simulation samples.  This method yields a cross-check of the calorimeter response to charged kaons at a precision of about 20\%.  To this level, the response is well modeled in the simulation.  The dominant uncertainties, however, are statistical and could be reduced in the future.

\subsubsection{Calorimeter response to additional species of particles and close-by particles}

\begin{figure}[htbp]
\centering
\includegraphics[width=.58\textwidth]{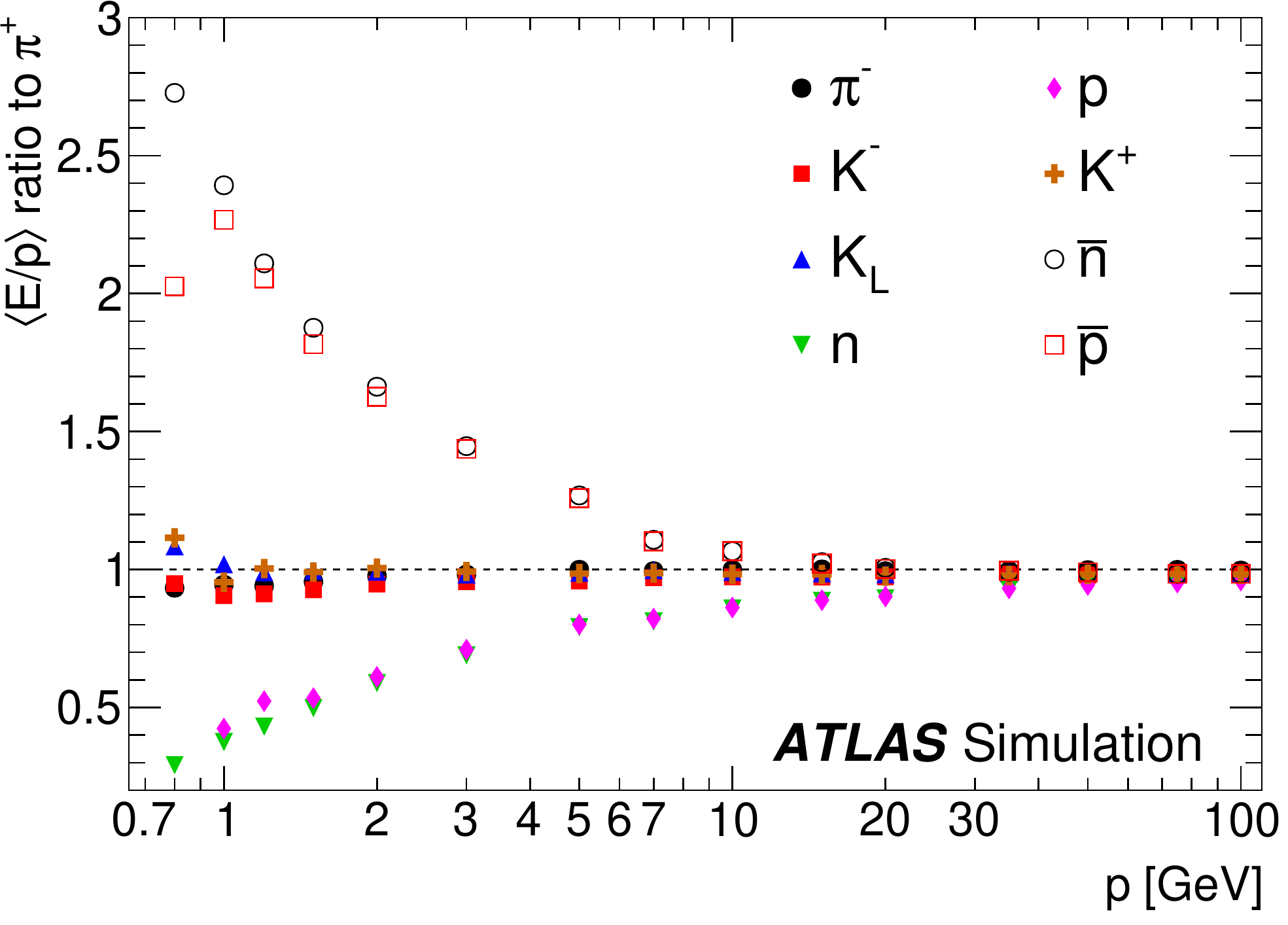}
\caption{The ratio of the calorimeter response to single particles of various species to the calorimeter response to $\pi^+$ with the \texttt{FTFP\_BERT} set of hadronic physics models.  Only statistical uncertainties are shown.}
\label{fig:singleparticle_summary}
\end{figure}

The calorimeter response to different particle species vary significantly at low energy.  Figure~\ref{fig:singleparticle_summary} shows the responses of various particle types using the \texttt{FTFP\_BERT} set of hadronic physics models.  Above 20~\GeV, the calorimeter response to all hadrons, charged or neutral, almost independent of species, is the same.  At lower energies, the response to protons and neutrons is significantly lower than to pions, and the response to anti-protons and anti-neutrons is significantly higher than the response to pions.  These differences are reflected in the different definitions of available energy used in this paper.

With the 2010 dataset, it is possible to use \pKS\ decays to test the hadronic shower widths and topological clustering effects using the response to nearby particles~\cite{PERF-2011-05}. The MC simulation was shown to be consistent with the data, albeit with sizeable statistical uncertainties. The 2012 dataset does not provide a sufficient number of events to test these effects with any additional accuracy.

\section{Extrapolation to jet energy response and uncertainty}
\label{sec-jets}
Reconstructed jets are formed from topological clusters of energy using the anti-$k_t$ algorithm~\cite{Cacciari:2008gp} with distance parameter $R=0.4$.
In simulated dijet events, the calibrated jet momenta are compared to particle jets formed using the same algorithm from particles with lifetimes greater than 15~ps, excluding muons and neutrinos.
The calorimeter response to jets can be calculated as the ratio of the $\pt$ of the reconstructed jet to that of the closest particle jet in $\eta$--$\phi$ space.
The calibration of the reconstructed jets involves compensation for all of the effects discussed in the previous sections of this paper.
The study of jet momenta in this section uses simulated dijet events generated using {\sc Pythia8} with the CT10 parton distribution set~\cite{CTEQ} and the AU2 tune~\cite{AU2}.
Only the 2012 data and 8~\TeV\ center-of-mass energy collisions are discussed in this section.  The 2010 data and 7~\TeV\ collisions show consistent features.

\subsection{Jet properties}

\begin{figure}[ht]
\centering
\subfloat[]{
  \includegraphics[width=0.48\textwidth]{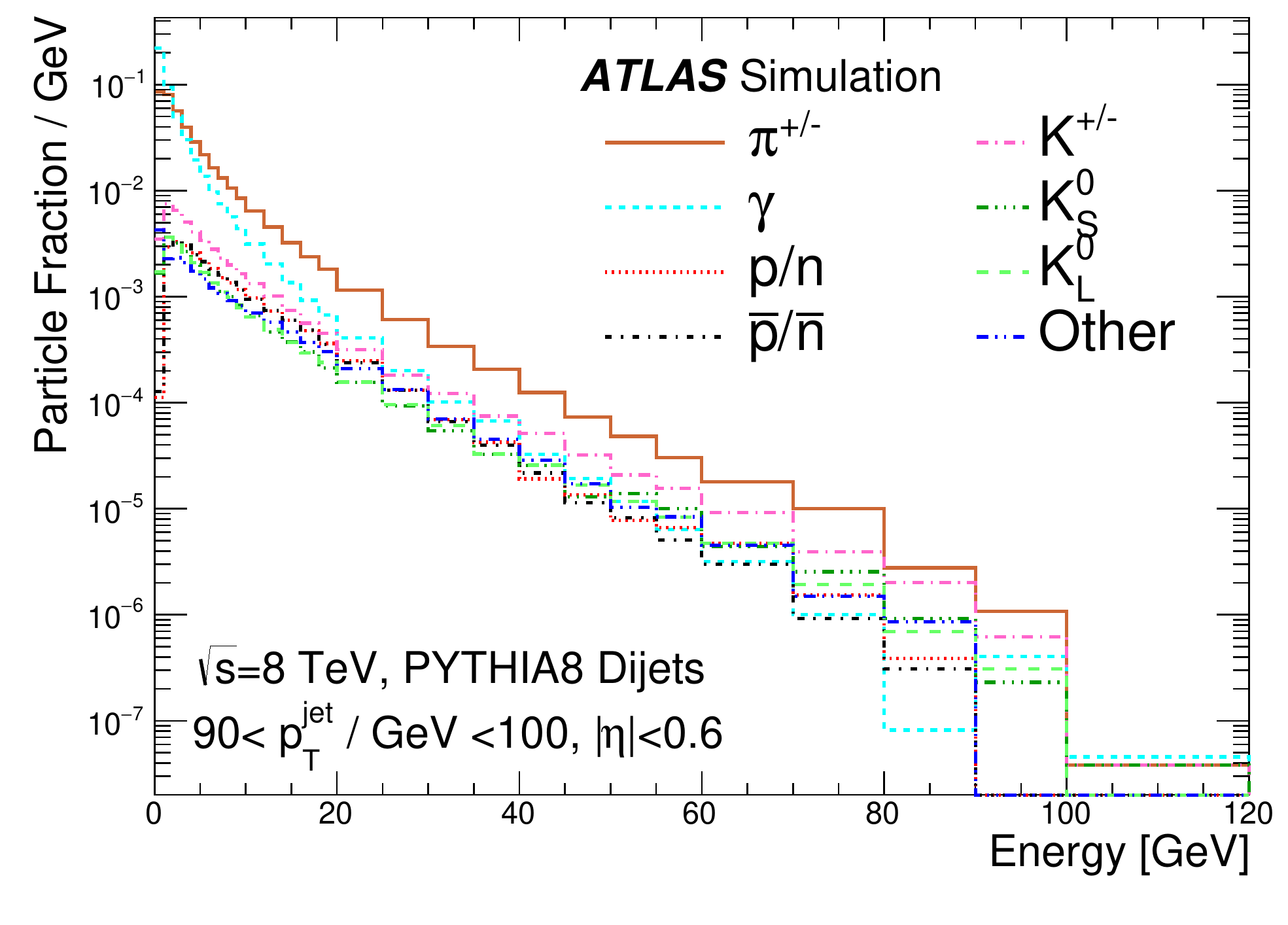}
}
\subfloat[]{
  \includegraphics[width=0.48\textwidth]{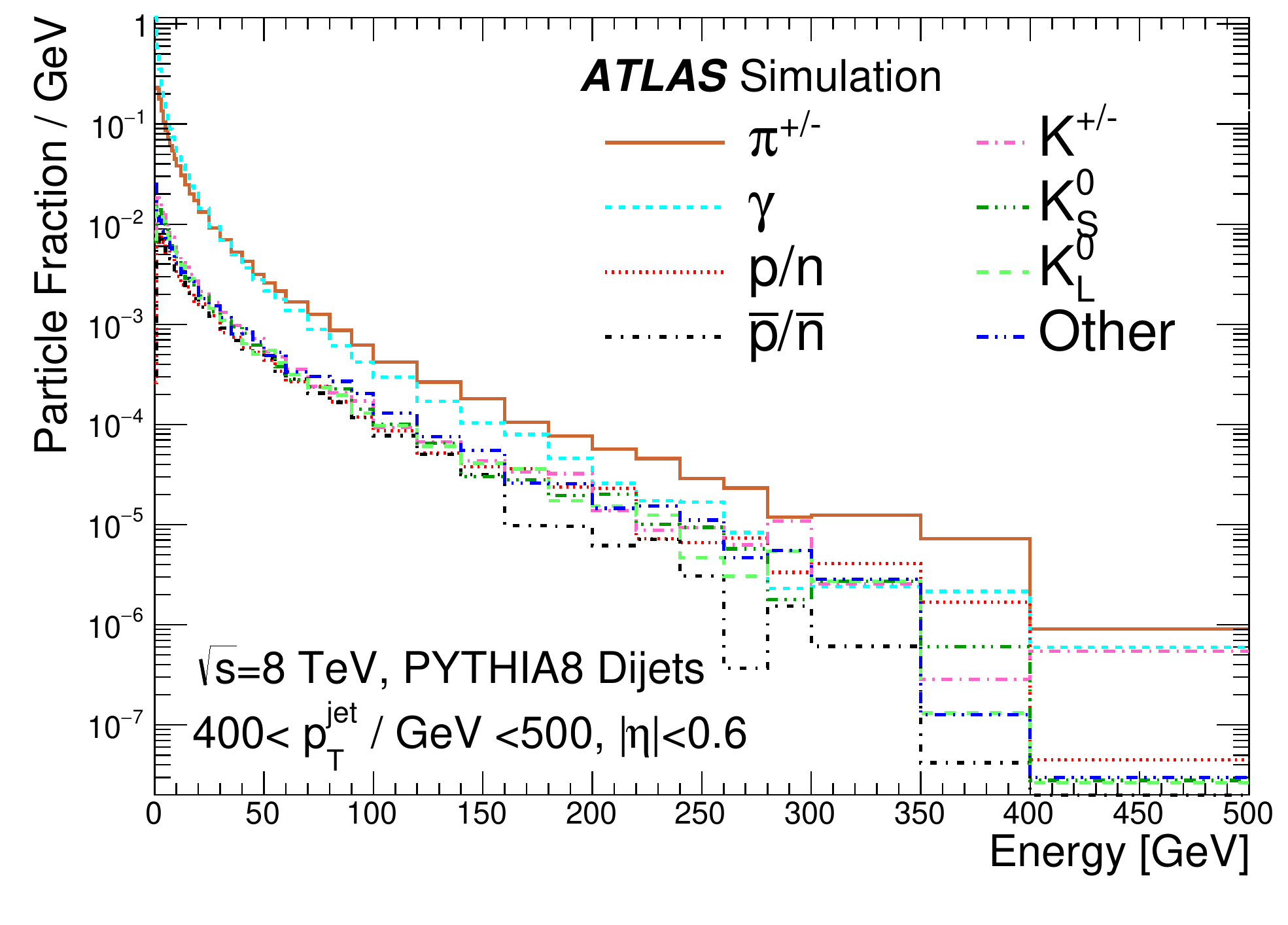}
}\\
\subfloat[]{
  \includegraphics[width=0.48\textwidth]{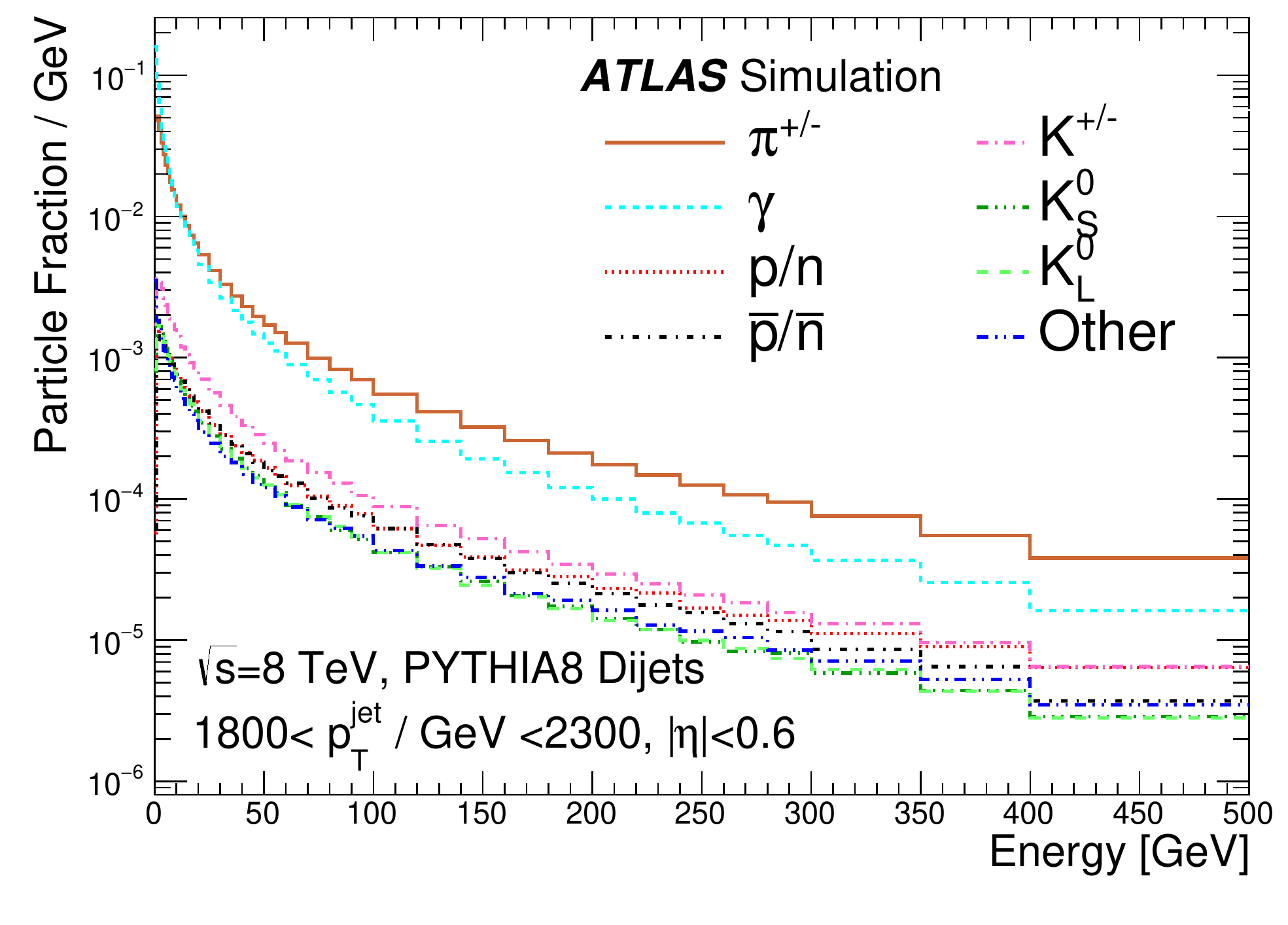}
}
\caption{The spectra of true particles inside anti-$k_t$, $R=0.4$ jets with (a) $90<\pt/\GeV<100$, (b) $400<\pt/\GeV<500$, and (c) $1800<\pt/\GeV<2300$. }
\label{fig-particle_spectra}
\end{figure}

Jet properties vary significantly from jet to jet, because of fragmentation and hadronisation effects~\cite{STDM-2015-12}.
The spectra of hadrons, both in terms of momentum and species, inside the jets differ, leading to differences in the calorimeter response to jets~\cite{PERF-2012-01}.
The spectra of particles inside jets are shown in Figure~\ref{fig-particle_spectra}.
The spectrum of photons is visibly softer than the spectrum of charged pions, owing to the fact that photons are coming predominantly from the decays of neutral pions.
If the charged and neutral hadrons have approximately the same spectrum, the photons from meson decays ought to be softer than the hadrons.
The spectrum is dominated by charged pions and photons.

\begin{figure}[ht]
\centering
\subfloat[]{
  \includegraphics[width=0.48\textwidth]{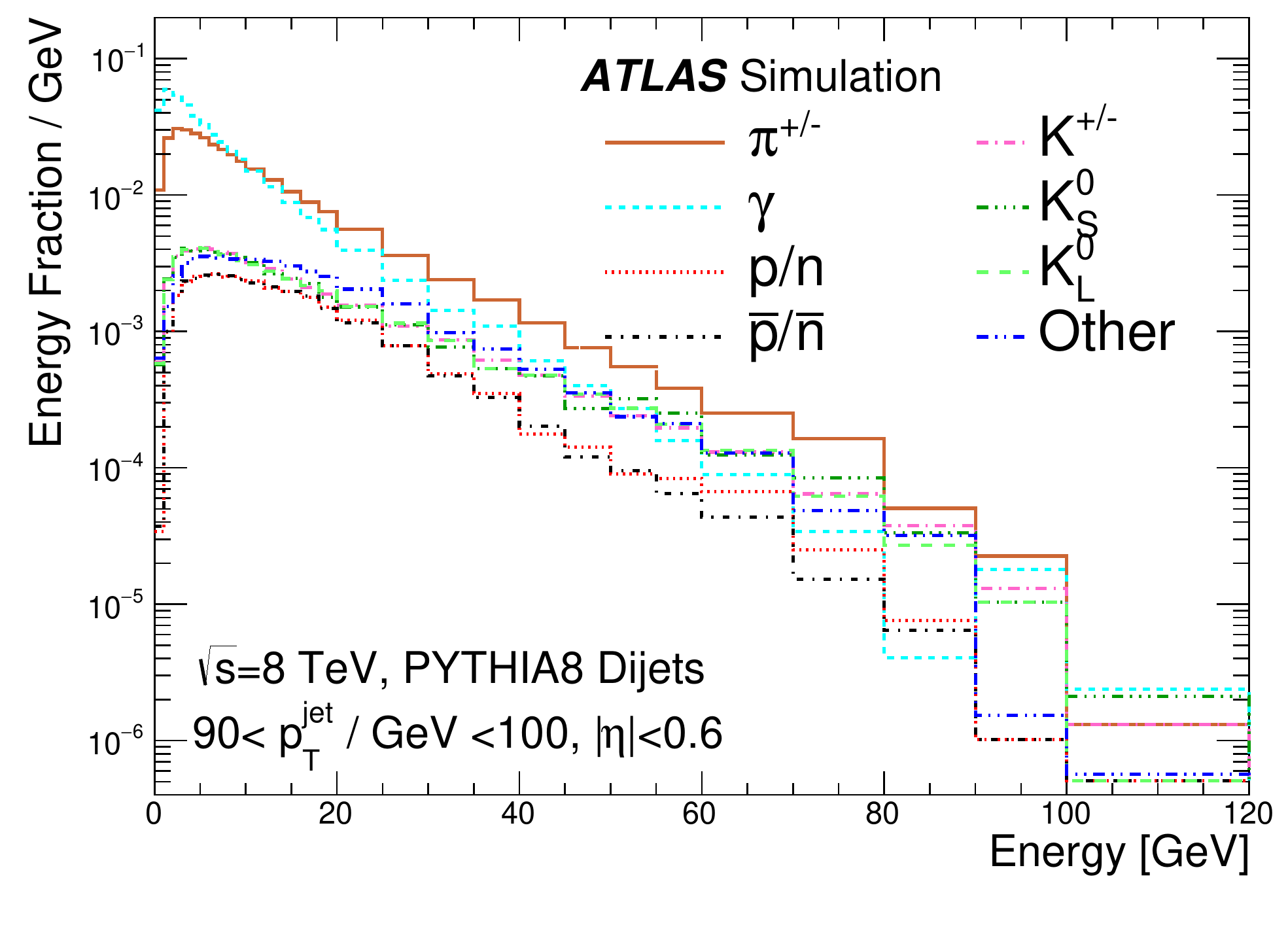}
}
\subfloat[]{
  \includegraphics[width=0.48\textwidth]{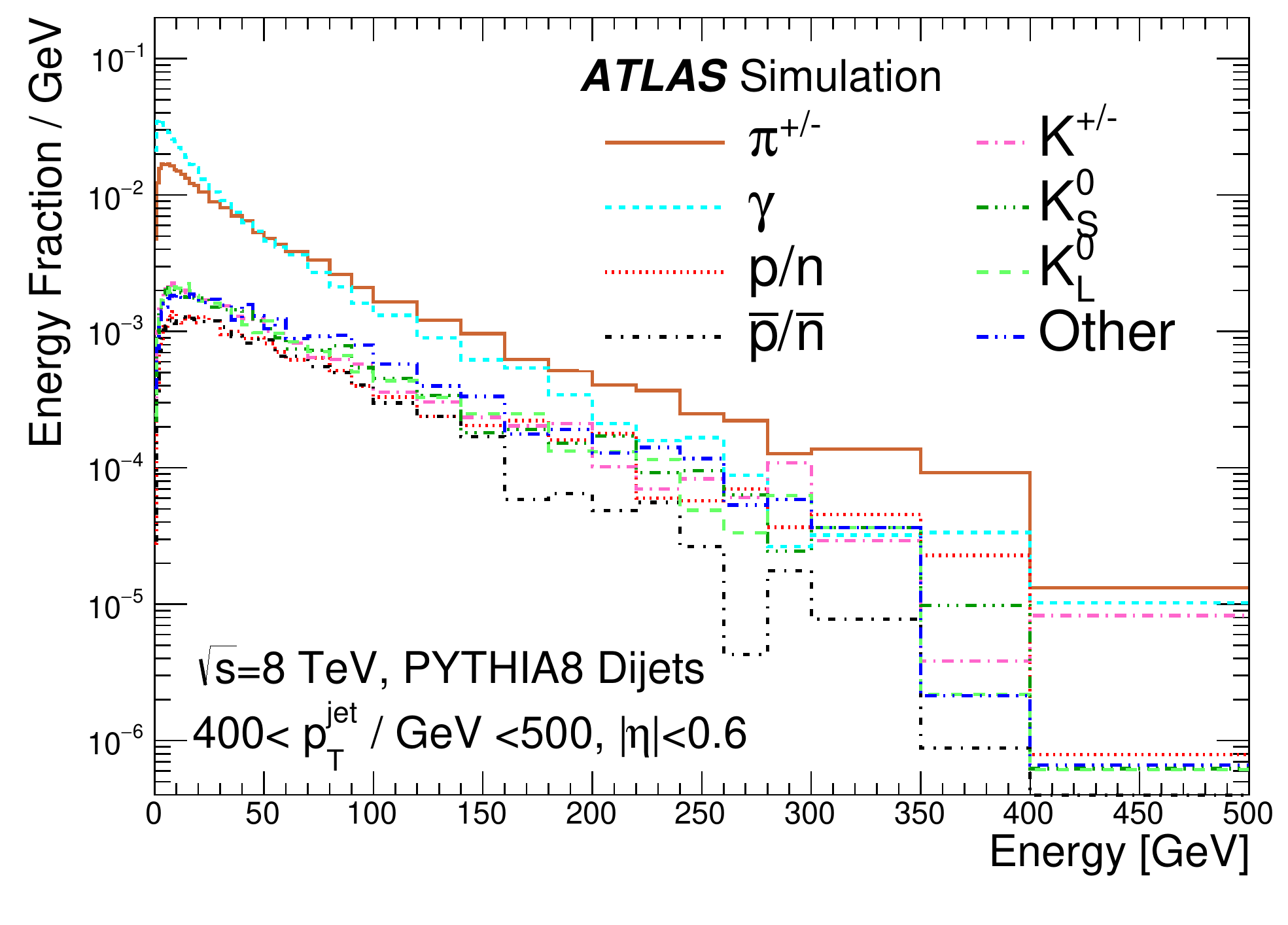}
}\\
\subfloat[]{
  \includegraphics[width=0.48\textwidth]{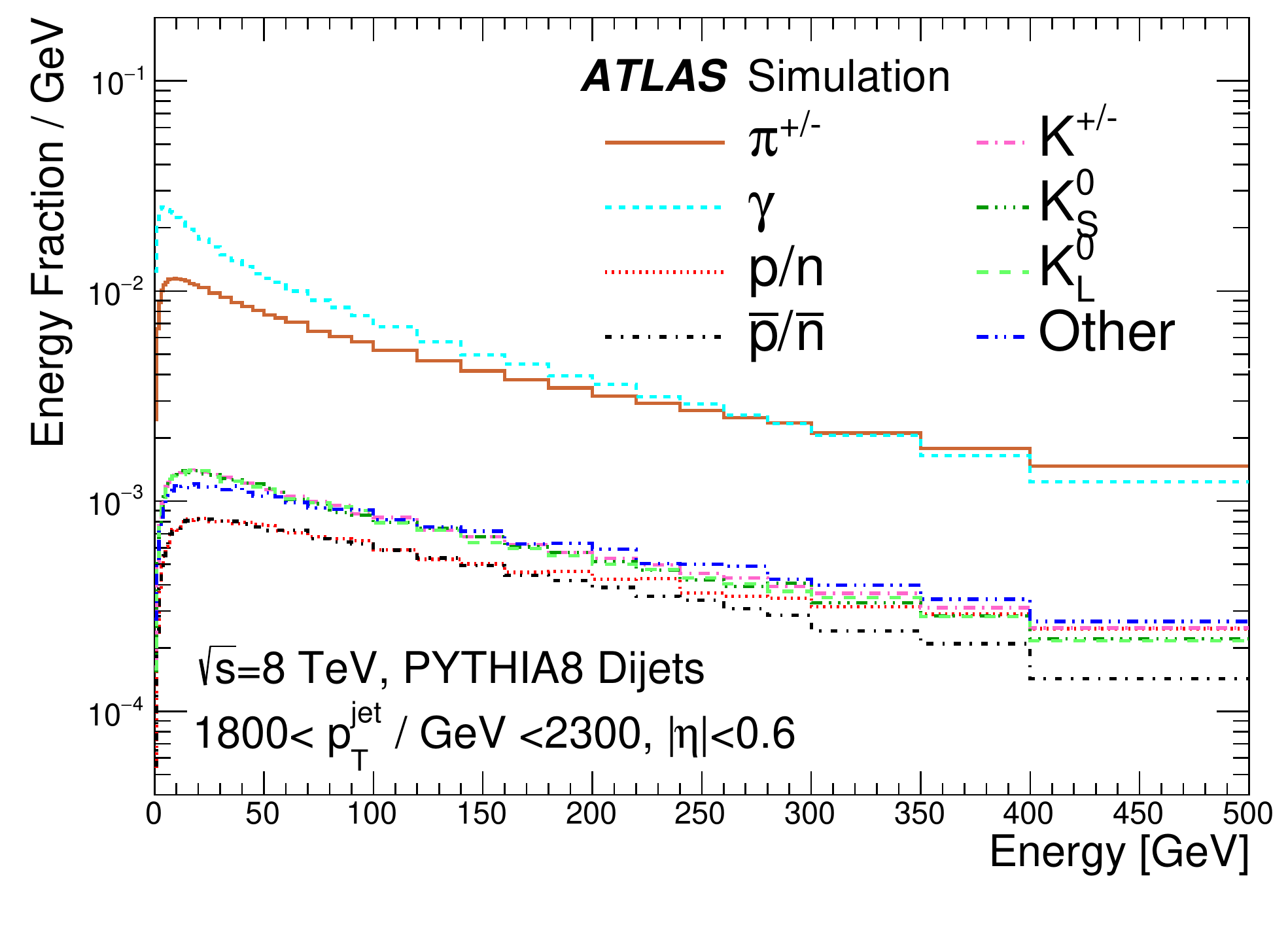}
}
\caption{The fractional contribution to the total jet energy of particles in a certain range of momenta, for particles inside anti-$k_t$, $R=0.4$ jets with (a) $90<\pt/\GeV<100$, (b) $400<\pt/\GeV<500$, and (c) $1800<\pt/\GeV<2300$. }
\label{fig-particle_spectra_contrib}
\end{figure}

Naturally, higher-energy particles contribute more energy to the total jet energy.
Thus, although the number of particles at higher energy is significantly lower than the number at low energy, their contribution to the total jet energy measured in the calorimeter is enhanced.
The contribution to the total jet energy of particles in specific momentum ranges is shown in Figure~\ref{fig-particle_spectra_contrib}. 
The fractional contribution of high-energy particles is, as expected, significant for jets with high energy.

\begin{figure}[ht]
\centering
\subfloat[]{
  \includegraphics[width=0.48\textwidth]{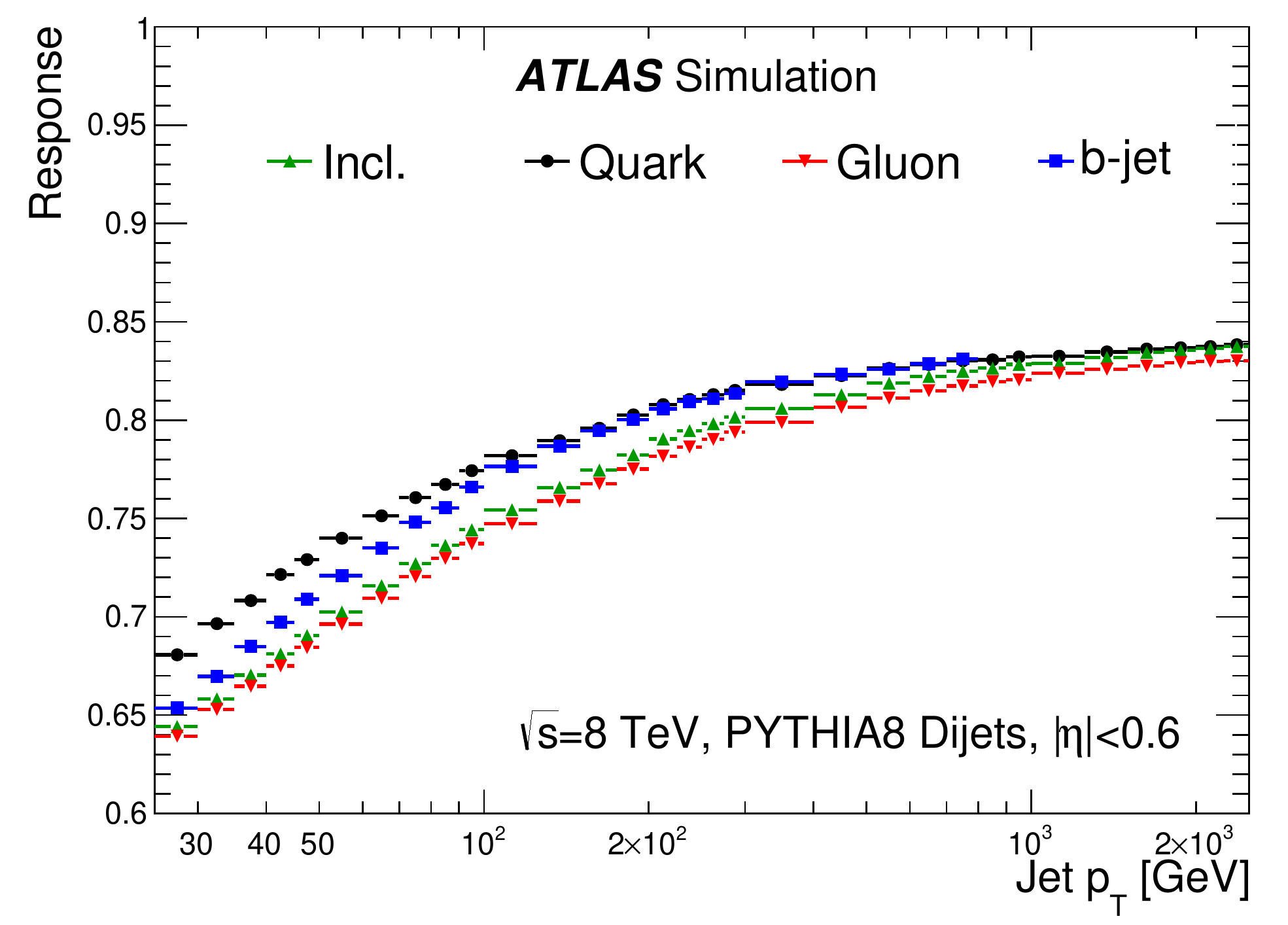}
}
\subfloat[]{
  \includegraphics[width=0.48\textwidth]{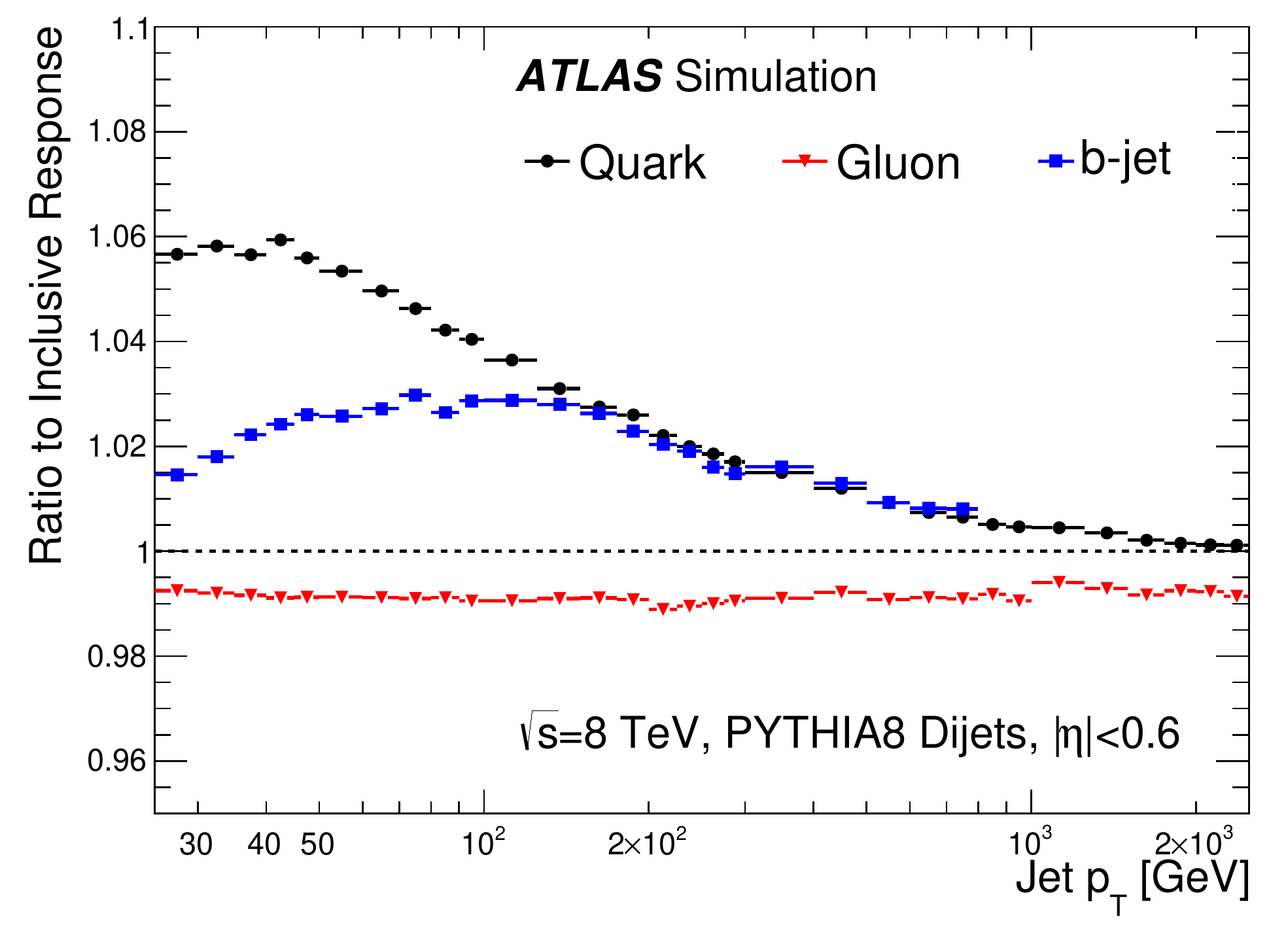}
}
\caption{(a) The energy response of anti-$k_t$, $R=0.4$ jets with $|\eta|<0.6$, based on a convolution of the response functions for isolated charged hadrons with the particle spectrum from {\sc Pythia8}.  (b) The ratio of the response of the categories of jets to the inclusive jet response.  Jets are separated into those containing a $b$-hadron ($b$-jets), those initiated by light quarks (quark), and those initiated by gluons (gluon).  The response distribution for $b$-jets is truncated at 800~\GeV.}
\label{fig-jet_flav_resp}
\end{figure}

As shown in Figure~\ref{fig-jet_flav_resp}, the calorimeter response to jets initiated by a light quark differs from that of jets initiated by a gluon.
This is due to the difference in particle spectrum, multiplicity, and composition.
In Figure~\ref{fig-jet_flav_resp}, the particle spectrum provided by {\sc Pythia8} is convolved with the calorimeter response to isolated charged hadrons measured in the data, extrapolated to higher energy where necessary.
As these measurements are at the EM scale, photons are assumed to have a calorimeter response of $1$ in this study.
Based on MC simulation, protons and neutrons at high energy (above 20~\GeV, cf. Figure~\ref{fig:singleparticle_summary}) are assumed to have the same response.
All hadrons at high energy are given a response of 0.78, in agreement with the combined test beam results at high particle energy~\cite{CTB}.
The jet energy response is then formed as the ratio of the jet energy after this convolution to the energy prior to convolution.
For this study, jets are labeled as light-quark-initiated (quark) or gluon-initiated (gluon) according to the highest-energy parton matched to the jet, following Ref.~\cite{PERF-2012-01}.
The difference in response between light-quark and gluon jets is 7\% at low jet $\pt$, falling to 1\% at high \pt.
This is consistent with the difference derived from the simulated calorimeter response to jets in Ref.~\cite{PERF-2012-01}.
Differences with jets containing a $b$-hadron ($b$-jets) are also visible in this figure: these jets appear more like gluons at low $\pt$ and more like light quarks at high \pt.
For simplicity, this study is restricted to jets with central $|\eta|$, but as the differences in jet properties persist at higher $|\eta|$, the differences in response discussed here also persist.

\begin{figure}[ht]
\centering
\includegraphics[width=0.95\textwidth]{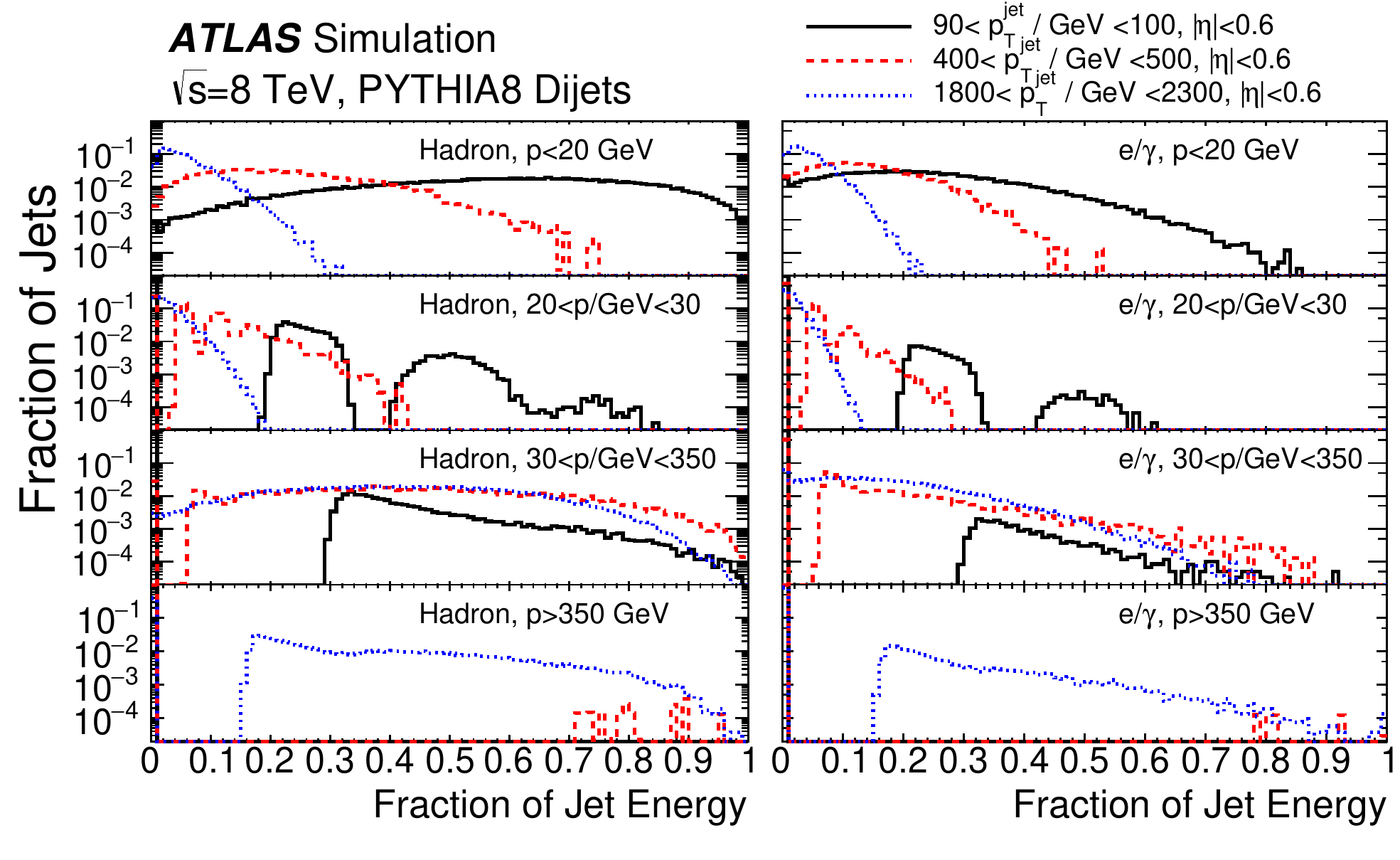}
\caption{The fraction of jet energy carried by particles in several categories (hadrons and EM particles, in several ranges of particle $p$), for anti-$k_t$, $R=0.4$ jets with $90<\pt/\GeV<100$, $400<\pt/\GeV<500$, or $1800<\pt/\GeV<2300$. }
\label{fig-particle_fractions}
\end{figure}

\begin{figure}[ht]
\centering
\includegraphics[width=0.68\textwidth]{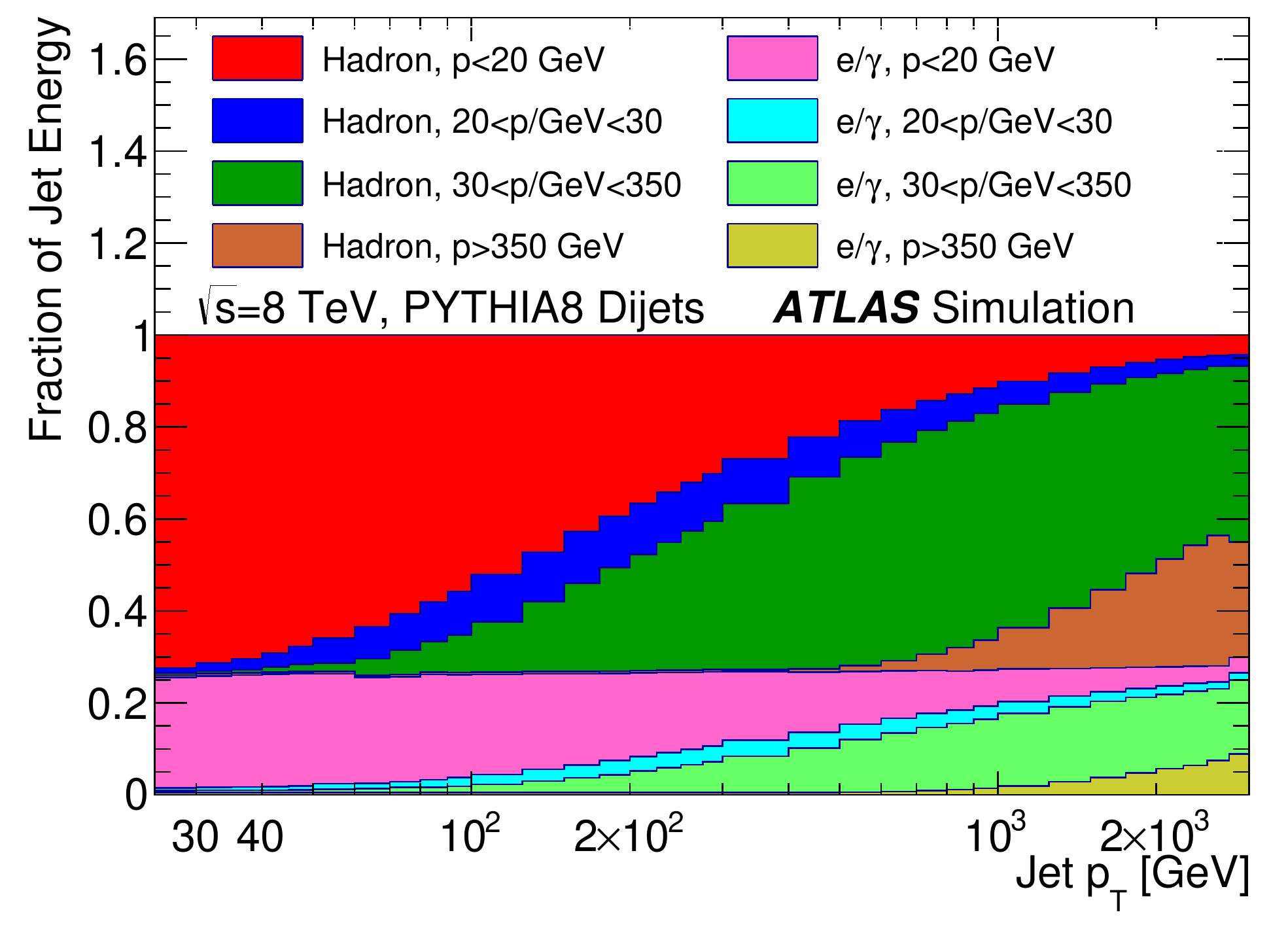}
\caption{The fraction of jet energy carried by particles in several categories (hadrons and EM particles, in several ranges of particle $p$), for anti-$k_t$, $R=0.4$ jets with $|\eta|<0.6$, as a function of jet $\pt$.}
\label{fig-particle_fractions_pt}
\end{figure}

For the jet response and uncertainty, the important quantity is the fraction of jet energy carried by particles in several momentum ranges.
This determines what the most important measurements are, through the convolution of particle-level response and uncertainties into jet-level response and uncertainties.
The fractions for several jet $\pt$ bins are shown in Figure~\ref{fig-particle_fractions}.
The $\pt$ bin ranges correspond to the region covered by only the \epcor\ studies in this paper ($\pt<20$~\GeV), the region covered by both these measurements and the combined test beam ($20<\pT/\GeV<30$), the region covered by only the combined test beam ($30<\pt/\GeV<350$), and the region that is uncovered by these measurements ($\pt>350$~\GeV).
Some structures are visible in this figure, in particular in the $20<p/\GeV<30$ bin of particle momentum.  The first structure corresponds to jets with exactly one such particle, the second structure to jets with two, and so on.
The narrow momentum range of that particular bin leads to a clearer structure than in the other cases.
The average fraction as a function of jet $\pt$ is shown in Figure~\ref{fig-particle_fractions_pt}.
As expected, the EM fraction of the jet is roughly constant as a function of the jet $\pt$, and the contribution from high-momentum particles increases with jet $\pt$.

\begin{figure}[ht]
\centering
\subfloat[]{
  \includegraphics[width=0.48\textwidth]{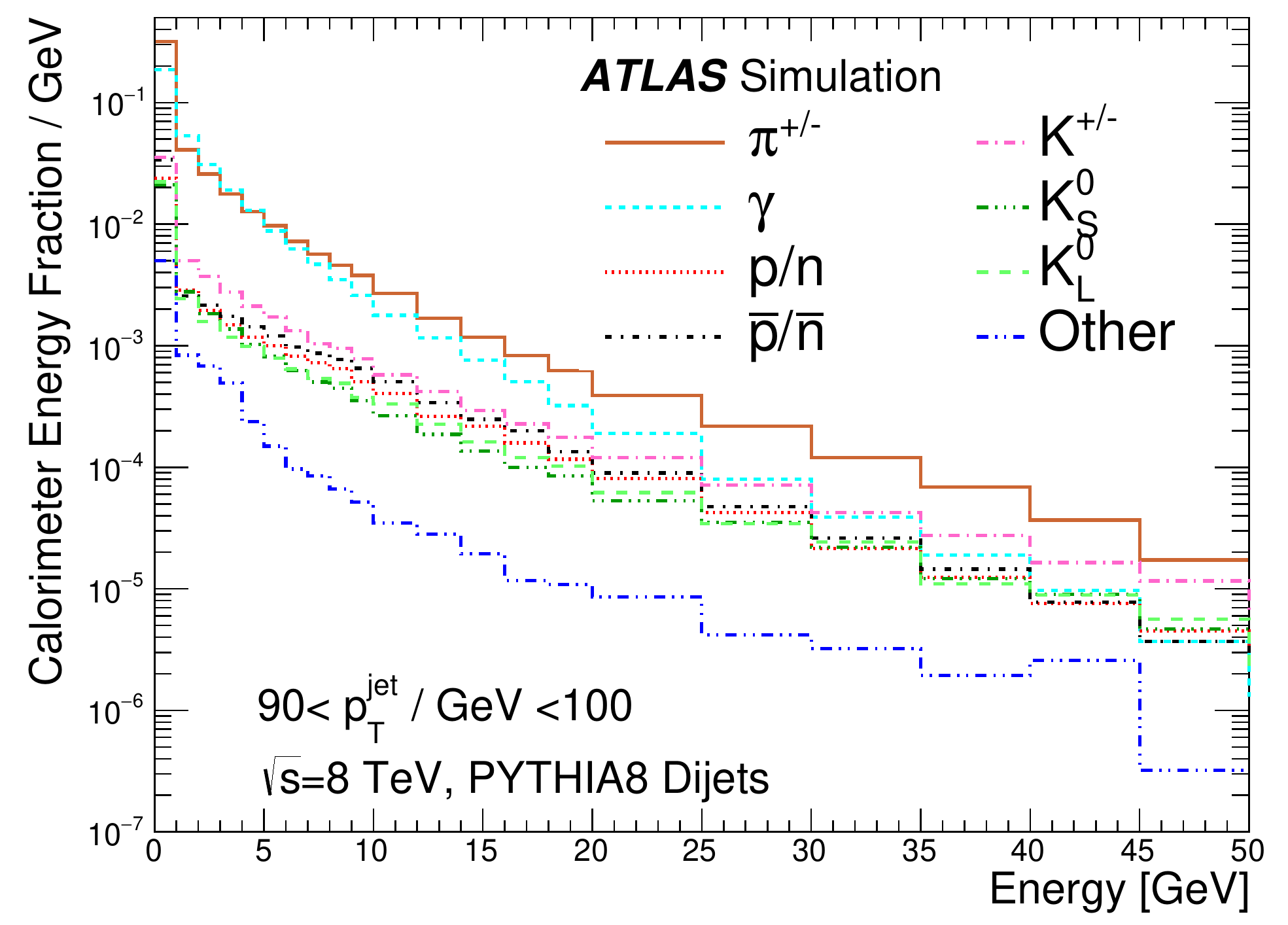}
}
\subfloat[]{
  \includegraphics[width=0.48\textwidth]{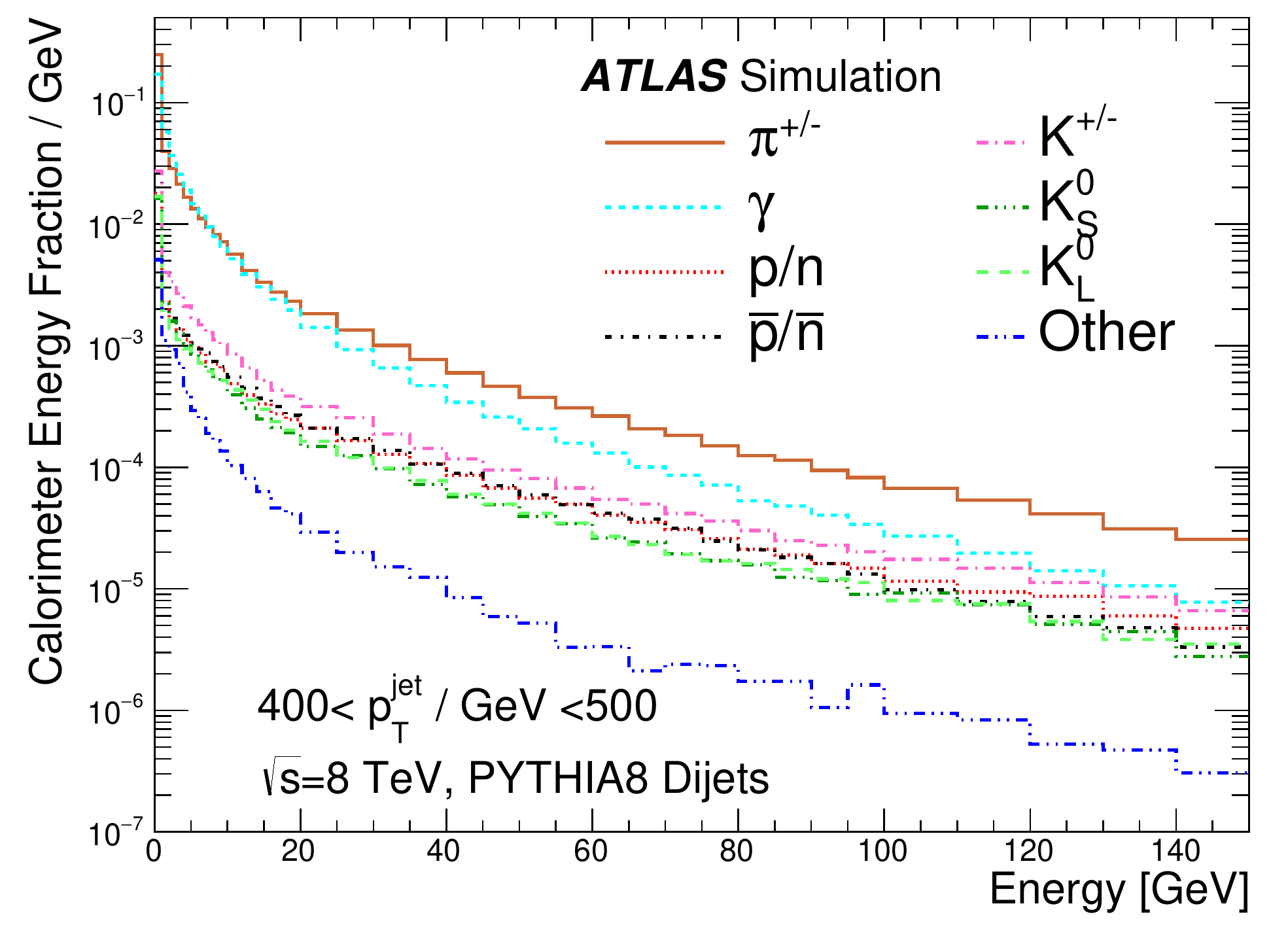}
}
\caption{The spectra of energy deposited in the calorimeter by particles inside anti-$k_t$, $R=0.4$ jets with $|\eta|<0.6$, for jets with (a) $90<\pt/\GeV<100$ and (b) $400<\pt/\GeV<500$.}
\label{fig-energy_spectra}
\end{figure}

Within the MC simulation, it is possible to isolate the contribution to the energy deposited in the calorimeter from each individual particle in the jet.
Figure~\ref{fig-energy_spectra} shows the spectrum of this energy deposited in the calorimeter within a jet, separated by the species of particle that caused the energy deposition.
In this figure, the entire energy from the shower of a positively charged pion is all considered part of the $\pi^{+}$ category, even if the energy was deposited by an electron that was created in the calorimeter, for example.  
The result is qualitatively similar to the spectrum from the generator-level particles in the jet (cf. Figure~\ref{fig-particle_spectra_contrib}), but this is a more accurate reflection of the particles that contribute to the jet's response.
Thus, it is this spectrum that is convoluted with the uncertainties described below in order to derive a jet energy scale uncertainty.

\FloatBarrier
\subsection{Jet energy scale and uncertainty}

In order to derive a jet energy scale response and uncertainty, the calorimeter response to single particles shown in this paper is combined with the response measured in the combined test beam~\cite{CTB}, and additional uncertainty terms are added from various effects that might not be well described by the MC simulation.  Each aspect of the energy scale and uncertainty can potentially include both an offset (a relative difference in the scale, on average, between the data and the MC simulation) and an uncertainty.  The individual terms included (with their labels used in the subsequent figures for the dominant terms) are:

\begin{itemize}
\item The statistical uncertainties in the main inclusive \epcor comparison, binned in $p$ and $|\eta|$, with a calorimeter noise threshold setting of $\langle\mu\rangle=30$,\footnote{The results with this calorimeter setting are consistent with the results with a calorimeter setting of $\langle\mu\rangle=0$, shown in Section~\ref{sec-inclusive}.} from 500~\MeV\ to 20~\GeV\ (``In situ $E/p$'', from Section~\ref{sec-final_eoverp}).
\item The uncertainty in \epcor at the EM scale at low momenta (5\% below 500~\GeV), where the full difference between data and MC simulation is taken as the uncertainty (from Section~\ref{sec-final_eoverp}).
\item The difference in the zero-fraction between data and MC simulation, with a calorimeter setting of $\langle\mu\rangle=30$, binned in $p$ and $|\eta|$, from 500~\MeV\ to 20~\GeV\ (``$E/p$ Zero Fraction'', from Section~\ref{sec:raw_ep}).
\item The uncertainty in the EM calorimeter response from the potential mis-modelling of threshold effects in topological clustering, derived using the comparison of response calculated with cells to that calculated using topological clusters ($0.3\times e^{-1.2\times p/\GeV}$ for $|\eta|<0.8$ and $0.09\times e^{-0.07 \times p/\GeV}$ for $|\eta|>0.8$, ``$E/p$ Threshold'', a parameterisation based on the studies in Section~\ref{sec-clusters_vs_cells}).
\item The uncertainty in the hadronic calorimeter response from the potential mis-modelling of threshold effects in topological clustering, taken as a flat 2\% uncertainty below 10~\GeV\ (from Section~\ref{sec-clusters_vs_cells}).
\item The electromagnetic scale and uncertainty in the tile calorimeter (3\% from Ref.~\cite{TCAL-2010-01}), LAr presampler (5\% from Ref.~\cite{PERF-2010-04}), LAr barrel and endcap EM calorimeters (1.5\% each from Ref.~\cite{PERF-2010-04}), and hadronic end cap calorimeter (3\% from Ref.~\cite{PERF-2010-04}).  This uncertainty is applied to particles at the EM scale or for which no measurement is available: neutral particles, $e^\pm$, and particles with $p>350$~\GeV.
\item The uncertainty in the calorimeter response to neutral hadrons based on studies of physics model variations in {\sc Geant4} (10\% for $p<3$~\GeV\ and 5\% above, ``Neutral'').
\item An additional uncertainty in the response to neutral \pKL in the calorimeter based on studies of physics model variations in {\sc Geant4} (20\%, ``$K_\text{L}$'').
\item The uncertainty in the background subtraction to the \epcor measurement at the EM scale (3\% for $p<2$~\GeV\ and 1\% above, from Section~\ref{sec:neutral_bg}).
\item An uncertainty derived from the difference in events with one and two reconstructed vertices, to account for possible pile-up effects (0.5\%).
\item The uncertainty in the $p$ measurement from misalignment of the ID (1\% above 5~\GeV, ``$E/p$ Misalignment'').
\item The main \epav comparison uncertainties, binned in $p$ and $|\eta|$, as derived from the combined test beam results, from 20 to 350~\GeV\ (``CTB'' from Ref.~\cite{CTB}).
\item The EM energy scale in the combined test beam of the LAr calorimeter (0.7\%) and tile calorimeter (0.5\%), for $20<p/\GeV<350$ (from Ref.~\cite{CTB}).
\item The response uniformity in the combined test beam of the LAr calorimeter (0.4\%) and tile calorimeter (1.5\%), for $20<p/\GeV<350$ (from Ref.~\cite{CTB}).
\item A flat 10\% uncertainty added to all particles above the energy range probed with the combined test beam (i.e. for $p/\GeV>350$) to conservatively cover the effects of saturation, punch-through, and non-linearity at high energy (included in ``Hadrons, $p>350$~\GeV'').
\item The same 10\% uncertainty is applied in the combined test beam momentum range when examining regions higher in $|\eta|$ where the response was not measured (also included in ``Hadrons, $p>350$~\GeV'').
\end{itemize}

No explicit uncertainty term is derived to account for possible dependence of the jet energy scale and uncertainty in the composition of the jets in the MC simulation, either in terms of partonic origin or particle content.  The uncertainty derived in this manner is applicable to data taken with a calorimeter threshold setting of $\langle\mu\rangle=30$, but without any pile-up.

\begin{figure}[ht]
\centering
\subfloat[]{
  \includegraphics[width=0.85\textwidth]{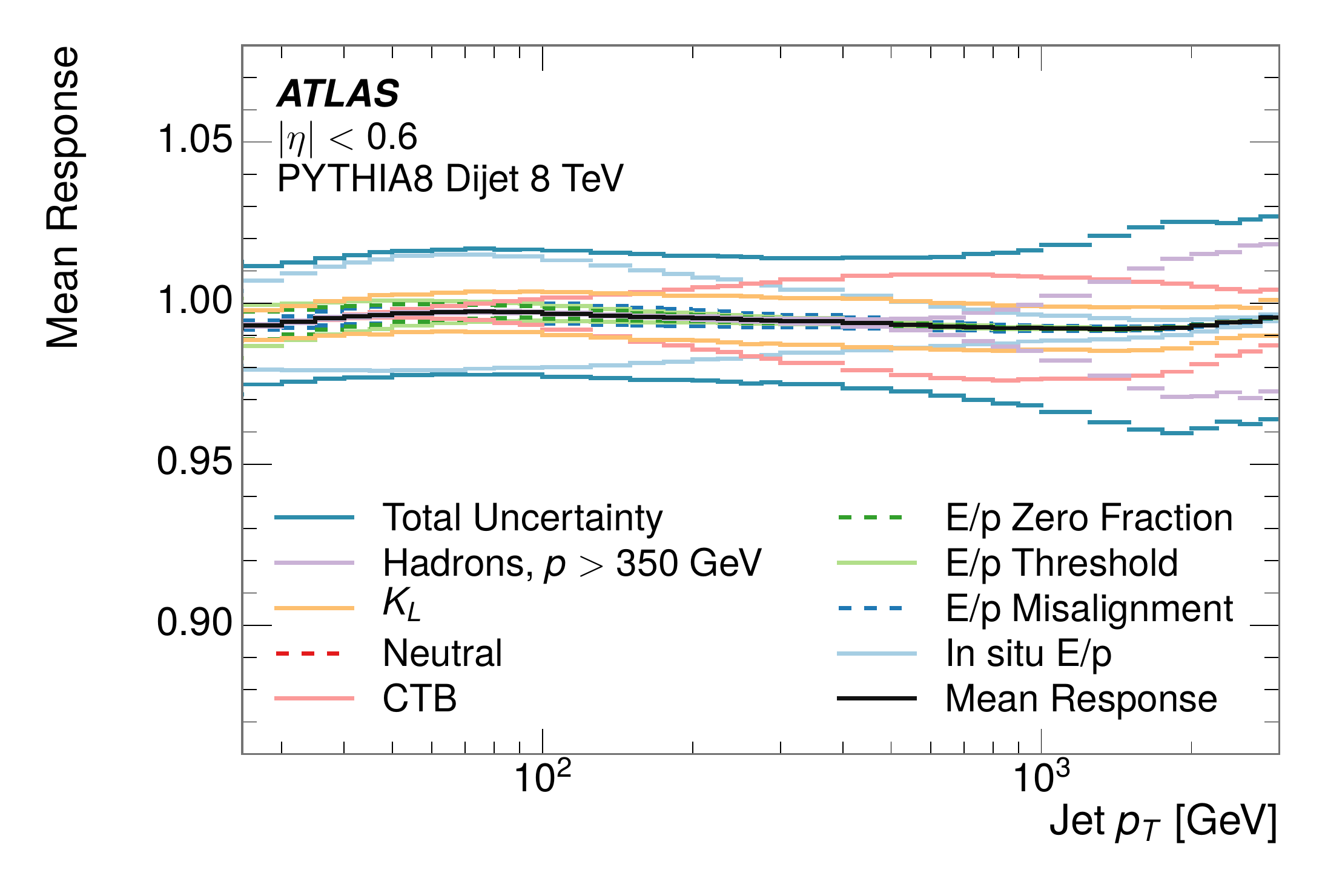}
} \\
\subfloat[]{
  \includegraphics[width=0.85\textwidth]{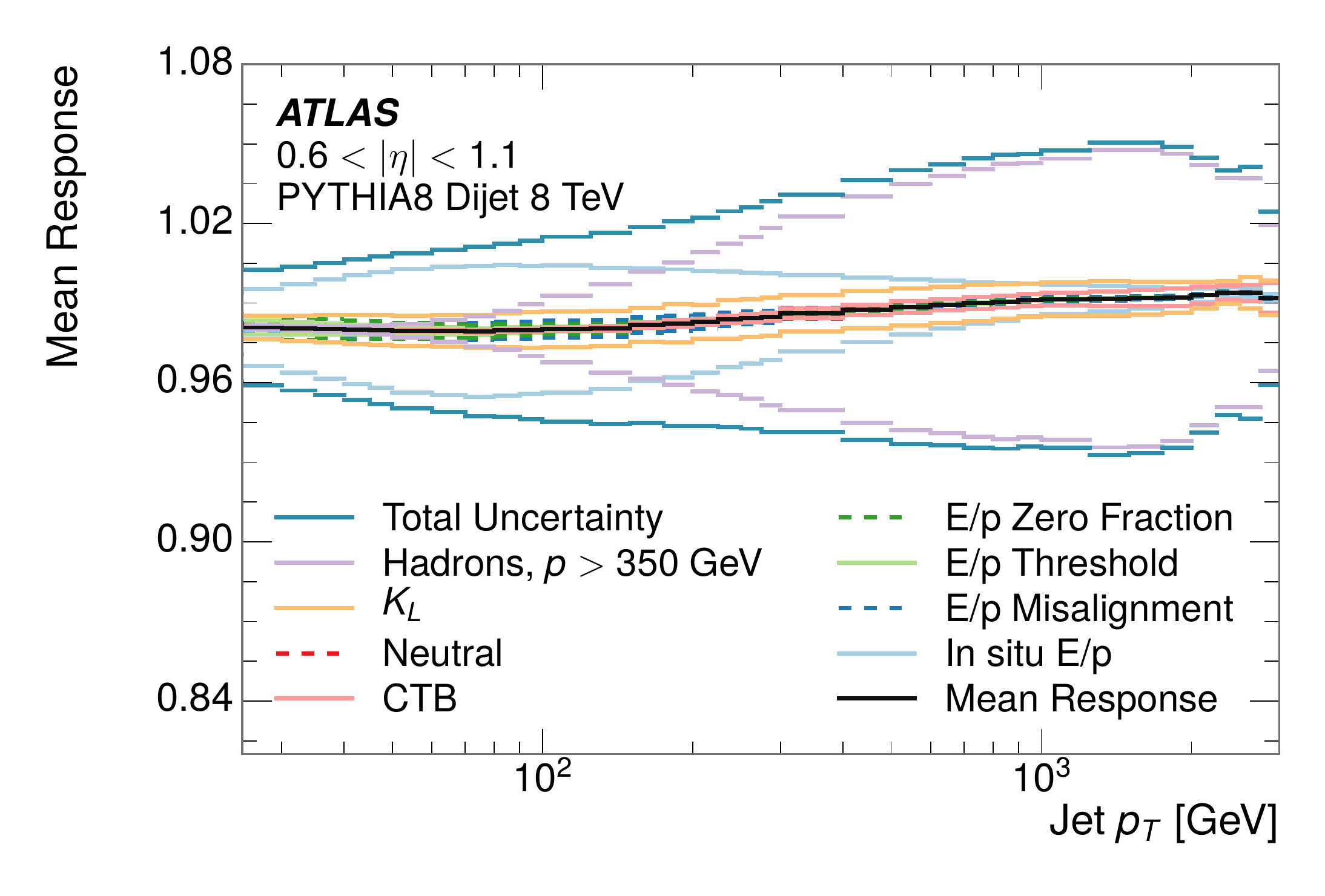}
}
\caption{ The jet energy scale uncertainty contributions, as well as the total jet energy scale uncertainty, as a function of jet $\pt$ for (a) $|\eta| < 0.6$ and (b) $0.6 < |\eta| < 1.1$.}
\label{fig-jes_uncertainty}
\end{figure}

Based on these effects, the jet energy scale uncertainty is extracted.  Each term is treated as an independent Gaussian-distributed uncertainty, and pseudo-data are used to determine both the full uncertainty and the size of the uncertainty correlations between jets with different $\pt$ and $|\eta|$.  The final jet energy scale uncertainty is shown in Figure~\ref{fig-jes_uncertainty}, with a detailed breakdown of the largest components of the jet energy scale uncertainty. The dominant components of the uncertainty, by far, are the uncertainty from the in situ \epcor measurement (for $\pt<600$~\GeV) and the uncertainty from particles that are outside the range probed by the test beam (for $\pt>600$~\GeV).  In the central region of the detector ($|\eta|<0.6$) and for jets of moderate $\pt$, the uncertainty derived in this manner is about twice as large as the uncertainty derived with in situ methods~\cite{PERF-2012-01}, though it is comparable to the uncertainty derived with MC simulation-based methods~\cite{PERF-2011-03}.\footnote{In comparing these uncertainties, it is important to remember that the full ATLAS jet energy scale uncertainty includes terms that may be analysis-dependent, for example uncertainties on the fragmentation of jets, that are not included here.  These uncertainties are derived separately and applied on top of the other uncertainties.  The uncertainty derived in this paper is most comparable to the ``in situ'' terms of the jet energy scale uncertainty from Ref.~\cite{PERF-2012-01}.}  However, this is the only estimate of the jet energy scale uncertainty at high energy ($\pt>1.8~\TeV$), and thus provides a critical component for many physics analyses.  It also serves as a complementary study of the in situ jet response and uncertainty that strengthens the understanding of the modelling of the measurement of hadronic showers by the MC simulation.

\begin{figure}[ht]
\centering
\includegraphics[width=0.7\textwidth]{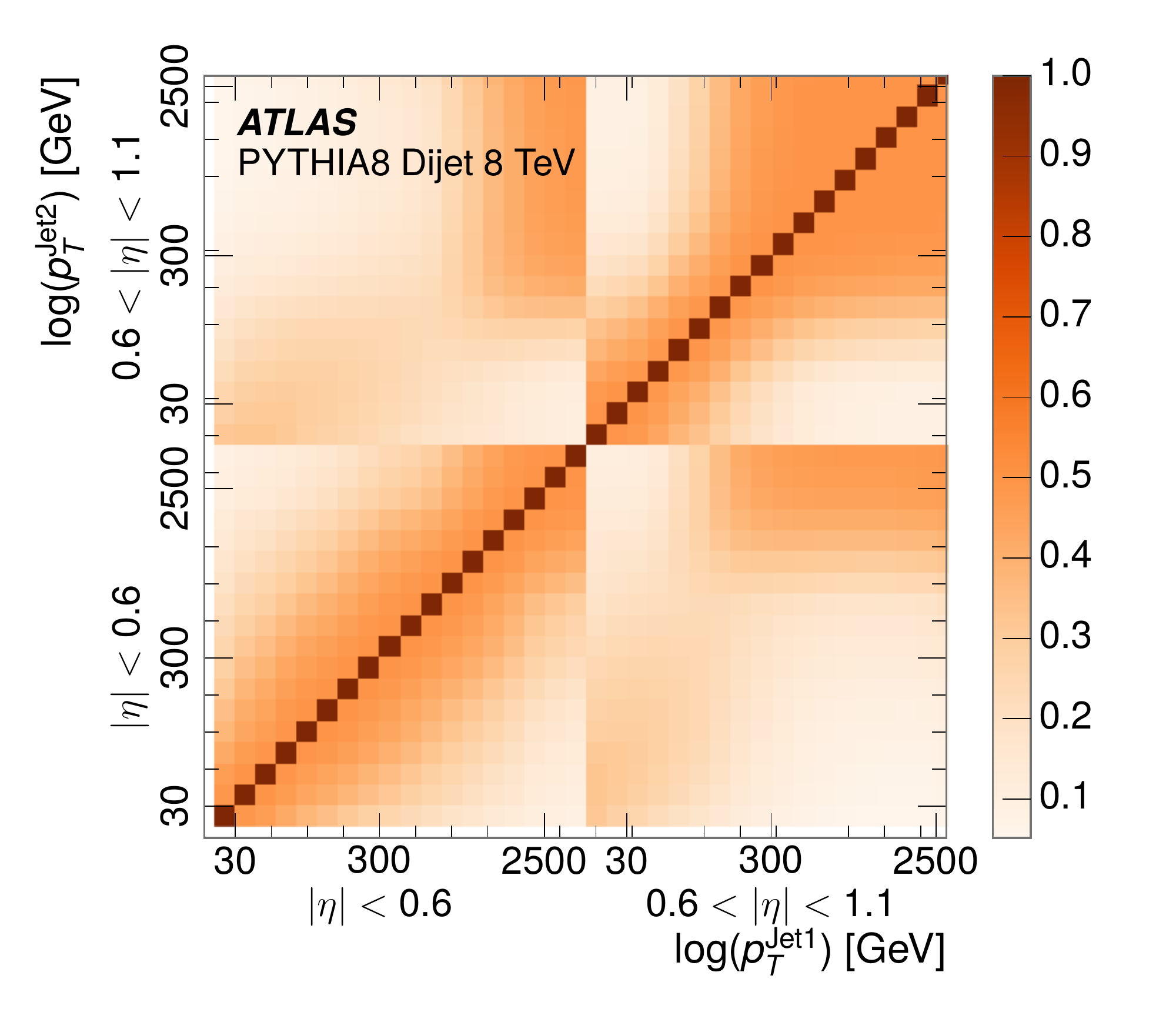}
\caption{ The jet energy scale correlations as a function of jet $\pt$ and $|\eta|$ for jets in the central region of the detector.}
\label{fig-jes_correlations}
\end{figure}

The pseudo-data are also used to explore the correlations in uncertainty of jets at different $\pt$ in the central region.  
The correlation is defined between the average reconstructed jet $\pt$ in a given bin of $\pt$ and $|\eta|$ and is shown in Figure~\ref{fig-jes_correlations}.
Jets at similar momenta are correlated, though the differences in average properties lessens these correlations.
At high $\pt$ and high $|\eta|$, because the jets are dominated by the ``Hadrons, $p>350$~\GeV'' term, the correlation becomes stronger.
These correlations are calculated using the average jet response in an MC simulation sample made using the \texttt{QGSP\_BERT} set of hadronic physics models.
The properties of the jets in this sample, for example the energy deposited in a given layer of the calorimeter, is dependent on the sample.
Therefore, the strength of these correlations may be different in a different MC simulation sample, or indeed in the data.

\FloatBarrier

\section{Conclusion}
\label{sec-conclusion}
A measurement of the calorimeter response to isolated single charged hadrons in the ATLAS detector with data at $\sqrt{s}=7$ and $8$~\TeV\ is presented.
This measurement is compared to the simulation that incorporates the best knowledge of the detector in 2010 and 2012.
After background subtraction, some discrepancy is observed in the response to charged hadrons in the central calorimeter at the level of 5\%.
In more forward regions the {\sc Geant4}-based MC simulation is consistent with the data.

Displaced decays are used to construct samples of pions, protons, and anti-protons.
These samples suggest that the description of response to anti-protons by the hadronic physics models in the \texttt{FTFP\_BERT} set of hadronic physics models is consistent with the data below 5~\GeV, while the description of \texttt{QGSP\_BERT} deviates from the measurement by 10--20\% at low momenta.
Both sets of hadronic physics models show discrepancies with the data in the response to low-energy pions, with the response to negatively charged pions in particular over-estimated by 10--20\% below a momentum of 2~\GeV.

The jet energy scale uncertainty is derived using these calorimeter response observables, along with results from the ATLAS test beam and additional MC simulation.
The uncertainty derived in this manner is 2--5\% for jets with $|\eta|<0.6$ across a broad range of the jet $\pt$ spectrum.
This uncertainty is somewhat larger than the in situ jet energy scale uncertainty, but at high $\pt$ it remains the only jet energy scale uncertainty available.
At high $\pt$ the jet energy scale uncertainty is dominated by the uncertainty in the response to particles above the momentum range probed by the test beam.

\section*{Acknowledgements}

We thank CERN for the very successful operation of the LHC, as well as the
support staff from our institutions without whom ATLAS could not be
operated efficiently.

We acknowledge the support of ANPCyT, Argentina; YerPhI, Armenia; ARC, Australia; BMWFW and FWF, Austria; ANAS, Azerbaijan; SSTC, Belarus; CNPq and FAPESP, Brazil; NSERC, NRC and CFI, Canada; CERN; CONICYT, Chile; CAS, MOST and NSFC, China; COLCIENCIAS, Colombia; MSMT CR, MPO CR and VSC CR, Czech Republic; DNRF and DNSRC, Denmark; IN2P3-CNRS, CEA-DSM/IRFU, France; GNSF, Georgia; BMBF, HGF, and MPG, Germany; GSRT, Greece; RGC, Hong Kong SAR, China; ISF, I-CORE and Benoziyo Center, Israel; INFN, Italy; MEXT and JSPS, Japan; CNRST, Morocco; FOM and NWO, Netherlands; RCN, Norway; MNiSW and NCN, Poland; FCT, Portugal; MNE/IFA, Romania; MES of Russia and NRC KI, Russian Federation; JINR; MESTD, Serbia; MSSR, Slovakia; ARRS and MIZ\v{S}, Slovenia; DST/NRF, South Africa; MINECO, Spain; SRC and Wallenberg Foundation, Sweden; SERI, SNSF and Cantons of Bern and Geneva, Switzerland; MOST, Taiwan; TAEK, Turkey; STFC, United Kingdom; DOE and NSF, United States of America. In addition, individual groups and members have received support from BCKDF, the Canada Council, CANARIE, CRC, Compute Canada, FQRNT, and the Ontario Innovation Trust, Canada; EPLANET, ERC, FP7, Horizon 2020 and Marie Sk{\l}odowska-Curie Actions, European Union; Investissements d'Avenir Labex and Idex, ANR, R{\'e}gion Auvergne and Fondation Partager le Savoir, France; DFG and AvH Foundation, Germany; Herakleitos, Thales and Aristeia programmes co-financed by EU-ESF and the Greek NSRF; BSF, GIF and Minerva, Israel; BRF, Norway; Generalitat de Catalunya, Generalitat Valenciana, Spain; the Royal Society and Leverhulme Trust, United Kingdom.

The crucial computing support from all WLCG partners is acknowledged gratefully, in particular from CERN, the ATLAS Tier-1 facilities at TRIUMF (Canada), NDGF (Denmark, Norway, Sweden), CC-IN2P3 (France), KIT/GridKA (Germany), INFN-CNAF (Italy), NL-T1 (Netherlands), PIC (Spain), ASGC (Taiwan), RAL (UK) and BNL (USA), the Tier-2 facilities worldwide and large non-WLCG resource providers. Major contributors of computing resources are listed in Ref.~\cite{ATL-GEN-PUB-2016-002}.

\printbibliography

\newpage 
\begin{flushleft}
{\Large The ATLAS Collaboration}

\bigskip

M.~Aaboud$^\textrm{\scriptsize 136d}$,
G.~Aad$^\textrm{\scriptsize 87}$,
B.~Abbott$^\textrm{\scriptsize 114}$,
J.~Abdallah$^\textrm{\scriptsize 65}$,
O.~Abdinov$^\textrm{\scriptsize 12}$,
B.~Abeloos$^\textrm{\scriptsize 118}$,
R.~Aben$^\textrm{\scriptsize 108}$,
O.S.~AbouZeid$^\textrm{\scriptsize 138}$,
N.L.~Abraham$^\textrm{\scriptsize 152}$,
H.~Abramowicz$^\textrm{\scriptsize 156}$
% ...
\bigskip
\\
% ...
$^{12}$ Institute of Physics, Azerbaijan Academy of Sciences, Baku, Azerbaijan\\
% ...
$^{65}$ University of Iowa, Iowa City IA, United States of America\\
% ...
$^{87}$ CPPM, Aix-Marseille Universit{\'e} and CNRS/IN2P3, Marseille, France\\
% ...
$^{108}$ Nikhef National Institute for Subatomic Physics and University of Amsterdam, Amsterdam, Netherlands\\
% ...
$^{114}$ Homer L. Dodge Department of Physics and Astronomy, University of Oklahoma, Norman OK, United States of America\\
% ...
$^{118}$ LAL, Univ. Paris-Sud, CNRS/IN2P3, Universit{\'e} Paris-Saclay, Orsay, France\\
% ...
$^{136}$ $^{(a)}$ Facult{\'e} des Sciences Ain Chock, R{\'e}seau Universitaire de Physique des Hautes Energies - Universit{\'e} Hassan II, Casablanca; $^{(b)}$ Centre National de l'Energie des Sciences Techniques Nucleaires, Rabat; $^{(c)}$ Facult{\'e} des Sciences Semlalia, Universit{\'e} Cadi Ayyad, LPHEA-Marrakech; $^{(d)}$ Facult{\'e} des Sciences, Universit{\'e} Mohamed Premier and LPTPM, Oujda; $^{(e)}$ Facult{\'e} des sciences, Universit{\'e} Mohammed V, Rabat, Morocco\\
% ...
$^{138}$ Santa Cruz Institute for Particle Physics, University of California Santa Cruz, Santa Cruz CA, United States of America\\
% ...
$^{152}$ Department of Physics and Astronomy, University of Sussex, Brighton, United Kingdom\\
% ...
$^{156}$ Raymond and Beverly Sackler School of Physics and Astronomy, Tel Aviv University, Tel Aviv, Israel\\
% ...
$^{a}$ Also at Department of Physics, King's College London, London, United Kingdom\\
$^{b}$ Also at Institute of Physics, Azerbaijan Academy of Sciences, Baku, Azerbaijan\\
$^{c}$ Also at Novosibirsk State University, Novosibirsk, Russia\\
$^{d}$ Also at TRIUMF, Vancouver BC, Canada\\
$^{e}$ Also at Department of Physics {\&} Astronomy, University of Louisville, Louisville, KY, United States of America\\
% ...
\end{flushleft}


\end{document}